\documentclass[aps,prb,twocolumn,superscriptaddress]{revtex4}
\usepackage{graphicx}
\usepackage{color}

\begin{document}

\title{Accessing the spectral function of \emph{in operando} devices by angle-resolved photoemission spectroscopy}
\author{Philip Hofmann}
\email{philip@phys.au.dk}
\affiliation{Department of Physics and Astronomy, Interdisciplinary Nanoscience Center (iNANO), Aarhus University, 8000 Aarhus C, Denmark}
\date{\today}
\begin{abstract}
Progress in performing angle-resolved photoemission spectroscopy (ARPES) with  high spatial resolution in the order of 1~$\mu$m or less (nanoARPES) has opened the possibility to map the spectral function of solids on this tiny scale and thereby obtain detailed information on the materials' \emph{local} electronic band structure and  many-body interactions. Recently, nanoARPES has been used to study simple electronic devices, based on two-dimensional materials, with the possibility of tuning the carrier type and density by field effect-gating, and while passing a current through the device. It was demonstrated that nanoARPES can detect possible changes in the materials' electronic structure in these situations and that it can map the local doping, conductance and mobility.  This article reviews these first  \emph{in operando} ARPES results  on devices, discusses the resulting new insights, as well as  the perspectives for future developments of the technique. 
\end{abstract}
\maketitle

\section{Introduction}

%ARPES introduction and development
Angle-resolved photoemission spectroscopy (ARPES) is an experimental approach to study the electronic properties of crystalline solids. ARPES is a sophisticated version of the photoelectric effect and was first developed in the 1970's and 80's. Initially, the main aim was to study the occupied band structure of solids but it was soon discovered that information about many-body effects such as electronic correlations or the electron-phonon interaction could be investigated as well. ARPES has seen an enormous technical improvement after the discovery of high-temperature superconductivity, as it was thought to be the technique of choice to characterise the interactions leading to this phenomenon. 

%historically spatially integrating but now high resolution
Over the last 40 years, ARPES experiments have almost exclusively been performed on nominally uniform millimetre-sized single crystals (the crystal has to be larger than the diameter of the UV light spot used for photoexcitation) \cite{Plummer:1982aa,Kevan:1992aa,Hufner:2003aa,Damascelli:2003aa,Hofmann:2009ab,Lu:2012ac,Sobota:2020aa}. Only very recently, it has become possible in an ARPES context to focus the UV light to a spot size of less than a few $\mu$m, opening the possibility to study a variation of electronic phenomena on this scale, as well as to map a sample's electronic structure by scanning the UV beam across its surface \cite{Rotenberg:2010ad}. 

%this will can be used for in operando
The high spatial resolution can also be exploited to study samples that were hitherto not accessible to ARPES, for example simple electronic devices made of two-dimensional (2D) materials such as graphene or single-layer MoS$_2$. Typically, such a device is built up as a field effect transistor, permitting the control of the carrier concentration in the 2D material via a gate voltage. The 2D material channel is only a few $\mu$m wide and it is thus extremely demanding to focus the UV spot on this channel only. Indeed, such systems cannot be studied using a standard ARPES beamline and the exploration of devices by ARPES has only become possible due to the aforementioned recent progress in focusing the UV beam to a size much smaller than the typical device dimensions.  This review describes the first ARPES experiments on such devices \emph{in operando} i.e., while the 2D material is exposed to gate fields and currents. 
%The spectroscopic information delivered by ARPES would also benefit a vast number of more ``traditional'' condensed matter transport setups, from the metal oxide field effect transistor to the quantum Hall effect, but we shall see that a 2D material exposed the surface of a device is much more accessible to ARPES than the electron gas at a buried interface.

The application of electric fields to a material may not only lead to a modified carrier density and electric currents, but potentially also alter many-body effects or even induce phase transition, e.g., from a Mott insulator to a conductor. The so-called ``quantum materials'' (QMs) are particularly prone to such transitions and the recent development of ARPES performed on devices opens the door to studying the electronic structure and many-body effects of QMs in field-induced states -- information that was  previously  inaccessible.

We briefly discuss why the application of electric fields and currents, along with many other tuning parameters such as strain, pressure and chemical composition, is particularly interesting in the context of QMs. We start with the question:  What are QMs? At present, there is no agreed-upon  definition of what  constitutes a QM. There is, however, broad agreement that QMs are typically characterised by strong electronic correlations or topological band structures, often combined with reduced dimensionality \cite{Keimer:2015aa,Keimer:2017aa,Basov:2017aa,Tokura:2017aa}. QMs feature extraordinary properties, such as high-temperature superconductivity and exotic quantum Hall effects. The field of QM research has evolved from the interest in correlated electronic materials but the class of QM now also contains materials that can be well-described by single-particle band structures, for example the Bi$_2$Se$_3$ group of topological insulators. 

%interactions of different strengths create equilibrium and emergent properties. Can be changed.

The state of many QMs is controlled by a subtle balance between different interactions of similar strength. This leads to  so-called ``emergent phenomena'' that arise due to the interplay of these interactions, entailing the QM with new properties. Moreover, a small stimulus can often be used to upset the ground state equilibrium and ``switch'' the QM between radically different states---e.g., from insulating to superconducting. Quantum phase transitions (not involving temperature) are a common phenomenon in these materials \cite{Vojta:2003aa} and so is the spontaneous spatial separation between coexisting different phases. Pushing QMs out of equilibrium can even give rise to entirely new forms of quantum matter \cite{Wang:2013ad,Rudner:2020aa}.

\begin{figure}
\includegraphics[width=0.48\textwidth]{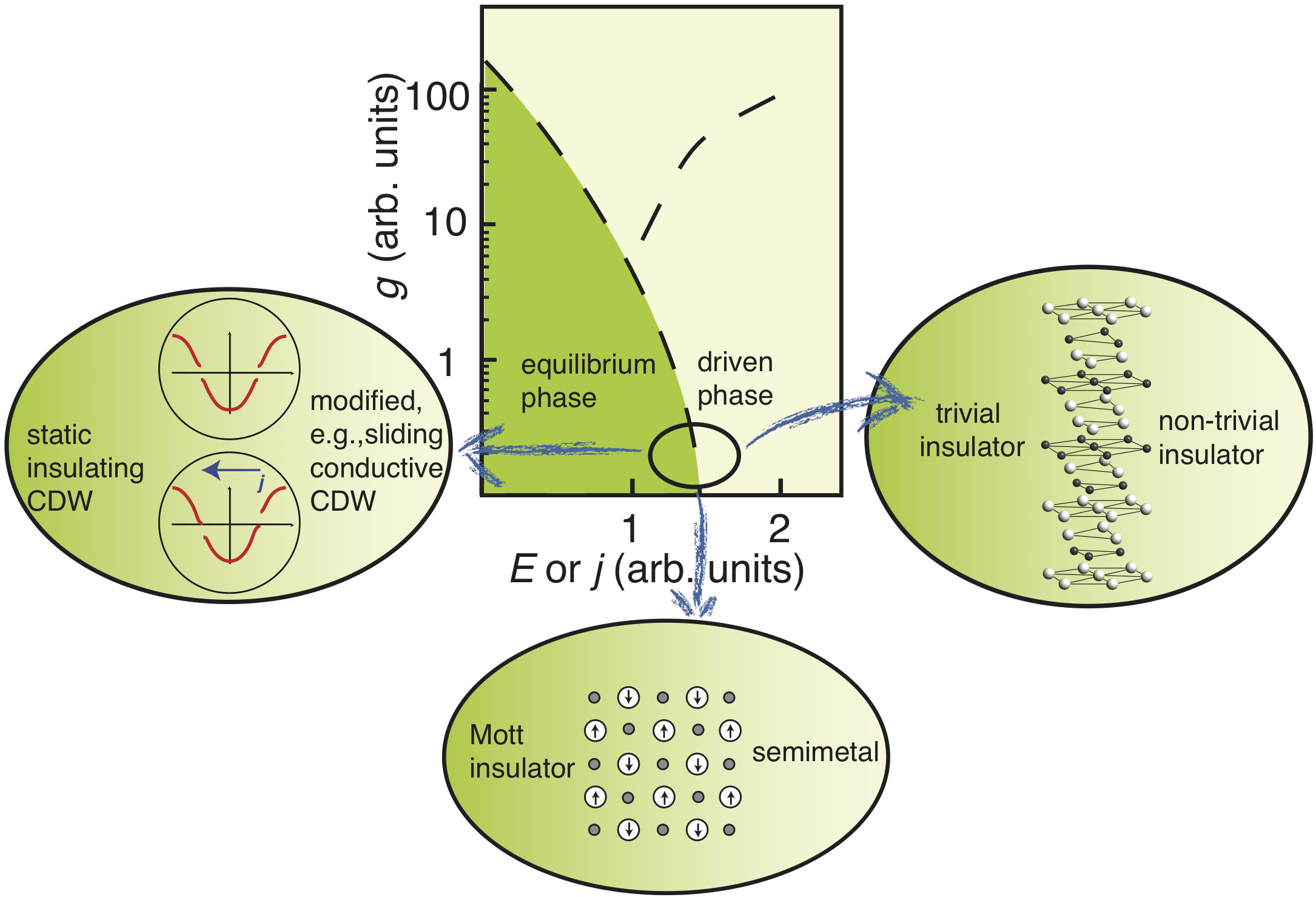}\\
\caption{Generic phase diagram for quantum materials as a function of a control parameter $g$ and a static electric field $E$ (steady state current density $j$). $g$ could be the temperature or some other parameter (chemical composition, pressure, strain, etc.). The three bubbles give some specific examples. From left to right: In charge density wave (CDW) materials, an electric field can induce a sliding of the CDW condensate, leading to an increased conductivity. In a Mott insulator, moderate electric fields can give rise to an insulator-to-metal transition, increasing the conductivity by many orders of magnitude. In topological materials, a high current density breaks crystalline symmetry, potentially leading to a transition in band topology. Note that the phase diagram is strongly simplified, not accounting for, e.g., additional phases in CDW systems and correlated materials.  The diagram could also have more than two relevant dimensions and the phases reached by a high $g$ and a high $j$ do not need to be the same. }
  \label{fig:intro}
\end{figure}

%transitions in QMs
Understanding and exploiting phase transitions in QMs is one of the major challenges in contemporary condensed matter physics. Much interest has been directed towards driving phase transitions in QMs by a range of stimuli such as ultrafast laser excitations  \cite{Cavalleri:2001aa,Schmitt:2008aa,Rohwer:2011aa,Fausti:2011aa,Wang:2013ad,Mitrano:2016aa,Caputo:2018aa}, or simply the application of electric fields or DC currents \cite{Taguchi:2000aa,Lee:2007aa,Vaju:2008aa,Monceau:2012aa,Stoliar:2013aa,Nakamura:2013aa,Rudner:2020aa,Bellec:2020aa}. An illustration of this is given in Figure~\ref{fig:intro} which shows a generic phase diagram of a QM as a function of an applied electric field and consequently a current density in the case of a conductive material, as well as an additional control parameter $g$. $g$ could be the sample temperature, for example in the case of insulator-to-metal transition in the Mott insulator Ca$_2$RuO$_4$ \cite{Nakamura:2013aa,Bertinshaw:2019aa,Cirillo:2019aa,Zhang:2019ac,Mattoni:2020aa} but it could also be some other control parameter. For Ca$_2$RuO$_4$, this could be chemical composition \cite{Nakatsuji:2000aa} (substitution of Ca by Sr), pressure \cite{Nakamura:2002aa} or strain \cite{Ricco:2018aa,Sunko:2019aa}. In fact, for many QMs the corresponding phase diagram is likely to be multi-dimensional. 

%For devices field and current interesting and this is also valuable from practical point of view
We highlight here the role of electric field and current-driven phase transitions in QMs because, in a practical device setting, electric fields are much easier to administer than, e.g., magnetic fields, strain or hydrostatic pressure. There is a large number of examples for current-driven transitions, such as sliding of charge density wave condensates, insulator-to-metal transitions in Mott insulators, the current-induced breakdown of superconductivity or field-induced changes in band topology \cite{Taguchi:2000aa,Lee:2007aa,Vaju:2008aa,Monceau:2012aa,Stoliar:2013aa,Nakamura:2013aa,Rudner:2020aa,Bellec:2020aa}. We stress that, with respect to applications of QMs in electronic devices, such phenomena imply that the QM's role in a device can be completely different from, e.g., silicon: Whereas an electric current leaves the properties of a normal material unaltered, it has the potential to change the state of some QMs, something that is not exploited in conventional electronic devices. 

The electronic structure of a material -- and its modification as a function of some external parameter or across a phase transition -- is closely related to its electrical transport properties. Transport measurements can be performed over a wide range of temperatures, in strong magnetic fields, for strained samples or during the exposure to light. While the primary quantity measured is the electrical conductivity of the sample, the simultaneous application of magnetic fields can give detailed information about the carrier density, the carrier's effective mass, the Berry phase, the presence of weak (anti)-localisation, quantum transport through edge states and many other phenomena. Magneto-transport measurements have therefore been of key-importance in the characterisation of solids for many years \cite{Ashcroft:1976aa}.

%the key to understanding the transitions is the spectral function
Ultimately, the transport properties and the complete electronic structure of a solid are encoded in its  spectral function $A(\hbar \omega, \mathbf{k})$. In a simple picture, $A(\hbar \omega, \mathbf{k})$ can be viewed as the probability of finding an electron with energy $\hbar \omega$ and wave vector $\mathbf{k}$. In a system of non-interacting electrons (not typically a good approximation in QMs),  $A(\hbar \omega, \mathbf{k})$ has the character of a $\delta$-function at the energy and $\mathbf{k}$ values given by the single particle dispersion $\epsilon(\mathbf{k})$. This changes when interactions are turned on. In the simplest situation described by a so-called quasiparticle picture, the maxima in $A(\hbar \omega, \mathbf{k})$ are broadened and shifted away from $\epsilon(\mathbf{k})$. For very strong many-body interactions, the quasiparticle picture  might even break down. This is reflected in more complex structures in  $A(\hbar \omega, \mathbf{k})$ that are not related to $\epsilon(\mathbf{k})$ in a simple way. Regardless, the spectral function is often proportional to the photoemission intensity observed in ARPES, so that the complex many-body interactions can be tracked by this technique.

%But ARPES doesn't work for all possible parameters
While ARPES has been extensively used to study QMs in equilibrium \cite{Damascelli:2003aa,Keimer:2015aa,Lu:2012ac}, its applicability to phenomena driven by external parameters, such as in  Figure~\ref{fig:intro}, can be restricted because such parameters need to be reconciled with the requirements of performing electron spectroscopy in ultra-high vacuum. While it is trivial to change the chemical composition by using different samples, and relatively simple to change the temperature using a cryostat, at least down to a few K, some highly interesting experimental conditions are difficult or even impossible to explore. The presence of electric and magnetic fields, for instance, is highly problematic because these change the trajectory of the photoemitted electrons. Also, hydrostatic pressure cannot be applied because the path of the photoelectrons to a detector is blocked in a diamond anvil cell. On the other hand, there has been recent progress with parameters such as strain to tune QMs and to trigger phase transitions while investigating the electronic structure using ARPES \cite{Ricco:2018aa,Flototto:2018aa,Lin:2020aa}.

%
%ARPES with spatial resolution
ARPES with spatial resolution, the key technique for this review,  is often called micro ($\mu$)-ARPES or nanoARPES. 
For simplicity, we will not make this distinction here and just refer to the technique as nanoARPES.
NanoARPES opens many interesting opportunities for the study of QMs because it not only gives access to intrinsically small samples but potentially also to phenomena such as phase separation in QMs and otherwise inhomogeneous systems. Indeed, while nanoARPES is a relatively recent technique, it has already given rise to a number of important results  \cite{Rotenberg:2010ad,Noguchi:2019aa,Kastl:2019aa,Ulstrup:2019aa,Utama:2020aa,Lisi:2020aa,Ulstrup:2020aa}.  Here we focus on the possibility to determine the spectral function of a QM in electronic devices in the presence field-induced doping and / or a finite current density. As we shall see, this opens a spectroscopic window on many phenomena that are usually inaccessible to ARPES investigations. Most importantly in connection with QMs, it should allow unique new insights in the modification of the materials' spectral function upon doping and under current-induced phase transitions.

% potential mapping
In addition to accessing the non-equilibrium spectral function of a QM in an operating device, we shall see that nanoARPES on current-carrying devices permits a contact-free mapping of the local potential, such that the spectroscopic measurements are accompanied by a transport measurement. This bridges the gap between the usually disparate fields of ARPES and transport measurements and it allows to link transport properties such as the local conductivity or mobility to local structural features in the investigated QM. 

%We need three things to actually do this: nanoARPES, 2D materials and devices
%Alfred criticises this by writing that the argument is the wrong way round, the question comes before the measurement.
%But I am not sure it really is in this case. Let's see if anybody else comes with the same criticism.
The advent of nanoARPES is only one of the necessary ingredients for mapping the spectral function in  operating  devices. As we shall see, ARPES is a very surface-sensitive technique and mapping the spectral function of a deeply buried conductive channel in a conventional silicon field effect transistor or a quantum Hall effect device would be extremely challenging. Therefore, other important conditions for the research reported here have been the discovery of (and the continued interest in) 2D materials, as well as the development of techniques that permit the construction 2D materials-based devices suitable for nanoARPES investigations. Indeed, there are several other experimental constraints that have so-far limited nanoARPES studies of functional devices to very simple situations involving 2D materials. The surface-sensitivity of the technique, for instance, does also imply that an atomically clean surface is required for nanoARPES investigations, along with need to place the sample in ultra-high vacuum for the surface to remain clean for an extended period of time. There are several other formidable experimental challenges that have so-far prohibited nanoARPES studies on many interesting nano-scaled devices exhibiting superconductivity, magnetism or spin-selection. These experimental constraints will be discussed in detail.

%structure of the review
This review is structured as follows: Following this introduction, section \ref{sec:II} will describe the necessary technical ingredients enabling the \emph{in operando} study of QM-based devices. These are 2D (mainly) QMs, nanoARPES, and 2D QM-based devices compatible with the requirements for nanoARPES. Section \ref{sec:III} illustrates how the spectral function can be mapped across a device in equilibrium and what kind of information can be obtained from such a measurement. Section \ref{sec:IV} reviews results obtained from field-doping 2D devices using a gate voltage. Section \ref{sec:V} describes the use of nanoARPES in the presence of a steady state current. Section \ref{sec:VI} discusses experimental challenges for nanoARPES on devices. Section \ref{sec:VII} is an extended outlook on future developments of the technique. The review is ended by a brief conclusion in Section \ref{sec:VIII}. 

\section{Experimental considerations}
\label{sec:II}

\subsection{Two-dimensional quantum materials}

%Low dimensionality is interesting but the 2D must be at the surface 

Low dimensionality is a frequently encountered feature of QMs, even if the materials are not outright 2D or one-dimensional (1D). Important examples of (quasi) 2D electronic systems arise in  the integer \cite{Klitzing:1980aa} and fractional \cite{Tsui:1982aa} quantum Hall effects, the quantum spin Hall effect \cite{Konig:2007aa}, the cuprate superconductors \cite{Bednorz:1986aa} and numerous CDW materials \cite{Gruner:1994aa}. As we shall see below,  ARPES is especially well-suited for determining the electronic structure of 2D electronic systems. On the other hand, it is a very surface-sensitive technique and this precludes access to the buried two-dimensional electron gas (2DEG) in a typical quantum Hall effect device or a metal oxide field effect transistor. The advent of 2D materials has been an essential condition for \emph{in operando} ARPES studies on devices because these materials can be placed right at the surface of the device. However, it should be mentioned that band bending-induced 2DEGs can also exist near the surfaces of many semiconductors and insulators and these have been studied by ARPES (see e.g. \cite{King:2010aa,Bianchi:2010ab,SantanderSyro:2011aa,Meevasana:2011aa,Kim:2015aa}).

%Discovery of 2D materials
The successful fabrication of  2D materials \cite{Novoselov:2005ab} was heralded by the discovery of graphene  \cite{Novoselov:2004aa,Zhang:2005ab} and the devices that are typically fabricated for transport experiments on such materials are already very close to meeting the requirements for being studied with surface-sensitive techniques such as ARPES. The field of 2D materials is vast and many excellent reviews about particular aspects and classes of materials are available \cite{Neto:2009aa,Kara:2012aa,Basov:2014aa,Das:2015ab,Avouris:2017aa,Manzeli:2017aa,Dong:2017aa,Zhao:2020aa,Andrei:2020aa}. We therefore only mention the most relevant 2D QMs for the use in ARPES investigations of \emph{in operando} devices.

%the zoo of 2D materials including $h$-BN
The vast majority of 2D materials is derived from layered bulk materials with weak van der Waals bonding between the layers but there are some for which this is not the case, such as silicene \cite{Kara:2012aa} and stanene \cite{Zhao:2020aa}, and some  do not even have a bulk counterpart of the same chemical composition \cite{Arnold:2018ab}. Single or few-layered 2D materials can often be extracted from their layered bulk counterpart by mechanical exfoliation and assembled into functional devices \cite{Avouris:2017aa}. The most important material classes in this context are graphene \cite{Neto:2009aa,Basov:2014aa} and the few-layer transition metal dichalcogenides (TMDCs) of the MX$_2$ type (and from this class  mostly the semiconducting and more inert compounds, such as MoS$_2$ \cite{Manzeli:2017aa,Dong:2017aa}). Simple field effect transistor-based devices made from these materials have  been at the heart of studying their transport properties \cite{Novoselov:2004aa,Zhang:2005ab,Radisavljevic:2011aa} and have illustrated many fascinating properties, such as the unconventional Hall effect for graphene \cite{Novoselov:2004aa,Zhang:2005ab}, the transition from a semiconductor to a superconductor in single layer (SL) MoS$_2$ \cite{Costanzo:2016aa}, as well as  a metal-insulator transition driven by electronic correlations  in the same material \cite{Radisavljevic:2013aa}. Another very important 2D version of a layered compound is thin hexagonal BN ($h$-BN). In fact, this highly insulating and inert material has a central  role in  2D materials devices. It is almost ubiquitously used as a protective layer or a gate dielectric and is the key to achieving ultra high mobility in graphene devices \cite{Dean:2010aa}. Its importance for 2D material-based devices can hardly be overstated \cite{Watanabe:2004aa}. 

%heterostructures
Currently, the main focus of 2D QMs research is no longer on SL materials, but on several such layers stacked on top of each other into heterostructures  \cite{Geim:2013aa,Novoselov:2016ab}. Naively, one could expect such stacks to inherit the properties of the individual layers but it turns out that this simple picture can break down, giving rise to combined materials with emergent properties. Examples for this are the formation of atomically thin $p-n$ junctions \cite{Cheng:2014ab}, hybridisation between 2D layers \cite{Wilson:2017aa,Zribi:2019aa}, as well as proximity effects (magnetic, spin-orbit, superconducting)  \cite{Avsar:2014aa,Li:2020ab,Lupke:2020aa,Zutic:2019aa}. Also, the exact stacking angle in a bilayer of 2D materials turns out to be an important tuning parameter for the combined system's properties. A particularly simple model system is twisted bilayer graphene \cite{Andrei:2020aa} which, depending on the twist angle, shows properties that are very different from those of the SL or the (usual) Bernal stacked bilayer, such as strong electronic correlations leading to an insulating state \cite{Cao:2018ad} or, upon slight electrostatic doping, superconductivity \cite{Cao:2018aa}. Remarkably, both phenomena occur for twist angles close to a so-called ``magic angle'' of $\approx 1.1^{\circ}$. This can be understood in terms of a long range moir\'e pattern forming at this twist angle, resulting  in a very small Brillouin zone. Hybridisation between the graphene sheets gives rise to nearly flat bands near the Fermi level throughout this Brillouin zone. The flat bands, in turn, lead to a high density of states and strong correlations. Note, however, that such phenomena can only be realised for a very accurate tuning of the twist angle and at low temperatures. 

%what has been done so far in terms of devices?
The functional devices investigated by nanoARPES so far have been built from graphene, TMDCs or combinations of these materials, always also using thin layers of $h$-BN. An important reason for the choice of these materials  was the ease with which these can be handled, their high stability and chemical inertness, which results in devices with reasonably clean surfaces, and the recent progress in assembling these materials with unprecedented controll and precision \cite{Frisenda:2018aa}. There is no reason why in principle devices for nanoARPES could not be extended to other classes of 2D materials \cite{Haastrup:2018aa},  for example 2D topological materials \cite{Hou:2020aa}, stanene \cite{Zhao:2020aa} or bismuthene \cite{Reis:2017aa}.  Indeed, several of the ideas used for 2D materials could also be explored for the 3D QMs discussed in the introduction. Electrostatic gating of 3D QMs is difficult unless they are non-metallic and very thin but exploring their response to a high current density close to the surface should be possible.

\subsection{Angle-resolved photoemission on the nano-scale}

\subsubsection{Very brief introduction to ARPES}
%mr introduction to ARPES and a bit of history
Angle-resolved photoemission spectroscopy is an experimental technique based on the photoelectric effect. It was developed in the 1970s to study the band structure of solids, building on tools developed for X-ray photoemission spectroscopy. First studies of many-body effects such as the electronic self-energy were also reported in the 1980s (see e.g. Ref. \cite{Jensen:1985aa}). The technique saw an enormous improvement of instrumentation \cite{Martensson:1994ab} following the discovery of high-temperature superconductivity \cite{Bednorz:1986aa}, since it promised a direct view on how many-body interactions would manifest themselves in the spectral function. Indeed, the large energy scale for high-temperature superconductivity ($\approx 100$~meV \cite{Lanzara:2001aa}) meant that relevant many-body effects were within reach of realistic energy resolution targets. Today, it is possible to achieve energy and angular resolutions better than 1~meV and 0.1$^{\circ}$ \cite{Kiss:2008aa}, as well as sample temperatures below 1~K \cite{Borisenko:2012aa}, opening the possibility to study even conventional superconductivity.

There are many excellent reviews and books published on ARPES (for a selection see Refs.  \cite{Plummer:1982aa,Kevan:1992aa,Hufner:2003aa}), even on more specialised issues such as ARPES on cuprate superconductors \cite{Damascelli:2003aa}, quantum materials \cite{Lu:2012ac,Sobota:2020aa}, the use of high photon energies \cite{Suga:2015aa,Strocov:2019aa}, the electron-phonon interaction \cite{Hofmann:2009ab} and ARPES for the investigation of many-body effects on the nano-scale \cite{Rotenberg:2010ad}. We therefore only summarise some of the most basic concepts here, aiming for an easily accessible level. 

\begin{figure}
\includegraphics[width=0.48\textwidth]{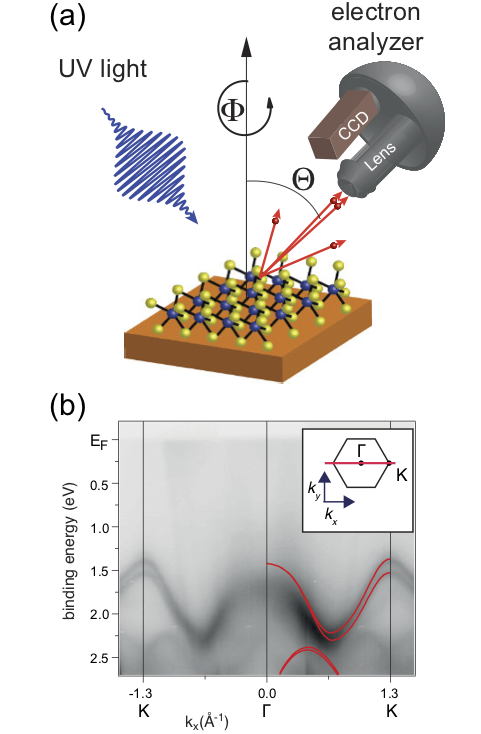}\\
\caption{(a) Principle of ARPES. A sample is exposed to photons with energies higher than the work function, resulting in the emission of photoelectrons. These are filtered in energy and detected using a two-dimensional detector that records photoemission intensity as a function of energy and for a small range of emission angles. (b) Photoemission intensity from a single layer of MoS$_2$ on Au(111), measured as a function of energy and crystal momentum parallel to the surface ($k_x$). The inset shows the path along $k_x$ relative to the hexagonal Brillouin zone of the material. The red lines in the image represented the calculated band structure of single layer MoS$_2$. Panel (b) reprinted with permission from J. A. Miwa, S. Ulstrup, S. G. S\o rensen, M. Dendzik, A. Grubi\v{s}i\'c~\v{C}abo, M. Bianchi, J. V. Lauritsen, and Ph. Hofmann, Phys. Rev. Lett.114, 046802 (2015). Copyright (2015) by the American Physical Society.}
  \label{fig:ARPESprinciple}
\end{figure}

%basic idea and kinematic theory: energy conservation
Figure~\ref{fig:ARPESprinciple}(a) illustrates the basic idea of ARPES. The surface of a material is exposed to monochromatic UV photons with an energy $h\nu$ exceeding the sample's work function $\Phi$, resulting in the emission of photoelectrons with a kinetic energy of 
\begin{equation}
E_{kin}=h\nu - \Phi - E_{b},
\label{equ:en}
\end{equation}
where $E_{b}$ is the electrons' binding energy, measured with respect to the Fermi energy $E_F$ of the sample. The measured $E_{b}$ of a spectral feature is the quantity of interest when considering the (single particle) band structure of the solid, $E_{b}(\mathbf{k})$. $E_{b}$ can be calculated using equation (\ref{equ:en}) or directly read from the photoemission spectra  because  $E_F$ is often clearly identifiable as a sharp intensity change (if the sample is not metallic, a metal in contact with the sample can be used for this energy calibration). Note that ARPES data is often displayed as a function of $\mathbf{k}$ and a positive binding energy $E_{b}$ but is also common to instead use the negative energy difference $E-E_F$. 

%momentum conservation, difficult but ok for 2D
While energy conservation is thus straight-forward, momentum conservation is not. The outgoing photoelectron has a momentum of $\hbar |\mathbf{k}|$ and the three dimensional $\mathbf{k}$ is completely determined by the measurement of the kinetic energy and the emission angles $\Phi$ and $\Theta$ in Figure  \ref{fig:ARPESprinciple}(a). The difficulty lies in working back to $\mathbf{k}$ inside the solid. The 2D component of $\mathbf{k}$ parallel to the surface, $\mathbf{k}_{\parallel}$, is conserved in the photoemission process because the system retains its translational symmetry parallel to the sample surface, so
\begin{equation}
\mathbf{k}_{\parallel}^{in}=\mathbf{k}_{\parallel}^{out},
\end{equation}
modulo a 2D (surface) reciprocal lattice vector. For the component perpendicular to the surface, $k_{\perp}$, the situation is more complicated. The translational symmetry in this direction is broken and $k_{\perp}$ is no longer a good quantum number. It is also poorly defined because the electrons of interest, those which escape the solid without inelastic scattering, stem from a region very close to the surface (see below), giving rise to a well-defined position and a therefore poorly defined (crystal) momentum $k_{\perp}$. Nevertheless, determinations of the 3D band structure of a solid are often possible by varying $k_{\perp}$ via changing the photon energy.  Fortunately, none of this is relevant for studying the (quasi) 2D materials we focus on here. For a  2D material, $\mathbf{k}_{\parallel}$ is the only relevant quantum number and the described procedure yields the desired initial state band structure $E_{b}(\mathbf{k}_{\parallel})$. 

%illustration of this for SL MoS2
A band structure determination along these lines is illustrated for the valence band of a SL of MoS$_2$, a  2D material, in Figure~\ref{fig:ARPESprinciple}(b) \cite{Miwa:2015aa}. The greyscale image shows the photoemission intensity as a function of $\mathbf{k}_{\parallel}=(k_x,0)$, following the red path in the inset of the figure, and the binding energy. The zero of the binding energy scale at $E_F$ is easily identified by the step-like change of the intensity. This is due to the fact that the SL MoS$_2$ is placed on a metallic gold substrate. The valence band is also clearly seen by an increased intensity, leading to particularly sharp features in the vicinity of the K point in the Brillouin zone. For comparison, the calculated dispersion for the valence band of free-standing SL MoS$_2$ is shown as a red line superimposed on the photoemission intensity.

\begin{figure}
\includegraphics[width=0.48\textwidth]{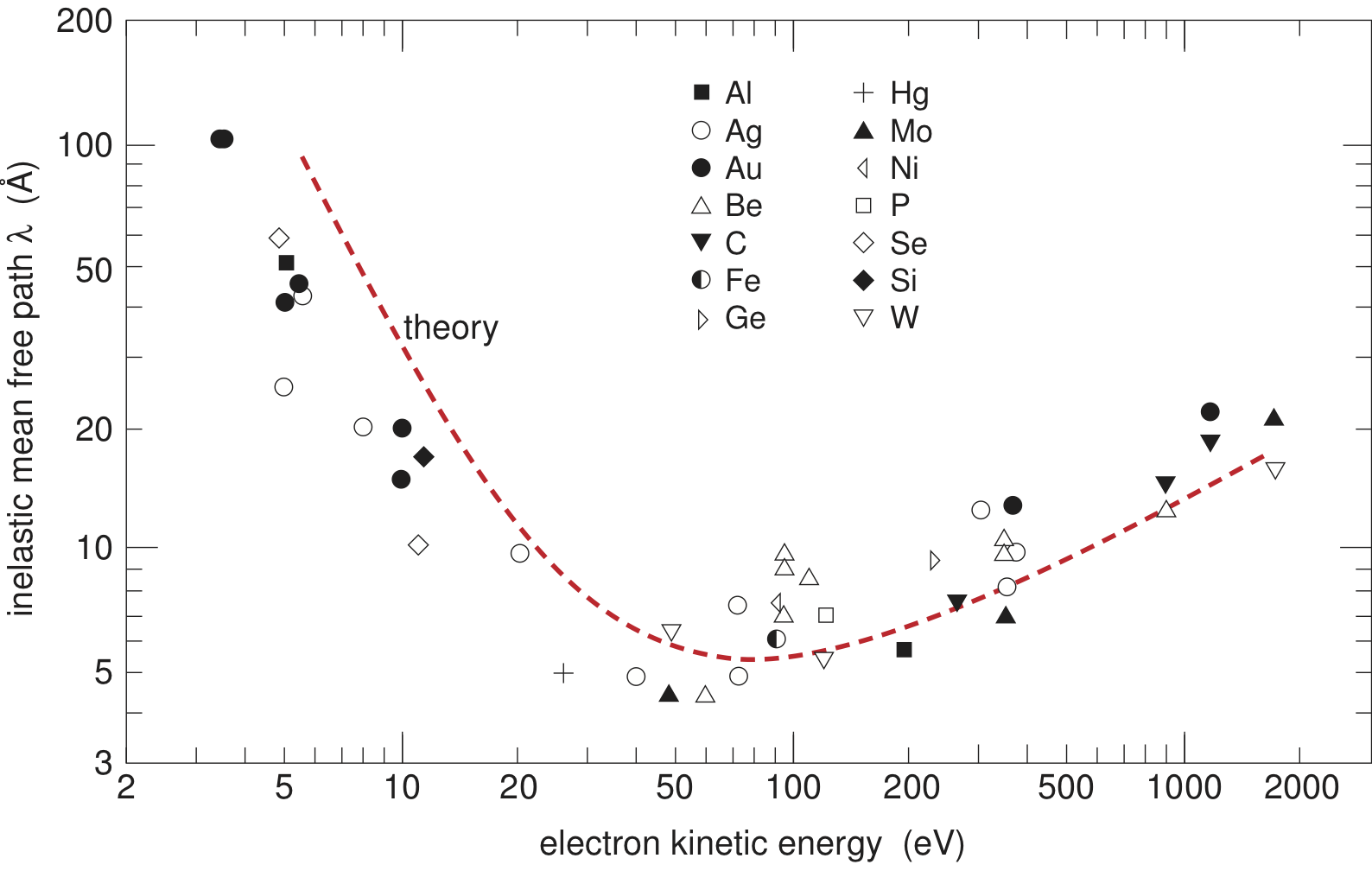}\\
\caption{Inelastic mean free path of electrons in solids. The dashed line is the calculated result for a free electron metal. Data taken from Refs.  \cite{Penn:1976aa,Zangwill:1988aa}.}
  \label{fig:electron_imfp}
\end{figure}

%surface sensitivity
A very important consideration in ARPES is the high surface sensitivity of the experiment that stems from the short inelastic mean free path of the excited photoelectrons in matter. Figure~\ref{fig:electron_imfp} shows the inelastic mean free path for some materials as a function of energy, along with a calculation for the free-electron model \cite{Penn:1976aa} (dashed line). Since this calculation represents the typical behaviour for many solids, the curve is often referred to as the ``universal curve'' for the electron inelastic mean free path but it needs to be noted that deviations from this generic behaviour have to be expected, in particular at low energies. The inelastic mean free path is rather short for all kinetic energies and especially for the energies most relevant to ARPES (between 10 and a 100~eV); the electrons that have escaped the solid without undergoing inelastic scattering processes are thus likely to stem from the first few layers of the material. Only these electrons are relevant for extracting band structure information in ARPES since information on the initial state energy and $\mathbf{k}$ is lost in an inelastic scattering process.

The short inelastic mean free path imposes some important restrictions on ARPES experiments in general and ARPES on devices in particular. All ARPES experiments are very surface-sensitive, implying that atomically clean surfaces are required for the measurements. This, in turn, calls for approaches to obtain  clean surfaces, such as the cleaving of bulk crystals, combinations of noble gas sputtering and thermal annealing or the use of very inert surfaces that do not need extra cleaning. Also, the surfaces need to remain clean during the time of the experiments. This requires carrying out ARPES experiments in ultra-high vacuum, i.e. at a pressure well below 1$\times 10^{-9}$~mbar. 
For ARPES experiments on devices, the short inelastic mean free path of the electrons also prevents spectroscopy on interfaces that are deeply buried under thick protective layers or top-electrodes for gating (approaches to circumvent this problem are discussed in Section \ref{sec:VII}). 

%more advanced: access to spectral function
The above considerations explain how ARPES can be used to measure the single particle band structure of crystalline materials but the enormous significance of the technique in modern condensed matter physics stems from the fact that it also opens a window onto the many-body effects governing the behaviour of complex QMs. In order to understand this, a slightly more advanced picture of the photoemission process is needed. Still in a simplified picture, and ignoring effects of finite energy and angular resolution, the measured photoemission intensity is given by
\begin{equation}
I(E_{b},\mathbf{k}) \propto |M_{if}|^2 f(E_{b},T)  \mathcal{A}(E_{b},\mathbf{k},T),
\end{equation}
where $|M_{if}|$ is the dipole matrix element for the excitation from initial state to final state in the photoemission process, $f(E_{b},T)$ is the Fermi-Dirac function ensuring that only occupied states can contribute to the photoemission intensity, and $ \mathcal{A}(E_{b},\mathbf{k},T)$ is the desired spectral function. We assume that the sample is a 2D material and $\mathbf{k}$ is a 2D vector. The matrix element $|M_{if}|$ for an initial state of given symmetry can strongly depend on parameters such as the photon energy or the light polarisation. However, when investigating the spectral function over a small range of $\mathbf{k}$, the $\mathbf{k}$-dependence of the matrix elements is often negligible.

%what is the spectral function?
The spectral function $\mathcal{A}$ describes the electronic structure
of a solid in the presence of many-body effects.   $\mathcal{A}(E_{b},\mathbf{k},T)$ can
be viewed as the probability of finding an electron with energy
$E_{b}$ and momentum $\mathbf{k}$ at a given temperature $T$. From now on, we drop the explicit reference to $T$  in the notation for convenience. In a quasiparticle picture, where the many-body interactions do not lead to a complete breakdown of the band structure idea but merely to the picture of quasiparticles instead of non-interacting electrons, $\mathcal{A}$ can be stated in terms of the 
``bare'' dispersion for non-interacting electrons
$\epsilon(\mathbf{k})$ and the complex self-energy $\Sigma$. It is usually assumed that $\Sigma$ is independent of  $\mathbf{k}$.  Then $\mathcal{A}$ has the form
 \begin{equation}
  \mathcal{A}(E_{b},\mathbf{k})=\frac{\pi^{-1}|\Sigma''(E_{b})|}{[E_{b}-\epsilon(\mathbf{k})-\Sigma'(E_{b})]^{2}+\Sigma''(E_{b})^{2}},
 \label{equ:p7}
 \end{equation}
where $\Sigma'$ and $\Sigma''$ are the real and imaginary parts of the self-energy. 

\begin{figure}
\includegraphics[width=0.48\textwidth]{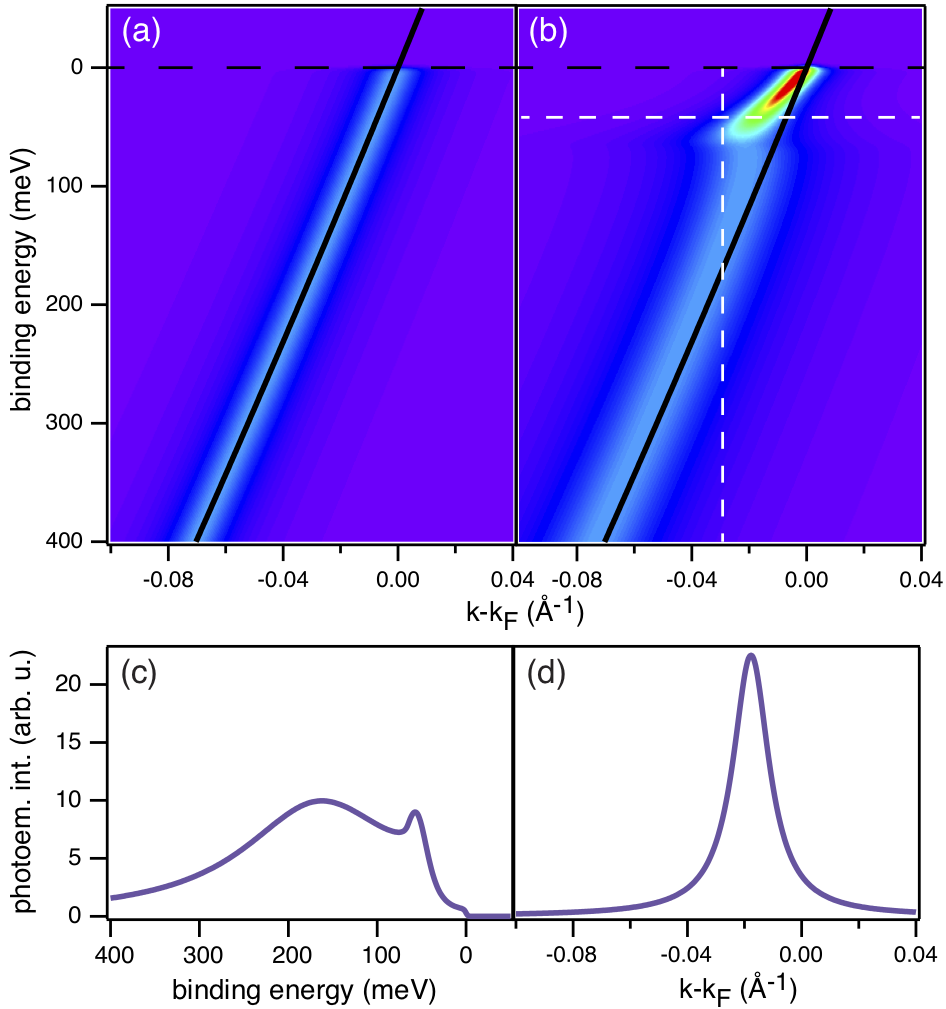}\\
\caption{Simulated spectral function according to equation (\ref{equ:p7}). The bare dispersion $\epsilon(k)$ is the solid black line. (a) Spectral function for a energy-independent lifetime broadening, as caused by electron-defect scattering, implemented by setting $\Sigma'=0$ and $\Sigma''=const$. (b) Spectral function in the presence of strong electron-phonon coupling \cite{Hofmann:2009ab}. The white dashed lines denote particular cuts through the spectral function given in panels (c) and (d). These panels show the cuts at  constant $k$ and constant energy, respectively.}
  \label{fig:specfunc}
\end{figure}

%what's the meaning of A?
A somewhat simplified interpretation of $\mathcal{A}$ in the form of equation (\ref{equ:p7}) is that the many-body effects, as encoded in $\Sigma$, lead to a spectral function that resembles that of the bare dispersion $\epsilon(\mathbf{k})$ with two important modifications: Instead of being a $\delta$-function tracking $\epsilon(\mathbf{k})$, the dispersion is broadened by $\Sigma''$, the imaginary part of $\Sigma$ and shifted by $\Sigma'$, the real part. The simplest situation is that of  an energy-independent scattering mechanism, for example defect scattering. In this case, $\Sigma$ is independent of energy and given by $\Sigma'=0$ and $\Sigma''=const.$ The resulting spectral function is shown in Figure~\ref{fig:specfunc}(a). The bare dispersion is given by the solid line and the maximum in the spectral function tracks this line.  The broadening corresponds to the quasiparticle now having a finite lifetime due to scattering and this lifetime is given by $\tau = \hbar / 2 \Sigma''$. In this simple case, the interpretation of the ARPES features tracking the single particle band structure $E_{b}(\mathbf{k}_{\parallel})$ is valid. The many-body effects merely lead to a broadening but do not affect the dispersion as such. 

The situation of a more complicated interaction with a bosonic mode (e.g., a phonon) is illustrated in Figure~\ref{fig:specfunc}(b). Here the dispersion changes in the vicinity of $E_F$. The position of the $E_F$ crossing, $k_F$, remains the same but the dispersion flattens out near $E_F$, leading to an increased effective mass of the quasiparticle which now is a mixture of an electron and a phonon. This is accompanied by a characteristic ``kink'' at a higher binding energy corresponding to the energy of the phonon. Such kinks are often observed in the case of systems with strong electron-boson coupling \cite{Hengsberger:1999ab,Valla:1999ab,Lanzara:2001aa,Hofmann:2009ab}.

%some idea about how to get the bare dispersion from this
The analysis of a measured spectral function should ideally yield the bare dispersion $\epsilon(\mathbf{k})$, as well as the complex self-energy $\Sigma$. Extracting these quantities is not trivial since $\epsilon(\mathbf{k})$ is usually not known. Commonly, 1D cuts through the spectral function are analysed, either at constant $\mathbf{k}$ (so-called energy distribution curves) or at constant $E_{b}$ (so-called momentum distribution curves). The latter approach has the advantage that, within the approximation of equation (\ref{equ:p7}) and assuming a locally linear dispersion, the momentum distribution curve can be described by a Lorentzian \cite{Valla:1999aa}. For every $E_{b}$, a single maximum in the spectral function can be found and a renormalised dispersion can be defined. In the case of energy distribution curves, on the other hand, several maxima can appear at a given $\mathbf{k}$, complicating the analysis \cite{Engelsberg:1963aa,Hofmann:2009ab}. This is illustrated with the dashed vertical (horizontal) line representing an energy (momentum) distribution curve in Figure~\ref{fig:specfunc}(b) and the corresponding intensity distributions in panels (c) and (d), respectively. Several schemes have been proposed to extract the bare dispersion from the data  \cite{Kordyuk:2005aa,Pletikosic:2012aa}. These make use of fact that $\Sigma'$ and $\Sigma''$ are related by a Kramers-Kronig transformation, that the result must be self-consistent and that the the Fermi wave vector $k_F$ is not altered by the many-body interaction \cite{Luttinger:1960ab} (something that is not strictly correct because only the overall volume (area) of the Fermi surface (contour) is unaffected by the interactions, not a particular $k_F$ value).  The quantities of interest ($\epsilon(\mathbf{k})$ and $\Sigma(E_b)$) can also be extracted by a fitting a simulated 2D spectral function to the data, doing away with the need to rely on 1D cuts \cite{Nechaev:2009aa,Mazzola:2013aa}. This has the advantage that all the information obtained in the measurement process is simultaneously exploited for the complex task of finding $\epsilon(\mathbf{k})$ and $\Sigma(E_b)$.

\begin{figure}
\includegraphics[width=0.48\textwidth]{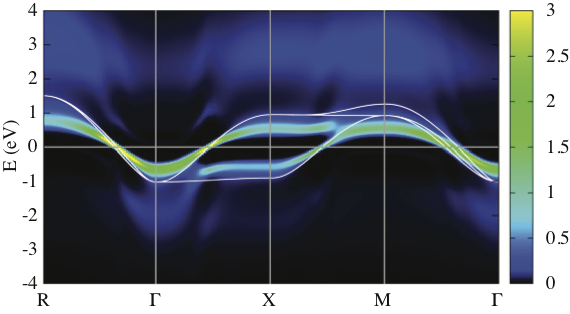}\\
\caption{Calculated spectral function for SrVO$_3$, a material with strong correlations. The energy scale denotes the difference to $E_F$. The white lines represent the band structure calculated within density functional theory. Reprinted with permission from M. Karolak, T. O. Wehling, F. Lechermann, and A. I. Lichtenstein, Journal of Physics: Condensed Matter 23, 085601 (2011). Copyright 2011, IOP Publishing.}
  \label{fig:correlations}
\end{figure}

In systems with strong electronic correlations, it may not be possible to describe the spectral function in the quasiparticle picture of equation (\ref{equ:p7}) and more complicated structures than in Figure~\ref{fig:specfunc} can arise in $\mathcal{A}$. A typical example is given in Figure~\ref{fig:correlations} which shows the calculated spectral function near $E_F$ for SrVO$_3$, a strongly correlated transition metal oxide, along with a band structure calculation using density functional theory (DFT) \cite{Karolak:2011aa}. This example shows some characteristic fingerprints of strong electronic correlations. The maxima of the spectral function still somewhat track the DFT band structure and the Fermi level crossings appear to be unaltered, as in the case of the strong electron-phonon interaction discussed above. On the other hand, the  correlations lead to an overall narrowing of the band. This is found in many correlated electron systems and can be understood as an increase of the electron effective mass that goes hand in hand with  the flatter bands. Moreover, additional rather diffuse bands are found for energies of around 2~eV below and 3~eV above the Fermi energy. These are the lower and upper Hubbard bands, respectively, representing localised electronic states. It is clear that this spectral function is rather complex and cannot be explained in terms of merely a shifted and broadened single-particle dispersion $\epsilon(\mathbf{k})$. Still, a comparison of ARPES data to calculations such as that in Figure~\ref{fig:correlations} remains meaningful. Note, of course, that ARPES can only provide information on the spectral function below $E_F$.  

\subsubsection{nanoARPES}

%history of ignoring position resolution - need for doing it especially for materials with nano-structuring
The advent of nanoARPES \cite{Rotenberg:2010ad,Usachov:2011aa}, i.e., 
ARPES with the capability of accurately positioning a focused light spot on a sample with a resolution of
 about 1~$\mu$m or below, has opened the route to studying a multitude of samples and addressing scientific questions that have been inaccessible to conventional ARPES  \cite{Rotenberg:2010ad}. 
 A few examples of first nanoARPES results are 
 the confirmation of predicted flat bands in very small samples of rhombohedral-stacked graphene \cite{Henck:2018ac},
 the identification of a weak topological insulator via the surface orientation-dependent electronic structure \cite{Noguchi:2019aa},
the effect of defects on the nano-scale electronic structure \cite{Kastl:2019aa},
the band alignment in nano-scale  heterostructures \cite{Wilson:2017aa,Kastl:2018aa,Ulstrup:2019aa},
the band structure of twisted bilayer graphene near the magic twist angle \cite{Utama:2020aa,Lisi:2020aa}
and the observation of minibands in twisted graphene/WS$_2$ heterostructures \cite{Ulstrup:2020aa}. 

%different surface termination
Interesting opportunities for nanoARPES investigations also arise for crystals with different possible terminations for a given surface orientation. For topological materials, the surface termination is crucial for the surface state dispersion \cite{Teo:2008aa,Wu:2020ab}, even for the simplest materials such as Bi$_{1-x}$Sb$_{x}$ \cite{Teo:2008aa,Zhu:2014ad}. Different surface termination can be simultaneously present, separated in domains or terraces, on a cleaved surface, either because their surface energy is very similar or because cleaving the bulk crystal along a plane of weak bonding necessarily produces two different terminations, e.g., when the cleavage plane is separating a quintuple and a septuple layer of the crystal structure \cite{Wu:2020ab}. In such cases, high spatial resolution is clearly desirable to determine the electronic structure of surface domains containing only a single termination. 

%but also good for nominally uniform samples because they aren't uniform
Even for nominally uniform sample surfaces, using a small light spot may have considerable advantages. Historically, the vast majority of ARPES investigations have  considered the sample under investigation as completely uniform, ignoring any microscopic variation or assuming that such a variation would average out and/or contribute to the incoherent background intensity. For many samples, this is not a bad approximation. After all, a background intensity can always be expected due to scattering by phonons, electrons or point defects. To illustrate this, consider the photoemission intensity near $E_F$ in Figure~\ref{fig:ARPESprinciple}(b). A clear cutoff due to the Fermi-Dirac distribution is visible in the entire $k_x$ range but, actually, this cutoff should be absent away from bulk band Fermi level crossings of the gold substrate, and especially in the projected bulk band gap of Au(111) around the $\bar{\mathrm{K}}$ point \cite{Takeuchi:1991aa}. An increased intensity of the background is not particularly important, as long as the band structure features of interest are sufficiently intense. It has even become a common practice to emphasize weakly dispersing features on a high background by displaying the second derivative of the photoemission intensity \cite{Zhang:2011aa}. On the other hand, a quantitative analysis in terms of a spectral function can be difficult in such a case. The assumption of a uniform sample surface can also lead to a misinterpretation of the data. For instance, local doping variations of a semiconductor with a length scale smaller than the integration area of the UV light spot would broaden spectral functions considerably and this could then mistakenly be interpreted as a defect-scattering induced lifetime reduction. This is an issue for samples showing electron and hole puddles such as graphene on SiO$_2$ \cite{Zhang:2009ab} or the surface states of topological insulators \cite{Beidenkopf:2011aa}. In short, there are many reasons to believe that investigations of nominally uniform samples could also strongly benefit from a high spatial resolution.

%there are basically two way to do this: electron optics and light spot size. Focus on the latter
There are essentially two approaches to obtaining spatial resolution in ARPES. One is to use the electron optics in a photoemission electron microscope (PEEM), a type of instrument that can routinely achieve a spatial resolution better than 100~nm. PEEM has the additional advantage of being a fast imaging technique, so that many different sample areas of interest can be explored. Conventional PEEM does not offer energy resolution and using a PEEM for ARPES requires the addition of an energy filter. Energy-filtered PEEM is a well-established technique for small-scale ARPES and has also been used to map the band structure of 2D materials, see e.g., Refs. \cite{Fujikawa:2009aa,Sutter:2009ac,Jin:2013aa,Cattelan:2018aa,Ulstrup:2019ab}. Integration over regions of interest with diameters of a few $\mu$m has been found sufficient for band structure mapping. The notable drawbacks of energy-filtered PEEM are image aberrations, the limited energy resolution in the order of 100~meV.  Both preclude the study of subtle many-body effects that are one of the main areas of interest in current ARPES experiments of QMs. Moreover, a PEEM setup requires the presence of a high voltage between sample and microscope lens, necessitating very flat sample surfaces that cannot usually be reconciled with the device designs explained below. We will thus focus on other techniques but note that varieties of energy-filtered PEEM (called nanoESCA or  momentum microscopy) are seeing strong instrumental improvements \cite{Escher:2005aa,Kromker:2008aa,Wiemann:2011aa,Schonhense:2015aa,Tusche:2019aa,Matsui:2020aa}. When time-of-flight strategies are used to obtain energy resolution in combination with pulsed sources, these instruments become particularly attractive for time-resolved experiments and we shall return to this point in Section VII. 

%small light spot, general considerations, diffraction limit, can't move the sample anymore
The other route to high spatial resolution is to focus the UV light spot to the nano-scale but to otherwise keep the conventional ARPES instrumentation of Figure~\ref{fig:ARPESprinciple}. In principle, this then permits experiments with the same energy and angular resolution as currently achieved in conventional ARPES. However, there are a number of important considerations. Concerning the focusing of the light beam, there is the principle diffraction limit of a few wavelengths $\lambda$ for the size of the light spot. For the low energies conventionally used in ARPES, this could be a serious limitation (e.g. $\lambda=62$~nm for $h\nu=$20~eV) but as most nanoARPES setups work with higher energies, it is not currently an important concern. As in any microscopy technique, the sample position needs to be stabilised against vibrations and drift far more rigorously than in conventional ARPES and it needs to be possible to move the sample very precisely to scan the light spot across the surface. It is also  desirable to combine this with efficient cooling to cryogenic temperatures. Finally, in current ARPES setups with large fixed electron analysers, a change of the emission angles $\Theta$ and $\Phi$ in Figure~\ref{fig:ARPESprinciple}(a) is often achieved by rotating the sample on a goniometer with respect to a fixed electron analyser. This is not a practical solution when using a strongly focused beam on a well-defined location because the light spot will inevitably move across the sample surface upon rotating the sample. Therefore, nanoARPES setups based on focusing the light rely either on rotating the electron analyser around a fixed sample, on electrostatic deflector plates in front of the analyser entrance in order to select different emission angles, or a combination of the two.  

\begin{figure}
\includegraphics[width=0.48\textwidth]{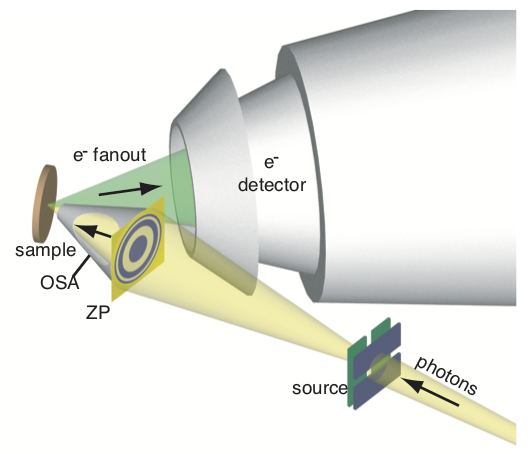}\\
\caption{NanoARPES setup based on focusing the light using a zone plate (ZP) and an order-sorting aperture (OSA). Reprinted with permission from E. Rotenberg and A. Bostwick, Journal of Synchrotron Radiation 21, 1048 (2014). Copyright 2014, International Union of Crystallography. }
  \label{fig:zpfocus}
\end{figure}

%focusing with Fresnel zone plates
For a sufficiently coherent light source, such as a 3rd or 4th generation synchrotron radiation source or a laser, focusing can be achieved by using diffractive optics such as a Fresnel zone plate combined with an order-sorting aperture, as shown in Figure~\ref{fig:zpfocus}. This approach is by now realised at several synchrotron radiation facilities (ALS, SOLEIL, DIAMOND)  \cite{Rotenberg:2010ad,Rotenberg:2014aa,Avila:2013aa,Avila:2014aa,Rosner:2019aa}. It has the practical drawback of requiring the order sorting aperture to be very close to the sample surface, while still permitting the photoelectrons to escape at normal emission (higher emission angles can simply be taken along the direction away from the focusing optics). Also, being a diffractive technique, zone plate focusing restricts the available photon energy range. This is not a very important limitation when investigating 2D materials because the most important reason for changing the photon energy is a variation in $k_{\perp}$ when studying 3D band structures (although changing the matrix elements via the photon energy can also be desirable). Finally, zone plates have a low efficiency of only about 10\% in the energy range of interest \cite{Rosner:2019aa}, Combined with the low coherent fraction of undulator radiation from 3rd generation synchrotron radiation sources, this gives rise to a very low photon flux at the sample, typically two orders of magnitude lower than in a conventional ARPES setup. This is a serious drawback because it requires a compromise in terms of resolution on the electron analyser side in order to collect usable data. On the other hand, using Fresnel zone plates is the light focusing approach that can currently achieve the highest spatial resolution of better than 200~nm \cite{Kastl:2019aa}.
The move to 4th generation synchrotron radiation sources with an increased coherent fraction of undulator radiation will further improve the spatial resolution and photon flux. 

\begin{figure}
\includegraphics[width=0.48\textwidth]{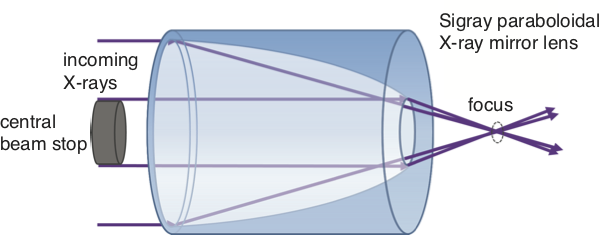}\\
\caption{Principle of light focusing using the inner paraboloid-shaped surface of a capillary (depending on the optical layout, ellipsoidal-shaped surfaces are also used but the overall concept is similar). The central beam is blocked. Image courtesy of Sigray Inc., used with permission. }
  \label{fig:capillary}
\end{figure}

%Alternative: Schwarzschild objectives, capillary sources
As an alternative, focusing can be achieved by non-diffractive optics. Using a Schwarzschild objective, a resolution of better than 1~$\mu$m has been achieved at ELETTRA \cite{Dudin:2010aa,Nguyen:2019aa}. At the ALS, a newly developed capillary focusing optics gives a resolution of around 1~$\mu$m \cite{Koch:2018ab,Ulstrup:2020aa} and a similar setup is under construction at DIAMOND, SOLEIL and ASTRID2. Figure~\ref{fig:capillary} illustrates the working principle of the device. An incoming beam is focused using the inner surface of a capillary. There is a requirement to block the central beam to eliminate the non-focused part of the beam but the device still gives a much higher photon flux compared to diffractive optics. It is also achromatic and operates at a larger working distance. For a low-energy synchrotron like ASTRID2, it is the focusing option of choice because the achievable light spot size is anyway limited by the long wavelength of the radiation. 

\subsection{Devices}

\begin{figure*}
\includegraphics[width=0.85\textwidth]{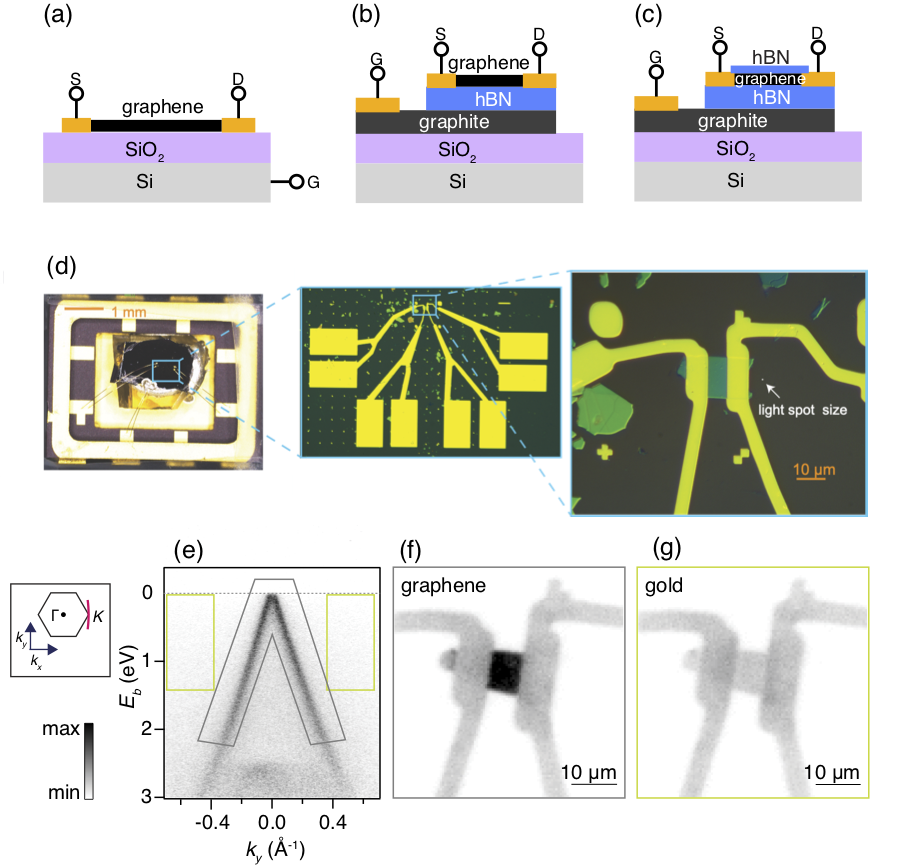}\\
\caption{Device designs for nanoARPES investigations. The ``graphene'' in the sketches could be exchanged for another 2D material, such as single layer MoS$_2$. (a) Source and drain contacts to graphene are fabricated on the surface of SiO$_2$/Si. A heavily doped Si substrate can be used as the gate material. (b) Highly conductive graphite is used as the gate and a thin layer of $h$-BN as a dielectric. This permits reduced operating voltages and gives rise to higher mobility graphene. (c) For extra protection of more sensitive 2D materials, the surface can be covered by an additional single layer of $h$-BN. (d) Actual device in different stages of magnification. The device is mounted on an ultra-high vacuum compatible chip carrier. Large gold pads on the chip carrier are used to connect to the device contacts via wire-bonding. The largest magnification on the right shows the device (graphene on $h$-BN on SiO$_2$/Si with two Au contacts) and the size of the light spot used in the nanoARPES experiment for comparison.  (e) Spectrum taken from the middle of the device (photoemission intensity as a function of binding energy $E_b$ and $k_y$), showing the Dirac cone of graphene and the valence band maximum of $h$-BN at $E_b \approx$2.7~eV.  To regions of interest are marked around the Dirac cone (grey) and around the Fermi level away from the graphene dispersion (green). The sketch to the left shows the Brillouin zone together with the scan direction (red line). (f) Integrated photoemission intensity in the grey window around the Dirac cone in panel (e), marking the location of graphene.  (g) Photoemission intensity from the green area in panel (e), emphasising the Au contacts.  Panels (d) - (g) reprinted with permission from D. Curcio, A. J. H. Jones, R. Muzzio, K. Volckaert, D. Biswas, C. E. Sanders, P. Dudin, C. Cacho, S. Singh, K. Watanabe, T. Taniguchi, J. A. Miwa, J. Katoch, S. Ulstrup and Ph. Hofmann, Phys. Rev. Lett. 125, 236403 (2020). Copyright (2020) by the American Physical Society.}
  \label{fig:devices}
\end{figure*}

%some general requirements
In order to be suitable for ARPES investigations, devices must meet a number of requirements that are not usually important in transport measurements \cite{Avouris:2017aa,Liu:2019ae,Fan:2020aa}. First of all, the device must not be too small. Typical devices made from 2D materials are only a few $\mu$m in size, strictly requiring nanoARPES to investigate their electronic structure, but device dimensions below $\approx 1$~$\mu$m would make the experiment very challenging.  Being an electron-based spectroscopy, ARPES measurements are very sensitive to stray electric and magnetic fields. In fact, in any ARPES setup, careful shielding of the earth's magnetic field must be ensured. Electric fields are usually not a concern for large flat single crystal surfaces but they are in connection with devices. Consider the simple graphene-based field effect transistor in Figure~\ref{fig:devices}(a) \cite{Novoselov:2004aa,Zhang:2005ab}. The device is fabricated on the thin oxide layer on a heavily doped Si substrate. A graphene flake is attached to two gold electrodes forming the source (S) and drain (D) of the transistor. A gate voltage is applied via the Si substrate. Due to the low gate capacitance, the operating voltage needs to be rather high, leading to a stray electric field perpendicular to the surface of graphene. Also there is a voltage difference between source and drain, giving rise to a field component parallel to the surface. Finally, even in the absence of any applied voltages, there are differences in work function and contact potentials between the different materials in the device. 
%doping and gate capacitance: Q=CU, n= C V / e
%so small capacitance will require high voltages
%what is the capacitance of the SiO2 typically? In Zhang 2014 it's 115 aF / mum^2 for a SiO2 device
%in arxiv1108.2021 it is 0.39 muF / cm2 for a $h$-BN transistor
%so then 1 cm^2 is equal to 10^8 mum^2, so then the capacitance is 3900 aF / mum^2 a lot higher than for SiO2

%generic device for nanoARPES
Because of the need to operate the device with a small gate voltage and issues with photodoping effects that compensate the gating in devices of the type shown in  Figure  \ref{fig:devices}(a) \cite{Ju:2014aa}, 
ARPES studies on graphene are  mostly based on devices similar to that in Figure  \ref{fig:devices}(b) where the gate dielectric is a thin flake of $h$-BN, placed on a piece of graphite that serves as the gate electrode. In addition to working at lower gate voltages, this has the advantage of resulting in graphene devices with much higher mobility \cite{Dean:2010aa}. Other high-$k$ gate dielectrics might be also be usable, such as SrTiO$_3$ which has been employed for gated scanning tunnelling microscopy experiments \cite{Zhang:2013ac}. 

%surface sensitivity. The surface needs to be clean.
A second important constraint for ARPES is the very high surface sensitivity of the technique, calling for an atomically clean surface. This is especially relevant when many-body effects such as electron-electron or electron-phonon scattering are to be investigated because even a small defect consideration can give rise to strong electron-defect scattering, resulting in short lifetimes and masking out the effects of interest \cite{Kevan:1986aa} (see Figure~\ref {fig:specfunc}).
One of the main practical reasons for using graphene in many of the experiments used below is thus the 
 need for a material that is easily cleaned and remains clean for many hours in ultra-high vacuum during the time-consuming data collection.  Still, it remains difficult to obtain a clean surface in the first place because approaches such as high-temperature annealing in vacuum, usually employed for graphene on metal surface or SiC, cannot be applied without damaging the devices. Semiconducting 2D TMDCs have the advantage of chemical stability similar to graphene and have also been used for nanoARPES experiments on devices \cite{Nguyen:2019aa}. Metallic 2D TMDCs, on the other hand, are typically too reactive. 

%protect surface by thin layer of $h$-BN
An elegant way of circumventing the problem of reactivity is to cover the 2D material of interest with a thin $h$-BN crystal. This so-called encapsulation technique is illustrated in Figure  \ref{fig:devices}(c). Such a $h$-BN layer is highly inert, protects the 2D material and can even reduce the number of defects at the surface \cite{Kretinin:2014aa}. The only drawback is that the $h$-BN layer needs to be extremely thin in order to observe the spectral features of the buried 2D material, due to the short inelastic mean free path of low-energy electrons in solids. By now, there have been multiple demonstrations that SLs of $h$-BN, graphene and TMDCs are sufficiently transparent to perform ARPES investigations of the material buried beneath these layers \cite{Ulstrup:2019aa,Nguyen:2019aa,Muzzio:2020aa}. However, devices that rely on top-gating or ionic liquid gating will not be easily accessible for ARPES investigations. 

%how many contacts does one need?
All the devices in Figure~\ref{fig:devices} are essentially back-gated field effect transistors, permitting both a change of the carrier density in the 2D material by application of a gate voltage and the passing of a current through the material. If only one of these properties is of interest, a simpler two-terminal device geometry can be used, as for example in the gating experiments of Ref. \cite{Nguyen:2019aa}.

%device needs to be UHV compatible
A final requirement for the device is that it must be compatible with maintaining ultra-high vacuum conditions in the experimental chamber used for ARPES. This limits the choice of materials to those with a low vapour pressure. Excessive amounts of plastic, glue etc. need to be avoided.  Figure~\ref{fig:devices}(d) shows a typical sample meeting these requirements in different magnifications. The device is  mounted on a chip carrier that is  attached to a standard transferrable ARPES sample holder with several electrical contacts (not shown). On the chip carrier, a  Si/SiO$_2$ wafer is mounted with gold tracks from the actual device area to larger gold pads, used for wire-bonding connections to the chip carrier. These wire bonds are  seen on the left hand image. The highest magnification on the right hand side shows the actual device, consisting of a SL of graphene on a thin sheet of $h$-BN, connected to the gold tracks \cite{Curcio:2020aa}.  

%basic characterisation by ARPES
%Alfred finds this too detailed but let's see what the others say. 
Before performing any detailed nanoARPES measurement on the device, it is necessary to find it on the SiO$_2$ substrate and to localise its basic components by nanoARPES. Finding the device turns out to be easily done by monitoring the photoemission intensity around $E_F$ or from a specific core level as a function of position on the substrate. While the SiO$_2$ is prone to charging and gives only a very weak signal, the gold contacts or the graphene in the device all have states around $E_F$ and are thus clearly identified. This is illustrated in  Figure~\ref{fig:devices}(e)-(g) for the same device as the one shown in Figure~\ref{fig:devices}(d). A spectrum from the graphene flake in the centre of the device is shown in Figure~\ref{fig:devices}(e). The Dirac cone of graphene is clearly visible, as well as the valence band maximum of the underlying $h$-BN at a binding energy $E_B$ of around 2.7~eV. Two spectral regions are marked in this figure: The grey and green windows enclose the Dirac cone and an area which is close to $E_F$ but does not contain any graphene features, respectively. The integrated intensity in the grey area  across the entire device is displayed in Figure~\ref{fig:devices}(f). The shape of the device is clearly seen and it is very similar to the optical image in Figure~\ref{fig:devices}(d). As expected, the highest intensity is found for the graphene flake in the centre. The gold contacts are polycrystalline and show an almost featureless photoemission intensity around $E_F$. They therefore also contribute to the intensity in the grey window but not as strongly. Figure~\ref{fig:devices}(g) finally shows the integrated intensity in the green window, emphasising the contribution of the gold. The  graphene states are not expected to show any photoemission intensity inside the green window but still the graphene is seen in  Figure~\ref{fig:devices}(g), presumably because of a weak background arising from electron scattering by, e.g., phonons. 
 
 %samples with a current 
 So far, all nanoARPES experiments with field-doping or currents have been performed on 2D materials in devices similar to those shown in Figure~\ref{fig:devices} but, as pointed out in the introduction,  there is a strong case to be made for performing similar experiments on 3D QMs. In particular, driving phase transitions by a current has many potential applications. In general, 3D crystals could have different conductivities at the surface and in the bulk, for example due to the presence of metallic (topological) surface states. For such situations, we can specify some requirements for the experiment in addition to the need for a small light spot. First of all, it is desirable to achieve a high current density near the surface of the sample while avoiding a high total current. This reduces both Joule heating and the magnetic field from the current. In experiments on the current-induced breakdown of superconductivity in high $T_C$ materials, avoiding a high total current has been achieved by using narrow bridges of thin films i.e., samples with a very small cross section \cite{Kaminski:2016aa,Naamneh:2016aa}.

\begin{figure}
\includegraphics[width=0.48\textwidth]{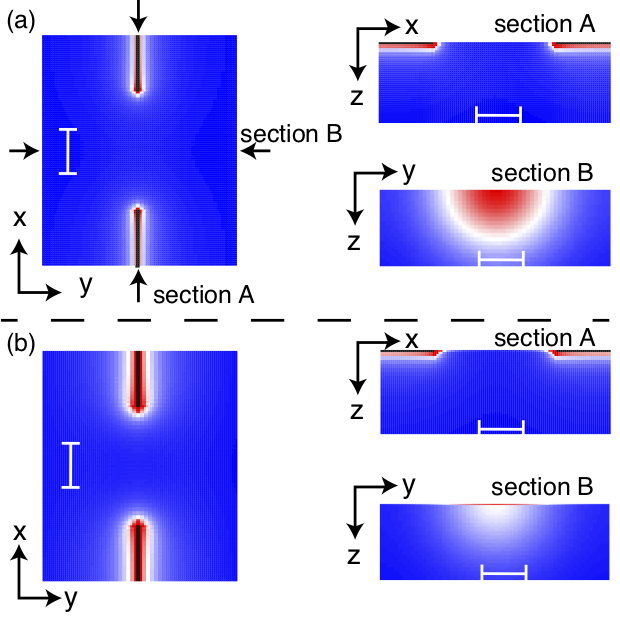}\\
\caption{Calculated current distribution between contacts placed on a 3D sample. The left hand panels show a top view of the sample. The black vertical lines are Au contacts. The colour scale indicates the current density from blue (low) to red (high).  The sections on the right are cuts through the data along the directions given by the arrows in the left-hand image. Note that a different maximum colour is used for the centre section B where the current density is much lower than close to the contacts.  (a) Situation for a sample with identical bulk and surface conductivity. (b) Current density for a sample with a surface conductance that is 20 times higher than the bulk conductance \cite{condnote}. The white scale bars have the same size in all panels. Note the strongly increased current density just at the surface in section B, as well as the stronger spreading along the surface in the top view.  }
  \label{fig:holder_current}
\end{figure}

%how would one construct this? 
Alternatively, high current densities in the region of interest could be achieved by placing electric contacts on the surface of a material in a suitable geometry.  We ignore the experimental difficulty of obtaining a clean surface and applying the electrical contacts and focus solely on the desirable geometry for such contacts.  Figure~\ref{fig:holder_current} shows two situations for the current density in the vicinity of  contact strips on the surface. In Figure~\ref{fig:holder_current}(a), it is assumed that the surface of the material is as conductive as the bulk. In Figure~\ref{fig:holder_current}(b), the surface conductance is 20 times higher \cite{condnote}. The resulting current density is obtained by numerically solving the Poisson equation \cite{Wells:2006aa,Hofmann:2009aa,Perkins:2013aa,Barreto:2014aa}. As might be expected, the resulting current distributions are drastically different: In Figure~\ref{fig:holder_current}(a), the current penetrates into the bulk with a depth similar to the contact distance (seen in the plane intersecting the contact-contact direction half-way between the contacts, section B). In Figure~\ref{fig:holder_current}(b) almost the entire current is concentrated at the surface, as seen in the corresponding cut. Moreover, the functional form of the potential and current density away from the contacts is different in the two situations (something well-known from transport measurements \cite{Hofmann:2009aa}) and this can be clearly seen in the top view current distributions. As expected, achieving a high current density near the surface for a material without a high density of metallic surface states requires a close proximity of the contacts. One should thus aim for a contact distance that is as small as possible but still significantly larger than the light spot. Moreover, one should consider the possibility of band bending near the contacts and ``blurred'' contact regions.

\section{Position-resolved spectral function}
\label{sec:III}

%need for nano-scale equilibrium property mapping and results from Curcio
Before applying gate voltages or a current, it is important to carefully characterise a device's equilibrium properties. Using nanoARPES, the local band structure variations in several materials systems have been explored \cite{Joucken:2016aa,Wilson:2017aa,Joucken:2019ab,Kastl:2019aa,Ulstrup:2019aa,Ulstrup:2020aa} and extending this to study more subtle details of the spectral function is straight-forward. In the following, we give a detailed account of the position-resolved spectral features of the graphene-based device shown in Figure~\ref{fig:devices}(d) \cite{Curcio:2020aa}. The analysis illustrates the nano-scale inhomogeneities in doping, domain  orientation and electronic self-energy in a typical graphene sample and how such information can be collected using nanoARPES. In section \ref{sec:V} we shall return to this particular device and relate the position-resolved static features to the local transport properties. 

\begin{figure}
\includegraphics[width=0.48\textwidth]{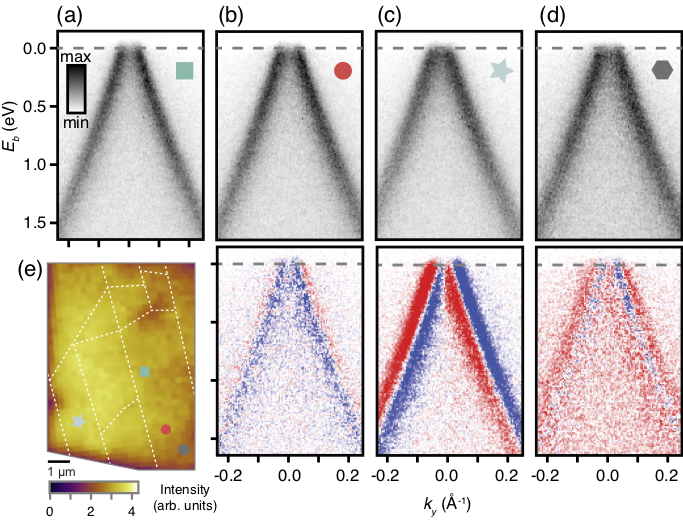}\\
\caption{Spectra of the Dirac cone of graphene,  taken in different locations on the device shown in Figure~\ref{fig:devices}(d). (a) Reference spectrum. (b)-(d) Spectra taken at different positions along with the difference to the reference spectrum below (blue negative, red positive). The markers in (a)-(d) reveal the position of the individual spectra in panel (e), which displays the intensity of the Dirac cone across the device. The dashed lines in (e) outline the position of defect lines in the sample, as observed by atomic force microscopy. Reprinted with permission from D. Curcio, A. J. H. Jones, R. Muzzio, K. Volckaert, D. Biswas, C. E. Sanders, P. Dudin, C. Cacho, S. Singh, K. Watanabe, T. Taniguchi, J. A. Miwa, J. Katoch, S. Ulstrup and Ph. Hofmann, Phys. Rev. Lett. 125, 236403 (2020). Copyright (2020) by the American Physical Society.
}
  \label{fig:specfunc1}
\end{figure}

The spectral function of the device in  Figure~\ref{fig:devices}(d)  has been mapped throughout the graphene flake by scanning the UV light spot in steps of 250~nm and collecting nanoARPES spectra in every position. Even a simple inspection of the raw data shows a considerable variation of properties across the device: Spectra from different locations are shown in Figure~\ref{fig:specfunc1}. A spectrum at an arbitrary position is taken as a reference (Figure~\ref{fig:specfunc1}(a)) and compared to spectra from other positions by calculating the difference to the reference spectrum ((Figure~\ref{fig:specfunc1}(b)-(d)). The difference plots from selected positions in the device are showing an energy shift in panel (b), a shift along $k_y$ in panel (c) and a broadening in panel (d). Similar variations in doping and alignment angle have also been observed for bilayer graphene on $h$-BN in Ref. \cite{Joucken:2019ab}.

\begin{figure}
\includegraphics[width=0.48\textwidth]{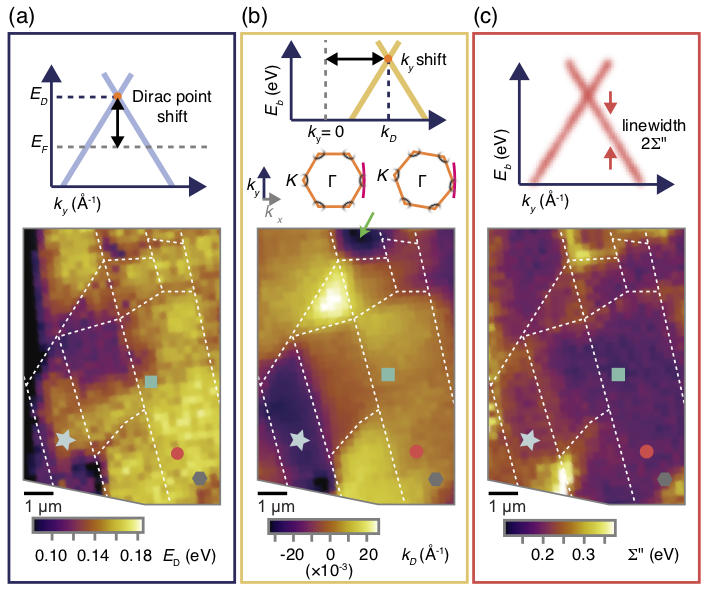}\\
\caption{Result of fitting photoemission spectra collected across the device in Figure \ref{fig:devices}(d) to a model spectral function. The upper parts of the sub-figures illustrate the role of the fitting parameter displayed. (a) Energy of the Dirac point $E_D$, mapping the local variation in doping across the device. The markers correspond to the locations for the spectra in Figure~\ref{fig:specfunc1}. (b) Position of the Dirac cone in $k_y$. Shifts in $k_y$ arise from azimuthal rotations between domains on the sample. The inset demonstrates how a domain rotation from a reference direction leads to a shift of the Dirac cone dispersion with respect to the fixed scan range in $k-$space (red line). (c) Imaginary part of the self energy $\Sigma''$ measuring the width of the spectra or, equivalently, the inverse lifetime of the state. Reprinted with permission from D. Curcio, A. J. H. Jones, R. Muzzio, K. Volckaert, D. Biswas, C. E. Sanders, P. Dudin, C. Cacho, S. Singh, K. Watanabe, T. Taniguchi, J. A. Miwa, J. Katoch, S. Ulstrup and Ph. Hofmann, Phys. Rev. Lett. 125, 236403 (2020). Copyright (2020) by the American Physical Society. }
  \label{fig:specfunc2}
\end{figure}

%fitting this
A more systematic analysis can be carried out by fitting the location-dependent spectral function across the device using a 2D model of the type given in equation (\ref{equ:p7}). Such a fit contains the bare dispersion of the Dirac fermions in graphene, as well as the self-energy $\Sigma$. For the analysis of the data in Figure~\ref{fig:specfunc1}, assuming an energy-independent broadening mechanism such as electron-defect scattering, encoded by $\Sigma'=0$ and $\Sigma''=const.$,  has been sufficient for obtaining a good fit. The most important parameters resulting from the fit are the energy of the Dirac point (i.e., the local doping of the graphene), the shift in $k_y$ already seen in Figure~\ref{fig:specfunc1}(c), and $\Sigma''$ which describes the linewidth of the bands. These fit parameters are mapped out across the device and displayed in Figure~\ref{fig:specfunc2}. The $k_y$ shift shown in Figure~\ref{fig:specfunc2}(b) can be shown to arise from slight relative rotations of crystalline domains. Actually, such rotations could be tracked more easily in ARPES by simply determining the precise location of the K point. However, this would require data collection as a function of the 2D crystal momentum ($k_x$, $k_y$) around the K point compared to the merely 1D cuts measured in Ref. \cite{Curcio:2020aa}. 
Boundaries between regions of different $k_y$ values appear to partly coincide with the dashed lines in the images which represent rows of defects observed in atomic force microscopy. There is also some correlation between the position of the dashed lines and an increased $\Sigma''$. This will be discussed in further detail below. 

%the result of this: 
Overall, mapping the spectral function across the  device in equilibrium reveals a detailed picture of doping and many-body effects on a scale of 500~nm that can be partly related to the defect structure observed by atomic force microscopy. Note that the local change of doping is also considerable, with the Dirac point energy showing a variation in the order of 50~meV across the device, reminiscent of the situation in the presence of charge puddles \cite{Zhang:2009ab}. In a position-integrated measurement, this variation would be interpreted in terms of an overall broadening, as for example caused by electron-defect scattering. 

\section{Electrostatic gating}
\label{sec:IV}

%why important?
One of the most important properties of a semiconductor is the possibility to locally change the carrier type and density by applying an electric field.  In 2D materials, high carrier densities can be reached in a field effect transistor-like geometry or by other means \cite{Ju:2014aa,Paradisi:2015aa,Rosenzweig:2020aa}, allowing a large tunability of properties. Single layer MoS$_2$, for instance, can be turned from being a semiconductor in equilibrium to a superconductor by gating \cite{Costanzo:2016aa}. While this requires very strong fields and can only be achieved by using ionic liquid gates, much weaker gate fields can be sufficient to completely change the properties of materials with correspondingly small energy scales, for example magic angle twisted bilayer graphene \cite{Cao:2018aa,Cao:2018ad}.  
%CCCC Check the eight bands again

%in situ gating observations with spectroscopic / local techniques
\emph{In situ} gating has been demonstrated in scanning tunnelling microscopy \cite{Zhang:2008ac} but in ARPES this is more challenging because of the photoelectrons' long path in a region of space that is potentially affected by stray fields (in contrast to a microscopic tunnelling junction). 

\begin{figure}
\includegraphics[width=0.48\textwidth]{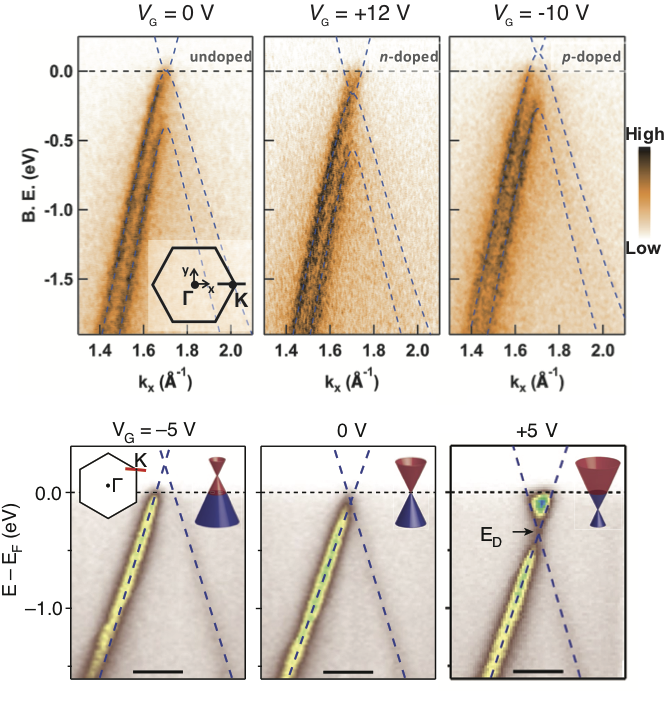}\\
\caption{First demonstrations of gating tracked by nanoARPES in graphene-based devices. Upper part: gate-induced doping of bilayer graphene after Ref. \cite{Joucken:2019aa}. Shown is a cut through the spectral function through the K point of the Brillouin zone, along the thick black line in the inset of the left panel. B. E. denotes the binding energy (defined as negative in contrast to the rest of this review). The dashed lines are the result of a tight-binding calculation. 
Lower part: corresponding doping for SL graphene after Ref.  \cite{Nguyen:2019aa}, showing the gate voltage-dependent spectral function of graphene near the K point of the Brillouin zone. The bottom axis is the crystal momentum. The direction of the scan is given in the inset showing the Brillouin zone of graphene. The scale bars have a length of 0.2~\AA$^{-1}$. The dashed lines are the result of a fit to the dispersion. Upper part adapted with permission from F. Joucken, J. Avila, Z. Ge, E. A. Quezada-Lopez, H. Yi, R. Le Goff, E. Baudin, J. L. Davenport, K. Watanabe, T. Taniguchi, M. C. Asensio and J.  Velasco, Nano Letters 19, 2682 (2019). Copyright 2019 American Chemical Society. Lower part adapted with permission from P. V. Nguyen, N. C. Teutsch, N. P. Wilson, J. Kahn, X. Xia, A. J. Graham, V. Kandyba, A. Giampietri, A. Barinov, G. C. Constantinescu, N. Yeung, N. D. M. Hine, X. Xu, D. H. Cobden and N. R. Wilson, Nature 572, 220 (2019). Copyright 2019, Springer Nature.}
  \label{fig:gating1}
\end{figure}

The first demonstration of gate-induced doping of 2D materials in ARPES was given for bilayer graphene  
and is presented in the upper part of  Figure~\ref{fig:gating1} \cite{Joucken:2019aa}. The data has been taken on a device corresponding to that of Figure~\ref{fig:devices}(b). For a gate voltage of 0~V, bilayer graphene on $h$-BN is essentially undoped. Clear electron and hole doping can be achieved by applying gate voltages of 12~V and -10~V, respectively. Breaking the symmetry in bilayer graphene by applying a vertical electric field has been predicted \cite{McCann:2006ab}
 and experimentally found \cite{Ohta:2006aa} to open a small band gap. Interestingly, no indications of such a band gap opening could be observed for the data in Figure~\ref{fig:gating1} within the experimental uncertainty. This was found to be in contradiction to a simple tight-binding model but consistent with a DFT-GW calculation.  
 
The lower part of Figure~\ref{fig:gating1} shows the results for a gated SL graphene device from Ref. \cite{Nguyen:2019aa}.
Actually, the graphene in this particular device is encapsulated between a $h$-BN dielectric and a protective $h$-BN layer on the surface, similar to what is shown in Fig \ref{fig:devices}(c). Figure~\ref{fig:gating1} demonstrates a wide gating range for the device, from strong $p$ to strong $n$-doping with a relatively small gate voltage. The $n$-doped dispersion shows what appears to be a small gap around the Dirac point but this is related to matrix element effects associated with the asymmetric scan direction and the fact that the spectra were not taken exactly through the K point of the Brillouin zone, such that the actual Dirac point is missed in the scan, leading to an apparent gap. Note that the doping-induced energy shift achieved for graphene in Ref. \cite{Nguyen:2019aa} is significantly larger than that reached for bilayer graphene in Ref. \cite{Joucken:2019aa} for similar gate voltages in similar devices. Ref. \cite{Nguyen:2019aa} also investigates the dependence of the Fermi velocity $v_F$ on the doping and finds no detectable change within the experimental uncertainties. 

%The right hand side of the figure shows the Dirac point energy $E_D$ and induced carrier concentration as a function of gate voltage. \mr{Maybe remove the doping level or comment on this.}

\begin{figure*}
\includegraphics[width=0.95\textwidth]{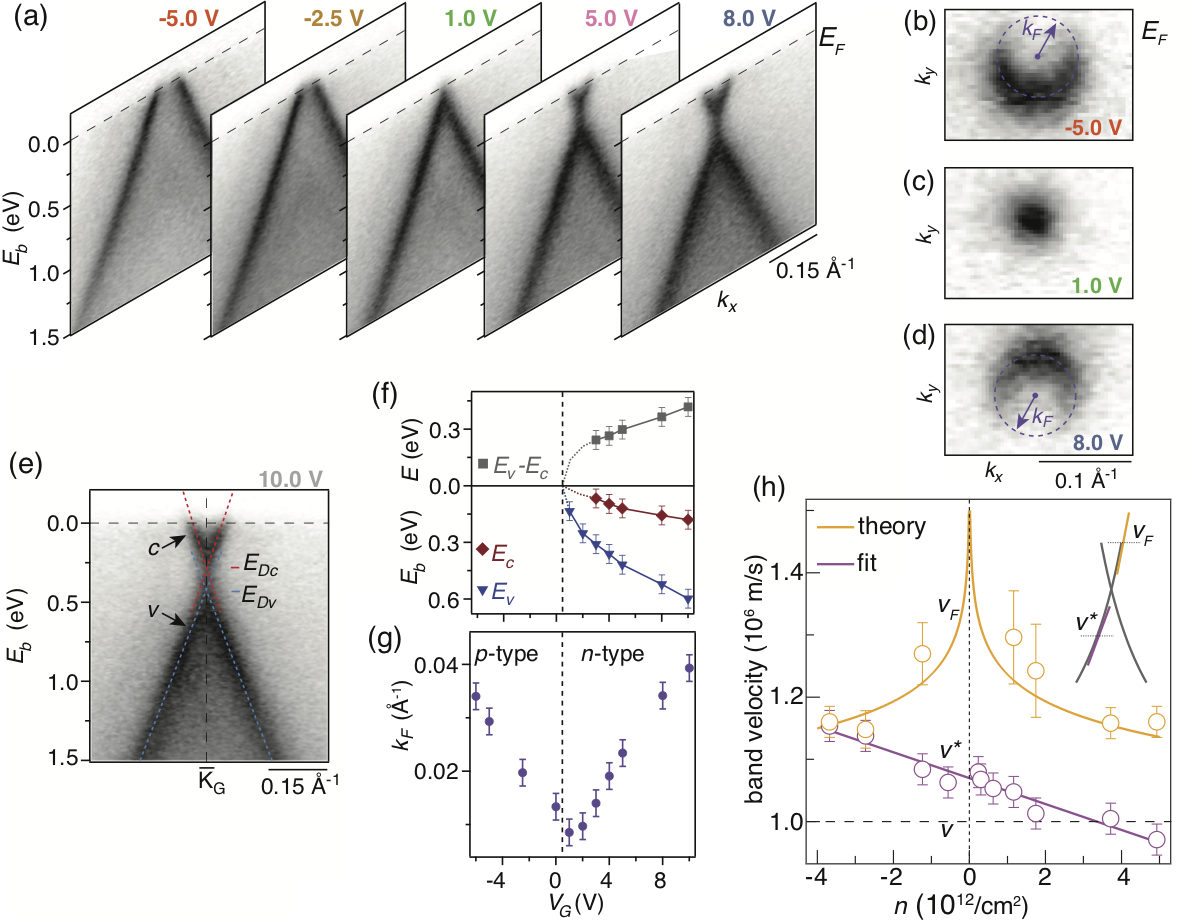}\\
\caption{Results for the spectral function of gated graphene on $h$-BN from Ref. \cite{Muzzio:2020aa}. (a) Dispersion through the K point and  (b)-(d) photoemission at the Fermi energy as a function of applied gate voltage given in the figure. $k_F$ is the radius of the Fermi circle. (e) Dispersion through the K point for the highest $n$-doping. The valence and conduction bands are marked by v and c, respectively. The blue (red) dashed lines mark the linear dispersion of the v (c) states. The observed crossing of the v (c) bands is denoted by $E_{Dv}$ and $E_{Dc}$, respectively. (f) Evolution of $E_V$ and $E_C$ (essentially the same as $E_{Dv}$ and $E_{Dc}$), as well as their difference, as a function of gate voltage. (g) Radius of the Fermi circle $k_F$. (h) Band velocities $v_F$ and $v^{\ast}$ measured at the Fermi level (orange circles) and 300~meV below $E_{Dv}$ (purple circles), respectively. The inset illustrates the definition of $v_F$ and $v^{\ast}$. The purple line is a fit to a linear dependence to guide the eye, and the orange curve is the analytic dependence of $v_F$ on doping for an effective Coulomb coupling constant given by $\alpha=0.5$. The bare velocity  $v$ (without many-body interactions) is indicated by a horizontal dashed line.  Adapted with permission from R. Muzzio, A. J. H. Jones, D. Curcio, D. Biswas, J. A. Miwa, Ph. Hofmann, K. Watanabe, T. Taniguchi, S. Singh, C. Jozwiak, E. Rotenberg, A. Bostwick, R.  Koch, S. Ulstrup and J. Katoch, Phys. Rev. B 101, 201409(R) (2020). Copyright (2020) by the American Physical Society.}
  \label{fig:gating2}
\end{figure*}

%result from graphene from Muzzio
Figure~\ref{fig:gating2} shows results from a similar device of graphene placed on $h$-BN but without an additional $h$-BN cover, reproduced from Ref. \cite{Muzzio:2020aa}. At first glance, the gate-dependent spectral function in Figure~\ref{fig:gating2}(a) is  quite similar to that shown in the lower part of Figure~\ref{fig:gating1} but there are some subtle differences. In particular, the results for high $n$-doping do not appear to show a distinct gap around the Dirac point but rather an elongated vertical intensity connecting an upper and a lower cone. An important experimental difference between Refs. \cite{Nguyen:2019aa} and \cite{Muzzio:2020aa} is that in the latter experiment the gate voltage-dependent spectral functions have been extracted from angle scans around the K point to represent cuts precisely through the Dirac point. Such scans can also be used to display the photoemission intensity at the Fermi level as a function of $\mathbf{k}_{\parallel}$ which, under certain conditions \cite{Kipp:1999aa}, can be viewed as an image of the 2D material's Fermi contour \cite{Gaylord:1989aa,Aebi:1994ab}, giving direct access to the electron or hole density in the device.  This is illustrated in Figure~\ref{fig:gating2}(b)-(d) together with a fit to a circular Fermi contour with radius $k_F$.  For the doping range explored in Ref. \cite{Muzzio:2020aa}, the carrier density varies between $\approx 4 \times 10^{12}$~holes and $\approx 5 \times 10^{12}$~electrons per cm$^2$.

%the issue of incomplete mapping
Note that the spatial resolution in the data in Figs. \ref{fig:gating1} and \ref{fig:gating2} is not sufficiently high to permit a systematic mapping of the spectral function including many-body effects and doping across the device, along the lines of Figure~\ref{fig:specfunc2}. The data used for the mapping in Figure~\ref{fig:specfunc2}, on the other hand, does not consist of full angular scans around the K point at each position on the sample but rather of a single spectrum near K. As a consequence, interpreting the $k_y$ displacement of the Dirac cone in Figure~\ref{fig:specfunc2}(b) in terms of domain rotations needs to be supported by a relatively involved argument based on the relative photoemission intensity between the two observed branches forming the lower half of the Dirac cone \cite{Shirley:1995aa,Mucha-Kruczynski:2008aa,Lizzit:2010aa,Gierz:2011ab}.

%description of distorted dispersion in Muzzio
Returning to the distorted dispersion around the Dirac point for high $n$-doping, Figure \ref{fig:gating2}(e) shows that both the valence and conduction band can be described by a linear dispersion (red and blue lines, respectively) but these do not meet in a single Dirac point. Instead, there appears to be a slight offset between the valence band and conduction band ``Dirac points'' $E_{Dv}$ and $E_{Dc}$. The gate dependence of  the Dirac point energies along with their difference and the Fermi circle radius $k_F$ are  shown in Figure~\ref{fig:gating2}(f) and (g). Note that the energies $E_V$ and $E_C$ shown in this figure are essentially the same as  $E_{Dv}$ and $E_{Dc}$, respectively, but based on a different type of analysis. For a gate voltage $V_G = 0$~V, corresponding to nearly undoped graphene on $h$-BN, $E_C$ and $E_V$ approach zero, such that the ideal Dirac cone dispersion is recovered. 

%this is known from the plasmarons
This observed distortion of the Dirac cone is not unexpected but rather a sign of the electron-plasmon interaction; and the observed separation between valence and conduction band can be used to determine the effective Coulomb coupling constant $\alpha$ in graphene \cite{Bostwick:2010ac,Walter:2011ac}. The knowledge of $\alpha$, in turn, can be used to predict the doping-dependent many-body renormalisation of the Fermi velocity $v_F$ \cite{DasSarma:2013aa}. The calculated result is shown as an orange curve in Figure~\ref{fig:gating2}(h) and it is consistent with the observed values of $v_F$. The observed behaviour is in sharp contrast to the constant (and lower) Fermi velocity in the absence of many-body effects (dashed line), as well as the monotonic decrease of band velocity below the Dirac point (purple line). 

%summary of these results
Overall, these results demonstrate that a very detailed picture of the many-body effects in graphene (or other 2D materials) can be gained with sufficient resolution. So far, ARPES investigations  of doping-dependent many-body effects in graphene required chemical doping via the intercalation or adsorption of chemical species (mainly alkali metal atoms, see e.g. Refs. \cite{Ulstrup:2016ab,Rosenzweig:2020aa} and references therein). This approach has several drawbacks: The ability to achieve $n$ or $p$-doping depends on the chemical properties of the substrate such that a wide range of doping can only be achieved by using a combination of metallic and semiconducting substrates. Moreover, the alkali adsorption process is irreversible. Finally, the presence of possibly disordered adsorbates is likely to increase the observed linewidth due to electron-defect scattering. 

%====================================================

\begin{figure*}
\includegraphics[width=0.85\textwidth]{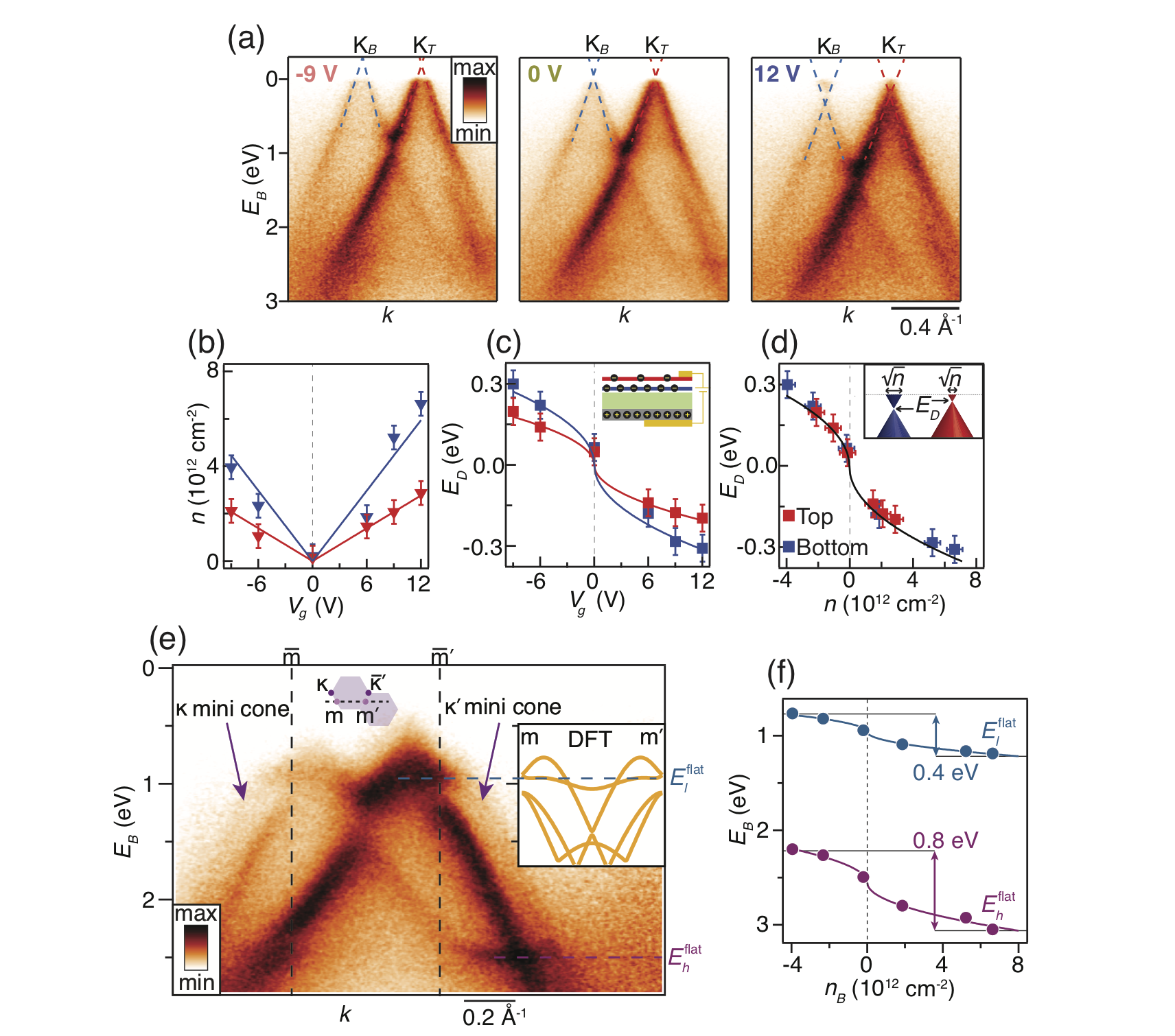}\\
\caption{Tuning the position of flat bands in twisted bilayer graphene after Ref.  \cite{Jones:2020aa}. (a) Gate voltage-dependent dispersion near $E_F$ along the $\mathbf{k}$ direction connecting the two K points K$_T$ and K$_B$ of the top and bottom graphene layer. (b) Gate voltage-dependent carrier density in the bottom (blue) and top (red) layer. (c)  Gate voltage-dependent position of the Dirac point for the two layers. The inset shows a sketch of the device interpreted as a parallel plate capacitor. (d) Dirac point energy as a function of carrier density, representing the expected $\sqrt{n}$ behaviour for both layers. (e) Detailed dispersion along a line between two m points of the mini Brillouin zone corresponding to the moir\'e lattice, showing the hybridisation between the main Dirac cones and the replica bands (so-called mini-cones) generated by the moir\'e lattice. The replica bands stem from Dirac cones centred around the $\kappa$ and $\kappa'$ points of the mini Brillouin zone. The energies of the hybridisation-derived flat bands are indicated by dashed horizontal lines for $E_l^\mathrm{flat}$ and $E_h^\mathrm{flat}$. (f) Binding energy of the flat bands as a function of doping-induced carrier concentration. Adapted with permission from A. J. H. Jones, R. Muzzio, P. Majchrzak, S. Pakdel, D. Curcio, K. Volckaert, D. Biswas, J. Gobbo, S. Singh, J. T. Robinson, K. Watanabe, T. Taniguchi, T. K. Kim,  C. Cacho, N.  Lanata, J. A. Miwa, Ph. Hofmann, J. Katoch, and S. Ulstrup, Advanced Materials 32, 2001656 (2020). Copyright 2020, John Wiley and Sons.}
  \label{fig:gating3}
\end{figure*}

%TBLG results from Jones et al.
The method of \emph{in situ} gating has also been applied to a device fabricated from twisted bilayer graphene by Jones \emph{et al.} \cite{Jones:2020aa}. The general device outline was similar to Figure~\ref{fig:devices}(b) and the twist angle between the graphene sheets in the device was 12.2$^{\circ}$ -- and thus far larger than the magic twist angle of 1.1$^{\circ}$ that leads to the observation of correlated electronic states and superconductivity \cite{Cao:2018ad,Cao:2018aa}. While choosing a large twist angle  has the obvious disadvantage of not reproducing the highly interesting flat bands at $E_F$ leading to these phenomena \cite{Utama:2020aa,Lisi:2020aa}, it is much easier to track the moir\'e-induced interaction of the Dirac cones and the effect of electrostatic gating on the band structure. As we shall see, the Dirac cones and flat bands are well-separated for this large twist angle whereas this is not the case near the magic twist angle (on an energy scale small compared to the typical linewidth broadening in ARPES), making it difficult to disentangle the four spin-degenerate bands near $E_F$ \cite{Utama:2020aa,Lisi:2020aa}.

%the doping dependency 
Figure~\ref{fig:gating3}(a) shows the band structure and gating dependence of twisted bilayer graphene near the Fermi energy. The cut in $\mathbf{k}$-space has been chosen such as to connect the K points of the bottom and top layer K$_B$ and K$_T$. Due  to the short inelastic mean free path of the photoelectrons (see Figure~\ref{fig:electron_imfp}), the photoemission intensity from the top layer is significantly higher than that from the bottom layer, allowing an easy identification of the spectral features. Near the Fermi energy, there is little interaction between the two Dirac cones. The spatial resolution in the data is about 700~nm and it was found necessary to carefully select the light spot position on the device in order to identify a region with a single twist angle and few defects. At first glance, the gate-induced doping follows the pattern already seen in the previous graphene examples: Using a relatively small gate voltage, it is possible to tune the device from clearly $p$-doped to clearly $n$-doped. However, it is already evident from the raw data that the bottom layer is always doped more strongly than the top layer. A more detailed analysis of this is shown in Figure~\ref{fig:gating3}(b)-(c) which shows the layer-resolved and gate voltage-dependent carrier density and Dirac point position, confirming the weaker doping of the top layer. The phenomenon can be understood in a simple parallel plate capacitor model by the fact that the charge accumulated of the bottom layer strongly -- but not completely \cite{Luryi:1988aa} --  screens the electric field affecting the top layer (see inset of Figure~\ref{fig:gating3}(c)).  Figure~\ref{fig:gating3}(d) finally shows the relation between Dirac point energy $E_D$ and carrier concentration. This is identical for the two layers, as expected due to the equal Dirac cone dispersion that, close to $E_F$, is unaffected by the bilayer. 

%formation of flat bands and different dependence on doping for different flat bands.
The formation of flat bands from the  interaction of the main Dirac cones and their moir\'e-induced replicas can also be observed at this large twist angle. Flat bands are seen at binding energies of 0.95 and 2.48~eV for zero gate voltage, as already evident in the overview of Figure~\ref{fig:gating3}(a). The flat bands are observed more clearly in the data of Figure~\ref{fig:gating3}(e) which shows the dispersion connecting two m points on the flat sides of the mini Brillouin zone formed by the moir\'e. The electronic structure in the mini Brillouin zone is governed by the original Dirac cones of the two layers along with their moir\'e-induced replicas at the $\kappa$ points of the mini Brillouin zone. The formation of flat bands can be understood particularly well at a binding energy of 0.95~eV because it can be related to the hybridisation-induced gap opening at the border of the mini Brillouin zone. When gating the device, band shifts similar to those near the Fermi energy are also observed for the flat bands and the detailed change of the flat band energies is given in Figure~\ref{fig:gating3}(f). Interestingly, the total shift of the flat band at $E_B$=0.95~eV is only half as big as for that at 2.48~eV, and also smaller than the shift of the Dirac cone of the bottom layer for the same range of doping. In fact, there are several effects contributing to the gating-dependent energy of the different flat bands: the fact that the gating affects the two graphene layers differently and, at higher binding energies, the non-linearity of the dispersion.

\begin{figure*}
\includegraphics[width=0.85\textwidth]{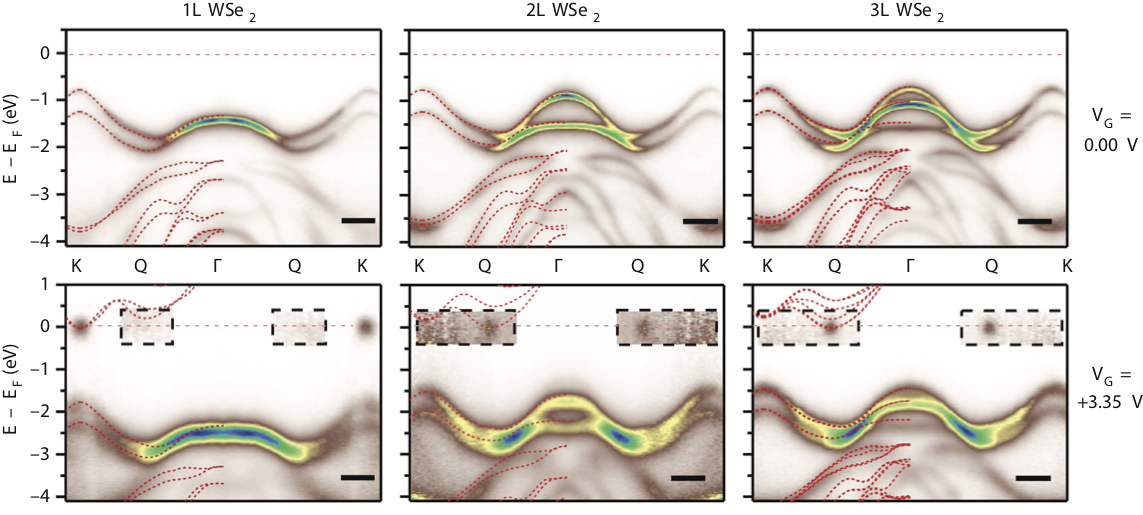}\\
\caption{Equilibrium (upper row) and gated (lower row) band dispersion of few layer WSe$_2$ reported in Ref.  \cite{Nguyen:2019aa}. The colour scale ranges from white (low) to blue (high) photoemission intensity. The intensity in the dashed boxes
is multiplied by a factor of 20. The length of the scale bar is 0.3~\AA$^{-1}$. The Brillouin zone is hexagonal and the Q point is on a line connecting the $\Gamma$ and K points. The calculated band structures in the GW approximation are overlayed as dashed lines. Adapted with permission from P. V. Nguyen, N. C. Teutsch, N. P. Wilson, J. Kahn, X. Xia, A. J. Graham, V. Kandyba, A. Giampietri, A. Barinov, G. C. Constantinescu, N. Yeung, N. D. M. Hine, X. Xu, D. H. Cobden and N. R. Wilson, Nature 572, 220 (2019). Copyright 2019, Springer Nature.}
  \label{fig:gating4}
\end{figure*}

%summary TBLG
These results on twisted bilayer graphene illustrate the detailed tunability of flat bands. A particular advantage of the approach lies in the ability to disentangle the different doping of the two layers, as well as the detailed interaction leading to the flat bands. In this context, the large twist angle in this experiment is an advantage because analysing similar details would be very difficult in the case of a sample with the magic twist angle where flat bands have also been observed by ARPES very recently \cite{Utama:2020aa,Lisi:2020aa}, but where it is  very challenging to reach the resolution, temperature and sample quality necessary to study the fine structure of the four spin-degenerate flat bands. 

%====================================================

%WSe2 on $h$-BN from Nguyen. Description of the materials and band structure
As a final example of gating-induced band shifts in devices, Figure~\ref{fig:gating4} illustrates the equilibrium and gated band structures of few layer WSe$_2$ on $h$-BN from Ref. \cite{Nguyen:2019aa}. Equilibrium band structure measurements for 1 to 3 layers of WSe$_2$ in the 2H structure on $h$-BN are given in the upper row of the figure along with the result of band structure calculations in the GW approximation. The layer thickness strongly affects the valence band maximum around $\Gamma$. For a SL, only one (spin degenerate) band is present at $\Gamma$ while the interaction of the out-of-plane orbitals forming this band leads to a strong bonding-antibonding splitting for a WSe$_2$ bilayer \cite{Cheiwchanchamnangij:2012aa,Cappelluti:2013aa}. This band structure change is what causes the SL to have a direct band gap. For three layers of WSe$_2$, a third band is clearly visible. In contrast to the $\Gamma$ point, the orbital character of the bands near K is in-plane and therefore the number of observable bands does not increase with the layer thickness. The splitting of the bands near K is due to the spin-orbit interaction. For a SL, the bands are spin polarised but for the bulk they are not because the spin ordering is inverted in neighbouring layers. There remains, however, a local spin polarisation in every layer \cite{Riley:2014aa}. 

%populating the CBM
ARPES can only probe occupied states and since WSe$_2$ is a semiconductor, there is no signal from the empty conduction band. However, when applying a sufficiently high gate voltage, the conduction band minimum of few layer WSe$_2$ can be populated, as shown in the lower panels of Figure~\ref{fig:gating4}. Note that the intensity of the conduction band features can be quite low, so that the scale for the photoemission intensity in the dashed boxes has been changed to emphasise weak features. For a SL of WSe$_2$, the conduction band minimum is clearly observed at K, confirming the direct band gap expected for this type of material \cite{Mak:2010aa,Splendiani:2010aa} and showing quantitative agreement with the calculated band gap. Very similar results have been obtained for SLs of WS$_2$ on $h$-BN by (irreversible) doping using alkali adsorption \cite{Katoch:2018aa}.  For two and three layers, the conduction band minimum switches over to the Q point of the Brillouin zone, where it is also found for the bulk crystal. Note that the gating does not merely lead to a rigid shift of the band structure but also to a broadening and deformation of the bands, something that has not been observed so clearly for the graphene-based devices. 

%why is this important? Normally hard to get access to CB
These results illustrate the ability of the technique to determine the character and size of the band gap in few layer TMDCs (assuming here that these band extrema are found along the K-$\Gamma$ direction, of course). This is an important result because obtaining such information can otherwise be difficult. Optical measurements suggest the presence of a direct band gap  \cite{Mak:2010aa,Splendiani:2010aa} but the absorption is governed by excitonic effects \cite{Qiu:2013aa}, masking the onset of the electronic band gap. Also, ARPES can potentially determine the effective masses of both holes and electrons because it is a $\mathbf{k}$-resolved techniques. In the present case, this is difficult for the electron effective masses because the conduction band population is too low to discern the band dispersion. 

%Alternative ways to determined this and what's the difference
The conduction band position in few layer (and bulk) TMDCs can also be determined by first populating the unoccupied states using an optical laser pulse and then performing (time-resolved) ARPES on this excited state \cite{Antonija-Grubisic-Cabo:2015aa,Ulstrup:2016aa,Bertoni:2016aa,Ulstrup:2017aa,Beyer:2019aa,Kutnyakhov:2020aa}.  Time-resolved ARPES has shown a strong band gap shrinkage for high carrier densities pumped into the conduction band \cite{Antonija-Grubisic-Cabo:2015aa}, as also found in optical experiments \cite{Chernikov:2015aa} and in the gated WSe$_2$ device in Ref. \cite{Nguyen:2019aa} (see Figure 4 in that paper). The position of the conduction band minimum in equilibrium can also be determined 
 by spin-polarised inverse photoemission spectroscopy, which has the additional advantage of being able to determine the spin texture of the conduction band minimum \cite{Eickholt:2018aa}. 
Alternatively, both valence band and conduction band edges can by determined by scanning tunnelling spectroscopy, albeit with the difficulty that this technique is very insensitive towards states far away from the centre of the Brillouin zone \cite{Zhang:2015aa,Bruix:2016aa}.

%for tmdcs the gating is particularly interesting for quantum capacitance, control of spin-splitting and negative electronic compressibility \cite{Riley:2015aa}

%electronic compressibility in bilayer graphene here: \cite{Young:2012ab}

%what can we learn from doping the semiconducting 2D materials: summary
The demonstration of nanoARPES experiments on gated 2D semiconductors opens a number of interesting research directions. It will, for instance, be possible to track the evaluation of the band width and Fermi contour as a function of doping, as well as the shift of higher binding energy features such as the valence band maximum. This can be used to obtain information about properties such as screening-induced band gap narrowing or (negative) electronic compressibility \cite{King:2010aa,Riley:2015aa,Chernikov:2015ab}. Studying the detailed dispersion of the conduction band as a function of electron density should reveal the evolution of many-body effects as the electron density is increased, shedding light on effects such as the superconductivity in few layer TMDCs at very high electron densities \cite{Costanzo:2016aa}. Another interesting aspect of the gating experiments for bilayer graphene is the possibility to separately study the carrier density in the two layers, giving a detailed insight into effects such as the electronic compressibility and quantum capacity in the device  \cite{Young:2012ab}. In SL TMDCs, it has recently been demonstrated that the valley polarisation can be switched by gating \cite{Li:2020ab} and this could also be tracked by nanoARPES.

\section{Current-carrying devices}
\label{sec:V}

%why currents are important
While gating can tune the carrier concentration in a QM, a device's functionality is closely related to passing a (controllable) current through the material or, even more interesting, to control the QMs properties by such a current. NanoARPES gives access to the electronic structure of the device while passing a current though it, as recently demonstrated in Ref. \cite{Curcio:2020aa} and discussed in detail below. Strictly speaking, of course, ARPES experiments  always result in a current density in the material under investigation to compensate for the photoemitted electrons. For a sufficiently resistive sample grounded from one side only, this effect causes an energy offset of $e I R$ in the spectra, where $I$ is the photoemission current and $R$ the resistance between the position of the light spot and the grounded part of the sample holder. This effect has been used to study the transport properties of graphene undergoing a metal-insulator transition \cite{Bostwick:2009aa}.

\begin{figure*}
\includegraphics[width=0.8\textwidth]{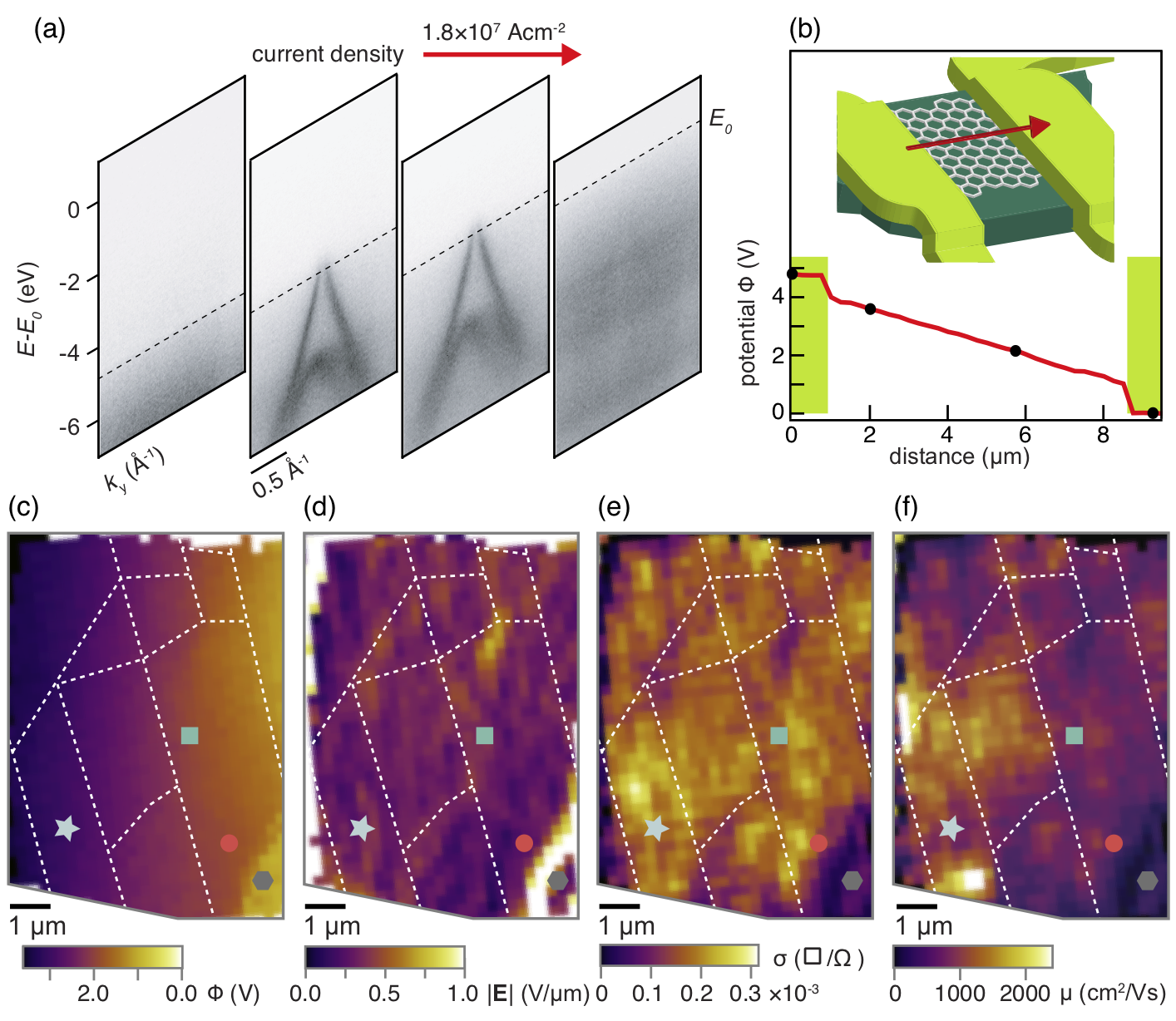}\\
\caption{Analysis of current-carrying device in Fig. \ref{fig:devices}(d) from Ref. \cite{Curcio:2020aa}. (a) Series of spectra along a line connecting the gold electrodes (see inset in panel (b)) for a device current of 0.5~mA. The Fermi level on the right hand electrode is used as a zero for the energy scale.  (b) Sketch of the device together with the potential $\phi$ along the red line, determined from the local $E_F$. The black markers give the location of the spectra shown in (a). (c) Map of the potential $\phi$ across the device. The energy zero is the Fermi energy on the right hand side. The markers correspond to the locations for the spectra in Figs. \ref{fig:specfunc1} and \ref{fig:specfunc2}.  (d) Magnitude of the electric field  across the device.  (e) Local 2D conductivity $\sigma$ calculated from $\phi$. (f) Local mobility $\mu$, calculated by combining $\sigma$ with the hole density derived from the Dirac point energy $E_D$ in Figure~\ref{fig:specfunc1}(a). Reprinted with permission from D. Curcio, A. J. H. Jones, R. Muzzio, K. Volckaert, D. Biswas, C. E. Sanders, P. Dudin, C. Cacho, S. Singh, K. Watanabe, T. Taniguchi, J. A. Miwa, J. Katoch, S. Ulstrup and Ph. Hofmann, Phys. Rev. Lett. 125, 236403 (2020). Copyright (2020) by the American Physical Society. }
  \label{fig:current1}
\end{figure*}

%why nanoARPES is needed for this. Problem with the size of the light spot
Here we are mostly concerned with nanoARPES on devices that are subject to an externally applied current, for example between the source and drain of the devices of Figure~\ref{fig:devices}, leading to a voltage drop across the sample. In conventional ARPES with a UV light spot size in the order of 100~$\mu$m, the voltage drop \emph{across the light spot} can be sufficiently large to have a detrimental effect on the energy resolution, making \emph{in operando} ARPES on current-carrying devices impossible. Indeed, this affect can even be significant when working with sub $\mu$m light spots and corresponding broadening due to the photocurrent has been linked to the spectral broadening in case of the few-layer WSe$_2$ experiments reported in Figure~\ref{fig:gating4}. The only class of sample where this problem does not occur are superconductors below $T_C$ and below the critical current density. ARPES on superconductors in the presence of a current density have been reported in Refs.  \cite{Kaminski:2016aa,Naamneh:2016aa}. Note, however, that even under these conditions, power dissipation and hence resistance can still arise due the current-induced motion of magnetic vortices for type II superconductors.

%results from Curcio: mapping spectral function and obtaining potential
The results  on a current-carrying graphene device by  Curcio \emph{et al.} are shown in Figure~\ref{fig:current1}. They are based on the device shown in Figure~\ref{fig:devices}(c) and the characteriasation of the device's equilibrium properties in Figure~\ref{fig:specfunc2} \cite{Curcio:2020aa}. Setting the device current to 0.5~mA (orders of magnitude higher than a typical photoemission current) leads to a current density of 1.8$\times$10$^7$~Acm$^{-2}$ through the graphene sample. The corresponding voltage drop can be seen as a shift of the photoemission spectra along a line connecting the two electrodes of the device in Figure~\ref{fig:current1}(a), where the first and the last spectrum are taken on the gold electrodes and the two spectra in the middle on the graphene flake. The potential along a line connecting the contacts, as obtained from the position-dependent Fermi energy of the local spectra, is shown in Figure~\ref{fig:current1}(b). The potential is essentially constant across the highly conductive gold contacts; it shows large steps at the gold-graphene interface due to contact potentials and it is linear across the graphene flake. For a detailed analysis across the graphene area, the spectra have been analysed by a fit to a 2D model spectral function in the same way as discussed in connection with Figure~\ref{fig:specfunc2}, but permitting a shift of the local  $E_F$. This results  in the potential landscape across the entire graphene flake shown in Figure~\ref{fig:current1}(c). One observes the expected smooth change except for region near the lower right corner (near the hexagonal marker) where the graphene flake is poorly connected to the electrode. 

%getting local conductivity 
While details are hard to make out in the potential due to the large variation across the device, they are more clearly visible in the local electric field $|\mathbf{E}|=|\nabla \phi |$ shown in Figure~\ref{fig:current1}(d). From this map of the local electric field magnitude, it is possible to calculate the local conductivity of the graphene sheet $\sigma$, following approaches developed for scanning tunnelling potentiometry and related techniques \cite{Muralt:1986aa,Zhang:2016ag,Tetienne:2017aa,Voigtlander:2018aa,Ella:2019aa}. Such a  map is given in Figure~\ref{fig:current1}(e). There appears to be a strong correlation between $\sigma$ and $1/|\mathbf{E}|$, as one might expect for a homogeneous situation where $\mathbf{E}=\mathbf{j}/\sigma$. More interestingly, there is a close correspondence between a low $\sigma$ and a high $\Sigma''$ (see Figure~\ref{fig:specfunc2}(c)) near the defect lines on the device represented by the dashed lines. This has the obvious interpretation that the defects reduce the carrier lifetime (increase $\Sigma''$)  and that this is then reflected in a low conductance. On the other hand, a certain amount of care is necessary because a broadening of the spectral function can occur when taking spectra at the boundary between two rotational domains because, in this situation, the photoemission signal contains contributions of the two Dirac cones in the different domains which, according to Figure  \ref{fig:specfunc2}(b), are slightly shifted against each other. The resulting broadening in the Dirac cone measured on the domain boundary could be interpreted as an increased $\Sigma''$ when it is really due to the superposition of two shifted Dirac cones. 

%... and mobility
Finally, it is possible to combine the results from the local doping, measured via the energy of the Dirac point $E_D$ in  Figure~\ref{fig:specfunc2}(a) with the local conductivity determined in Figure~\ref{fig:current1}(e) to calculate the local mobility in the device. The result is displayed in Figure~\ref{fig:current1}(f). There are some clear qualitative differences between the conductivity and the mobility maps. For instance, the region that contains the square and circle markers represents a relatively uniform high conductivity but a low mobility because it is also quite strongly and uniformly doped, as can be seen in Figure~\ref{fig:specfunc2}(a). 

%summary of this 

Overall, the small light spot in nanoARPES permits meaningful measurements of the spectral function in the presence of a transport current, simply by avoiding energy broadening from the potential change over the large diameter of the light spot in conventional ARPES. This is a rather trivial point but it has the power to integrate the normally disjunct areas of transport measurements and spectroscopy, at least to some degree. As demonstrated above, the device's spectral function can be mapped while simultaneously obtaining the type of information that is otherwise extracted from a transport experiment, such as the sample's conductivity and carrier mobility. This is particularly attractive since the transport measurements are both position-resolved and contact-free. The very useful application of strong magnetic fields in transport experiments will probably remain impossible in ARPES, precluding the study of phenomena such as the Hall effect and the sample's magnetoresistance. On the other hand, information such as the carrier density and type, normally  obtained by Hall measurements, can now be extracted from the local Fermi contour obtained from the spectral function. 

Note that the results of Ref.  \cite{Curcio:2020aa} are a proof of principle, demonstrating the power of the technique and illustrating effects like the correlation between defects on the sample, a low conductivity and a high imaginary part of the self energy. Even at the high current density used for obtaining the results in Figure~\ref{fig:current1} ($\approx2 \times 10^7$~Acm$^{-2}$), one does neither expect a fundamental change in the spectral function of graphene, nor an asymmetry in the population of the electronic states around the Fermi level. The latter effect can easily be estimated to be very small \cite{Kaminski:2016aa} and it is only observed when exciting a two-dimensional electronic system transiently with much stronger electric fields  \cite{Gudde:2007aa,Reimann:2018aa}. 

The most interesting consequence of a current density in a QM is an accompanying phase transition. As discussed in the introduction, there are many examples of this in 3D QMs such as insulator-to-metal transitions in Mott insulators, field-driven currents in CDW systems, the destruction of superconductivity and the time-reversal symmetry breaking of the current as such that might lead to changes in band topology. NanoARPES studies of such transitions could significantly contribute to their understanding.

\section{Practical challenges}
\label{sec:VI}

%\mr{make sure that this is all cited and mentioned} Achieving high spatial resolution by focusing the light spot to a sub-$\mu$m spot, no matter by which approach, can lead to a number of challenges such as beam damage and space charge broadening, as discussed in detail in Refs.  \cite{Hellmann:2012ac,Rotenberg:2014aa}. We shall return to these issues in the end of the review.

%don't know very much yet and other problems might arise
\emph{In operando} nanoARPES investigations on QM-based devices are very much in their infancy and a lot remains to be learned about the opportunities and challenges. A few practical limitations are already apparent and we will briefly discuss them here.

%general question if the nanoARPES and the transport measurement is the same
An important question is to what degree the operating device investigated by nanoARPES can be probed under the same conditions as in a transport experiments.  In a conventional transport setup, it is possible to reach much lower temperatures and to operate in strong magnetic fields to measure (quantum) Hall effects, magnetoresistance, quantum oscillations and similar phenomena. In ARPES, temperatures below 1~K can be reached \cite{Borisenko:2012aa} but the mK regime of transport setups is not accessible. The application of magnetic fields is not possible. To some degree, this can be compensated by other measurements. As we have seen, the carrier concentration in a sample and the type of carriers can be inferred from the measured Fermi contours. The carrier effective mass can be inferred from the measured dispersion and it can be even tracked across the Fermi surface. 

%damage by the light spot?
An important difference between ARPES and transport measurements is the presence of the small and very intense UV light spot and the resulting photocurrent. Exposing the sample to the UV light can severely alter its properties and focusing the photon beam to a nano-scale spot amplifies this problem. In order to estimate the photon density in a typical nanoARPES setup (the MAESTRO beamline at the ALS), we can follow the considerations in Ref. \cite{Rotenberg:2014aa}. Achieving a 10~$\mu$m size spot with a photon energy of 80~eV and a resolving power of 20,000 results in a photon flux  $\approx 3\times 10^{11}$~s$^{-1}$ in this spot. Using a zone plate for further focusing would reduce the flux by more than a factor of 10. This flux is at least two orders of magnitude smaller compared to conventional high resolution ARPES with a spot size of more than 50~$\mu$m but it is focused to an area that is 10,000 times smaller. Using a focusing capillary as a final optical element instead would  reduce the photon flux from the beamline only by a factor of two or so but the spot area would still be more than 1,000 times smaller than in conventional ARPES. 

\begin{figure}
\includegraphics[width=0.35\textwidth]{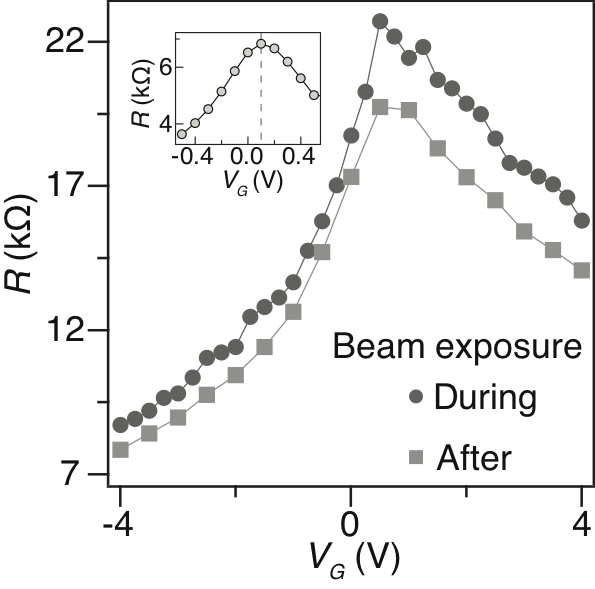}\\
\caption{Consequence of beam-induced damage in a device: Resistance of the conductive channel as a function of gate voltage for the device used in Ref.  \cite{Muzzio:2020aa} during and after the nanoARPES experiment. The inset shows the corresponding curve before exposing the device to the high-intensity UV light. Adapted with permission from R. Muzzio, A. J. H. Jones, D. Curcio, D. Biswas, J. A. Miwa, Ph. Hofmann, K. Watanabe, T. Taniguchi, S. Singh, C. Jozwiak, E. Rotenberg, A. Bostwick, R.  Koch, S. Ulstrup and J. Katoch, Phys. Rev. B 101, 201409(R) (2020). Copyright (2020) by the American Physical Society.}
  \label{fig:damage}
\end{figure}

An effect of the intense light spot from a focusing capillary is demonstrated in Figure~\ref{fig:damage} which shows the gate voltage-dependent resistance of the device used in Ref. \cite{Muzzio:2020aa} during and after exposure to the UV beam. Compared to the  resistance measured before the nanoARPES experiment (shown as an inset), the resistance is much higher and the resistance curves are rather asymmetric. This suggests that the light spot induces some permanent damage in the device. On the other hand, the ARPES spectra collected from the graphene flake do not significantly degrade over the measurement time of many hours \cite{Muzzio:2020aa}. There could be several explanations for the increased resistance. The intense photon beam could result in photodoping \cite{Ju:2014aa} or even produce defects in the SiO$_2$ or $h$-BN layers that affect the screening properties of these materials. The beam could also decompose (and chemically activate) residual material from the graphene and $h$-BN transfer processes that remained on the graphene surface or trapped between the layers of the devices. 

Beam-induced damage in materials investigated by photoemission and similar techniques is, of course, not new, and  especially prevalent in organic materials  rarely used in the devices discussed here \cite{Nakayama:2020aa}.
On the other hand, it needs to be stressed that the fluence increase in future nanoARPES experiments could be dramatic. The light spot area in nanoARPES is more than four orders of magnitude smaller than in conventional ARPES. Currently, this does not have such dramatic consequences because of the very inefficient focusing using zone plates. However, once this sees the expected improvements and if a similar photon flux as in conventional ARPES is desired, the high fluence is going to be an issue.  

%photo-induced conductance by the light spot?
A non-destructive effect of the light spot is the photoionisation of the materials in the device and the resulting currents. This has been discussed in detail by Nguyen \emph{et al.}, in particular for a device consisting of a non-metallic WSe$_2$ layer on top of a $h$-BN dielectric \cite{Nguyen:2019aa} for which three currents  are considered: The current of photoemitted electrons $I_{PE}$ and two currents that result in order to compensate for the total photoemitted charge; the first, $I_C$, flows between the lateral contacts and the position of the light spot via the 2D material and the second, $I_G$, is a leakage current between the gate and the position of the light spot, passing through the dielectric. Obviously, $I_{PE}=I_C + I_G$ but the relative importance of $I_C$ and $I_G$ varies. $h$-BN is a highly insulating material but the intense UV light spot can lead to an increased quantity of mobile carriers  and an accompanying leakage current. For a small gate voltage, the chemical potential of the semiconducting SL WSe$_2$ is found in the band gap, implying a very low conductivity. When performing nanoARPES far away from the source or drain, $I_C$ is therefore negligible. In this situation $I_{PE} \approx I_G$. If, on the other hand, the gating is sufficiently strong to populate the conduction band of WSe$_2$ or in the case of a conductive 2D material such as graphene, most of the photoemission current can be supplied from the lateral contacts via the 2D material and $I_{PE} \approx I_C$. In this case,  Nguyen \emph{et al.} have pointed out that the lateral current and accompanying voltage drop \emph{within the probing area of the nanoARPES light spot} can be responsible for broadening of spectral features in the 2D material, as observed in their WSe$_2$ samples (see Figure \ref{fig:gating4}). For similar experiments on graphene samples, such broadening effects have not been observed \cite{Jones:2020aa,Joucken:2019aa,Muzzio:2020aa}. The expected voltage drop, or at least its order of magnitude, and thus the energy broadening over the area of the light spot can be estimated from $I_{PE}\approx$1~nA, the size of the light spot $\approx$1~$\mu$m, the size of the device $\approx$10~$\mu$m and the resistance $R$ of the 2D material channel. For graphene,  $R\approx$10~k$\Omega$ or smaller. Assuming that therefore $I_C \approx I_{PE}$, the spectral broadening remains below 10~meV,  and thus negligible. Indeed, similar broadening effects could potentially manifest themselves also for graphene on insulating substrates, such as graphene on SiC, especially when probing these systems with large light spots in conventional ARPES \cite{Bostwick:2007aa,Bostwick:2010ac} but graphene is too conductive for this to play a role.  For much more resistive 2D materials, the discussed broadening effect could play a role but, on the other hand, $I_G$ will dominate over $I_C$ for a sufficiently resistive 2D material. For gated SL TMDCs such as MoS$_2$, the resistance of the conductivity channel can become as low as $\approx$100~k$\Omega$. Overall, the interplay between the tuneable conductivity of the 2D material, the currents $I_G$ and $I_C$ and the resulting broadening effects in gated materials deserves further scrutiny.

%How big can the broadening effects be:
%Photocurrent is 1e-9 A
%resistance of path light spot to contact in graphene is about 1e6 ohm
%So voltage drop along this path is  1e-3 V
%length of the path is 10e-6m
%light spot diameter is 1/10 of this, so voltage drop is also 1/10 of this. So 1e-4V. 
%
%What about SL TMDCs? For gated MoS2 one finds for example a 10-7 A for a source drain voltage of 10-2 V. \cite{Radisavljevic:2011aa}
%So in this case R is 100 kOhm

%space charge broadening
An important potential broadening mechanism in nanoARPES arises from the  Coulomb-repulsion within the dense cloud of photoemitted electrons. This so-called space charge effect can alter the direction and kinetic energy of the photoelectrons and thereby broaden the energy and $\mathbf{k}$-distribution of the photoelectrons -- the two quantities of interest in high resolution ARPES. The space charge issue becomes especially acute for a high density of photoemitted electrons above the sample surface, resulting from an excitation  with a small focus, as in nanoARPES or for a short pulse duration of the UV light, or both. Light sources useful for nanoARPES are likely to be high harmonic lasers or synchrotrons and thus always pulsed. The space charge problem in ARPES has been thoroughly investigated (see e.g. Refs. \cite{Zhou:2005ab,Passlack:2006aa,Hellmann:2009aa,Hellmann:2012ac,Rotenberg:2014aa}) and its influence has been evaluated in particular for nanoARPES  \cite{Hellmann:2012ac}. 

A specific simulation of space charge effects in nanoARPES for the ALS has been performed in Ref. \cite{Rotenberg:2014aa}.  With the currently reachable focus and photon flux in this setup, space charge-induced energy and angular broadening are a minor concern, with energy broadening well below 10~meV. However,  as in the discussion of beam damage above, one needs to keep in mind that a much higher photon flux could be achievable on fourth generation synchrotron radiation sources due to the increased coherent fraction of the undulator radiation or  when using non-diffractive focusing optics. 

%deflection and fields, magnetic fields
Another concern for nanoARPES from operating devices is the deflection of photoelectrons by current-induced magnetic fields and stray electric fields. The influence of stray magnetic fields has been investigated by Naamneh \emph{et al.}  on superconducting samples for rather high absolute currents in the order of 10~mA \cite{Naamneh:2016aa}. While a deflection from the magnetic field does, of course, occur, it is very small, even for photoelectrons with only a few eV kinetic energy. In Ref. \cite{Kaminski:2016aa}, the authors have designed the sample mounting such that the current through the material of interest is compensated by a current in the opposite direction on the back of the sample holder, minimising the effecting magnetic field. 

%electric fields
The role of electric fields is difficult to evaluate, especially for the rather complicated devices shown in Figure~\ref{fig:devices}. Holding source and drain at a different potential results in an electric field parallel to the surface. However, fields perpendicular to the surface are also possible. Consider the simple situation of a device as in Figure~\ref{fig:devices}(b) or (c) with source and drain contacts both held on ground potential but a finite gate voltage with respect to ground is applied. Such a setup is usually modelled as a parallel plate capacitor but when a 2D material is used as a top electrode, this is not sufficient to screen the field as efficiently as an ideal metallic electrode, leading to a remaining vertical field above the device\cite{Luryi:1988aa}. This happens even for highly conductive 2D materials such as graphene. Indeed, the effect is immediately evident from Figure~\ref{fig:gating3} where the presence of the bottom graphene layer does not prevent field-doping of the top layer. 

Without a voltage applied between source and drain, the main stray electric field will therefore be roughly perpendicular to the surface, at least far away from the contacts. It will thus affect the kinetic energy of photoelectrons in normal emission, resulting in a shift of the photoemission spectrum but not in an angular deflection. For off-normal emission, however, the direction of the photoemitted electrons will also be affected and this is relevant for, e.g., graphene where the interesting part of the electronic structure is found at K, such that examining it necessitates a large $\mathbf{k}_{\parallel}$ and an emission direction far away from the surface normal, at least for low photon energies. In order to ensure that the desired features in the spectral function are captured, it is thus necessary to map an entire range of $\mathbf{k}_{\parallel}$ as a function of gate voltage, greatly increasing the required data collection time (see also next section). 

%more complicated stray fields, not only because incomplete gate screening
Stray electric fields affecting the outgoing photoelectrons can also arise due to reasons other than incomplete screening of the gate field. Alone the use of different materials in the device results in work function differences, contact potentials and accompanying stray fields. Irregularities in the 2D material itself can give rise to stray fields. Consider for example the region near the bottom right corner  of the device (close to the hexagonal marker) in Figure~\ref{fig:current1}(d). In this region, the graphene flake appears to be connected to the gold electrode but not to the rest of the graphene, as indicated by the fact that the potential on this part of the flake is constant (unaffected by the current) and equal to that of the electrode. A strong electric field arises thus between the flake connected to the gold and the rest of the graphene, as seen in Figure~\ref{fig:current1}(d). 

%the issue of Joule heating
Finally, an important issue when studying current-induced phase transition in QMs is that similar transitions can often be caused by raising the temperature (see Figure~\ref{fig:intro}) and one thus needs to worry about inducing a given transition by Joule heating instead of directly by the electric field or current density. While this is not a problem that is particular to nanoARPES experiments, we mention it anyway briefly because device design can be influenced by the need to measure the energy dissipation in the sample. The first experiments combining ARPES with a transport current were performed on superconductors in order to study the current-induced breakdown of superconductivity \cite{Kaminski:2016aa,Naamneh:2016aa}. At first glance, the current-induced voltage drop within the area of the light spot can expected to vanish for a superconductor but phenomena such as vortex de-pinning can still give rise to energy dissipation. Therefore, great care was taken in Refs. \cite{Kaminski:2016aa,Naamneh:2016aa} to ensure that the current does not simply heat the sample above $T_C$ due to energy dissipation. In order to know the dissipated power, it is useful to measure the voltage drop across the device in a four point geometry, such as realised in Ref. \cite{Naamneh:2016aa}. In nanoARPES experiments, of the type shown in Figure~\ref{fig:current1}, this is not required because the potential mapping across the device can be used to determine the voltage drop and power dissipation (even locally, if needed). In principle, it should also be possible to use the width of the Fermi-Dirac distribution for a local temperature measurement but this approach is to be used with great care \cite{Kroger:2001aa,Ulstrup:2014aa}.The question of Joule heating is not only important for superconductors but for many current-induced phase transitions, for example in the case of the insulator-to-metal transition in Ca$_2$RuO$_4$ \cite{Nakamura:2013aa,Zhang:2019ac,Mattoni:2020aa}

\section{Future directions}
\label{sec:VII}

%some developments are challenging but obvious
Opportunities for many developments in the young research field of \emph{in operando} nanoARPES are rather obvious, especially when bearing in mind that most of the results reported in this review have been obtained from (bilayer) graphene or semiconducting TMDCs. Clearly, an extension of the approach to other QMs is needed, including 3D materials as in the pioneering studies on current-carrying cuprates \cite{Kaminski:2016aa,Naamneh:2016aa}. Also, the relevant effects in QMs are often happening at small energy and length scales and there is much room for improvement in energy and spatial resolution.

 For materials showing magnetism, topological band structures or strong spin-orbit interaction combined with inversion symmetry breaking, adding spin resolution to the detector would be an interesting opportunity but this will be  challenging because it combines nanoARPES -- which already suffers from low count rates due to inefficient focusing -- with inefficient electron detection. 
 
 In the following, we  briefly elaborate on the subject of sample temperature and spectral resolution. We then discuss some other, maybe less obvious, future opportunities in more detail.

\subsection{Higher resolution and extended data sets}

%access to many body effects still needs better resolution
The overall  objective of nanoARPES on QM devices and nanoARPES in general is to gain access to many-body interactions in the spectral function on a relevant length scale, preferably maintaining the energy resolution achievable in conventional ARPES.  This is extremely challenging. In conventional ARPES, the current sub-meV energy resolution is sufficient to detect signatures of many-body effects such as small gaps on parts of the Fermi contour, not only in high $T_C$ materials but in conventional superconductors or CDW  materials. In nanoARPES, the currently reachable energy resolution is not nearly as high as in conventional ARPES and this is caused by the inefficient light focusing by the zone plates currently used (the electron analysers are essentially the same as those used in conventional ARPES). As already discussed, this is bound to improve with the fourth generation, diffraction-limited synchrotron radiation sources or non-diffractive approaches to focusing the beam. 

The state-of-the-art spectral resolution of a few hundred $\mu$eV in conventional ARPES \cite{Kiss:2008aa} is roughly matched by the lowest currently reachable sample temperature of $\approx 1$~K \cite{Borisenko:2012aa} but this is far better than what is achievable in a typical nanoARPES setup. Moreover, even in conventional ARPES, the sample temperature is much higher than what is typically needed to reveal phenomena such as the recently discovered complex many-body states in twisted bilayer graphene \cite{Cao:2018ad,Cao:2018aa}. Sample cooling is even more challenging in nanoARPES because it has to be combined with a highly precise motion control of the sample. 

%hard with energy resolution because of low photon flux

%much bigger data sets needed because now we have two dimensions more
Comparing nanoARPES with conventional ARPES, a  severe practical issue is that a common data set in conventional ARPES consists of the photoemission intensity as a function of binding energy $E_{b}$ and the two-dimensional $\mathbf{k}_{\parallel}$  in a three dimensional region of interest. The scan in $\mathbf{k}_{\parallel}$ is performed by either scanning the two emission angles $\Theta$ and $\Phi$ in Figure~\ref{fig:ARPESprinciple}(a) or by scanning corresponding deflector voltages in the lens of the electron analyser. In nanoARPES on a QM-based device, it is usually not only desirable to perform a corresponding scan on the ``best'' position of a sample but for every position on the device, leading to the need of measuring a five dimensional instead of a three dimensional data set. Collecting data on every position of the device,  sampled with a fine grid corresponding to the spatial resolution, hugely increases the time needed for the experiments (there are 1600 points in the grid used for scanning the graphene device in Figure~\ref{fig:specfunc2})  and this comes on top of the already existing challenge of a low photon flux. 

Being able to measure a full $E_{b}$, $\mathbf{k}_{\parallel}$ scan  matters: In Figure~\ref{fig:gating1}, only a single cut in $\mathbf{k}_{\parallel}$ was measured as a function of gate voltage and only at a single position. Since the gating leads to stray electric fields affecting the emission angle of the photoelectrons, it is not possible to ensure that the cut is always covering exactly the same $\mathbf{k}_{\parallel}$ line, for example that including the Dirac point of graphene, as also pointed out by the authors of Ref. \cite{Nguyen:2019aa}. The position-resolved data in Figure~\ref{fig:specfunc2} has a similar issue because here, too, only a single cut in  $E_{b}$, $\mathbf{k}_{\parallel}$ was collected for every position in space, necessitating a complex and indirect analysis to determine parameters such as the local azimuthal rotation of the graphene flake. By contrast, the analysis in Figure~\ref{fig:gating2} is based on full $E_{b}$, $\mathbf{k}_{\parallel}$ scans for every gate voltage and this, in addition to the high spectral resolution, is the key for the accurate determination of the doping-dependent many-body effects. However, this data set has been collected at a single position of the sample. 

%how to handle the data that comes out of this
Once an efficient measurement of photoemission intensity as a function of $E_{b}$, $\mathbf{k}_{\parallel}$ and position on the sample is achieved, it will be challenging to handle and analyse such multi-dimensional data sets. This is not helped by the fact that collecting the data as a function of additional parameters can be imagined,  such as the photon energy, sample temperature, gate voltage, source-drain current, the photoelectrons' spin polarisation and so on. On the other hand, this is an issue facing many experimental approaches and efficient data-handling schemes are developed by the community \cite{Xian:2019aa}.

\subsection{High photon energies}

%surface sensitivity not good because restricts device geometry very severely
A serious drawback in ARPES is the high surface sensitivity caused by the short inelastic mean free path of the photoelectrons illustrated in Figure~\ref{fig:electron_imfp}. As far as nanoARPES on devices is concerned, this is one of the reasons for mainly studying 2D materials where this is not a major issue (at least not for SL materials exposed to the surface). In fact, already when it comes to graphene covered by a protective layer of $h$-BN or another layer of graphene, a clear reduction of the signal from the lower layer is seen, as illustrated in Figure~\ref{fig:gating3}(a). For a buried 2DEG, ARPES experiments are usually not possible except, in some cases, by  choosing photon energies for a resonant excitation \cite{Miwa:2013aa}. In general, the short inelastic mean free path puts severe constraints on the device design shown in Figure~\ref{fig:devices}. Adding an additional top gate on the device, for example, is  out of the question because the conductive channel would no longer be accessible in ARPES.

%high energy soft X-ray would work
A possible way to circumvent this problem is to use high photon energies, leading to correspondingly high kinetic energies of the electrons photoemitted from the valence band, and thus to a longer inelastic mean free path \cite{Suga:2004aa,Suga:2015aa,Strocov:2019aa}. According to Figure~\ref{fig:electron_imfp}, the increase of the inelastic mean free path is modest, even when using photon energies of more than 1~keV. Still, ARPES studies of materials covered by protective layers  have, in fact, been reported using the technique of soft X-ray ARPES \cite{Kobayashi:2012aa}. As far as nanoARPES is concerned, using high photon energies has the additional advantage of lessening the focusing constraints imposed by the diffraction limit. 

Soft X-ray nanoARPES would also open the possibility of accessing core level binding energies. This would allow to perform local core level spectroscopy and structural determination via X-ray photoelectron diffraction \cite{Woodruff:2002aa}. Given the well-defined energy of core level lines and their high photoemission cross section, core level lines can also be used to track the local potential across a device with much higher precision and accuracy than in the data of  Figure~\ref{fig:specfunc1} (for a more detailed discussion, see below). Finally, access to inner shell energies  can be used to gain chemical sensitivity in valence band studies via resonant photoemission effects \cite{Molodtsov:1997aa}. 

On the other hand, using soft X-rays also drastically reduces the photoemission cross section for the valence band and requires a higher angular resolution to achieve the same $\mathbf{k}_{\parallel}$ resolution, turning it into a highly photon hungry technique \cite{Strocov:2019aa} which, of course, is an unfortunate combination with the inefficient focusing of photons in nanoARPES. Indeed, it is questionable if the high resolution for studying many-body effects in the spectral function of QMs will be achievable at very high photon energies. Moreover, Debye Waller-like broadening effects become more important, especially for 3D electronic bulk states  \cite{Hofmann:2002aa}. Nevertheless, soft X-ray nanoARPES is a promising direction for studying QM-based devices, especially when performed using one of the up-coming high-energy fourth generation synchrotron radiation sources. As far as device construction is concerned, top dielectric and gate materials should preferably be amorphous, such that their photoemission signal merely contributes to the incoherent background in the spectra.

\subsection{Time-resolved studies}

%%Chances emerging through TR and nanoARPES
While spatial resolution in ARPES studies is still in its infancy, time-resolved (TR) ARPES has seen enormous progress over the last decade, such that the non-equilibrium electronic structure of solids can now be routinely studied on an ultrafast time scale. In most TR-ARPES experiments, non-equilibrium situations are created in a pump-probe type of experiment. An ultrashort pump pulse (typically in the infrared or optical regime) drives a sample out of equilibrium. Following a variable time-delay, a high energy UV or X-ray pulse is used to drive the photoemission process in ARPES. This approach can be used to study processes such as the dynamics of excited carriers in QMs \cite{Perfetti:2007aa,Johannsen:2013ab,Gierz:2013aa,Ulstrup:2014ac,Rohde:2018ab}, the ultrafast melting of CDWs \cite{Perfetti:2006aa,Rohwer:2011aa,Hellmann:2012aa}, the generation of coherent phonons and the electron-phonon interaction \cite{Leuenberger:2013aa,Sobota:2014aa,Gerber:2017aa,Hein:2020aa}, as well as the response of an electron gas to a  strong electric field \cite{Gudde:2007aa,Reimann:2018aa}. In fact, very strong electric fields, pressures or electronic temperatures can be tolerated  because these are applied only for a very short time. TR-ARPES also opens the possibility to study normally unoccupied states \cite{Bertoni:2016aa,Reimann:2014aa} and to monitor the size and renormalisation of band gaps \cite{Antonija-Grubisic-Cabo:2015aa,Ulstrup:2016aa}. In general, it permits ARPES investigations of transient states of matter created by the pump pulse.

%TR ARPES is cool for all kinds of this.

It is easy to build a strong scientific case for the combination of nanoARPES (on devices) and TR-ARPES. For a 2D semiconductor placed on a patterned substrate \cite{Rosner:2016aa}, this would not only allow static properties such as stacking-dependent band offsets \cite{Rosner:2016aa,Ulstrup:2019aa} to be measured but also the influence of such local properties on the electron dynamics in the 2D system. Indeed, in a more general picture, all the phenomena that give rise to spontaneous or imposed nano structuring that can be investigated with nanoARPES could also be studied with respect to their unoccupied band structure or electron dynamics. TR-ARPES could also be combined with nanoARPES to perform the the local transport measurements described in the next section. One could, for instance, envision to create a state of transient superconductivity \cite{Mitrano:2016aa} and monitor the local conductivity using nanoARPES on a current-carrying sample.

%What are the problems? Mainly space charge
While the scientific prospect of combining spatial and time resolution is thus highly attractive, the technical realisation faces formidable difficulties, the most important one being that of space charge effects. Space charge is already a potential issue for nanoARPES using a synchrotron radiation source because the photoemitted electrons are compressed in a cloud that is small parallel to the sample surface due to the tight focus, and also small perpendicular to the surface due to the short pulse length of the synchrotron light \cite{Zhou:2005ab,Passlack:2006aa,Hellmann:2009aa,Hellmann:2012ac,Rotenberg:2014aa}. This pulse length is tens or hundreds of picoseconds and therefore much longer than the $<100$~fs required for meaningful TR-ARPES experiments. Ultrafast TR-ARPES experiments do thus lead to a severe further compression of the electron cloud perpendicular to the surface and will therefore suffer more from space charge broadening. This is not helped by the fact that high-harmonic laser sources tend to have a low repetition rates, and a high number of photons per pulse is thus required to achieve acceptable count rates. Moreover, TR-ARPES experiments suffer from additional space charge effects due to the interaction of the photoemitted electrons with the dense cloud of electrons created by the intense \emph{pump} pulse \cite{Oloff:2014aa,Ulstrup:2015ab}. In fact, this is currently the most severe space charge limitation for this technique. This will, at least, not necessarily have to worsen when combining TR-ARPES with nanoARPES because there is no \emph{a priori} need to focus the pump pulse as tightly as the probe pulse. 

%Maybe energy resolved PEEM anyway?
For combining TR-ARPES with nanoARPES, other techniques than focusing the light may be more promising, especially energy-filtered PEEM. This is particularly true if the achievable energy resolution is anyway limited because of space charge effects. Moreover, when ultrafast lasers are used as a light source, highly efficient time-of-flight approaches become attractive for the energy filtering in a PEEM \cite{Spiecker:1998aa,Chernov:2015aa}. Indeed, very recently a TR-ARPES from a small (6$\times$40~$\mu$m) area of a SL WSe$_2$ flake on $h$-BN was reported \cite{Madeo:2020aa}. On the other hand, the application of energy-filtered PEEM on actual devices is yet to be demonstrated. Using this technique on a structured device faces the difficulty that PEEM requires the application of a high electric field between the sample and entrance lens of the PEEM, making it mainly suitable for studying flat samples. 

\subsection{Contact-free transport measurements}

%integrating spectroscopy and transport
As discussed in connection with Figure~\ref{fig:current1}, mapping the spectral function of a current-carrying device simultaneously provides information about the local potential and electric field, integrating spectroscopic and transport measurements. This opens up a number of opportunities for characterising QMs close to phase transitions, as briefly discussed in the following. 

\begin{figure}
\includegraphics[width=0.48\textwidth]{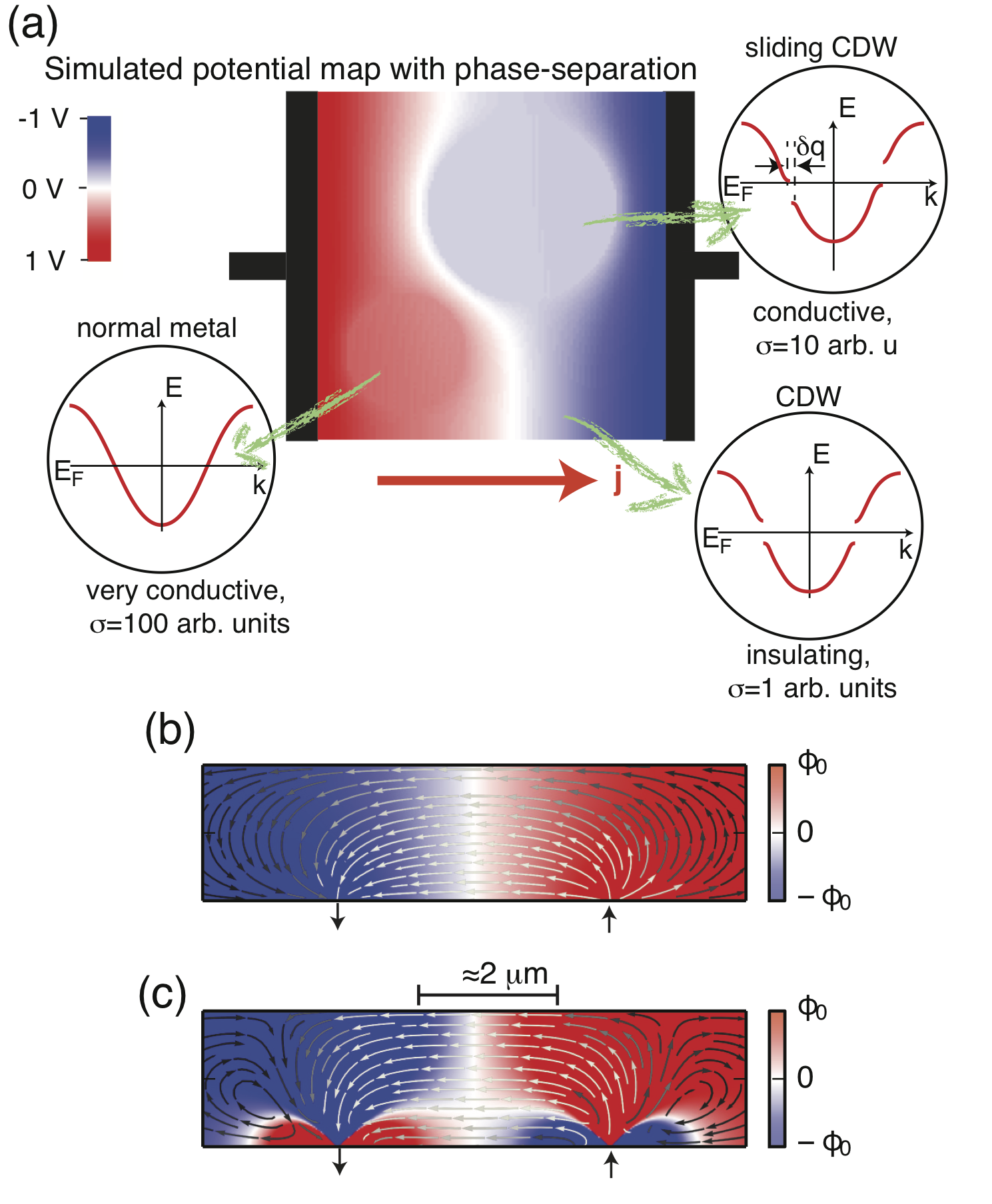}\\
\caption{(a) Qualitative picture of what could be the potential in a CDW material in the presence of a current density $\mathbf{j}$. It is assumed that a separation takes place into regions with static CDW, sliding CDW and the un-gapped metallic state, and that these phases have different conductivities. \emph{In operando} nanoARPES will be in a position to determine both the electronic structure and the local conductivity of all the phases in the system. (b) and (c) Nano-scale potential landscape (colour) and current density (arrows) in a graphene device in the absence and the presence of viscous electron flow, respectively, adapted from Ref.  \cite{Torre:2015aa}. The arrows outside the box represent the current injection points. \emph{In operando} nanoARPES could  permit a direct mapping of such a potential landscape instead of an indirect measurement based on potential differences between contacts mounted on the side. The scale bar gives merely an order of magnitude for realistic devices inspired by Ref. \cite{Bandurin:2016aa}. Lower part adapted with permission from I. Torre, A. Tomadin, A. K. Geim, and M. Polini, Physical Review B 92, 165433 (2015). Copyright (2015) by the American Physical Society. }
  \label{fig:potentialmapping}
\end{figure}

%why would this be cool. Local conductivity map
Figure~\ref{fig:potentialmapping}(a) illustrates the general idea of combining spectroscopy and transport properties, using a  CDW system as an example.  Consider a materials with a phase diagram like that of Figure  \ref{fig:intro}, with the parameter $g$ being the sample temperature \cite{Monceau:2012aa}. The system shall be at finite temperature and a transport current shall be present. Due to fluctuations of properties, it is conceivable that three phases can coexist: the sliding CDW, the static CDW and the ``normal'' metallic phase. If these phases are spatially separated, they could be singled out by their characteristic spectral function in ARPES, as illustrated in Figure~\ref{fig:potentialmapping}(a). Moreover, the different conductivity of the phases would show up in the potential map, as shown qualitatively in the simulation (calculated by solving the Poisson equation with variable conductivity for this 2D situation). From such data, it would not only be possible to obtain detailed spectroscopic information on the different phases but also on the local conductivity of each phase and the conductivity of the boundaries between the phases. Following the proof-of-principle demonstration in Figure~\ref{fig:current1}, data of this type can be analysed using the same approach as in scanning tunnelling potentiometry and similar techniques \cite{Muralt:1986aa,Zhang:2016ag,Tetienne:2017aa,Voigtlander:2018aa,Ella:2019aa}.

%one should think of this as a generic QM with different local spectral properties and conductivities.
 While this example is inspired by a driven CDW system, the generic idea behind the figure is that of a QM with a tendency towards local phase separation, exposed to a current that leads to an insulator-to-metal transition or some other transition influencing the material's conductivity. Areas of different conductivity could, for instance, also arise because of the formation of conductive filaments in current-driven Mott insulators or other types of phase separation \cite{Kumai:1999aa,Lang:2002aa,Qazilbash:2007aa,Vaju:2008aa,Lai:2010aa}. As already mentioned, mapping the local potential and inferring the conductivity and current is also important in order to evaluate local Joule heating effects in QMs, to exclude that the sample temperature rather than another parameter drives a phase transition.

%another example could be viscous flow
The combination of spectroscopy and local potential mapping could also give access to a number of other phenomena. A particularly interesting example is the recently discovered viscous current flow in graphene \cite{Torre:2015aa,Bandurin:2016aa,Krishna-Kumar:2017aa,Ella:2019aa,Sulpizio:2019aa,Polini:2020aa}. This unusual hydrodynamic flow regime of the electron liquid can only be reached when the electron-electron scattering length becomes significantly shorter than the sample dimensions and the electron-phonon and electron-defect scattering length \cite{Torre:2015aa}. In this regime, which has otherwise only been realised in (Al,Ga)As heterostructures \cite{Jong:1995aa}, phenomena such as electron ``whirlpools'' appear in potential maps calculated from the hydrodynamic transport equations. Examples of potential maps with and without viscosity are given in Figure~\ref{fig:potentialmapping}(b) and (c), respectively (from Ref.  \cite{Torre:2015aa}). In transport experiments \cite{Bandurin:2016aa}, the existence of current vortices and electron ``back-flow'' can only be inferred indirectly from voltage measurements at the side contacts of a device, even though it has recently become possible to image such situations directly by scanning-sensor techniques \cite{Ella:2019aa,Sulpizio:2019aa}.  \emph{In operando} nanoARPES could complement these techniques by simultaneously probing  the local spectral function and the potential in the device.

%The idea of determining surface and bulk conductivity separately. 
Local potential mapping can also be used to disentangle bulk and surface contributions to transport, an aspect which is particularly relevant to topological insulator research. Topological insulators are expected to be bulk insulators or semiconductors supporting metallic surface states \cite{Hasan:2010aa,Ando:2013aa}. At sufficiently low temperature, the bulk carriers are frozen out and the surface states should thus give rise to a 2D conductance. Unfortunately, this is rarely the case. In fact, topological insulators are plagued by bulk-dominated conductance due to unintended doping and / or small band gaps. In a given transport situation, it is not trivial to determine if the conductance is through surface or bulk states even though 2D vs. 3D transport can be distinguished by collinear four point probes with variable spacing (or movable contacts) \cite{Barreto:2014aa}, due to the different functional form of the potential decay in a 2D plane and on the surface of a semi-infinite bulk ($\ln(x)$  vs $1/x$) \cite{Hofmann:2009aa,Perkins:2013aa}. Mapping the potential in nanoARPES can directly yield this functional form and help to disentangle surface and bulk contributions to transport. This is illustrated in Figure~\ref{fig:holder_current} where not only the penetration of the current into the bulk is different but also the distribution along the surface. 
Even when the detected transport regime is found to be 2D rather than 3D in a transport measurement, it cannot be concluded that it is only the topological surface state that gives rise to metallic transport since several topological insulator materials support band bending-induced (and topologically trivial) two-dimensional metallic states at the surface \cite{Bianchi:2010ab,Benia:2011aa,King:2011aa}. The simultaneous mapping of the spectral function by nanoARPES, however, would ensure that all contributions to metallic transport can be identified. 

%use core leves
Some of the above effects are quite subtle and put high demands on the local potential mapping. In the proof-of-principle from Ref. \cite{Curcio:2020aa} illustrated in Figure~\ref{fig:current1}, the local potential mapping was carried out by fitting a 2D model spectral function to the data. The model is constructed from a Dirac-cone dispersion with the possibility of shifts in energy and $k$, as well as broadening, combined with a Fermi-Dirac distribution. The local Fermi energy from the fit is then used as a map of the potential. 
This is not a particularly good way of obtaining a potential map because, in the case of almost un-doped graphene, only a vanishingly small spectral feature is present at $E_F$, resulting in a low accuracy of the $E_F$ value obtained from a fit. A far better approach to measuring such maps would be to use a sharp core level and map the core level binding energy across the sample. This would greatly increase both accuracy and precision. Note that when fitting core level peaks, the resulting uncertainty of the energy position can be very small, in the order of 1~meV, even if the peaks are broad (in the order of 100 meV). 

%========================================================

\section{Conclusions}
\label{sec:VIII}

The combination of recent advances in nanoARPES, 2D materials and device fabrication is now permitting  \emph{in operando} studies of field effect transistor-like devices made of 2D QMs, such as graphene or few layer TMDCs. We have discussed the first experimental demonstrations of tracking a material's doping,  spectral function and many body-effects upon gating and while passing a current through the material. While this technique is still  in its infancy, it provides a unique new insight into the properties of operating devices and it should be possible to extend the approach to a wide range of QMs showing electric field and current-induced phenomena, such as doping-induced transitions to superconducting or correlated states, field-induced insulator-to-metal transitions in Mott insulators or the field-induced sliding of charge density waves. A side product of mapping the spectral function of a current-carrying device is a position-resolved map of the potential. This can be used to combine spectroscopic and contact-free transport measurements on  the nano-scale, determining properties such as the local conductivity or carrier mobility.

\section{acknowledgement}
The author acknowledges inspiring discussions with S\o ren Ulstrup, Jill A. Miwa, Marco Bianchi, Jyoti Katoch, Alfred J. H. Jones, Davide Curcio, Deepnarayan Biswas, Paulina Majchrzak, Charlotte E. Sanders, Klara Volckaert, Eli Rotenberg, Aaron Bostwick, Cephise Cacho, Pavel Dudin, Kai Rossnagel and Frederick Joucken. I especially thank SU, AJHJ, JAM, DC, KR and PD for critical comments on the manuscript. This work was supported by VILLUM FONDEN via the Centre of Excellence for Dirac Materials (Grant No. 11744).

\vspace{2mm}
This paper is dedicated to the memory of E. Ward Plummer (30.10.1940 - 23.7.2020).

\section{data availability}

The data that support the findings of this study are available from the corresponding author upon reasonable request.

%\bibliographystyle{apsrev_ph}
%\bibliography{phref,local}

\begin{thebibliography}{247}
\expandafter\ifx\csname natexlab\endcsname\relax\def\natexlab#1{#1}\fi
\expandafter\ifx\csname bibnamefont\endcsname\relax
  \def\bibnamefont#1{#1}\fi
\expandafter\ifx\csname bibfnamefont\endcsname\relax
  \def\bibfnamefont#1{#1}\fi
\expandafter\ifx\csname citenamefont\endcsname\relax
  \def\citenamefont#1{#1}\fi
\expandafter\ifx\csname url\endcsname\relax
  \def\url#1{\texttt{#1}}\fi
\expandafter\ifx\csname urlprefix\endcsname\relax\def\urlprefix{URL }\fi
\providecommand{\bibinfo}[2]{#2}
\providecommand{\eprint}[2][]{\url{#2}}

\bibitem[{\citenamefont{Plummer and Eberhardt}(1982)}]{Plummer:1982aa}
\bibinfo{author}{\bibfnamefont{E.~W.} \bibnamefont{Plummer}} \bibnamefont{and}
  \bibinfo{author}{\bibfnamefont{W.}~\bibnamefont{Eberhardt}},
  \bibinfo{journal}{Advances in Chemical Physics}
  \textbf{\bibinfo{volume}{49}}, \bibinfo{pages}{533} (\bibinfo{year}{1982}).

\bibitem[{\citenamefont{Kevan}(1992)}]{Kevan:1992aa}
\bibinfo{editor}{\bibfnamefont{S.~D.} \bibnamefont{Kevan}}, ed.,
  \emph{\bibinfo{title}{Angle-resolved photoemission}},
  vol.~\bibinfo{volume}{74} of \emph{\bibinfo{series}{Studies in Surface
  Chemistry and Catalysis}} (\bibinfo{publisher}{Elsevier},
  \bibinfo{address}{Amsterdam}, \bibinfo{year}{1992}).

\bibitem[{\citenamefont{H{{\"u}}fner}(2003)}]{Hufner:2003aa}
\bibinfo{author}{\bibfnamefont{S.}~\bibnamefont{H{{\"u}}fner}},
  \emph{\bibinfo{title}{Photoelectron spectroscopy}}
  (\bibinfo{publisher}{Springer}, \bibinfo{address}{Berlin},
  \bibinfo{year}{2003}), \bibinfo{edition}{3rd} ed.

\bibitem[{\citenamefont{Damascelli et~al.}(2003)\citenamefont{Damascelli,
  Hussain, and Shen}}]{Damascelli:2003aa}
\bibinfo{author}{\bibfnamefont{A.}~\bibnamefont{Damascelli}},
  \bibinfo{author}{\bibfnamefont{Z.}~\bibnamefont{Hussain}}, \bibnamefont{and}
  \bibinfo{author}{\bibfnamefont{Z.-X.} \bibnamefont{Shen}},
  \bibinfo{journal}{Reviews of Modern Physics} \textbf{\bibinfo{volume}{75}},
  \bibinfo{pages}{473} (\bibinfo{year}{2003}).

\bibitem[{\citenamefont{Hofmann et~al.}(2009)\citenamefont{Hofmann, Sklyadneva,
  Rienks, and Chulkov}}]{Hofmann:2009ab}
\bibinfo{author}{\bibfnamefont{P.}~\bibnamefont{Hofmann}},
  \bibinfo{author}{\bibfnamefont{I.~Y.} \bibnamefont{Sklyadneva}},
  \bibinfo{author}{\bibfnamefont{E.~D.~L.} \bibnamefont{Rienks}},
  \bibnamefont{and} \bibinfo{author}{\bibfnamefont{E.~V.}
  \bibnamefont{Chulkov}}, \bibinfo{journal}{New Journal of Physics}
  \textbf{\bibinfo{volume}{11}}, \bibinfo{pages}{125005}
  (\bibinfo{year}{2009}).

\bibitem[{\citenamefont{Lu et~al.}(2012)\citenamefont{Lu, Vishik, Yi, Chen,
  Moore, and Shen}}]{Lu:2012ac}
\bibinfo{author}{\bibfnamefont{D.}~\bibnamefont{Lu}},
  \bibinfo{author}{\bibfnamefont{I.~M.} \bibnamefont{Vishik}},
  \bibinfo{author}{\bibfnamefont{M.}~\bibnamefont{Yi}},
  \bibinfo{author}{\bibfnamefont{Y.}~\bibnamefont{Chen}},
  \bibinfo{author}{\bibfnamefont{R.~G.} \bibnamefont{Moore}}, \bibnamefont{and}
  \bibinfo{author}{\bibfnamefont{Z.-X.} \bibnamefont{Shen}},
  \bibinfo{journal}{Annual Review of Condensed Matter Physics}
  \textbf{\bibinfo{volume}{3}}, \bibinfo{pages}{129} (\bibinfo{year}{2012}).

\bibitem[{\citenamefont{Sobota et~al.}(2020)\citenamefont{Sobota, He, and
  Shen}}]{Sobota:2020aa}
\bibinfo{author}{\bibfnamefont{J.~A.} \bibnamefont{Sobota}},
  \bibinfo{author}{\bibfnamefont{Y.}~\bibnamefont{He}}, \bibnamefont{and}
  \bibinfo{author}{\bibfnamefont{Z.-X.} \bibnamefont{Shen}}
  (\bibinfo{year}{2020}), \eprint{arXiv.2008.02378}.

\bibitem[{\citenamefont{Rotenberg}(2010)}]{Rotenberg:2010ad}
\bibinfo{author}{\bibfnamefont{E.}~\bibnamefont{Rotenberg}}, in
  \emph{\bibinfo{booktitle}{Many-Body Interactions in Nanoscale Materials by
  Angle-Resolved Photoemission Spectroscopy}}, edited by
  \bibinfo{editor}{\bibfnamefont{J.~a.} \bibnamefont{Guo}}
  (\bibinfo{publisher}{Wiley-VCH}, \bibinfo{year}{2010}), p.
  \bibinfo{pages}{169}.

\bibitem[{\citenamefont{Keimer et~al.}(2015)\citenamefont{Keimer, Kivelson,
  Norman, Uchida, and Zaanen}}]{Keimer:2015aa}
\bibinfo{author}{\bibfnamefont{B.}~\bibnamefont{Keimer}},
  \bibinfo{author}{\bibfnamefont{S.~A.} \bibnamefont{Kivelson}},
  \bibinfo{author}{\bibfnamefont{M.~R.} \bibnamefont{Norman}},
  \bibinfo{author}{\bibfnamefont{S.}~\bibnamefont{Uchida}}, \bibnamefont{and}
  \bibinfo{author}{\bibfnamefont{J.}~\bibnamefont{Zaanen}},
  \bibinfo{journal}{Nature} \textbf{\bibinfo{volume}{518}},
  \bibinfo{pages}{179} (\bibinfo{year}{2015}).

\bibitem[{\citenamefont{Keimer and Moore}(2017)}]{Keimer:2017aa}
\bibinfo{author}{\bibfnamefont{B.}~\bibnamefont{Keimer}} \bibnamefont{and}
  \bibinfo{author}{\bibfnamefont{J.~E.} \bibnamefont{Moore}},
  \bibinfo{journal}{Nature Physics} \textbf{\bibinfo{volume}{13}},
  \bibinfo{pages}{1045} (\bibinfo{year}{2017}).

\bibitem[{\citenamefont{Basov et~al.}(2017)\citenamefont{Basov, Averitt, and
  Hsieh}}]{Basov:2017aa}
\bibinfo{author}{\bibfnamefont{D.~N.} \bibnamefont{Basov}},
  \bibinfo{author}{\bibfnamefont{R.~D.} \bibnamefont{Averitt}},
  \bibnamefont{and} \bibinfo{author}{\bibfnamefont{D.}~\bibnamefont{Hsieh}},
  \bibinfo{journal}{Nature Materials} \textbf{\bibinfo{volume}{16}},
  \bibinfo{pages}{1077} (\bibinfo{year}{2017}).

\bibitem[{\citenamefont{Tokura et~al.}(2017)\citenamefont{Tokura, Kawasaki, and
  Nagaosa}}]{Tokura:2017aa}
\bibinfo{author}{\bibfnamefont{Y.}~\bibnamefont{Tokura}},
  \bibinfo{author}{\bibfnamefont{M.}~\bibnamefont{Kawasaki}}, \bibnamefont{and}
  \bibinfo{author}{\bibfnamefont{N.}~\bibnamefont{Nagaosa}},
  \bibinfo{journal}{Nature Physics} \textbf{\bibinfo{volume}{13}},
  \bibinfo{pages}{1056} (\bibinfo{year}{2017}).

\bibitem[{\citenamefont{Vojta}(2003)}]{Vojta:2003aa}
\bibinfo{author}{\bibfnamefont{M.}~\bibnamefont{Vojta}},
  \bibinfo{journal}{Reports on Progress in Physics}
  \textbf{\bibinfo{volume}{66}}, \bibinfo{pages}{2069} (\bibinfo{year}{2003}).

\bibitem[{\citenamefont{Wang et~al.}(2013)\citenamefont{Wang, Steinberg,
  Jarillo-Herrero, and Gedik}}]{Wang:2013ad}
\bibinfo{author}{\bibfnamefont{Y.~H.} \bibnamefont{Wang}},
  \bibinfo{author}{\bibfnamefont{H.}~\bibnamefont{Steinberg}},
  \bibinfo{author}{\bibfnamefont{P.}~\bibnamefont{Jarillo-Herrero}},
  \bibnamefont{and} \bibinfo{author}{\bibfnamefont{N.}~\bibnamefont{Gedik}},
  \bibinfo{journal}{Science} \textbf{\bibinfo{volume}{342}},
  \bibinfo{pages}{453} (\bibinfo{year}{2013}).

\bibitem[{\citenamefont{Rudner and Lindner}(2020)}]{Rudner:2020aa}
\bibinfo{author}{\bibfnamefont{M.~S.} \bibnamefont{Rudner}} \bibnamefont{and}
  \bibinfo{author}{\bibfnamefont{N.~H.} \bibnamefont{Lindner}},
  \bibinfo{journal}{Nature Reviews Physics} \textbf{\bibinfo{volume}{2}},
  \bibinfo{pages}{229} (\bibinfo{year}{2020}).

\bibitem[{\citenamefont{Cavalleri et~al.}(2001)\citenamefont{Cavalleri,
  T{\'{o}}th, Siders, Squier, R{\'{a}}ksi, Forget, and
  Kieffer}}]{Cavalleri:2001aa}
\bibinfo{author}{\bibfnamefont{A.}~\bibnamefont{Cavalleri}},
  \bibinfo{author}{\bibfnamefont{C.}~\bibnamefont{T{\'{o}}th}},
  \bibinfo{author}{\bibfnamefont{C.~W.} \bibnamefont{Siders}},
  \bibinfo{author}{\bibfnamefont{J.~A.} \bibnamefont{Squier}},
  \bibinfo{author}{\bibfnamefont{F.}~\bibnamefont{R{\'{a}}ksi}},
  \bibinfo{author}{\bibfnamefont{P.}~\bibnamefont{Forget}}, \bibnamefont{and}
  \bibinfo{author}{\bibfnamefont{J.~C.} \bibnamefont{Kieffer}},
  \bibinfo{journal}{Physical Review Letters} \textbf{\bibinfo{volume}{87}}
  (\bibinfo{year}{2001}).

\bibitem[{\citenamefont{Schmitt et~al.}(2008)\citenamefont{Schmitt, Kirchmann,
  Bovensiepen, Moore, Rettig, Krenz, Chu, Ru, Perfetti, Lu
  et~al.}}]{Schmitt:2008aa}
\bibinfo{author}{\bibfnamefont{F.}~\bibnamefont{Schmitt}},
  \bibinfo{author}{\bibfnamefont{P.~S.} \bibnamefont{Kirchmann}},
  \bibinfo{author}{\bibfnamefont{U.}~\bibnamefont{Bovensiepen}},
  \bibinfo{author}{\bibfnamefont{R.~G.} \bibnamefont{Moore}},
  \bibinfo{author}{\bibfnamefont{L.}~\bibnamefont{Rettig}},
  \bibinfo{author}{\bibfnamefont{M.}~\bibnamefont{Krenz}},
  \bibinfo{author}{\bibfnamefont{J.~H.} \bibnamefont{Chu}},
  \bibinfo{author}{\bibfnamefont{N.}~\bibnamefont{Ru}},
  \bibinfo{author}{\bibfnamefont{L.}~\bibnamefont{Perfetti}},
  \bibinfo{author}{\bibfnamefont{D.~H.} \bibnamefont{Lu}},
  \bibnamefont{et~al.}, \bibinfo{journal}{Science}
  \textbf{\bibinfo{volume}{321}}, \bibinfo{pages}{1649} (\bibinfo{year}{2008}).

\bibitem[{\citenamefont{Rohwer et~al.}(2011)\citenamefont{Rohwer, Hellmann,
  Wiesenmayer, Sohrt, Stange, Slomski, Carr, Liu, Avila, Kallane
  et~al.}}]{Rohwer:2011aa}
\bibinfo{author}{\bibfnamefont{T.}~\bibnamefont{Rohwer}},
  \bibinfo{author}{\bibfnamefont{S.}~\bibnamefont{Hellmann}},
  \bibinfo{author}{\bibfnamefont{M.}~\bibnamefont{Wiesenmayer}},
  \bibinfo{author}{\bibfnamefont{C.}~\bibnamefont{Sohrt}},
  \bibinfo{author}{\bibfnamefont{A.}~\bibnamefont{Stange}},
  \bibinfo{author}{\bibfnamefont{B.}~\bibnamefont{Slomski}},
  \bibinfo{author}{\bibfnamefont{A.}~\bibnamefont{Carr}},
  \bibinfo{author}{\bibfnamefont{Y.}~\bibnamefont{Liu}},
  \bibinfo{author}{\bibfnamefont{L.~M.} \bibnamefont{Avila}},
  \bibinfo{author}{\bibfnamefont{M.}~\bibnamefont{Kallane}},
  \bibnamefont{et~al.}, \bibinfo{journal}{Nature}
  \textbf{\bibinfo{volume}{471}}, \bibinfo{pages}{490} (\bibinfo{year}{2011}).

\bibitem[{\citenamefont{Fausti et~al.}(2011)\citenamefont{Fausti, Tobey, Dean,
  Kaiser, Dienst, Hoffmann, Pyon, Takayama, Takagi, and
  Cavalleri}}]{Fausti:2011aa}
\bibinfo{author}{\bibfnamefont{D.}~\bibnamefont{Fausti}},
  \bibinfo{author}{\bibfnamefont{R.~I.} \bibnamefont{Tobey}},
  \bibinfo{author}{\bibfnamefont{N.}~\bibnamefont{Dean}},
  \bibinfo{author}{\bibfnamefont{S.}~\bibnamefont{Kaiser}},
  \bibinfo{author}{\bibfnamefont{A.}~\bibnamefont{Dienst}},
  \bibinfo{author}{\bibfnamefont{M.~C.} \bibnamefont{Hoffmann}},
  \bibinfo{author}{\bibfnamefont{S.}~\bibnamefont{Pyon}},
  \bibinfo{author}{\bibfnamefont{T.}~\bibnamefont{Takayama}},
  \bibinfo{author}{\bibfnamefont{H.}~\bibnamefont{Takagi}}, \bibnamefont{and}
  \bibinfo{author}{\bibfnamefont{A.}~\bibnamefont{Cavalleri}},
  \bibinfo{journal}{Science} \textbf{\bibinfo{volume}{331}},
  \bibinfo{pages}{189} (\bibinfo{year}{2011}).

\bibitem[{\citenamefont{Mitrano et~al.}(2016)\citenamefont{Mitrano, Cantaluppi,
  Nicoletti, Kaiser, Perucchi, Lupi, Di~Pietro, Pontiroli, Ricc{\`o}, Clark
  et~al.}}]{Mitrano:2016aa}
\bibinfo{author}{\bibfnamefont{M.}~\bibnamefont{Mitrano}},
  \bibinfo{author}{\bibfnamefont{A.}~\bibnamefont{Cantaluppi}},
  \bibinfo{author}{\bibfnamefont{D.}~\bibnamefont{Nicoletti}},
  \bibinfo{author}{\bibfnamefont{S.}~\bibnamefont{Kaiser}},
  \bibinfo{author}{\bibfnamefont{A.}~\bibnamefont{Perucchi}},
  \bibinfo{author}{\bibfnamefont{S.}~\bibnamefont{Lupi}},
  \bibinfo{author}{\bibfnamefont{P.}~\bibnamefont{Di~Pietro}},
  \bibinfo{author}{\bibfnamefont{D.}~\bibnamefont{Pontiroli}},
  \bibinfo{author}{\bibfnamefont{M.}~\bibnamefont{Ricc{\`o}}},
  \bibinfo{author}{\bibfnamefont{S.~R.} \bibnamefont{Clark}},
  \bibnamefont{et~al.}, \bibinfo{journal}{Nature}
  \textbf{\bibinfo{volume}{530}}, \bibinfo{pages}{461} (\bibinfo{year}{2016}).

\bibitem[{\citenamefont{Caputo et~al.}(2018)\citenamefont{Caputo, Khalil,
  Papalazarou, Nilforoushan, Perfetti, Taleb-Ibrahimi, Gibson, Cava, and
  Marsi}}]{Caputo:2018aa}
\bibinfo{author}{\bibfnamefont{M.}~\bibnamefont{Caputo}},
  \bibinfo{author}{\bibfnamefont{L.}~\bibnamefont{Khalil}},
  \bibinfo{author}{\bibfnamefont{E.}~\bibnamefont{Papalazarou}},
  \bibinfo{author}{\bibfnamefont{N.}~\bibnamefont{Nilforoushan}},
  \bibinfo{author}{\bibfnamefont{L.}~\bibnamefont{Perfetti}},
  \bibinfo{author}{\bibfnamefont{A.}~\bibnamefont{Taleb-Ibrahimi}},
  \bibinfo{author}{\bibfnamefont{Q.~D.} \bibnamefont{Gibson}},
  \bibinfo{author}{\bibfnamefont{R.~J.} \bibnamefont{Cava}}, \bibnamefont{and}
  \bibinfo{author}{\bibfnamefont{M.}~\bibnamefont{Marsi}},
  \bibinfo{journal}{Physical Review B} \textbf{\bibinfo{volume}{97}}
  (\bibinfo{year}{2018}).

\bibitem[{\citenamefont{Taguchi et~al.}(2000)\citenamefont{Taguchi, Matsumoto,
  and Tokura}}]{Taguchi:2000aa}
\bibinfo{author}{\bibfnamefont{Y.}~\bibnamefont{Taguchi}},
  \bibinfo{author}{\bibfnamefont{T.}~\bibnamefont{Matsumoto}},
  \bibnamefont{and} \bibinfo{author}{\bibfnamefont{Y.}~\bibnamefont{Tokura}},
  \bibinfo{journal}{Phys. Rev. B} \textbf{\bibinfo{volume}{62}},
  \bibinfo{pages}{7015} (\bibinfo{year}{2000}).

\bibitem[{\citenamefont{Lee et~al.}(2007)\citenamefont{Lee, Fursina, Mayo,
  Yavuz, Colvin, Sumesh~Sofin, Shvets, and Natelson}}]{Lee:2007aa}
\bibinfo{author}{\bibfnamefont{S.}~\bibnamefont{Lee}},
  \bibinfo{author}{\bibfnamefont{A.}~\bibnamefont{Fursina}},
  \bibinfo{author}{\bibfnamefont{J.~T.} \bibnamefont{Mayo}},
  \bibinfo{author}{\bibfnamefont{C.~T.} \bibnamefont{Yavuz}},
  \bibinfo{author}{\bibfnamefont{V.~L.} \bibnamefont{Colvin}},
  \bibinfo{author}{\bibfnamefont{R.~G.} \bibnamefont{Sumesh~Sofin}},
  \bibinfo{author}{\bibfnamefont{I.~V.} \bibnamefont{Shvets}},
  \bibnamefont{and} \bibinfo{author}{\bibfnamefont{D.}~\bibnamefont{Natelson}},
  \bibinfo{journal}{Nature Materials} \textbf{\bibinfo{volume}{7}},
  \bibinfo{pages}{130} (\bibinfo{year}{2007}).

\bibitem[{\citenamefont{Vaju et~al.}(2008)\citenamefont{Vaju, Cario, Corraze,
  Janod, Dubost, Cren, Roditchev, Braithwaite, and Chauvet}}]{Vaju:2008aa}
\bibinfo{author}{\bibfnamefont{C.}~\bibnamefont{Vaju}},
  \bibinfo{author}{\bibfnamefont{L.}~\bibnamefont{Cario}},
  \bibinfo{author}{\bibfnamefont{B.}~\bibnamefont{Corraze}},
  \bibinfo{author}{\bibfnamefont{E.}~\bibnamefont{Janod}},
  \bibinfo{author}{\bibfnamefont{V.}~\bibnamefont{Dubost}},
  \bibinfo{author}{\bibfnamefont{T.}~\bibnamefont{Cren}},
  \bibinfo{author}{\bibfnamefont{D.}~\bibnamefont{Roditchev}},
  \bibinfo{author}{\bibfnamefont{D.}~\bibnamefont{Braithwaite}},
  \bibnamefont{and} \bibinfo{author}{\bibfnamefont{O.}~\bibnamefont{Chauvet}},
  \bibinfo{journal}{Advanced Materials} \textbf{\bibinfo{volume}{20}},
  \bibinfo{pages}{2760} (\bibinfo{year}{2008}).

\bibitem[{\citenamefont{Monceau}(2012)}]{Monceau:2012aa}
\bibinfo{author}{\bibfnamefont{P.}~\bibnamefont{Monceau}},
  \bibinfo{journal}{Advances in Physics} \textbf{\bibinfo{volume}{61}},
  \bibinfo{pages}{325} (\bibinfo{year}{2012}).

\bibitem[{\citenamefont{Stoliar et~al.}(2013)\citenamefont{Stoliar, Cario,
  Janod, Corraze, Guillot-Deudon, Salmon-Bourmand, Guiot, Tranchant, and
  Rozenberg}}]{Stoliar:2013aa}
\bibinfo{author}{\bibfnamefont{P.}~\bibnamefont{Stoliar}},
  \bibinfo{author}{\bibfnamefont{L.}~\bibnamefont{Cario}},
  \bibinfo{author}{\bibfnamefont{E.}~\bibnamefont{Janod}},
  \bibinfo{author}{\bibfnamefont{B.}~\bibnamefont{Corraze}},
  \bibinfo{author}{\bibfnamefont{C.}~\bibnamefont{Guillot-Deudon}},
  \bibinfo{author}{\bibfnamefont{S.}~\bibnamefont{Salmon-Bourmand}},
  \bibinfo{author}{\bibfnamefont{V.}~\bibnamefont{Guiot}},
  \bibinfo{author}{\bibfnamefont{J.}~\bibnamefont{Tranchant}},
  \bibnamefont{and}
  \bibinfo{author}{\bibfnamefont{M.}~\bibnamefont{Rozenberg}},
  \bibinfo{journal}{Advanced Materials} \textbf{\bibinfo{volume}{25}},
  \bibinfo{pages}{3222} (\bibinfo{year}{2013}).

\bibitem[{\citenamefont{Nakamura et~al.}(2013)\citenamefont{Nakamura, Sakaki,
  Yamanaka, Tamaru, Suzuki, and Maeno}}]{Nakamura:2013aa}
\bibinfo{author}{\bibfnamefont{F.}~\bibnamefont{Nakamura}},
  \bibinfo{author}{\bibfnamefont{M.}~\bibnamefont{Sakaki}},
  \bibinfo{author}{\bibfnamefont{Y.}~\bibnamefont{Yamanaka}},
  \bibinfo{author}{\bibfnamefont{S.}~\bibnamefont{Tamaru}},
  \bibinfo{author}{\bibfnamefont{T.}~\bibnamefont{Suzuki}}, \bibnamefont{and}
  \bibinfo{author}{\bibfnamefont{Y.}~\bibnamefont{Maeno}},
  \bibinfo{journal}{Scientific Reports} \textbf{\bibinfo{volume}{3}},
  \bibinfo{pages}{2536} (\bibinfo{year}{2013}).

\bibitem[{\citenamefont{Bellec et~al.}(2020)\citenamefont{Bellec,
  Gonzalez-Vallejo, Jacques, Sinchenko, Orlov, Monceau, Leake, and
  Le~Bolloc'h}}]{Bellec:2020aa}
\bibinfo{author}{\bibfnamefont{E.}~\bibnamefont{Bellec}},
  \bibinfo{author}{\bibfnamefont{I.}~\bibnamefont{Gonzalez-Vallejo}},
  \bibinfo{author}{\bibfnamefont{V.~L.~R.} \bibnamefont{Jacques}},
  \bibinfo{author}{\bibfnamefont{A.~A.} \bibnamefont{Sinchenko}},
  \bibinfo{author}{\bibfnamefont{A.~P.} \bibnamefont{Orlov}},
  \bibinfo{author}{\bibfnamefont{P.}~\bibnamefont{Monceau}},
  \bibinfo{author}{\bibfnamefont{S.~J.} \bibnamefont{Leake}}, \bibnamefont{and}
  \bibinfo{author}{\bibfnamefont{D.}~\bibnamefont{Le~Bolloc'h}},
  \bibinfo{journal}{Phys. Rev. B} \textbf{\bibinfo{volume}{101}},
  \bibinfo{pages}{125122} (\bibinfo{year}{2020}).

\bibitem[{\citenamefont{Bertinshaw et~al.}(2019)\citenamefont{Bertinshaw,
  Gurung, Jorba, Liu, Schmid, Mantadakis, Daghofer, Krautloher, Jain, Ryu
  et~al.}}]{Bertinshaw:2019aa}
\bibinfo{author}{\bibfnamefont{J.}~\bibnamefont{Bertinshaw}},
  \bibinfo{author}{\bibfnamefont{N.}~\bibnamefont{Gurung}},
  \bibinfo{author}{\bibfnamefont{P.}~\bibnamefont{Jorba}},
  \bibinfo{author}{\bibfnamefont{H.}~\bibnamefont{Liu}},
  \bibinfo{author}{\bibfnamefont{M.}~\bibnamefont{Schmid}},
  \bibinfo{author}{\bibfnamefont{D.~T.} \bibnamefont{Mantadakis}},
  \bibinfo{author}{\bibfnamefont{M.}~\bibnamefont{Daghofer}},
  \bibinfo{author}{\bibfnamefont{M.}~\bibnamefont{Krautloher}},
  \bibinfo{author}{\bibfnamefont{A.}~\bibnamefont{Jain}},
  \bibinfo{author}{\bibfnamefont{G.~H.} \bibnamefont{Ryu}},
  \bibnamefont{et~al.}, \bibinfo{journal}{Physical Review Letters}
  \textbf{\bibinfo{volume}{123}} (\bibinfo{year}{2019}).

\bibitem[{\citenamefont{Cirillo et~al.}(2019)\citenamefont{Cirillo, Granata,
  Avallone, Fittipaldi, Attanasio, Avella, and Vecchione}}]{Cirillo:2019aa}
\bibinfo{author}{\bibfnamefont{C.}~\bibnamefont{Cirillo}},
  \bibinfo{author}{\bibfnamefont{V.}~\bibnamefont{Granata}},
  \bibinfo{author}{\bibfnamefont{G.}~\bibnamefont{Avallone}},
  \bibinfo{author}{\bibfnamefont{R.}~\bibnamefont{Fittipaldi}},
  \bibinfo{author}{\bibfnamefont{C.}~\bibnamefont{Attanasio}},
  \bibinfo{author}{\bibfnamefont{A.}~\bibnamefont{Avella}}, \bibnamefont{and}
  \bibinfo{author}{\bibfnamefont{A.}~\bibnamefont{Vecchione}},
  \bibinfo{journal}{Physical Review B} \textbf{\bibinfo{volume}{100}}
  (\bibinfo{year}{2019}).

\bibitem[{\citenamefont{Zhang et~al.}(2019)\citenamefont{Zhang, McLeod, Han,
  Chen, Bechtel, Yao, Gilbert~Corder, Ciavatti, Tao, Aronson
  et~al.}}]{Zhang:2019ac}
\bibinfo{author}{\bibfnamefont{J.}~\bibnamefont{Zhang}},
  \bibinfo{author}{\bibfnamefont{A.~S.} \bibnamefont{McLeod}},
  \bibinfo{author}{\bibfnamefont{Q.}~\bibnamefont{Han}},
  \bibinfo{author}{\bibfnamefont{X.}~\bibnamefont{Chen}},
  \bibinfo{author}{\bibfnamefont{H.~A.} \bibnamefont{Bechtel}},
  \bibinfo{author}{\bibfnamefont{Z.}~\bibnamefont{Yao}},
  \bibinfo{author}{\bibfnamefont{S.~N.} \bibnamefont{Gilbert~Corder}},
  \bibinfo{author}{\bibfnamefont{T.}~\bibnamefont{Ciavatti}},
  \bibinfo{author}{\bibfnamefont{T.~H.} \bibnamefont{Tao}},
  \bibinfo{author}{\bibfnamefont{M.}~\bibnamefont{Aronson}},
  \bibnamefont{et~al.}, \bibinfo{journal}{Physical Review X}
  \textbf{\bibinfo{volume}{9}} (\bibinfo{year}{2019}).

\bibitem[{\citenamefont{Mattoni et~al.}(2020)\citenamefont{Mattoni, Yonezawa,
  Nakamura, and Maeno}}]{Mattoni:2020aa}
\bibinfo{author}{\bibfnamefont{G.}~\bibnamefont{Mattoni}},
  \bibinfo{author}{\bibfnamefont{S.}~\bibnamefont{Yonezawa}},
  \bibinfo{author}{\bibfnamefont{F.}~\bibnamefont{Nakamura}}, \bibnamefont{and}
  \bibinfo{author}{\bibfnamefont{Y.}~\bibnamefont{Maeno}},
  \bibinfo{journal}{arXiv}  (\bibinfo{year}{2020}), \eprint{arXiv.2007.06885}.

\bibitem[{\citenamefont{Nakatsuji and Maeno}(2000)}]{Nakatsuji:2000aa}
\bibinfo{author}{\bibfnamefont{S.}~\bibnamefont{Nakatsuji}} \bibnamefont{and}
  \bibinfo{author}{\bibfnamefont{Y.}~\bibnamefont{Maeno}},
  \bibinfo{journal}{Phys. Rev. Lett.} \textbf{\bibinfo{volume}{84}},
  \bibinfo{pages}{2666} (\bibinfo{year}{2000}).

\bibitem[{\citenamefont{Nakamura et~al.}(2002)\citenamefont{Nakamura, Goko,
  Ito, Fujita, Nakatsuji, Fukazawa, Maeno, Alireza, Forsythe, and
  Julian}}]{Nakamura:2002aa}
\bibinfo{author}{\bibfnamefont{F.}~\bibnamefont{Nakamura}},
  \bibinfo{author}{\bibfnamefont{T.}~\bibnamefont{Goko}},
  \bibinfo{author}{\bibfnamefont{M.}~\bibnamefont{Ito}},
  \bibinfo{author}{\bibfnamefont{T.}~\bibnamefont{Fujita}},
  \bibinfo{author}{\bibfnamefont{S.}~\bibnamefont{Nakatsuji}},
  \bibinfo{author}{\bibfnamefont{H.}~\bibnamefont{Fukazawa}},
  \bibinfo{author}{\bibfnamefont{Y.}~\bibnamefont{Maeno}},
  \bibinfo{author}{\bibfnamefont{P.}~\bibnamefont{Alireza}},
  \bibinfo{author}{\bibfnamefont{D.}~\bibnamefont{Forsythe}}, \bibnamefont{and}
  \bibinfo{author}{\bibfnamefont{S.~R.} \bibnamefont{Julian}},
  \bibinfo{journal}{Phys. Rev. B} \textbf{\bibinfo{volume}{65}},
  \bibinfo{pages}{220402} (\bibinfo{year}{2002}).

\bibitem[{\citenamefont{Ricc{\`o} et~al.}(2018)\citenamefont{Ricc{\`o}, Kim,
  Tamai, McKeown~Walker, Bruno, Cucchi, Cappelli, Besnard, Kim, Dudin
  et~al.}}]{Ricco:2018aa}
\bibinfo{author}{\bibfnamefont{S.}~\bibnamefont{Ricc{\`o}}},
  \bibinfo{author}{\bibfnamefont{M.}~\bibnamefont{Kim}},
  \bibinfo{author}{\bibfnamefont{A.}~\bibnamefont{Tamai}},
  \bibinfo{author}{\bibfnamefont{S.}~\bibnamefont{McKeown~Walker}},
  \bibinfo{author}{\bibfnamefont{F.~Y.} \bibnamefont{Bruno}},
  \bibinfo{author}{\bibfnamefont{I.}~\bibnamefont{Cucchi}},
  \bibinfo{author}{\bibfnamefont{E.}~\bibnamefont{Cappelli}},
  \bibinfo{author}{\bibfnamefont{C.}~\bibnamefont{Besnard}},
  \bibinfo{author}{\bibfnamefont{T.~K.} \bibnamefont{Kim}},
  \bibinfo{author}{\bibfnamefont{P.}~\bibnamefont{Dudin}},
  \bibnamefont{et~al.}, \bibinfo{journal}{Nature Communications}
  \textbf{\bibinfo{volume}{9}}, \bibinfo{pages}{4535} (\bibinfo{year}{2018}).

\bibitem[{\citenamefont{Sunko et~al.}(2019)\citenamefont{Sunko, Morales,
  Markovi{\'{c}}, Barber, Milosavljevi{\'{c}}, Mazzola, Sokolov, Kikugawa,
  Cacho, Dudin et~al.}}]{Sunko:2019aa}
\bibinfo{author}{\bibfnamefont{V.}~\bibnamefont{Sunko}},
  \bibinfo{author}{\bibfnamefont{E.~A.} \bibnamefont{Morales}},
  \bibinfo{author}{\bibfnamefont{I.}~\bibnamefont{Markovi{\'{c}}}},
  \bibinfo{author}{\bibfnamefont{M.~E.} \bibnamefont{Barber}},
  \bibinfo{author}{\bibfnamefont{D.}~\bibnamefont{Milosavljevi{\'{c}}}},
  \bibinfo{author}{\bibfnamefont{F.}~\bibnamefont{Mazzola}},
  \bibinfo{author}{\bibfnamefont{D.~A.} \bibnamefont{Sokolov}},
  \bibinfo{author}{\bibfnamefont{N.}~\bibnamefont{Kikugawa}},
  \bibinfo{author}{\bibfnamefont{C.}~\bibnamefont{Cacho}},
  \bibinfo{author}{\bibfnamefont{P.}~\bibnamefont{Dudin}},
  \bibnamefont{et~al.}, \bibinfo{journal}{npj Quantum Materials}
  \textbf{\bibinfo{volume}{4}} (\bibinfo{year}{2019}).

\bibitem[{\citenamefont{Ashcroft and Mermin}(1976)}]{Ashcroft:1976aa}
\bibinfo{author}{\bibfnamefont{N.~E.} \bibnamefont{Ashcroft}} \bibnamefont{and}
  \bibinfo{author}{\bibfnamefont{N.~D.} \bibnamefont{Mermin}},
  \emph{\bibinfo{title}{Solid state physics}} (\bibinfo{publisher}{Saunders
  College}, \bibinfo{address}{Philadelphia}, \bibinfo{year}{1976}),
  \bibinfo{edition}{international} ed.

\bibitem[{\citenamefont{Fl{\"o}totto et~al.}(2018)\citenamefont{Fl{\"o}totto,
  Bai, Chan, Chen, Wang, Rossi, Xu, Zhang, Hlevyack, Denlinger
  et~al.}}]{Flototto:2018aa}
\bibinfo{author}{\bibfnamefont{D.}~\bibnamefont{Fl{\"o}totto}},
  \bibinfo{author}{\bibfnamefont{Y.}~\bibnamefont{Bai}},
  \bibinfo{author}{\bibfnamefont{Y.-H.} \bibnamefont{Chan}},
  \bibinfo{author}{\bibfnamefont{P.}~\bibnamefont{Chen}},
  \bibinfo{author}{\bibfnamefont{X.}~\bibnamefont{Wang}},
  \bibinfo{author}{\bibfnamefont{P.}~\bibnamefont{Rossi}},
  \bibinfo{author}{\bibfnamefont{C.-Z.} \bibnamefont{Xu}},
  \bibinfo{author}{\bibfnamefont{C.}~\bibnamefont{Zhang}},
  \bibinfo{author}{\bibfnamefont{J.~A.} \bibnamefont{Hlevyack}},
  \bibinfo{author}{\bibfnamefont{J.~D.} \bibnamefont{Denlinger}},
  \bibnamefont{et~al.}, \bibinfo{journal}{Nano Letters}
  \textbf{\bibinfo{volume}{18}}, \bibinfo{pages}{5628} (\bibinfo{year}{2018}).

\bibitem[{\citenamefont{Lin et~al.}(2020)\citenamefont{Lin, Ochi, Noguchi,
  Kuroda, Sakoda, Nomura, Tsubota, Zhang, Bareille, Kurokawa
  et~al.}}]{Lin:2020aa}
\bibinfo{author}{\bibfnamefont{C.}~\bibnamefont{Lin}},
  \bibinfo{author}{\bibfnamefont{M.}~\bibnamefont{Ochi}},
  \bibinfo{author}{\bibfnamefont{R.}~\bibnamefont{Noguchi}},
  \bibinfo{author}{\bibfnamefont{K.}~\bibnamefont{Kuroda}},
  \bibinfo{author}{\bibfnamefont{M.}~\bibnamefont{Sakoda}},
  \bibinfo{author}{\bibfnamefont{A.}~\bibnamefont{Nomura}},
  \bibinfo{author}{\bibfnamefont{M.}~\bibnamefont{Tsubota}},
  \bibinfo{author}{\bibfnamefont{P.}~\bibnamefont{Zhang}},
  \bibinfo{author}{\bibfnamefont{C.}~\bibnamefont{Bareille}},
  \bibinfo{author}{\bibfnamefont{K.}~\bibnamefont{Kurokawa}},
  \bibnamefont{et~al.} (\bibinfo{year}{2020}), \eprint{arXiv.2009.06353}.

\bibitem[{\citenamefont{Noguchi et~al.}(2019)\citenamefont{Noguchi, Takahashi,
  Kuroda, Ochi, Shirasawa, Sakano, Bareille, Nakayama, Watson, Yaji
  et~al.}}]{Noguchi:2019aa}
\bibinfo{author}{\bibfnamefont{R.}~\bibnamefont{Noguchi}},
  \bibinfo{author}{\bibfnamefont{T.}~\bibnamefont{Takahashi}},
  \bibinfo{author}{\bibfnamefont{K.}~\bibnamefont{Kuroda}},
  \bibinfo{author}{\bibfnamefont{M.}~\bibnamefont{Ochi}},
  \bibinfo{author}{\bibfnamefont{T.}~\bibnamefont{Shirasawa}},
  \bibinfo{author}{\bibfnamefont{M.}~\bibnamefont{Sakano}},
  \bibinfo{author}{\bibfnamefont{C.}~\bibnamefont{Bareille}},
  \bibinfo{author}{\bibfnamefont{M.}~\bibnamefont{Nakayama}},
  \bibinfo{author}{\bibfnamefont{M.~D.} \bibnamefont{Watson}},
  \bibinfo{author}{\bibfnamefont{K.}~\bibnamefont{Yaji}}, \bibnamefont{et~al.},
  \bibinfo{journal}{Nature} \textbf{\bibinfo{volume}{566}},
  \bibinfo{pages}{518} (\bibinfo{year}{2019}).

\bibitem[{\citenamefont{Kastl et~al.}(2019)\citenamefont{Kastl, Koch, Chen,
  Eichhorn, Ulstrup, Bostwick, Jozwiak, Kuykendall, Borys, Toma
  et~al.}}]{Kastl:2019aa}
\bibinfo{author}{\bibfnamefont{C.}~\bibnamefont{Kastl}},
  \bibinfo{author}{\bibfnamefont{R.~J.} \bibnamefont{Koch}},
  \bibinfo{author}{\bibfnamefont{C.~T.} \bibnamefont{Chen}},
  \bibinfo{author}{\bibfnamefont{J.}~\bibnamefont{Eichhorn}},
  \bibinfo{author}{\bibfnamefont{S.}~\bibnamefont{Ulstrup}},
  \bibinfo{author}{\bibfnamefont{A.}~\bibnamefont{Bostwick}},
  \bibinfo{author}{\bibfnamefont{C.}~\bibnamefont{Jozwiak}},
  \bibinfo{author}{\bibfnamefont{T.~R.} \bibnamefont{Kuykendall}},
  \bibinfo{author}{\bibfnamefont{N.~J.} \bibnamefont{Borys}},
  \bibinfo{author}{\bibfnamefont{F.~M.} \bibnamefont{Toma}},
  \bibnamefont{et~al.}, \bibinfo{journal}{{ACS} Nano}
  \textbf{\bibinfo{volume}{13}}, \bibinfo{pages}{1284} (\bibinfo{year}{2019}).

\bibitem[{\citenamefont{Ulstrup
  et~al.}(2019{\natexlab{a}})\citenamefont{Ulstrup, Giusca, Miwa, Sanders,
  Browning, Dudin, Cacho, Kazakova, Gaskill, Myers-Ward
  et~al.}}]{Ulstrup:2019aa}
\bibinfo{author}{\bibfnamefont{S.}~\bibnamefont{Ulstrup}},
  \bibinfo{author}{\bibfnamefont{C.~E.} \bibnamefont{Giusca}},
  \bibinfo{author}{\bibfnamefont{J.~A.} \bibnamefont{Miwa}},
  \bibinfo{author}{\bibfnamefont{C.~E.} \bibnamefont{Sanders}},
  \bibinfo{author}{\bibfnamefont{A.}~\bibnamefont{Browning}},
  \bibinfo{author}{\bibfnamefont{P.}~\bibnamefont{Dudin}},
  \bibinfo{author}{\bibfnamefont{C.}~\bibnamefont{Cacho}},
  \bibinfo{author}{\bibfnamefont{O.}~\bibnamefont{Kazakova}},
  \bibinfo{author}{\bibfnamefont{D.~K.} \bibnamefont{Gaskill}},
  \bibinfo{author}{\bibfnamefont{R.~L.} \bibnamefont{Myers-Ward}},
  \bibnamefont{et~al.}, \bibinfo{journal}{Nature Communications}
  \textbf{\bibinfo{volume}{10}}, \bibinfo{pages}{3283}
  (\bibinfo{year}{2019}{\natexlab{a}}).

\bibitem[{\citenamefont{Utama et~al.}(2020)\citenamefont{Utama, Koch, Lee,
  Leconte, Li, Zhao, Jiang, Zhu, Watanabe, Taniguchi et~al.}}]{Utama:2020aa}
\bibinfo{author}{\bibfnamefont{M.~I.~B.} \bibnamefont{Utama}},
  \bibinfo{author}{\bibfnamefont{R.~J.} \bibnamefont{Koch}},
  \bibinfo{author}{\bibfnamefont{K.}~\bibnamefont{Lee}},
  \bibinfo{author}{\bibfnamefont{N.}~\bibnamefont{Leconte}},
  \bibinfo{author}{\bibfnamefont{H.}~\bibnamefont{Li}},
  \bibinfo{author}{\bibfnamefont{S.}~\bibnamefont{Zhao}},
  \bibinfo{author}{\bibfnamefont{L.}~\bibnamefont{Jiang}},
  \bibinfo{author}{\bibfnamefont{J.}~\bibnamefont{Zhu}},
  \bibinfo{author}{\bibfnamefont{K.}~\bibnamefont{Watanabe}},
  \bibinfo{author}{\bibfnamefont{T.}~\bibnamefont{Taniguchi}},
  \bibnamefont{et~al.}, \bibinfo{journal}{Nature Physics}
  (\bibinfo{year}{2020}).

\bibitem[{\citenamefont{Lisi et~al.}(2020)\citenamefont{Lisi, Lu, Benschop,
  de~Jong, Stepanov, Duran, Margot, Cucchi, Cappelli, Hunter
  et~al.}}]{Lisi:2020aa}
\bibinfo{author}{\bibfnamefont{S.}~\bibnamefont{Lisi}},
  \bibinfo{author}{\bibfnamefont{X.}~\bibnamefont{Lu}},
  \bibinfo{author}{\bibfnamefont{T.}~\bibnamefont{Benschop}},
  \bibinfo{author}{\bibfnamefont{T.~A.} \bibnamefont{de~Jong}},
  \bibinfo{author}{\bibfnamefont{P.}~\bibnamefont{Stepanov}},
  \bibinfo{author}{\bibfnamefont{J.~R.} \bibnamefont{Duran}},
  \bibinfo{author}{\bibfnamefont{F.}~\bibnamefont{Margot}},
  \bibinfo{author}{\bibfnamefont{I.}~\bibnamefont{Cucchi}},
  \bibinfo{author}{\bibfnamefont{E.}~\bibnamefont{Cappelli}},
  \bibinfo{author}{\bibfnamefont{A.}~\bibnamefont{Hunter}},
  \bibnamefont{et~al.}, \bibinfo{journal}{Nature Physics}
  (\bibinfo{year}{2020}).

\bibitem[{\citenamefont{Ulstrup et~al.}(2020)\citenamefont{Ulstrup, Koch,
  Singh, McCreary, Jonker, Robinson, Jozwiak, Rotenberg, Bostwick, Katoch
  et~al.}}]{Ulstrup:2020aa}
\bibinfo{author}{\bibfnamefont{S.}~\bibnamefont{Ulstrup}},
  \bibinfo{author}{\bibfnamefont{R.~J.} \bibnamefont{Koch}},
  \bibinfo{author}{\bibfnamefont{S.}~\bibnamefont{Singh}},
  \bibinfo{author}{\bibfnamefont{K.~M.} \bibnamefont{McCreary}},
  \bibinfo{author}{\bibfnamefont{B.~T.} \bibnamefont{Jonker}},
  \bibinfo{author}{\bibfnamefont{J.~T.} \bibnamefont{Robinson}},
  \bibinfo{author}{\bibfnamefont{C.}~\bibnamefont{Jozwiak}},
  \bibinfo{author}{\bibfnamefont{E.}~\bibnamefont{Rotenberg}},
  \bibinfo{author}{\bibfnamefont{A.}~\bibnamefont{Bostwick}},
  \bibinfo{author}{\bibfnamefont{J.}~\bibnamefont{Katoch}},
  \bibnamefont{et~al.}, \bibinfo{journal}{Science Advances}
  \textbf{\bibinfo{volume}{6}}, \bibinfo{pages}{eaay6104}
  (\bibinfo{year}{2020}).

\bibitem[{\citenamefont{Klitzing et~al.}(1980)\citenamefont{Klitzing, Dorda,
  and Pepper}}]{Klitzing:1980aa}
\bibinfo{author}{\bibfnamefont{K.~v.} \bibnamefont{Klitzing}},
  \bibinfo{author}{\bibfnamefont{G.}~\bibnamefont{Dorda}}, \bibnamefont{and}
  \bibinfo{author}{\bibfnamefont{M.}~\bibnamefont{Pepper}},
  \bibinfo{journal}{Phys. Rev. Lett.} \textbf{\bibinfo{volume}{45}},
  \bibinfo{pages}{494} (\bibinfo{year}{1980}).

\bibitem[{\citenamefont{Tsui et~al.}(1982)\citenamefont{Tsui, Stormer, and
  Gossard}}]{Tsui:1982aa}
\bibinfo{author}{\bibfnamefont{D.~C.} \bibnamefont{Tsui}},
  \bibinfo{author}{\bibfnamefont{H.~L.} \bibnamefont{Stormer}},
  \bibnamefont{and} \bibinfo{author}{\bibfnamefont{A.~C.}
  \bibnamefont{Gossard}}, \bibinfo{journal}{Phys. Rev. Lett.}
  \textbf{\bibinfo{volume}{48}}, \bibinfo{pages}{1559} (\bibinfo{year}{1982}).

\bibitem[{\citenamefont{Konig et~al.}(2007)\citenamefont{Konig, Wiedmann,
  Brune, Roth, Buhmann, Molenkamp, Qi, and Zhang}}]{Konig:2007aa}
\bibinfo{author}{\bibfnamefont{M.}~\bibnamefont{Konig}},
  \bibinfo{author}{\bibfnamefont{S.}~\bibnamefont{Wiedmann}},
  \bibinfo{author}{\bibfnamefont{C.}~\bibnamefont{Brune}},
  \bibinfo{author}{\bibfnamefont{A.}~\bibnamefont{Roth}},
  \bibinfo{author}{\bibfnamefont{H.}~\bibnamefont{Buhmann}},
  \bibinfo{author}{\bibfnamefont{L.~W.} \bibnamefont{Molenkamp}},
  \bibinfo{author}{\bibfnamefont{X.-L.} \bibnamefont{Qi}}, \bibnamefont{and}
  \bibinfo{author}{\bibfnamefont{S.-C.} \bibnamefont{Zhang}},
  \bibinfo{journal}{Science} \textbf{\bibinfo{volume}{318}},
  \bibinfo{pages}{766} (\bibinfo{year}{2007}).

\bibitem[{\citenamefont{Bednorz and M{\"u}ller}(1986)}]{Bednorz:1986aa}
\bibinfo{author}{\bibfnamefont{J.~G.} \bibnamefont{Bednorz}} \bibnamefont{and}
  \bibinfo{author}{\bibfnamefont{K.~A.} \bibnamefont{M{\"u}ller}},
  \bibinfo{journal}{Zeitschrift f{\"u}r Physik B Condensed Matter}
  \textbf{\bibinfo{volume}{64}}, \bibinfo{pages}{189} (\bibinfo{year}{1986}).

\bibitem[{\citenamefont{Gr{{\"u}}ner}(1994)}]{Gruner:1994aa}
\bibinfo{author}{\bibfnamefont{G.}~\bibnamefont{Gr{{\"u}}ner}},
  \emph{\bibinfo{title}{Density waves in solids}}, vol.~\bibinfo{volume}{89} of
  \emph{\bibinfo{series}{Frontiers in physics}} (\bibinfo{publisher}{Perseus
  Publishing}, \bibinfo{address}{Cambridge, Massachusetts},
  \bibinfo{year}{1994}).

\bibitem[{\citenamefont{King et~al.}(2010)\citenamefont{King, Veal, McConville,
  Z\'u\~niga P\'erez, Mu\~noz Sanjos\'e, Hopkinson, Rienks, Jensen, and
  Hofmann}}]{King:2010aa}
\bibinfo{author}{\bibfnamefont{P.~D.~C.} \bibnamefont{King}},
  \bibinfo{author}{\bibfnamefont{T.~D.} \bibnamefont{Veal}},
  \bibinfo{author}{\bibfnamefont{C.~F.} \bibnamefont{McConville}},
  \bibinfo{author}{\bibfnamefont{J.}~\bibnamefont{Z\'u\~niga P\'erez}},
  \bibinfo{author}{\bibfnamefont{V.}~\bibnamefont{Mu\~noz Sanjos\'e}},
  \bibinfo{author}{\bibfnamefont{M.}~\bibnamefont{Hopkinson}},
  \bibinfo{author}{\bibfnamefont{E.~D.~L.} \bibnamefont{Rienks}},
  \bibinfo{author}{\bibfnamefont{M.~F.} \bibnamefont{Jensen}},
  \bibnamefont{and} \bibinfo{author}{\bibfnamefont{P.}~\bibnamefont{Hofmann}},
  \bibinfo{journal}{Phys. Rev. Lett.} \textbf{\bibinfo{volume}{104}},
  \bibinfo{pages}{256803} (\bibinfo{year}{2010}).

\bibitem[{\citenamefont{Bianchi et~al.}(2010)\citenamefont{Bianchi, Guan, Bao,
  Mi, Iversen, King, and Hofmann}}]{Bianchi:2010ab}
\bibinfo{author}{\bibfnamefont{M.}~\bibnamefont{Bianchi}},
  \bibinfo{author}{\bibfnamefont{D.}~\bibnamefont{Guan}},
  \bibinfo{author}{\bibfnamefont{S.}~\bibnamefont{Bao}},
  \bibinfo{author}{\bibfnamefont{J.}~\bibnamefont{Mi}},
  \bibinfo{author}{\bibfnamefont{B.~B.} \bibnamefont{Iversen}},
  \bibinfo{author}{\bibfnamefont{P.~D.~C.} \bibnamefont{King}},
  \bibnamefont{and} \bibinfo{author}{\bibfnamefont{P.}~\bibnamefont{Hofmann}},
  \bibinfo{journal}{Nature Communications} \textbf{\bibinfo{volume}{1}},
  \bibinfo{pages}{128} (\bibinfo{year}{2010}).

\bibitem[{\citenamefont{Santander-Syro
  et~al.}(2011)\citenamefont{Santander-Syro, Copie, Kondo, Fortuna,
  Pailh{\`e}s, Weht, Qiu, Bertran, Nicolaou, Taleb-Ibrahimi
  et~al.}}]{SantanderSyro:2011aa}
\bibinfo{author}{\bibfnamefont{A.~F.} \bibnamefont{Santander-Syro}},
  \bibinfo{author}{\bibfnamefont{O.}~\bibnamefont{Copie}},
  \bibinfo{author}{\bibfnamefont{T.}~\bibnamefont{Kondo}},
  \bibinfo{author}{\bibfnamefont{F.}~\bibnamefont{Fortuna}},
  \bibinfo{author}{\bibfnamefont{S.}~\bibnamefont{Pailh{\`e}s}},
  \bibinfo{author}{\bibfnamefont{R.}~\bibnamefont{Weht}},
  \bibinfo{author}{\bibfnamefont{X.~G.} \bibnamefont{Qiu}},
  \bibinfo{author}{\bibfnamefont{F.}~\bibnamefont{Bertran}},
  \bibinfo{author}{\bibfnamefont{A.}~\bibnamefont{Nicolaou}},
  \bibinfo{author}{\bibfnamefont{A.}~\bibnamefont{Taleb-Ibrahimi}},
  \bibnamefont{et~al.}, \bibinfo{journal}{Nature}
  \textbf{\bibinfo{volume}{469}}, \bibinfo{pages}{189} (\bibinfo{year}{2011}).

\bibitem[{\citenamefont{Meevasana et~al.}(2011)\citenamefont{Meevasana, King,
  He, Mo, Hashimoto, Tamai, Songsiriritthigul, Baumberger, and
  Shen}}]{Meevasana:2011aa}
\bibinfo{author}{\bibfnamefont{W.}~\bibnamefont{Meevasana}},
  \bibinfo{author}{\bibfnamefont{P.~D.~C.} \bibnamefont{King}},
  \bibinfo{author}{\bibfnamefont{R.~H.} \bibnamefont{He}},
  \bibinfo{author}{\bibfnamefont{S.-K.} \bibnamefont{Mo}},
  \bibinfo{author}{\bibfnamefont{M.}~\bibnamefont{Hashimoto}},
  \bibinfo{author}{\bibfnamefont{A.}~\bibnamefont{Tamai}},
  \bibinfo{author}{\bibfnamefont{P.}~\bibnamefont{Songsiriritthigul}},
  \bibinfo{author}{\bibfnamefont{F.}~\bibnamefont{Baumberger}},
  \bibnamefont{and} \bibinfo{author}{\bibfnamefont{Z.-X.} \bibnamefont{Shen}},
  \bibinfo{journal}{Nature Materials} \textbf{\bibinfo{volume}{10}},
  \bibinfo{pages}{114} (\bibinfo{year}{2011}).

\bibitem[{\citenamefont{Kim et~al.}(2015)\citenamefont{Kim, Baik, Ryu, Sohn,
  Park, Park, Denlinger, Yi, Choi, and Kim}}]{Kim:2015aa}
\bibinfo{author}{\bibfnamefont{J.}~\bibnamefont{Kim}},
  \bibinfo{author}{\bibfnamefont{S.~S.} \bibnamefont{Baik}},
  \bibinfo{author}{\bibfnamefont{S.~H.} \bibnamefont{Ryu}},
  \bibinfo{author}{\bibfnamefont{Y.}~\bibnamefont{Sohn}},
  \bibinfo{author}{\bibfnamefont{S.}~\bibnamefont{Park}},
  \bibinfo{author}{\bibfnamefont{B.-G.} \bibnamefont{Park}},
  \bibinfo{author}{\bibfnamefont{J.}~\bibnamefont{Denlinger}},
  \bibinfo{author}{\bibfnamefont{Y.}~\bibnamefont{Yi}},
  \bibinfo{author}{\bibfnamefont{H.~J.} \bibnamefont{Choi}}, \bibnamefont{and}
  \bibinfo{author}{\bibfnamefont{K.~S.} \bibnamefont{Kim}},
  \bibinfo{journal}{Science} \textbf{\bibinfo{volume}{349}},
  \bibinfo{pages}{723} (\bibinfo{year}{2015}).

\bibitem[{\citenamefont{Novoselov et~al.}(2005)\citenamefont{Novoselov, Jiang,
  Schedin, Booth, Khotkevich, Morozov, and Geim}}]{Novoselov:2005ab}
\bibinfo{author}{\bibfnamefont{K.~S.} \bibnamefont{Novoselov}},
  \bibinfo{author}{\bibfnamefont{D.}~\bibnamefont{Jiang}},
  \bibinfo{author}{\bibfnamefont{F.}~\bibnamefont{Schedin}},
  \bibinfo{author}{\bibfnamefont{T.~J.} \bibnamefont{Booth}},
  \bibinfo{author}{\bibfnamefont{V.~V.} \bibnamefont{Khotkevich}},
  \bibinfo{author}{\bibfnamefont{S.~V.} \bibnamefont{Morozov}},
  \bibnamefont{and} \bibinfo{author}{\bibfnamefont{A.~K.} \bibnamefont{Geim}},
  \bibinfo{journal}{Proceedings of the National Academy of Sciences of the
  United States of America} \textbf{\bibinfo{volume}{102}},
  \bibinfo{pages}{10451} (\bibinfo{year}{2005}).

\bibitem[{\citenamefont{Novoselov et~al.}(2004)\citenamefont{Novoselov, Geim,
  Morozov, Jiang, Zhang, Dubonos, Grigorieva, and Firsov}}]{Novoselov:2004aa}
\bibinfo{author}{\bibfnamefont{K.~S.} \bibnamefont{Novoselov}},
  \bibinfo{author}{\bibfnamefont{A.~K.} \bibnamefont{Geim}},
  \bibinfo{author}{\bibfnamefont{S.~V.} \bibnamefont{Morozov}},
  \bibinfo{author}{\bibfnamefont{D.}~\bibnamefont{Jiang}},
  \bibinfo{author}{\bibfnamefont{Y.}~\bibnamefont{Zhang}},
  \bibinfo{author}{\bibfnamefont{S.~V.} \bibnamefont{Dubonos}},
  \bibinfo{author}{\bibfnamefont{I.~V.} \bibnamefont{Grigorieva}},
  \bibnamefont{and} \bibinfo{author}{\bibfnamefont{A.~A.}
  \bibnamefont{Firsov}}, \bibinfo{journal}{Science}
  \textbf{\bibinfo{volume}{306}}, \bibinfo{pages}{666} (\bibinfo{year}{2004}).

\bibitem[{\citenamefont{Zhang et~al.}(2005)\citenamefont{Zhang, Tan, Stormer,
  and Kim}}]{Zhang:2005ab}
\bibinfo{author}{\bibfnamefont{Y.~B.} \bibnamefont{Zhang}},
  \bibinfo{author}{\bibfnamefont{Y.~W.} \bibnamefont{Tan}},
  \bibinfo{author}{\bibfnamefont{H.~L.} \bibnamefont{Stormer}},
  \bibnamefont{and} \bibinfo{author}{\bibfnamefont{P.}~\bibnamefont{Kim}},
  \bibinfo{journal}{Nature} \textbf{\bibinfo{volume}{438}},
  \bibinfo{pages}{201} (\bibinfo{year}{2005}).

\bibitem[{\citenamefont{Neto et~al.}(2009)\citenamefont{Neto, Guinea, Peres,
  Novoselov, and Geim}}]{Neto:2009aa}
\bibinfo{author}{\bibfnamefont{A.~H.~C.} \bibnamefont{Neto}},
  \bibinfo{author}{\bibfnamefont{F.}~\bibnamefont{Guinea}},
  \bibinfo{author}{\bibfnamefont{N.~M.~R.} \bibnamefont{Peres}},
  \bibinfo{author}{\bibfnamefont{K.~S.} \bibnamefont{Novoselov}},
  \bibnamefont{and} \bibinfo{author}{\bibfnamefont{A.~K.} \bibnamefont{Geim}},
  \bibinfo{journal}{Reviews of Modern Physics} \textbf{\bibinfo{volume}{81}},
  \bibinfo{eid}{109} (\bibinfo{year}{2009}).

\bibitem[{\citenamefont{Kara et~al.}(2012)\citenamefont{Kara, Enriquez,
  Seitsonen, {Lew Yan Voon}, Vizzini, Aufray, and Oughaddou}}]{Kara:2012aa}
\bibinfo{author}{\bibfnamefont{A.}~\bibnamefont{Kara}},
  \bibinfo{author}{\bibfnamefont{H.}~\bibnamefont{Enriquez}},
  \bibinfo{author}{\bibfnamefont{A.~P.} \bibnamefont{Seitsonen}},
  \bibinfo{author}{\bibfnamefont{L.}~\bibnamefont{{Lew Yan Voon}}},
  \bibinfo{author}{\bibfnamefont{S.}~\bibnamefont{Vizzini}},
  \bibinfo{author}{\bibfnamefont{B.}~\bibnamefont{Aufray}}, \bibnamefont{and}
  \bibinfo{author}{\bibfnamefont{H.}~\bibnamefont{Oughaddou}},
  \bibinfo{journal}{Surface Science Reports} \textbf{\bibinfo{volume}{67}},
  \bibinfo{pages}{1 } (\bibinfo{year}{2012}).

\bibitem[{\citenamefont{Basov et~al.}(2014)\citenamefont{Basov, Fogler,
  Lanzara, Wang, and Zhang}}]{Basov:2014aa}
\bibinfo{author}{\bibfnamefont{D.~N.} \bibnamefont{Basov}},
  \bibinfo{author}{\bibfnamefont{M.~M.} \bibnamefont{Fogler}},
  \bibinfo{author}{\bibfnamefont{A.}~\bibnamefont{Lanzara}},
  \bibinfo{author}{\bibfnamefont{F.}~\bibnamefont{Wang}}, \bibnamefont{and}
  \bibinfo{author}{\bibfnamefont{Y.}~\bibnamefont{Zhang}},
  \bibinfo{journal}{Rev. Mod. Phys.} \textbf{\bibinfo{volume}{86}},
  \bibinfo{pages}{959} (\bibinfo{year}{2014}).

\bibitem[{\citenamefont{Das et~al.}(2015)\citenamefont{Das, Robinson, Dubey,
  Terrones, and Terrones}}]{Das:2015ab}
\bibinfo{author}{\bibfnamefont{S.}~\bibnamefont{Das}},
  \bibinfo{author}{\bibfnamefont{J.~A.} \bibnamefont{Robinson}},
  \bibinfo{author}{\bibfnamefont{M.}~\bibnamefont{Dubey}},
  \bibinfo{author}{\bibfnamefont{H.}~\bibnamefont{Terrones}}, \bibnamefont{and}
  \bibinfo{author}{\bibfnamefont{M.}~\bibnamefont{Terrones}},
  \bibinfo{journal}{Annual Review of Materials Research}
  \textbf{\bibinfo{volume}{45}}, \bibinfo{pages}{1} (\bibinfo{year}{2015}).

\bibitem[{\citenamefont{Avouris et~al.}(2017)\citenamefont{Avouris, Heinz, and
  Low}}]{Avouris:2017aa}
\bibinfo{editor}{\bibfnamefont{P.}~\bibnamefont{Avouris}},
  \bibinfo{editor}{\bibfnamefont{T.~F.} \bibnamefont{Heinz}}, \bibnamefont{and}
  \bibinfo{editor}{\bibfnamefont{T.}~\bibnamefont{Low}}, eds.,
  \emph{\bibinfo{title}{2D Materials: Properties and Devices}}
  (\bibinfo{publisher}{Cambridge University Press}, \bibinfo{year}{2017}).

\bibitem[{\citenamefont{Manzeli et~al.}(2017)\citenamefont{Manzeli,
  Ovchinnikov, Pasquier, Yazyev, and Kis}}]{Manzeli:2017aa}
\bibinfo{author}{\bibfnamefont{S.}~\bibnamefont{Manzeli}},
  \bibinfo{author}{\bibfnamefont{D.}~\bibnamefont{Ovchinnikov}},
  \bibinfo{author}{\bibfnamefont{D.}~\bibnamefont{Pasquier}},
  \bibinfo{author}{\bibfnamefont{O.~V.} \bibnamefont{Yazyev}},
  \bibnamefont{and} \bibinfo{author}{\bibfnamefont{A.}~\bibnamefont{Kis}},
  \bibinfo{journal}{Nature Reviews Materials} \textbf{\bibinfo{volume}{2}},
  \bibinfo{pages}{17033} (\bibinfo{year}{2017}).

\bibitem[{\citenamefont{Dong and Kuljanishvili}(2017)}]{Dong:2017aa}
\bibinfo{author}{\bibfnamefont{R.}~\bibnamefont{Dong}} \bibnamefont{and}
  \bibinfo{author}{\bibfnamefont{I.}~\bibnamefont{Kuljanishvili}},
  \bibinfo{journal}{Journal of Vacuum Science and Technology B, Nanotechnology
  and Microelectronics: Materials, Processing, Measurement, and Phenomena}
  \textbf{\bibinfo{volume}{35}}, \bibinfo{pages}{030803}
  (\bibinfo{year}{2017}).

\bibitem[{\citenamefont{Zhao and Jia}(2020)}]{Zhao:2020aa}
\bibinfo{author}{\bibfnamefont{C.-X.} \bibnamefont{Zhao}} \bibnamefont{and}
  \bibinfo{author}{\bibfnamefont{J.-F.} \bibnamefont{Jia}},
  \bibinfo{journal}{Frontiers of Physics} \textbf{\bibinfo{volume}{15}}
  (\bibinfo{year}{2020}).

\bibitem[{\citenamefont{Andrei and MacDonald}(2020)}]{Andrei:2020aa}
\bibinfo{author}{\bibfnamefont{E.~Y.} \bibnamefont{Andrei}} \bibnamefont{and}
  \bibinfo{author}{\bibfnamefont{A.~H.} \bibnamefont{MacDonald}}
  (\bibinfo{year}{2020}), \eprint{arXiv.2008.08129}.

\bibitem[{\citenamefont{Arnold et~al.}(2018)\citenamefont{Arnold, Stan,
  Mahatha, Lund, Curcio, Dendzik, Bana, Travaglia, Bignardi, Lacovig
  et~al.}}]{Arnold:2018ab}
\bibinfo{author}{\bibfnamefont{F.}~\bibnamefont{Arnold}},
  \bibinfo{author}{\bibfnamefont{R.-M.} \bibnamefont{Stan}},
  \bibinfo{author}{\bibfnamefont{S.~K.} \bibnamefont{Mahatha}},
  \bibinfo{author}{\bibfnamefont{H.~E.} \bibnamefont{Lund}},
  \bibinfo{author}{\bibfnamefont{D.}~\bibnamefont{Curcio}},
  \bibinfo{author}{\bibfnamefont{M.}~\bibnamefont{Dendzik}},
  \bibinfo{author}{\bibfnamefont{H.}~\bibnamefont{Bana}},
  \bibinfo{author}{\bibfnamefont{E.}~\bibnamefont{Travaglia}},
  \bibinfo{author}{\bibfnamefont{L.}~\bibnamefont{Bignardi}},
  \bibinfo{author}{\bibfnamefont{P.}~\bibnamefont{Lacovig}},
  \bibnamefont{et~al.}, \bibinfo{journal}{2D Materials}
  \textbf{\bibinfo{volume}{5}}, \bibinfo{pages}{045009} (\bibinfo{year}{2018}).

\bibitem[{\citenamefont{Radisavljevic et~al.}(2011)\citenamefont{Radisavljevic,
  Radenovic, Brivio, Giacometti, and Kis}}]{Radisavljevic:2011aa}
\bibinfo{author}{\bibfnamefont{B.}~\bibnamefont{Radisavljevic}},
  \bibinfo{author}{\bibfnamefont{A.}~\bibnamefont{Radenovic}},
  \bibinfo{author}{\bibfnamefont{J.}~\bibnamefont{Brivio}},
  \bibinfo{author}{\bibfnamefont{V.}~\bibnamefont{Giacometti}},
  \bibnamefont{and} \bibinfo{author}{\bibfnamefont{A.}~\bibnamefont{Kis}},
  \bibinfo{journal}{Nature Nanotechnology} \textbf{\bibinfo{volume}{6}},
  \bibinfo{pages}{147} (\bibinfo{year}{2011}).

\bibitem[{\citenamefont{Costanzo et~al.}(2016)\citenamefont{Costanzo, Jo,
  Berger, and Morpurgo}}]{Costanzo:2016aa}
\bibinfo{author}{\bibfnamefont{D.}~\bibnamefont{Costanzo}},
  \bibinfo{author}{\bibfnamefont{S.}~\bibnamefont{Jo}},
  \bibinfo{author}{\bibfnamefont{H.}~\bibnamefont{Berger}}, \bibnamefont{and}
  \bibinfo{author}{\bibfnamefont{A.~F.} \bibnamefont{Morpurgo}},
  \bibinfo{journal}{Nat Nano} \textbf{\bibinfo{volume}{11}},
  \bibinfo{pages}{339} (\bibinfo{year}{2016}).

\bibitem[{\citenamefont{Radisavljevic and Kis}(2013)}]{Radisavljevic:2013aa}
\bibinfo{author}{\bibfnamefont{B.}~\bibnamefont{Radisavljevic}}
  \bibnamefont{and} \bibinfo{author}{\bibfnamefont{A.}~\bibnamefont{Kis}},
  \bibinfo{journal}{Nature Materials} \textbf{\bibinfo{volume}{12}},
  \bibinfo{pages}{815} (\bibinfo{year}{2013}).

\bibitem[{\citenamefont{Dean et~al.}(2010)\citenamefont{Dean, Young, Meric,
  Lee, Wang, Sorgenfrei, Watanabe, Taniguchi, Kim, Shepard
  et~al.}}]{Dean:2010aa}
\bibinfo{author}{\bibfnamefont{C.~R.} \bibnamefont{Dean}},
  \bibinfo{author}{\bibfnamefont{A.~F.} \bibnamefont{Young}},
  \bibinfo{author}{\bibfnamefont{I.}~\bibnamefont{Meric}},
  \bibinfo{author}{\bibfnamefont{C.}~\bibnamefont{Lee}},
  \bibinfo{author}{\bibfnamefont{L.}~\bibnamefont{Wang}},
  \bibinfo{author}{\bibfnamefont{S.}~\bibnamefont{Sorgenfrei}},
  \bibinfo{author}{\bibfnamefont{K.}~\bibnamefont{Watanabe}},
  \bibinfo{author}{\bibfnamefont{T.}~\bibnamefont{Taniguchi}},
  \bibinfo{author}{\bibfnamefont{P.}~\bibnamefont{Kim}},
  \bibinfo{author}{\bibfnamefont{K.~L.} \bibnamefont{Shepard}},
  \bibnamefont{et~al.}, \bibinfo{journal}{Nature Nanotechnology}
  \textbf{\bibinfo{volume}{5}}, \bibinfo{pages}{722} (\bibinfo{year}{2010}).

\bibitem[{\citenamefont{Watanabe et~al.}(2004)\citenamefont{Watanabe,
  Taniguchi, and Kanda}}]{Watanabe:2004aa}
\bibinfo{author}{\bibfnamefont{K.}~\bibnamefont{Watanabe}},
  \bibinfo{author}{\bibfnamefont{T.}~\bibnamefont{Taniguchi}},
  \bibnamefont{and} \bibinfo{author}{\bibfnamefont{H.}~\bibnamefont{Kanda}},
  \bibinfo{journal}{Nature Materials} \textbf{\bibinfo{volume}{3}},
  \bibinfo{pages}{404} (\bibinfo{year}{2004}).

\bibitem[{\citenamefont{Geim and Grigorieva}(2013)}]{Geim:2013aa}
\bibinfo{author}{\bibfnamefont{A.~K.} \bibnamefont{Geim}} \bibnamefont{and}
  \bibinfo{author}{\bibfnamefont{I.~V.} \bibnamefont{Grigorieva}},
  \bibinfo{journal}{Nature} \textbf{\bibinfo{volume}{499}},
  \bibinfo{pages}{419} (\bibinfo{year}{2013}).

\bibitem[{\citenamefont{Novoselov et~al.}(2016)\citenamefont{Novoselov,
  Mishchenko, Carvalho, and Neto}}]{Novoselov:2016ab}
\bibinfo{author}{\bibfnamefont{K.~S.} \bibnamefont{Novoselov}},
  \bibinfo{author}{\bibfnamefont{A.}~\bibnamefont{Mishchenko}},
  \bibinfo{author}{\bibfnamefont{A.}~\bibnamefont{Carvalho}}, \bibnamefont{and}
  \bibinfo{author}{\bibfnamefont{A.~H.~C.} \bibnamefont{Neto}},
  \bibinfo{journal}{Science} \textbf{\bibinfo{volume}{353}},
  \bibinfo{pages}{aac9439} (\bibinfo{year}{2016}).

\bibitem[{\citenamefont{Cheng et~al.}(2014)\citenamefont{Cheng, Li, Zhou, Wang,
  Yin, Jiang, Liu, Chen, Huang, and Duan}}]{Cheng:2014ab}
\bibinfo{author}{\bibfnamefont{R.}~\bibnamefont{Cheng}},
  \bibinfo{author}{\bibfnamefont{D.}~\bibnamefont{Li}},
  \bibinfo{author}{\bibfnamefont{H.}~\bibnamefont{Zhou}},
  \bibinfo{author}{\bibfnamefont{C.}~\bibnamefont{Wang}},
  \bibinfo{author}{\bibfnamefont{A.}~\bibnamefont{Yin}},
  \bibinfo{author}{\bibfnamefont{S.}~\bibnamefont{Jiang}},
  \bibinfo{author}{\bibfnamefont{Y.}~\bibnamefont{Liu}},
  \bibinfo{author}{\bibfnamefont{Y.}~\bibnamefont{Chen}},
  \bibinfo{author}{\bibfnamefont{Y.}~\bibnamefont{Huang}}, \bibnamefont{and}
  \bibinfo{author}{\bibfnamefont{X.}~\bibnamefont{Duan}},
  \bibinfo{journal}{Nano Letters} \textbf{\bibinfo{volume}{14}},
  \bibinfo{pages}{5590} (\bibinfo{year}{2014}).

\bibitem[{\citenamefont{Wilson et~al.}(2017)\citenamefont{Wilson, Nguyen,
  Seyler, Rivera, Marsden, Laker, Constantinescu, Kandyba, Barinov, Hine
  et~al.}}]{Wilson:2017aa}
\bibinfo{author}{\bibfnamefont{N.~R.} \bibnamefont{Wilson}},
  \bibinfo{author}{\bibfnamefont{P.~V.} \bibnamefont{Nguyen}},
  \bibinfo{author}{\bibfnamefont{K.}~\bibnamefont{Seyler}},
  \bibinfo{author}{\bibfnamefont{P.}~\bibnamefont{Rivera}},
  \bibinfo{author}{\bibfnamefont{A.~J.} \bibnamefont{Marsden}},
  \bibinfo{author}{\bibfnamefont{Z.~P.~L.} \bibnamefont{Laker}},
  \bibinfo{author}{\bibfnamefont{G.~C.} \bibnamefont{Constantinescu}},
  \bibinfo{author}{\bibfnamefont{V.}~\bibnamefont{Kandyba}},
  \bibinfo{author}{\bibfnamefont{A.}~\bibnamefont{Barinov}},
  \bibinfo{author}{\bibfnamefont{N.~D.~M.} \bibnamefont{Hine}},
  \bibnamefont{et~al.}, \bibinfo{journal}{Science Advances}
  \textbf{\bibinfo{volume}{3}} (\bibinfo{year}{2017}).

\bibitem[{\citenamefont{Zribi et~al.}(2019)\citenamefont{Zribi, Khalil, Zheng,
  Avila, Pierucci, Brul{\'{e}}, Chaste, Lhuillier, Asensio, Pan
  et~al.}}]{Zribi:2019aa}
\bibinfo{author}{\bibfnamefont{J.}~\bibnamefont{Zribi}},
  \bibinfo{author}{\bibfnamefont{L.}~\bibnamefont{Khalil}},
  \bibinfo{author}{\bibfnamefont{B.}~\bibnamefont{Zheng}},
  \bibinfo{author}{\bibfnamefont{J.}~\bibnamefont{Avila}},
  \bibinfo{author}{\bibfnamefont{D.}~\bibnamefont{Pierucci}},
  \bibinfo{author}{\bibfnamefont{T.}~\bibnamefont{Brul{\'{e}}}},
  \bibinfo{author}{\bibfnamefont{J.}~\bibnamefont{Chaste}},
  \bibinfo{author}{\bibfnamefont{E.}~\bibnamefont{Lhuillier}},
  \bibinfo{author}{\bibfnamefont{M.~C.} \bibnamefont{Asensio}},
  \bibinfo{author}{\bibfnamefont{A.}~\bibnamefont{Pan}}, \bibnamefont{et~al.},
  \bibinfo{journal}{npj 2D Materials and Applications}
  \textbf{\bibinfo{volume}{3}} (\bibinfo{year}{2019}).

\bibitem[{\citenamefont{Avsar et~al.}(2014)\citenamefont{Avsar, Tan,
  Taychatanapat, Balakrishnan, Koon, Yeo, Lahiri, Carvalho, Rodin, O'Farrell
  et~al.}}]{Avsar:2014aa}
\bibinfo{author}{\bibfnamefont{A.}~\bibnamefont{Avsar}},
  \bibinfo{author}{\bibfnamefont{J.~Y.} \bibnamefont{Tan}},
  \bibinfo{author}{\bibfnamefont{T.}~\bibnamefont{Taychatanapat}},
  \bibinfo{author}{\bibfnamefont{J.}~\bibnamefont{Balakrishnan}},
  \bibinfo{author}{\bibfnamefont{G.~K.~W.} \bibnamefont{Koon}},
  \bibinfo{author}{\bibfnamefont{Y.}~\bibnamefont{Yeo}},
  \bibinfo{author}{\bibfnamefont{J.}~\bibnamefont{Lahiri}},
  \bibinfo{author}{\bibfnamefont{A.}~\bibnamefont{Carvalho}},
  \bibinfo{author}{\bibfnamefont{A.~S.} \bibnamefont{Rodin}},
  \bibinfo{author}{\bibfnamefont{E.~C.~T.} \bibnamefont{O'Farrell}},
  \bibnamefont{et~al.}, \bibinfo{journal}{Nature Communications}
  \textbf{\bibinfo{volume}{5}}, \bibinfo{pages}{4875} (\bibinfo{year}{2014}).

\bibitem[{\citenamefont{Li et~al.}(2020)\citenamefont{Li, Jiang, Wang,
  Watanabe, Taniguchi, Shan, and Mak}}]{Li:2020ab}
\bibinfo{author}{\bibfnamefont{L.}~\bibnamefont{Li}},
  \bibinfo{author}{\bibfnamefont{S.}~\bibnamefont{Jiang}},
  \bibinfo{author}{\bibfnamefont{Z.}~\bibnamefont{Wang}},
  \bibinfo{author}{\bibfnamefont{K.}~\bibnamefont{Watanabe}},
  \bibinfo{author}{\bibfnamefont{T.}~\bibnamefont{Taniguchi}},
  \bibinfo{author}{\bibfnamefont{J.}~\bibnamefont{Shan}}, \bibnamefont{and}
  \bibinfo{author}{\bibfnamefont{K.~F.} \bibnamefont{Mak}},
  \bibinfo{journal}{Phys. Rev. Materials} \textbf{\bibinfo{volume}{4}},
  \bibinfo{pages}{104005} (\bibinfo{year}{2020}).

\bibitem[{\citenamefont{L{\"u}pke et~al.}(2020)\citenamefont{L{\"u}pke, Waters,
  de~la Barrera, Widom, Mandrus, Yan, Feenstra, and Hunt}}]{Lupke:2020aa}
\bibinfo{author}{\bibfnamefont{F.}~\bibnamefont{L{\"u}pke}},
  \bibinfo{author}{\bibfnamefont{D.}~\bibnamefont{Waters}},
  \bibinfo{author}{\bibfnamefont{S.~C.} \bibnamefont{de~la Barrera}},
  \bibinfo{author}{\bibfnamefont{M.}~\bibnamefont{Widom}},
  \bibinfo{author}{\bibfnamefont{D.~G.} \bibnamefont{Mandrus}},
  \bibinfo{author}{\bibfnamefont{J.}~\bibnamefont{Yan}},
  \bibinfo{author}{\bibfnamefont{R.~M.} \bibnamefont{Feenstra}},
  \bibnamefont{and} \bibinfo{author}{\bibfnamefont{B.~M.} \bibnamefont{Hunt}},
  \bibinfo{journal}{Nature Physics} \textbf{\bibinfo{volume}{16}},
  \bibinfo{pages}{526} (\bibinfo{year}{2020}).

\bibitem[{\citenamefont{{\v Z}uti{\'c} et~al.}(2019)\citenamefont{{\v
  Z}uti{\'c}, Matos-Abiague, Scharf, Dery, and Belashchenko}}]{Zutic:2019aa}
\bibinfo{author}{\bibfnamefont{I.}~\bibnamefont{{\v Z}uti{\'c}}},
  \bibinfo{author}{\bibfnamefont{A.}~\bibnamefont{Matos-Abiague}},
  \bibinfo{author}{\bibfnamefont{B.}~\bibnamefont{Scharf}},
  \bibinfo{author}{\bibfnamefont{H.}~\bibnamefont{Dery}}, \bibnamefont{and}
  \bibinfo{author}{\bibfnamefont{K.}~\bibnamefont{Belashchenko}},
  \bibinfo{journal}{Materials Today} \textbf{\bibinfo{volume}{22}},
  \bibinfo{pages}{85 } (\bibinfo{year}{2019}).

\bibitem[{\citenamefont{Cao et~al.}(2018{\natexlab{a}})\citenamefont{Cao,
  Fatemi, Demir, Fang, Tomarken, Luo, Sanchez-Yamagishi, Watanabe, Taniguchi,
  Kaxiras et~al.}}]{Cao:2018ad}
\bibinfo{author}{\bibfnamefont{Y.}~\bibnamefont{Cao}},
  \bibinfo{author}{\bibfnamefont{V.}~\bibnamefont{Fatemi}},
  \bibinfo{author}{\bibfnamefont{A.}~\bibnamefont{Demir}},
  \bibinfo{author}{\bibfnamefont{S.}~\bibnamefont{Fang}},
  \bibinfo{author}{\bibfnamefont{S.~L.} \bibnamefont{Tomarken}},
  \bibinfo{author}{\bibfnamefont{J.~Y.} \bibnamefont{Luo}},
  \bibinfo{author}{\bibfnamefont{J.~D.} \bibnamefont{Sanchez-Yamagishi}},
  \bibinfo{author}{\bibfnamefont{K.}~\bibnamefont{Watanabe}},
  \bibinfo{author}{\bibfnamefont{T.}~\bibnamefont{Taniguchi}},
  \bibinfo{author}{\bibfnamefont{E.}~\bibnamefont{Kaxiras}},
  \bibnamefont{et~al.}, \bibinfo{journal}{Nature}
  \textbf{\bibinfo{volume}{556}}, \bibinfo{pages}{80}
  (\bibinfo{year}{2018}{\natexlab{a}}).

\bibitem[{\citenamefont{Cao et~al.}(2018{\natexlab{b}})\citenamefont{Cao,
  Fatemi, Fang, Watanabe, Taniguchi, Kaxiras, and
  Jarillo-Herrero}}]{Cao:2018aa}
\bibinfo{author}{\bibfnamefont{Y.}~\bibnamefont{Cao}},
  \bibinfo{author}{\bibfnamefont{V.}~\bibnamefont{Fatemi}},
  \bibinfo{author}{\bibfnamefont{S.}~\bibnamefont{Fang}},
  \bibinfo{author}{\bibfnamefont{K.}~\bibnamefont{Watanabe}},
  \bibinfo{author}{\bibfnamefont{T.}~\bibnamefont{Taniguchi}},
  \bibinfo{author}{\bibfnamefont{E.}~\bibnamefont{Kaxiras}}, \bibnamefont{and}
  \bibinfo{author}{\bibfnamefont{P.}~\bibnamefont{Jarillo-Herrero}},
  \bibinfo{journal}{Nature} \textbf{\bibinfo{volume}{556}}, \bibinfo{pages}{43}
  (\bibinfo{year}{2018}{\natexlab{b}}).

\bibitem[{\citenamefont{Frisenda et~al.}(2018)\citenamefont{Frisenda,
  Navarro-Moratalla, Gant, P{\'e}rez De~Lara, Jarillo-Herrero, Gorbachev, and
  Castellanos-Gomez}}]{Frisenda:2018aa}
\bibinfo{author}{\bibfnamefont{R.}~\bibnamefont{Frisenda}},
  \bibinfo{author}{\bibfnamefont{E.}~\bibnamefont{Navarro-Moratalla}},
  \bibinfo{author}{\bibfnamefont{P.}~\bibnamefont{Gant}},
  \bibinfo{author}{\bibfnamefont{D.}~\bibnamefont{P{\'e}rez De~Lara}},
  \bibinfo{author}{\bibfnamefont{P.}~\bibnamefont{Jarillo-Herrero}},
  \bibinfo{author}{\bibfnamefont{R.~V.} \bibnamefont{Gorbachev}},
  \bibnamefont{and}
  \bibinfo{author}{\bibfnamefont{A.}~\bibnamefont{Castellanos-Gomez}},
  \bibinfo{journal}{Chem. Soc. Rev.} \textbf{\bibinfo{volume}{47}},
  \bibinfo{pages}{53} (\bibinfo{year}{2018}).

\bibitem[{\citenamefont{Haastrup et~al.}(2018)\citenamefont{Haastrup, Strange,
  Pandey, Deilmann, Schmidt, Hinsche, Gjerding, Torelli, Larsen, Riis-Jensen
  et~al.}}]{Haastrup:2018aa}
\bibinfo{author}{\bibfnamefont{S.}~\bibnamefont{Haastrup}},
  \bibinfo{author}{\bibfnamefont{M.}~\bibnamefont{Strange}},
  \bibinfo{author}{\bibfnamefont{M.}~\bibnamefont{Pandey}},
  \bibinfo{author}{\bibfnamefont{T.}~\bibnamefont{Deilmann}},
  \bibinfo{author}{\bibfnamefont{P.~S.} \bibnamefont{Schmidt}},
  \bibinfo{author}{\bibfnamefont{N.~F.} \bibnamefont{Hinsche}},
  \bibinfo{author}{\bibfnamefont{M.~N.} \bibnamefont{Gjerding}},
  \bibinfo{author}{\bibfnamefont{D.}~\bibnamefont{Torelli}},
  \bibinfo{author}{\bibfnamefont{P.~M.} \bibnamefont{Larsen}},
  \bibinfo{author}{\bibfnamefont{A.~C.} \bibnamefont{Riis-Jensen}},
  \bibnamefont{et~al.}, \bibinfo{journal}{2D Materials}
  \textbf{\bibinfo{volume}{5}}, \bibinfo{pages}{042002} (\bibinfo{year}{2018}).

\bibitem[{\citenamefont{Hou et~al.}(2020)\citenamefont{Hou, Zhang, Sun, Liu,
  Yao, and Wang}}]{Hou:2020aa}
\bibinfo{author}{\bibfnamefont{Y.}~\bibnamefont{Hou}},
  \bibinfo{author}{\bibfnamefont{T.}~\bibnamefont{Zhang}},
  \bibinfo{author}{\bibfnamefont{J.}~\bibnamefont{Sun}},
  \bibinfo{author}{\bibfnamefont{L.}~\bibnamefont{Liu}},
  \bibinfo{author}{\bibfnamefont{Y.}~\bibnamefont{Yao}}, \bibnamefont{and}
  \bibinfo{author}{\bibfnamefont{Y.}~\bibnamefont{Wang}}
  (\bibinfo{year}{2020}), \eprint{arXiv.2008.04150}.

\bibitem[{\citenamefont{Reis et~al.}(2017)\citenamefont{Reis, Li, Dudy,
  Bauernfeind, Glass, Hanke, Thomale, Sch{\"a}fer, and Claessen}}]{Reis:2017aa}
\bibinfo{author}{\bibfnamefont{F.}~\bibnamefont{Reis}},
  \bibinfo{author}{\bibfnamefont{G.}~\bibnamefont{Li}},
  \bibinfo{author}{\bibfnamefont{L.}~\bibnamefont{Dudy}},
  \bibinfo{author}{\bibfnamefont{M.}~\bibnamefont{Bauernfeind}},
  \bibinfo{author}{\bibfnamefont{S.}~\bibnamefont{Glass}},
  \bibinfo{author}{\bibfnamefont{W.}~\bibnamefont{Hanke}},
  \bibinfo{author}{\bibfnamefont{R.}~\bibnamefont{Thomale}},
  \bibinfo{author}{\bibfnamefont{J.}~\bibnamefont{Sch{\"a}fer}},
  \bibnamefont{and} \bibinfo{author}{\bibfnamefont{R.}~\bibnamefont{Claessen}},
  \bibinfo{journal}{Science} \textbf{\bibinfo{volume}{357}},
  \bibinfo{pages}{287} (\bibinfo{year}{2017}).

\bibitem[{\citenamefont{Jensen and Plummer}(1985)}]{Jensen:1985aa}
\bibinfo{author}{\bibfnamefont{E.}~\bibnamefont{Jensen}} \bibnamefont{and}
  \bibinfo{author}{\bibfnamefont{E.~W.} \bibnamefont{Plummer}},
  \bibinfo{journal}{Physical Review Letters} \textbf{\bibinfo{volume}{55}},
  \bibinfo{pages}{1912} (\bibinfo{year}{1985}).

\bibitem[{\citenamefont{M{\aa}rtensson
  et~al.}(1994)\citenamefont{M{\aa}rtensson, Baltzer, Br{\"u}hwiler, Forsell,
  Nilsson, Stenborg, and Wannberg}}]{Martensson:1994ab}
\bibinfo{author}{\bibfnamefont{N.}~\bibnamefont{M{\aa}rtensson}},
  \bibinfo{author}{\bibfnamefont{P.}~\bibnamefont{Baltzer}},
  \bibinfo{author}{\bibfnamefont{P.}~\bibnamefont{Br{\"u}hwiler}},
  \bibinfo{author}{\bibfnamefont{J.-O.} \bibnamefont{Forsell}},
  \bibinfo{author}{\bibfnamefont{A.}~\bibnamefont{Nilsson}},
  \bibinfo{author}{\bibfnamefont{A.}~\bibnamefont{Stenborg}}, \bibnamefont{and}
  \bibinfo{author}{\bibfnamefont{B.}~\bibnamefont{Wannberg}},
  \bibinfo{journal}{Journal of Electron Spectroscopy and Related Phenomena}
  \textbf{\bibinfo{volume}{70}}, \bibinfo{pages}{117} (\bibinfo{year}{1994}).

\bibitem[{\citenamefont{Lanzara et~al.}(2001)\citenamefont{Lanzara, Bogdanov,
  Zhou, Kellar, Feng, Lu, Yoshida, Eisaki, Fujimori, Kishio
  et~al.}}]{Lanzara:2001aa}
\bibinfo{author}{\bibfnamefont{A.}~\bibnamefont{Lanzara}},
  \bibinfo{author}{\bibfnamefont{P.~V.} \bibnamefont{Bogdanov}},
  \bibinfo{author}{\bibfnamefont{X.~J.} \bibnamefont{Zhou}},
  \bibinfo{author}{\bibfnamefont{S.~A.} \bibnamefont{Kellar}},
  \bibinfo{author}{\bibfnamefont{D.~L.} \bibnamefont{Feng}},
  \bibinfo{author}{\bibfnamefont{E.~D.} \bibnamefont{Lu}},
  \bibinfo{author}{\bibfnamefont{T.}~\bibnamefont{Yoshida}},
  \bibinfo{author}{\bibfnamefont{H.}~\bibnamefont{Eisaki}},
  \bibinfo{author}{\bibfnamefont{A.}~\bibnamefont{Fujimori}},
  \bibinfo{author}{\bibfnamefont{K.}~\bibnamefont{Kishio}},
  \bibnamefont{et~al.}, \bibinfo{journal}{Nature}
  \textbf{\bibinfo{volume}{412}}, \bibinfo{pages}{510} (\bibinfo{year}{2001}).

\bibitem[{\citenamefont{Kiss et~al.}(2008)\citenamefont{Kiss, Shimojima,
  Ishizaka, Chainani, Togashi, Kanai, Wang, Chen, Watanabe, and
  Shin}}]{Kiss:2008aa}
\bibinfo{author}{\bibfnamefont{T.}~\bibnamefont{Kiss}},
  \bibinfo{author}{\bibfnamefont{T.}~\bibnamefont{Shimojima}},
  \bibinfo{author}{\bibfnamefont{K.}~\bibnamefont{Ishizaka}},
  \bibinfo{author}{\bibfnamefont{A.}~\bibnamefont{Chainani}},
  \bibinfo{author}{\bibfnamefont{T.}~\bibnamefont{Togashi}},
  \bibinfo{author}{\bibfnamefont{T.}~\bibnamefont{Kanai}},
  \bibinfo{author}{\bibfnamefont{X.-Y.} \bibnamefont{Wang}},
  \bibinfo{author}{\bibfnamefont{C.-T.} \bibnamefont{Chen}},
  \bibinfo{author}{\bibfnamefont{S.}~\bibnamefont{Watanabe}}, \bibnamefont{and}
  \bibinfo{author}{\bibfnamefont{S.}~\bibnamefont{Shin}},
  \bibinfo{journal}{Review of Scientific Instruments}
  \textbf{\bibinfo{volume}{79}}, \bibinfo{pages}{023106}
  (\bibinfo{year}{2008}).

\bibitem[{\citenamefont{Borisenko}(2012)}]{Borisenko:2012aa}
\bibinfo{author}{\bibfnamefont{S.~V.} \bibnamefont{Borisenko}},
  \bibinfo{journal}{Synchrotron Radiation News} \textbf{\bibinfo{volume}{25}},
  \bibinfo{pages}{6} (\bibinfo{year}{2012}).

\bibitem[{\citenamefont{Suga and Tusche}(2015)}]{Suga:2015aa}
\bibinfo{author}{\bibfnamefont{S.}~\bibnamefont{Suga}} \bibnamefont{and}
  \bibinfo{author}{\bibfnamefont{C.}~\bibnamefont{Tusche}},
  \bibinfo{journal}{Journal of Electron Spectroscopy and Related Phenomena}
  \textbf{\bibinfo{volume}{200}}, \bibinfo{pages}{119} (\bibinfo{year}{2015}).

\bibitem[{\citenamefont{Strocov et~al.}(2019)\citenamefont{Strocov, Lev,
  Kobayashi, Cancellieri, Husanu, Chikina, Schr{\"o}ter, Wang, Krieger, and
  Salman}}]{Strocov:2019aa}
\bibinfo{author}{\bibfnamefont{V.~N.} \bibnamefont{Strocov}},
  \bibinfo{author}{\bibfnamefont{L.~L.} \bibnamefont{Lev}},
  \bibinfo{author}{\bibfnamefont{M.}~\bibnamefont{Kobayashi}},
  \bibinfo{author}{\bibfnamefont{C.}~\bibnamefont{Cancellieri}},
  \bibinfo{author}{\bibfnamefont{M.~A.} \bibnamefont{Husanu}},
  \bibinfo{author}{\bibfnamefont{A.}~\bibnamefont{Chikina}},
  \bibinfo{author}{\bibfnamefont{N.~B.~M.} \bibnamefont{Schr{\"o}ter}},
  \bibinfo{author}{\bibfnamefont{X.}~\bibnamefont{Wang}},
  \bibinfo{author}{\bibfnamefont{J.~A.} \bibnamefont{Krieger}},
  \bibnamefont{and} \bibinfo{author}{\bibfnamefont{Z.}~\bibnamefont{Salman}},
  \emph{\bibinfo{title}{k-resolved electronic structure of buried
  heterostructure and impurity systems by soft-x-ray arpes}}
  (\bibinfo{year}{2019}), \eprint{arXiv.1906.11025}.

\bibitem[{\citenamefont{Miwa et~al.}(2015)\citenamefont{Miwa, Ulstrup,
  S\o{}rensen, Dendzik, \ifmmode\check{C}\else\v{C}\fi{}abo, Bianchi,
  Lauritsen, and Hofmann}}]{Miwa:2015aa}
\bibinfo{author}{\bibfnamefont{J.~A.} \bibnamefont{Miwa}},
  \bibinfo{author}{\bibfnamefont{S.}~\bibnamefont{Ulstrup}},
  \bibinfo{author}{\bibfnamefont{S.~G.} \bibnamefont{S\o{}rensen}},
  \bibinfo{author}{\bibfnamefont{M.}~\bibnamefont{Dendzik}},
  \bibinfo{author}{\bibfnamefont{A.~G.}
  \bibnamefont{\ifmmode\check{C}\else\v{C}\fi{}abo}},
  \bibinfo{author}{\bibfnamefont{M.}~\bibnamefont{Bianchi}},
  \bibinfo{author}{\bibfnamefont{J.~V.} \bibnamefont{Lauritsen}},
  \bibnamefont{and} \bibinfo{author}{\bibfnamefont{P.}~\bibnamefont{Hofmann}},
  \bibinfo{journal}{Phys. Rev. Lett.} \textbf{\bibinfo{volume}{114}},
  \bibinfo{pages}{046802} (\bibinfo{year}{2015}).

\bibitem[{\citenamefont{Penn}(1976)}]{Penn:1976aa}
\bibinfo{author}{\bibfnamefont{D.~R.} \bibnamefont{Penn}},
  \bibinfo{journal}{Physical Review B} \textbf{\bibinfo{volume}{13}},
  \bibinfo{pages}{5248} (\bibinfo{year}{1976}).

\bibitem[{\citenamefont{Zangwill}(1988)}]{Zangwill:1988aa}
\bibinfo{author}{\bibfnamefont{A.}~\bibnamefont{Zangwill}},
  \emph{\bibinfo{title}{Physics at Surfaces (and references therein).}}
  (\bibinfo{publisher}{Cambridge University Press}, \bibinfo{year}{1988}).

\bibitem[{\citenamefont{Hengsberger et~al.}(1999)\citenamefont{Hengsberger,
  Purdie, Segovia, Garnier, and Baer}}]{Hengsberger:1999ab}
\bibinfo{author}{\bibfnamefont{M.}~\bibnamefont{Hengsberger}},
  \bibinfo{author}{\bibfnamefont{D.}~\bibnamefont{Purdie}},
  \bibinfo{author}{\bibfnamefont{P.}~\bibnamefont{Segovia}},
  \bibinfo{author}{\bibfnamefont{M.}~\bibnamefont{Garnier}}, \bibnamefont{and}
  \bibinfo{author}{\bibfnamefont{Y.}~\bibnamefont{Baer}},
  \bibinfo{journal}{Physical Review Letters} \textbf{\bibinfo{volume}{83}},
  \bibinfo{pages}{592} (\bibinfo{year}{1999}).

\bibitem[{\citenamefont{Valla et~al.}(1999{\natexlab{a}})\citenamefont{Valla,
  Fedorov, Johnson, and Hulbert}}]{Valla:1999ab}
\bibinfo{author}{\bibfnamefont{T.}~\bibnamefont{Valla}},
  \bibinfo{author}{\bibfnamefont{A.~V.} \bibnamefont{Fedorov}},
  \bibinfo{author}{\bibfnamefont{P.~D.} \bibnamefont{Johnson}},
  \bibnamefont{and} \bibinfo{author}{\bibfnamefont{S.~L.}
  \bibnamefont{Hulbert}}, \bibinfo{journal}{Physical Review Letters}
  \textbf{\bibinfo{volume}{83}}, \bibinfo{pages}{2085}
  (\bibinfo{year}{1999}{\natexlab{a}}).

\bibitem[{\citenamefont{Valla et~al.}(1999{\natexlab{b}})\citenamefont{Valla,
  Fedorov, Johnson, Wells, Hulbert, Li, Gu, and Koshizuka}}]{Valla:1999aa}
\bibinfo{author}{\bibfnamefont{T.}~\bibnamefont{Valla}},
  \bibinfo{author}{\bibfnamefont{A.~V.} \bibnamefont{Fedorov}},
  \bibinfo{author}{\bibfnamefont{P.~D.} \bibnamefont{Johnson}},
  \bibinfo{author}{\bibfnamefont{B.~O.} \bibnamefont{Wells}},
  \bibinfo{author}{\bibfnamefont{S.~L.} \bibnamefont{Hulbert}},
  \bibinfo{author}{\bibfnamefont{Q.}~\bibnamefont{Li}},
  \bibinfo{author}{\bibfnamefont{G.~D.} \bibnamefont{Gu}}, \bibnamefont{and}
  \bibinfo{author}{\bibfnamefont{N.}~\bibnamefont{Koshizuka}},
  \bibinfo{journal}{Science} \textbf{\bibinfo{volume}{285}},
  \bibinfo{pages}{2110} (\bibinfo{year}{1999}{\natexlab{b}}).

\bibitem[{\citenamefont{Engelsberg and Schrieffer}(1963)}]{Engelsberg:1963aa}
\bibinfo{author}{\bibfnamefont{S.}~\bibnamefont{Engelsberg}} \bibnamefont{and}
  \bibinfo{author}{\bibfnamefont{J.~R.} \bibnamefont{Schrieffer}},
  \bibinfo{journal}{Physical Review} \textbf{\bibinfo{volume}{131}},
  \bibinfo{pages}{993} (\bibinfo{year}{1963}).

\bibitem[{\citenamefont{Kordyuk et~al.}(2005)\citenamefont{Kordyuk, Borisenko,
  Koitzsch, Fink, Knupfer, and Berger}}]{Kordyuk:2005aa}
\bibinfo{author}{\bibfnamefont{A.~A.} \bibnamefont{Kordyuk}},
  \bibinfo{author}{\bibfnamefont{S.~V.} \bibnamefont{Borisenko}},
  \bibinfo{author}{\bibfnamefont{A.}~\bibnamefont{Koitzsch}},
  \bibinfo{author}{\bibfnamefont{J.}~\bibnamefont{Fink}},
  \bibinfo{author}{\bibfnamefont{M.}~\bibnamefont{Knupfer}}, \bibnamefont{and}
  \bibinfo{author}{\bibfnamefont{H.}~\bibnamefont{Berger}},
  \bibinfo{journal}{Physical Review B} \textbf{\bibinfo{volume}{71}},
  \bibinfo{pages}{214513} (\bibinfo{year}{2005}).

\bibitem[{\citenamefont{Pletikosi\ifmmode~\acute{c}\else \'{c}\fi{}
  et~al.}(2012)\citenamefont{Pletikosi\ifmmode~\acute{c}\else \'{c}\fi{},
  Kralj, Milun, and Pervan}}]{Pletikosic:2012aa}
\bibinfo{author}{\bibfnamefont{I.}~\bibnamefont{Pletikosi\ifmmode~\acute{c}\else
  \'{c}\fi{}}}, \bibinfo{author}{\bibfnamefont{M.}~\bibnamefont{Kralj}},
  \bibinfo{author}{\bibfnamefont{M.}~\bibnamefont{Milun}}, \bibnamefont{and}
  \bibinfo{author}{\bibfnamefont{P.}~\bibnamefont{Pervan}},
  \bibinfo{journal}{Phys. Rev. B} \textbf{\bibinfo{volume}{85}},
  \bibinfo{pages}{155447} (\bibinfo{year}{2012}).

\bibitem[{\citenamefont{Luttinger and Ward}(1960)}]{Luttinger:1960ab}
\bibinfo{author}{\bibfnamefont{J.~M.} \bibnamefont{Luttinger}}
  \bibnamefont{and} \bibinfo{author}{\bibfnamefont{J.~C.} \bibnamefont{Ward}},
  \bibinfo{journal}{Phys. Rev.} \textbf{\bibinfo{volume}{118}},
  \bibinfo{pages}{1417} (\bibinfo{year}{1960}).

\bibitem[{\citenamefont{Nechaev et~al.}(2009)\citenamefont{Nechaev, Jensen,
  Rienks, Silkin, Echenique, Chulkov, and Hofmann}}]{Nechaev:2009aa}
\bibinfo{author}{\bibfnamefont{I.~A.} \bibnamefont{Nechaev}},
  \bibinfo{author}{\bibfnamefont{M.~F.} \bibnamefont{Jensen}},
  \bibinfo{author}{\bibfnamefont{E.~D.~L.} \bibnamefont{Rienks}},
  \bibinfo{author}{\bibfnamefont{V.~M.} \bibnamefont{Silkin}},
  \bibinfo{author}{\bibfnamefont{P.~M.} \bibnamefont{Echenique}},
  \bibinfo{author}{\bibfnamefont{E.~V.} \bibnamefont{Chulkov}},
  \bibnamefont{and} \bibinfo{author}{\bibfnamefont{P.}~\bibnamefont{Hofmann}},
  \bibinfo{journal}{Physical Review B} \textbf{\bibinfo{volume}{80}},
  \bibinfo{eid}{113402} (\bibinfo{year}{2009}).

\bibitem[{\citenamefont{Mazzola et~al.}(2013)\citenamefont{Mazzola, Wells,
  Yakimova, Ulstrup, Miwa, Balog, Bianchi, Leandersson, Adell, Hofmann
  et~al.}}]{Mazzola:2013aa}
\bibinfo{author}{\bibfnamefont{F.}~\bibnamefont{Mazzola}},
  \bibinfo{author}{\bibfnamefont{J.~W.} \bibnamefont{Wells}},
  \bibinfo{author}{\bibfnamefont{R.}~\bibnamefont{Yakimova}},
  \bibinfo{author}{\bibfnamefont{S.}~\bibnamefont{Ulstrup}},
  \bibinfo{author}{\bibfnamefont{J.~A.} \bibnamefont{Miwa}},
  \bibinfo{author}{\bibfnamefont{R.}~\bibnamefont{Balog}},
  \bibinfo{author}{\bibfnamefont{M.}~\bibnamefont{Bianchi}},
  \bibinfo{author}{\bibfnamefont{M.}~\bibnamefont{Leandersson}},
  \bibinfo{author}{\bibfnamefont{J.}~\bibnamefont{Adell}},
  \bibinfo{author}{\bibfnamefont{P.}~\bibnamefont{Hofmann}},
  \bibnamefont{et~al.}, \bibinfo{journal}{Phys. Rev. Lett.}
  \textbf{\bibinfo{volume}{111}}, \bibinfo{pages}{216806}
  (\bibinfo{year}{2013}).

\bibitem[{\citenamefont{Karolak et~al.}(2011)\citenamefont{Karolak, Wehling,
  Lechermann, and Lichtenstein}}]{Karolak:2011aa}
\bibinfo{author}{\bibfnamefont{M.}~\bibnamefont{Karolak}},
  \bibinfo{author}{\bibfnamefont{T.~O.} \bibnamefont{Wehling}},
  \bibinfo{author}{\bibfnamefont{F.}~\bibnamefont{Lechermann}},
  \bibnamefont{and} \bibinfo{author}{\bibfnamefont{A.~I.}
  \bibnamefont{Lichtenstein}}, \bibinfo{journal}{Journal of Physics: Condensed
  Matter} \textbf{\bibinfo{volume}{23}}, \bibinfo{pages}{085601}
  (\bibinfo{year}{2011}).

\bibitem[{\citenamefont{Usachov et~al.}(2011)\citenamefont{Usachov, Vilkov,
  Gr{\"u}neis, Haberer, Fedorov, Adamchuk, Preobrajenski, Dudin, Barinov,
  Oehzelt et~al.}}]{Usachov:2011aa}
\bibinfo{author}{\bibfnamefont{D.}~\bibnamefont{Usachov}},
  \bibinfo{author}{\bibfnamefont{O.}~\bibnamefont{Vilkov}},
  \bibinfo{author}{\bibfnamefont{A.}~\bibnamefont{Gr{\"u}neis}},
  \bibinfo{author}{\bibfnamefont{D.}~\bibnamefont{Haberer}},
  \bibinfo{author}{\bibfnamefont{A.}~\bibnamefont{Fedorov}},
  \bibinfo{author}{\bibfnamefont{V.~K.} \bibnamefont{Adamchuk}},
  \bibinfo{author}{\bibfnamefont{A.~B.} \bibnamefont{Preobrajenski}},
  \bibinfo{author}{\bibfnamefont{P.}~\bibnamefont{Dudin}},
  \bibinfo{author}{\bibfnamefont{A.}~\bibnamefont{Barinov}},
  \bibinfo{author}{\bibfnamefont{M.}~\bibnamefont{Oehzelt}},
  \bibnamefont{et~al.}, \bibinfo{journal}{Nano Letters}
  \textbf{\bibinfo{volume}{11}}, \bibinfo{pages}{5401} (\bibinfo{year}{2011}).

\bibitem[{\citenamefont{Henck et~al.}(2018)\citenamefont{Henck, Avila,
  Ben~Aziza, Pierucci, Baima, Pamuk, Chaste, Utt, Bartos, Nogajewski
  et~al.}}]{Henck:2018ac}
\bibinfo{author}{\bibfnamefont{H.}~\bibnamefont{Henck}},
  \bibinfo{author}{\bibfnamefont{J.}~\bibnamefont{Avila}},
  \bibinfo{author}{\bibfnamefont{Z.}~\bibnamefont{Ben~Aziza}},
  \bibinfo{author}{\bibfnamefont{D.}~\bibnamefont{Pierucci}},
  \bibinfo{author}{\bibfnamefont{J.}~\bibnamefont{Baima}},
  \bibinfo{author}{\bibfnamefont{B.}~\bibnamefont{Pamuk}},
  \bibinfo{author}{\bibfnamefont{J.}~\bibnamefont{Chaste}},
  \bibinfo{author}{\bibfnamefont{D.}~\bibnamefont{Utt}},
  \bibinfo{author}{\bibfnamefont{M.}~\bibnamefont{Bartos}},
  \bibinfo{author}{\bibfnamefont{K.}~\bibnamefont{Nogajewski}},
  \bibnamefont{et~al.}, \bibinfo{journal}{Phys. Rev. B}
  \textbf{\bibinfo{volume}{97}}, \bibinfo{pages}{245421}
  (\bibinfo{year}{2018}).

\bibitem[{\citenamefont{Kastl et~al.}(2018)\citenamefont{Kastl, Chen, Koch,
  Schuler, Kuykendall, Bostwick, Jozwiak, Seyller, Rotenberg, Weber-Bargioni
  et~al.}}]{Kastl:2018aa}
\bibinfo{author}{\bibfnamefont{C.}~\bibnamefont{Kastl}},
  \bibinfo{author}{\bibfnamefont{C.~T.} \bibnamefont{Chen}},
  \bibinfo{author}{\bibfnamefont{R.~J.} \bibnamefont{Koch}},
  \bibinfo{author}{\bibfnamefont{B.}~\bibnamefont{Schuler}},
  \bibinfo{author}{\bibfnamefont{T.~R.} \bibnamefont{Kuykendall}},
  \bibinfo{author}{\bibfnamefont{A.}~\bibnamefont{Bostwick}},
  \bibinfo{author}{\bibfnamefont{C.}~\bibnamefont{Jozwiak}},
  \bibinfo{author}{\bibfnamefont{T.}~\bibnamefont{Seyller}},
  \bibinfo{author}{\bibfnamefont{E.}~\bibnamefont{Rotenberg}},
  \bibinfo{author}{\bibfnamefont{A.}~\bibnamefont{Weber-Bargioni}},
  \bibnamefont{et~al.}, \bibinfo{journal}{2D Materials}
  \textbf{\bibinfo{volume}{5}}, \bibinfo{pages}{045010} (\bibinfo{year}{2018}).

\bibitem[{\citenamefont{Teo et~al.}(2008)\citenamefont{Teo, Fu, and
  Kane}}]{Teo:2008aa}
\bibinfo{author}{\bibfnamefont{J.~C.~Y.} \bibnamefont{Teo}},
  \bibinfo{author}{\bibfnamefont{L.}~\bibnamefont{Fu}}, \bibnamefont{and}
  \bibinfo{author}{\bibfnamefont{C.~L.} \bibnamefont{Kane}},
  \bibinfo{journal}{Physical Review B} \textbf{\bibinfo{volume}{78}},
  \bibinfo{eid}{045426} (\bibinfo{year}{2008}).

\bibitem[{\citenamefont{Wu et~al.}(2020)\citenamefont{Wu, Li, Ma, Zhang, Liu,
  Zhou, Shao, Wang, Hao, Feng et~al.}}]{Wu:2020ab}
\bibinfo{author}{\bibfnamefont{X.}~\bibnamefont{Wu}},
  \bibinfo{author}{\bibfnamefont{J.}~\bibnamefont{Li}},
  \bibinfo{author}{\bibfnamefont{X.-M.} \bibnamefont{Ma}},
  \bibinfo{author}{\bibfnamefont{Y.}~\bibnamefont{Zhang}},
  \bibinfo{author}{\bibfnamefont{Y.}~\bibnamefont{Liu}},
  \bibinfo{author}{\bibfnamefont{C.-S.} \bibnamefont{Zhou}},
  \bibinfo{author}{\bibfnamefont{J.}~\bibnamefont{Shao}},
  \bibinfo{author}{\bibfnamefont{Q.}~\bibnamefont{Wang}},
  \bibinfo{author}{\bibfnamefont{Y.-J.} \bibnamefont{Hao}},
  \bibinfo{author}{\bibfnamefont{Y.}~\bibnamefont{Feng}}, \bibnamefont{et~al.},
  \bibinfo{journal}{Physical Review X} \textbf{\bibinfo{volume}{10}}
  (\bibinfo{year}{2020}).

\bibitem[{\citenamefont{Zhu and Hofmann}(2014)}]{Zhu:2014ad}
\bibinfo{author}{\bibfnamefont{X.-G.} \bibnamefont{Zhu}} \bibnamefont{and}
  \bibinfo{author}{\bibfnamefont{P.}~\bibnamefont{Hofmann}},
  \bibinfo{journal}{Phys. Rev. B} \textbf{\bibinfo{volume}{89}},
  \bibinfo{pages}{125402} (\bibinfo{year}{2014}).

\bibitem[{\citenamefont{Takeuchi et~al.}(1991)\citenamefont{Takeuchi, Chan, and
  Ho}}]{Takeuchi:1991aa}
\bibinfo{author}{\bibfnamefont{N.}~\bibnamefont{Takeuchi}},
  \bibinfo{author}{\bibfnamefont{C.~T.} \bibnamefont{Chan}}, \bibnamefont{and}
  \bibinfo{author}{\bibfnamefont{K.~M.} \bibnamefont{Ho}},
  \bibinfo{journal}{Phys. Rev. B} \textbf{\bibinfo{volume}{43}},
  \bibinfo{pages}{13899} (\bibinfo{year}{1991}).

\bibitem[{\citenamefont{Zhang et~al.}(2011)\citenamefont{Zhang, Richard, Qian,
  Xu, Dai, and Ding}}]{Zhang:2011aa}
\bibinfo{author}{\bibfnamefont{P.}~\bibnamefont{Zhang}},
  \bibinfo{author}{\bibfnamefont{P.}~\bibnamefont{Richard}},
  \bibinfo{author}{\bibfnamefont{T.}~\bibnamefont{Qian}},
  \bibinfo{author}{\bibfnamefont{Y.-M.} \bibnamefont{Xu}},
  \bibinfo{author}{\bibfnamefont{X.}~\bibnamefont{Dai}}, \bibnamefont{and}
  \bibinfo{author}{\bibfnamefont{H.}~\bibnamefont{Ding}},
  \bibinfo{journal}{Review of Scientific Instruments}
  \textbf{\bibinfo{volume}{82}}, \bibinfo{eid}{043712} (\bibinfo{year}{2011}).

\bibitem[{\citenamefont{Zhang et~al.}(2009)\citenamefont{Zhang, Brar, Girit,
  Zettl, and Crommie}}]{Zhang:2009ab}
\bibinfo{author}{\bibfnamefont{Y.}~\bibnamefont{Zhang}},
  \bibinfo{author}{\bibfnamefont{V.~W.} \bibnamefont{Brar}},
  \bibinfo{author}{\bibfnamefont{C.}~\bibnamefont{Girit}},
  \bibinfo{author}{\bibfnamefont{A.}~\bibnamefont{Zettl}}, \bibnamefont{and}
  \bibinfo{author}{\bibfnamefont{M.~F.} \bibnamefont{Crommie}},
  \bibinfo{journal}{Nature Physics} \textbf{\bibinfo{volume}{5}},
  \bibinfo{pages}{722} (\bibinfo{year}{2009}).

\bibitem[{\citenamefont{Beidenkopf et~al.}(2011)\citenamefont{Beidenkopf,
  Roushan, Seo, Gorman, Drozdov, Hor, Cava, and Yazdani}}]{Beidenkopf:2011aa}
\bibinfo{author}{\bibfnamefont{H.}~\bibnamefont{Beidenkopf}},
  \bibinfo{author}{\bibfnamefont{P.}~\bibnamefont{Roushan}},
  \bibinfo{author}{\bibfnamefont{J.}~\bibnamefont{Seo}},
  \bibinfo{author}{\bibfnamefont{L.}~\bibnamefont{Gorman}},
  \bibinfo{author}{\bibfnamefont{I.}~\bibnamefont{Drozdov}},
  \bibinfo{author}{\bibfnamefont{Y.~S.} \bibnamefont{Hor}},
  \bibinfo{author}{\bibfnamefont{R.~J.} \bibnamefont{Cava}}, \bibnamefont{and}
  \bibinfo{author}{\bibfnamefont{A.}~\bibnamefont{Yazdani}},
  \bibinfo{journal}{Nature Physics} \textbf{\bibinfo{volume}{7}},
  \bibinfo{pages}{939} (\bibinfo{year}{2011}).

\bibitem[{\citenamefont{Fujikawa et~al.}(2009)\citenamefont{Fujikawa, Sakurai,
  and Tromp}}]{Fujikawa:2009aa}
\bibinfo{author}{\bibfnamefont{Y.}~\bibnamefont{Fujikawa}},
  \bibinfo{author}{\bibfnamefont{T.}~\bibnamefont{Sakurai}}, \bibnamefont{and}
  \bibinfo{author}{\bibfnamefont{R.~M.} \bibnamefont{Tromp}},
  \bibinfo{journal}{Physical Review B} \textbf{\bibinfo{volume}{79}}
  (\bibinfo{year}{2009}).

\bibitem[{\citenamefont{Sutter et~al.}(2009)\citenamefont{Sutter, Hybertsen,
  Sadowski, and Sutter}}]{Sutter:2009ac}
\bibinfo{author}{\bibfnamefont{P.}~\bibnamefont{Sutter}},
  \bibinfo{author}{\bibfnamefont{M.~S.} \bibnamefont{Hybertsen}},
  \bibinfo{author}{\bibfnamefont{J.~T.} \bibnamefont{Sadowski}},
  \bibnamefont{and} \bibinfo{author}{\bibfnamefont{E.}~\bibnamefont{Sutter}},
  \bibinfo{journal}{Nano Letters} \textbf{\bibinfo{volume}{9}},
  \bibinfo{pages}{2654} (\bibinfo{year}{2009}).

\bibitem[{\citenamefont{Jin et~al.}(2013)\citenamefont{Jin, Yeh, Zaki, Zhang,
  Sadowski, Al-Mahboob, van~der Zande, Chenet, Dadap, Herman
  et~al.}}]{Jin:2013aa}
\bibinfo{author}{\bibfnamefont{W.}~\bibnamefont{Jin}},
  \bibinfo{author}{\bibfnamefont{P.-C.} \bibnamefont{Yeh}},
  \bibinfo{author}{\bibfnamefont{N.}~\bibnamefont{Zaki}},
  \bibinfo{author}{\bibfnamefont{D.}~\bibnamefont{Zhang}},
  \bibinfo{author}{\bibfnamefont{J.~T.} \bibnamefont{Sadowski}},
  \bibinfo{author}{\bibfnamefont{A.}~\bibnamefont{Al-Mahboob}},
  \bibinfo{author}{\bibfnamefont{A.~M.} \bibnamefont{van~der Zande}},
  \bibinfo{author}{\bibfnamefont{D.~A.} \bibnamefont{Chenet}},
  \bibinfo{author}{\bibfnamefont{J.~I.} \bibnamefont{Dadap}},
  \bibinfo{author}{\bibfnamefont{I.~P.} \bibnamefont{Herman}},
  \bibnamefont{et~al.}, \bibinfo{journal}{Phys. Rev. Lett.}
  \textbf{\bibinfo{volume}{111}}, \bibinfo{pages}{106801}
  (\bibinfo{year}{2013}).

\bibitem[{\citenamefont{Cattelan and Fox}(2018)}]{Cattelan:2018aa}
\bibinfo{author}{\bibfnamefont{M.}~\bibnamefont{Cattelan}} \bibnamefont{and}
  \bibinfo{author}{\bibfnamefont{N.}~\bibnamefont{Fox}},
  \bibinfo{journal}{Nanomaterials} \textbf{\bibinfo{volume}{8}},
  \bibinfo{pages}{284} (\bibinfo{year}{2018}).

\bibitem[{\citenamefont{Ulstrup
  et~al.}(2019{\natexlab{b}})\citenamefont{Ulstrup, Koch, Schwarz, McCreary,
  Jonker, Singh, Bostwick, Rotenberg, Jozwiak, and Katoch}}]{Ulstrup:2019ab}
\bibinfo{author}{\bibfnamefont{S.}~\bibnamefont{Ulstrup}},
  \bibinfo{author}{\bibfnamefont{R.~J.} \bibnamefont{Koch}},
  \bibinfo{author}{\bibfnamefont{D.}~\bibnamefont{Schwarz}},
  \bibinfo{author}{\bibfnamefont{K.~M.} \bibnamefont{McCreary}},
  \bibinfo{author}{\bibfnamefont{B.~T.} \bibnamefont{Jonker}},
  \bibinfo{author}{\bibfnamefont{S.}~\bibnamefont{Singh}},
  \bibinfo{author}{\bibfnamefont{A.}~\bibnamefont{Bostwick}},
  \bibinfo{author}{\bibfnamefont{E.}~\bibnamefont{Rotenberg}},
  \bibinfo{author}{\bibfnamefont{C.}~\bibnamefont{Jozwiak}}, \bibnamefont{and}
  \bibinfo{author}{\bibfnamefont{J.}~\bibnamefont{Katoch}},
  \bibinfo{journal}{Applied Physics Letters} \textbf{\bibinfo{volume}{114}},
  \bibinfo{pages}{151601} (\bibinfo{year}{2019}{\natexlab{b}}).

\bibitem[{\citenamefont{Escher et~al.}(2005)\citenamefont{Escher, Weber,
  Merkel, Ziethen, Bernhard, Sch{\"o}nhense, Schmidt, Forster, Reinert,
  Kr{\"o}mker et~al.}}]{Escher:2005aa}
\bibinfo{author}{\bibfnamefont{M.}~\bibnamefont{Escher}},
  \bibinfo{author}{\bibfnamefont{N.}~\bibnamefont{Weber}},
  \bibinfo{author}{\bibfnamefont{M.}~\bibnamefont{Merkel}},
  \bibinfo{author}{\bibfnamefont{C.}~\bibnamefont{Ziethen}},
  \bibinfo{author}{\bibfnamefont{P.}~\bibnamefont{Bernhard}},
  \bibinfo{author}{\bibfnamefont{G.}~\bibnamefont{Sch{\"o}nhense}},
  \bibinfo{author}{\bibfnamefont{S.}~\bibnamefont{Schmidt}},
  \bibinfo{author}{\bibfnamefont{F.}~\bibnamefont{Forster}},
  \bibinfo{author}{\bibfnamefont{F.}~\bibnamefont{Reinert}},
  \bibinfo{author}{\bibfnamefont{B.}~\bibnamefont{Kr{\"o}mker}},
  \bibnamefont{et~al.}, \bibinfo{journal}{Journal of Physics: Condensed Matter}
  \textbf{\bibinfo{volume}{17}}, \bibinfo{pages}{S1329} (\bibinfo{year}{2005}).

\bibitem[{\citenamefont{Kr{\"o}mker et~al.}(2008)\citenamefont{Kr{\"o}mker,
  Escher, Funnemann, Hartung, Engelhard, and Kirschner}}]{Kromker:2008aa}
\bibinfo{author}{\bibfnamefont{B.}~\bibnamefont{Kr{\"o}mker}},
  \bibinfo{author}{\bibfnamefont{M.}~\bibnamefont{Escher}},
  \bibinfo{author}{\bibfnamefont{D.}~\bibnamefont{Funnemann}},
  \bibinfo{author}{\bibfnamefont{D.}~\bibnamefont{Hartung}},
  \bibinfo{author}{\bibfnamefont{H.}~\bibnamefont{Engelhard}},
  \bibnamefont{and}
  \bibinfo{author}{\bibfnamefont{J.}~\bibnamefont{Kirschner}},
  \bibinfo{journal}{Review of Scientific Instruments}
  \textbf{\bibinfo{volume}{79}}, \bibinfo{pages}{053702}
  (\bibinfo{year}{2008}).

\bibitem[{\citenamefont{Wiemann et~al.}(2011)\citenamefont{Wiemann, Patt, Krug,
  Weber, Escher, Merkel, and Schneider}}]{Wiemann:2011aa}
\bibinfo{author}{\bibfnamefont{C.}~\bibnamefont{Wiemann}},
  \bibinfo{author}{\bibfnamefont{M.}~\bibnamefont{Patt}},
  \bibinfo{author}{\bibfnamefont{I.~P.} \bibnamefont{Krug}},
  \bibinfo{author}{\bibfnamefont{N.~B.} \bibnamefont{Weber}},
  \bibinfo{author}{\bibfnamefont{M.}~\bibnamefont{Escher}},
  \bibinfo{author}{\bibfnamefont{M.}~\bibnamefont{Merkel}}, \bibnamefont{and}
  \bibinfo{author}{\bibfnamefont{C.~M.} \bibnamefont{Schneider}},
  \bibinfo{journal}{e-Journal of Surface Science and Nanotechnology}
  \textbf{\bibinfo{volume}{9}}, \bibinfo{pages}{395} (\bibinfo{year}{2011}).

\bibitem[{\citenamefont{Sch{\"o}nhense
  et~al.}(2015)\citenamefont{Sch{\"o}nhense, Medjanik, and
  Elmers}}]{Schonhense:2015aa}
\bibinfo{author}{\bibfnamefont{G.}~\bibnamefont{Sch{\"o}nhense}},
  \bibinfo{author}{\bibfnamefont{K.}~\bibnamefont{Medjanik}}, \bibnamefont{and}
  \bibinfo{author}{\bibfnamefont{H.-J.} \bibnamefont{Elmers}},
  \bibinfo{journal}{Journal of Electron Spectroscopy and Related Phenomena}
  \textbf{\bibinfo{volume}{200}}, \bibinfo{pages}{94} (\bibinfo{year}{2015}).

\bibitem[{\citenamefont{Tusche et~al.}(2019)\citenamefont{Tusche, Chen,
  Schneider, and Kirschner}}]{Tusche:2019aa}
\bibinfo{author}{\bibfnamefont{C.}~\bibnamefont{Tusche}},
  \bibinfo{author}{\bibfnamefont{Y.-J.} \bibnamefont{Chen}},
  \bibinfo{author}{\bibfnamefont{C.~M.} \bibnamefont{Schneider}},
  \bibnamefont{and}
  \bibinfo{author}{\bibfnamefont{J.}~\bibnamefont{Kirschner}},
  \bibinfo{journal}{Ultramicroscopy} \textbf{\bibinfo{volume}{206}},
  \bibinfo{pages}{112815} (\bibinfo{year}{2019}).

\bibitem[{\citenamefont{Matsui et~al.}(2020)\citenamefont{Matsui, Makita,
  Matsuda, Yano, Nakamura, Tanaka, Suga, and Kera}}]{Matsui:2020aa}
\bibinfo{author}{\bibfnamefont{F.}~\bibnamefont{Matsui}},
  \bibinfo{author}{\bibfnamefont{S.}~\bibnamefont{Makita}},
  \bibinfo{author}{\bibfnamefont{H.}~\bibnamefont{Matsuda}},
  \bibinfo{author}{\bibfnamefont{T.}~\bibnamefont{Yano}},
  \bibinfo{author}{\bibfnamefont{E.}~\bibnamefont{Nakamura}},
  \bibinfo{author}{\bibfnamefont{K.}~\bibnamefont{Tanaka}},
  \bibinfo{author}{\bibfnamefont{S.}~\bibnamefont{Suga}}, \bibnamefont{and}
  \bibinfo{author}{\bibfnamefont{S.}~\bibnamefont{Kera}},
  \bibinfo{journal}{Japanese Journal of Applied Physics}
  \textbf{\bibinfo{volume}{59}}, \bibinfo{pages}{067001}
  (\bibinfo{year}{2020}).

\bibitem[{\citenamefont{Rotenberg and Bostwick}(2014)}]{Rotenberg:2014aa}
\bibinfo{author}{\bibfnamefont{E.}~\bibnamefont{Rotenberg}} \bibnamefont{and}
  \bibinfo{author}{\bibfnamefont{A.}~\bibnamefont{Bostwick}},
  \bibinfo{journal}{Journal of Synchrotron Radiation}
  \textbf{\bibinfo{volume}{21}}, \bibinfo{pages}{1048} (\bibinfo{year}{2014}).

\bibitem[{\citenamefont{Avila et~al.}(2013)\citenamefont{Avila, Razado-Colambo,
  Lorcy, Lagarde, Giorgetta, Polack, and Asensio}}]{Avila:2013aa}
\bibinfo{author}{\bibfnamefont{J.}~\bibnamefont{Avila}},
  \bibinfo{author}{\bibfnamefont{I.}~\bibnamefont{Razado-Colambo}},
  \bibinfo{author}{\bibfnamefont{S.}~\bibnamefont{Lorcy}},
  \bibinfo{author}{\bibfnamefont{B.}~\bibnamefont{Lagarde}},
  \bibinfo{author}{\bibfnamefont{J.-L.} \bibnamefont{Giorgetta}},
  \bibinfo{author}{\bibfnamefont{F.}~\bibnamefont{Polack}}, \bibnamefont{and}
  \bibinfo{author}{\bibfnamefont{M.~C.} \bibnamefont{Asensio}},
  \bibinfo{journal}{Journal of Physics: Conference Series}
  \textbf{\bibinfo{volume}{425}}, \bibinfo{pages}{192023}
  (\bibinfo{year}{2013}).

\bibitem[{\citenamefont{Avila and Asensio}(2014)}]{Avila:2014aa}
\bibinfo{author}{\bibfnamefont{J.}~\bibnamefont{Avila}} \bibnamefont{and}
  \bibinfo{author}{\bibfnamefont{M.~C.} \bibnamefont{Asensio}},
  \bibinfo{journal}{Synchrotron Radiation News} \textbf{\bibinfo{volume}{27}},
  \bibinfo{pages}{24} (\bibinfo{year}{2014}).

\bibitem[{\citenamefont{R{\"{o}}sner et~al.}(2019)\citenamefont{R{\"{o}}sner,
  Dudin, Bosgra, Hoesch, and David}}]{Rosner:2019aa}
\bibinfo{author}{\bibfnamefont{B.}~\bibnamefont{R{\"{o}}sner}},
  \bibinfo{author}{\bibfnamefont{P.}~\bibnamefont{Dudin}},
  \bibinfo{author}{\bibfnamefont{J.}~\bibnamefont{Bosgra}},
  \bibinfo{author}{\bibfnamefont{M.}~\bibnamefont{Hoesch}}, \bibnamefont{and}
  \bibinfo{author}{\bibfnamefont{C.}~\bibnamefont{David}},
  \bibinfo{journal}{Journal of Synchrotron Radiation}
  \textbf{\bibinfo{volume}{26}}, \bibinfo{pages}{467} (\bibinfo{year}{2019}).

\bibitem[{\citenamefont{Dudin et~al.}(2010)\citenamefont{Dudin, Lacovig, Fava,
  Nicolini, Bianco, Cautero, and Barinov}}]{Dudin:2010aa}
\bibinfo{author}{\bibfnamefont{P.}~\bibnamefont{Dudin}},
  \bibinfo{author}{\bibfnamefont{P.}~\bibnamefont{Lacovig}},
  \bibinfo{author}{\bibfnamefont{C.}~\bibnamefont{Fava}},
  \bibinfo{author}{\bibfnamefont{E.}~\bibnamefont{Nicolini}},
  \bibinfo{author}{\bibfnamefont{A.}~\bibnamefont{Bianco}},
  \bibinfo{author}{\bibfnamefont{G.}~\bibnamefont{Cautero}}, \bibnamefont{and}
  \bibinfo{author}{\bibfnamefont{A.}~\bibnamefont{Barinov}},
  \bibinfo{journal}{Journal of Synchrotron Radiation}
  \textbf{\bibinfo{volume}{17}}, \bibinfo{pages}{445} (\bibinfo{year}{2010}).

\bibitem[{\citenamefont{Nguyen et~al.}(2019)\citenamefont{Nguyen, Teutsch,
  Wilson, Kahn, Xia, Graham, Kandyba, Giampietri, Barinov, Constantinescu
  et~al.}}]{Nguyen:2019aa}
\bibinfo{author}{\bibfnamefont{P.~V.} \bibnamefont{Nguyen}},
  \bibinfo{author}{\bibfnamefont{N.~C.} \bibnamefont{Teutsch}},
  \bibinfo{author}{\bibfnamefont{N.~P.} \bibnamefont{Wilson}},
  \bibinfo{author}{\bibfnamefont{J.}~\bibnamefont{Kahn}},
  \bibinfo{author}{\bibfnamefont{X.}~\bibnamefont{Xia}},
  \bibinfo{author}{\bibfnamefont{A.~J.} \bibnamefont{Graham}},
  \bibinfo{author}{\bibfnamefont{V.}~\bibnamefont{Kandyba}},
  \bibinfo{author}{\bibfnamefont{A.}~\bibnamefont{Giampietri}},
  \bibinfo{author}{\bibfnamefont{A.}~\bibnamefont{Barinov}},
  \bibinfo{author}{\bibfnamefont{G.~C.} \bibnamefont{Constantinescu}},
  \bibnamefont{et~al.}, \bibinfo{journal}{Nature}
  \textbf{\bibinfo{volume}{572}}, \bibinfo{pages}{220} (\bibinfo{year}{2019}).

\bibitem[{\citenamefont{Koch et~al.}(2018)\citenamefont{Koch, Jozwiak,
  Bostwick, Stripe, Cordier, Hussain, Yun, and Rotenberg}}]{Koch:2018ab}
\bibinfo{author}{\bibfnamefont{R.~J.} \bibnamefont{Koch}},
  \bibinfo{author}{\bibfnamefont{C.}~\bibnamefont{Jozwiak}},
  \bibinfo{author}{\bibfnamefont{A.}~\bibnamefont{Bostwick}},
  \bibinfo{author}{\bibfnamefont{B.}~\bibnamefont{Stripe}},
  \bibinfo{author}{\bibfnamefont{M.}~\bibnamefont{Cordier}},
  \bibinfo{author}{\bibfnamefont{Z.}~\bibnamefont{Hussain}},
  \bibinfo{author}{\bibfnamefont{W.}~\bibnamefont{Yun}}, \bibnamefont{and}
  \bibinfo{author}{\bibfnamefont{E.}~\bibnamefont{Rotenberg}},
  \bibinfo{journal}{Synchrotron Radiation News} \textbf{\bibinfo{volume}{31}},
  \bibinfo{pages}{50} (\bibinfo{year}{2018}).

\bibitem[{\citenamefont{Curcio et~al.}(2020)\citenamefont{Curcio, Jones,
  Muzzio, Volckaert, Biswas, Sanders, Dudin, Cacho, Singh, Watanabe
  et~al.}}]{Curcio:2020aa}
\bibinfo{author}{\bibfnamefont{D.}~\bibnamefont{Curcio}},
  \bibinfo{author}{\bibfnamefont{A.~J.~H.} \bibnamefont{Jones}},
  \bibinfo{author}{\bibfnamefont{R.}~\bibnamefont{Muzzio}},
  \bibinfo{author}{\bibfnamefont{K.}~\bibnamefont{Volckaert}},
  \bibinfo{author}{\bibfnamefont{D.}~\bibnamefont{Biswas}},
  \bibinfo{author}{\bibfnamefont{C.~E.} \bibnamefont{Sanders}},
  \bibinfo{author}{\bibfnamefont{P.}~\bibnamefont{Dudin}},
  \bibinfo{author}{\bibfnamefont{C.}~\bibnamefont{Cacho}},
  \bibinfo{author}{\bibfnamefont{S.}~\bibnamefont{Singh}},
  \bibinfo{author}{\bibfnamefont{K.}~\bibnamefont{Watanabe}},
  \bibnamefont{et~al.}, \bibinfo{journal}{Phys. Rev. Lett.}
  \textbf{\bibinfo{volume}{125}}, \bibinfo{pages}{236403} (\bibinfo{year}{2020}).
  
\bibitem[{\citenamefont{Liu et~al.}(2019)\citenamefont{Liu, Zhang, He, Wang,
  and Liu}}]{Liu:2019ae}
\bibinfo{author}{\bibfnamefont{Y.}~\bibnamefont{Liu}},
  \bibinfo{author}{\bibfnamefont{S.}~\bibnamefont{Zhang}},
  \bibinfo{author}{\bibfnamefont{J.}~\bibnamefont{He}},
  \bibinfo{author}{\bibfnamefont{Z.~M.} \bibnamefont{Wang}}, \bibnamefont{and}
  \bibinfo{author}{\bibfnamefont{Z.}~\bibnamefont{Liu}},
  \bibinfo{journal}{Nano-Micro Letters} \textbf{\bibinfo{volume}{11}}
  (\bibinfo{year}{2019}).

\bibitem[{\citenamefont{Fan et~al.}(2020)\citenamefont{Fan, Vu, Tran, Adhikari,
  and Lee}}]{Fan:2020aa}
\bibinfo{author}{\bibfnamefont{S.}~\bibnamefont{Fan}},
  \bibinfo{author}{\bibfnamefont{Q.~A.} \bibnamefont{Vu}},
  \bibinfo{author}{\bibfnamefont{M.~D.} \bibnamefont{Tran}},
  \bibinfo{author}{\bibfnamefont{S.}~\bibnamefont{Adhikari}}, \bibnamefont{and}
  \bibinfo{author}{\bibfnamefont{Y.~H.} \bibnamefont{Lee}},
  \bibinfo{journal}{2D Materials} \textbf{\bibinfo{volume}{7}},
  \bibinfo{pages}{022005} (\bibinfo{year}{2020}).

\bibitem[{\citenamefont{Ju et~al.}(2014)\citenamefont{Ju, Velasco, Huang, Kahn,
  Nosiglia, Tsai, Yang, Taniguchi, Watanabe, Zhang et~al.}}]{Ju:2014aa}
\bibinfo{author}{\bibfnamefont{L.}~\bibnamefont{Ju}},
  \bibinfo{author}{\bibfnamefont{J.}~\bibnamefont{Velasco}},
  \bibinfo{author}{\bibfnamefont{E.}~\bibnamefont{Huang}},
  \bibinfo{author}{\bibfnamefont{S.}~\bibnamefont{Kahn}},
  \bibinfo{author}{\bibfnamefont{C.}~\bibnamefont{Nosiglia}},
  \bibinfo{author}{\bibfnamefont{H.-Z.} \bibnamefont{Tsai}},
  \bibinfo{author}{\bibfnamefont{W.}~\bibnamefont{Yang}},
  \bibinfo{author}{\bibfnamefont{T.}~\bibnamefont{Taniguchi}},
  \bibinfo{author}{\bibfnamefont{K.}~\bibnamefont{Watanabe}},
  \bibinfo{author}{\bibfnamefont{Y.}~\bibnamefont{Zhang}},
  \bibnamefont{et~al.}, \bibinfo{journal}{Nature Nanotechnology}
  \textbf{\bibinfo{volume}{9}}, \bibinfo{pages}{348} (\bibinfo{year}{2014}).

\bibitem[{\citenamefont{Zhang et~al.}(2013)\citenamefont{Zhang, Levy, Ha, Kuk,
  and Stroscio}}]{Zhang:2013ac}
\bibinfo{author}{\bibfnamefont{T.}~\bibnamefont{Zhang}},
  \bibinfo{author}{\bibfnamefont{N.}~\bibnamefont{Levy}},
  \bibinfo{author}{\bibfnamefont{J.}~\bibnamefont{Ha}},
  \bibinfo{author}{\bibfnamefont{Y.}~\bibnamefont{Kuk}}, \bibnamefont{and}
  \bibinfo{author}{\bibfnamefont{J.~A.} \bibnamefont{Stroscio}},
  \bibinfo{journal}{Phys. Rev. B} \textbf{\bibinfo{volume}{87}},
  \bibinfo{pages}{115410} (\bibinfo{year}{2013}).

\bibitem[{\citenamefont{Kevan}(1986)}]{Kevan:1986aa}
\bibinfo{author}{\bibfnamefont{S.}~\bibnamefont{Kevan}},
  \bibinfo{journal}{Physical Review B} \textbf{\bibinfo{volume}{33}},
  \bibinfo{pages}{4364} (\bibinfo{year}{1986}).

\bibitem[{\citenamefont{Kretinin et~al.}(2014)\citenamefont{Kretinin, Cao, Tu,
  Yu, Jalil, Novoselov, Haigh, Gholinia, Mishchenko, Lozada
  et~al.}}]{Kretinin:2014aa}
\bibinfo{author}{\bibfnamefont{A.~V.} \bibnamefont{Kretinin}},
  \bibinfo{author}{\bibfnamefont{Y.}~\bibnamefont{Cao}},
  \bibinfo{author}{\bibfnamefont{J.~S.} \bibnamefont{Tu}},
  \bibinfo{author}{\bibfnamefont{G.~L.} \bibnamefont{Yu}},
  \bibinfo{author}{\bibfnamefont{R.}~\bibnamefont{Jalil}},
  \bibinfo{author}{\bibfnamefont{K.~S.} \bibnamefont{Novoselov}},
  \bibinfo{author}{\bibfnamefont{S.~J.} \bibnamefont{Haigh}},
  \bibinfo{author}{\bibfnamefont{A.}~\bibnamefont{Gholinia}},
  \bibinfo{author}{\bibfnamefont{A.}~\bibnamefont{Mishchenko}},
  \bibinfo{author}{\bibfnamefont{M.}~\bibnamefont{Lozada}},
  \bibnamefont{et~al.}, \bibinfo{journal}{Nano Letters}
  \textbf{\bibinfo{volume}{14}}, \bibinfo{pages}{3270} (\bibinfo{year}{2014}).

\bibitem[{\citenamefont{Muzzio et~al.}(2020)\citenamefont{Muzzio, Jones,
  Curcio, Biswas, Miwa, Hofmann, Watanabe, Taniguchi, Singh, Jozwiak
  et~al.}}]{Muzzio:2020aa}
\bibinfo{author}{\bibfnamefont{R.}~\bibnamefont{Muzzio}},
  \bibinfo{author}{\bibfnamefont{A.~J.~H.} \bibnamefont{Jones}},
  \bibinfo{author}{\bibfnamefont{D.}~\bibnamefont{Curcio}},
  \bibinfo{author}{\bibfnamefont{D.}~\bibnamefont{Biswas}},
  \bibinfo{author}{\bibfnamefont{J.~A.} \bibnamefont{Miwa}},
  \bibinfo{author}{\bibfnamefont{P.}~\bibnamefont{Hofmann}},
  \bibinfo{author}{\bibfnamefont{K.}~\bibnamefont{Watanabe}},
  \bibinfo{author}{\bibfnamefont{T.}~\bibnamefont{Taniguchi}},
  \bibinfo{author}{\bibfnamefont{S.}~\bibnamefont{Singh}},
  \bibinfo{author}{\bibfnamefont{C.}~\bibnamefont{Jozwiak}},
  \bibnamefont{et~al.}, \bibinfo{journal}{Physical Review B}
  \textbf{\bibinfo{volume}{101}}, \bibinfo{pages}{201409}
  (\bibinfo{year}{2020}).

\bibitem[{\citenamefont{Kaminski et~al.}(2016)\citenamefont{Kaminski,
  Rosenkranz, Norman, Randeria, Li, Raffy, and Campuzano}}]{Kaminski:2016aa}
\bibinfo{author}{\bibfnamefont{A.}~\bibnamefont{Kaminski}},
  \bibinfo{author}{\bibfnamefont{S.}~\bibnamefont{Rosenkranz}},
  \bibinfo{author}{\bibfnamefont{M.~R.} \bibnamefont{Norman}},
  \bibinfo{author}{\bibfnamefont{M.}~\bibnamefont{Randeria}},
  \bibinfo{author}{\bibfnamefont{Z.~Z.} \bibnamefont{Li}},
  \bibinfo{author}{\bibfnamefont{H.}~\bibnamefont{Raffy}}, \bibnamefont{and}
  \bibinfo{author}{\bibfnamefont{J.~C.} \bibnamefont{Campuzano}},
  \bibinfo{journal}{Phys. Rev. X} \textbf{\bibinfo{volume}{6}},
  \bibinfo{pages}{031040} (\bibinfo{year}{2016}).

\bibitem[{\citenamefont{{Naamneh} et~al.}(2016)\citenamefont{{Naamneh},
  {Campuzano}, and {Kanigel}}}]{Naamneh:2016aa}
\bibinfo{author}{\bibfnamefont{M.}~\bibnamefont{{Naamneh}}},
  \bibinfo{author}{\bibfnamefont{J.~C.} \bibnamefont{{Campuzano}}},
  \bibnamefont{and} \bibinfo{author}{\bibfnamefont{A.}~\bibnamefont{{Kanigel}}}
  (\bibinfo{year}{2016}), \eprint{arXiv.1607.02901}.

\bibitem[{con()}]{condnote}
\bibinfo{note}{By the surface conductance being 20 times higher than the bulk
  conductance, we do not refer to a 2D surface conductance compared to a 3D
  bulk conductance. We merely mean that the top layer voxels in the simulation
  have a conductivity that is 20 times higher than that of all the other
  voxels. The depth of the first layer voxels is 60 times smaller than the
  distance between the contacts.}

\bibitem[{\citenamefont{Wells et~al.}(2006)\citenamefont{Wells, Kallehauge,
  Hansen, and Hofmann}}]{Wells:2006aa}
\bibinfo{author}{\bibfnamefont{J.~W.} \bibnamefont{Wells}},
  \bibinfo{author}{\bibfnamefont{J.~F.} \bibnamefont{Kallehauge}},
  \bibinfo{author}{\bibfnamefont{T.~M.} \bibnamefont{Hansen}},
  \bibnamefont{and} \bibinfo{author}{\bibfnamefont{P.}~\bibnamefont{Hofmann}},
  \bibinfo{journal}{Physical Review Letters} \textbf{\bibinfo{volume}{97}},
  \bibinfo{pages}{206803} (\bibinfo{year}{2006}).

\bibitem[{\citenamefont{Hofmann and Wells}(2009)}]{Hofmann:2009aa}
\bibinfo{author}{\bibfnamefont{P.}~\bibnamefont{Hofmann}} \bibnamefont{and}
  \bibinfo{author}{\bibfnamefont{J.~W.} \bibnamefont{Wells}},
  \bibinfo{journal}{Journal of Physics: Condensed Matter}
  \textbf{\bibinfo{volume}{21}}, \bibinfo{pages}{013003}
  (\bibinfo{year}{2009}).

\bibitem[{\citenamefont{Perkins et~al.}(2013)\citenamefont{Perkins, Barreto,
  Wells, and Hofmann}}]{Perkins:2013aa}
\bibinfo{author}{\bibfnamefont{E.}~\bibnamefont{Perkins}},
  \bibinfo{author}{\bibfnamefont{L.}~\bibnamefont{Barreto}},
  \bibinfo{author}{\bibfnamefont{J.}~\bibnamefont{Wells}}, \bibnamefont{and}
  \bibinfo{author}{\bibfnamefont{P.}~\bibnamefont{Hofmann}},
  \bibinfo{journal}{Review of Scientific Instruments}
  \textbf{\bibinfo{volume}{84}}, \bibinfo{eid}{033901} (\bibinfo{year}{2013}).

\bibitem[{\citenamefont{Barreto et~al.}(2014)\citenamefont{Barreto,
  K{\"u}hnemund, Edler, Tegenkamp, Mi, Bremholm, Iversen, Frydendahl, Bianchi,
  and Hofmann}}]{Barreto:2014aa}
\bibinfo{author}{\bibfnamefont{L.}~\bibnamefont{Barreto}},
  \bibinfo{author}{\bibfnamefont{L.}~\bibnamefont{K{\"u}hnemund}},
  \bibinfo{author}{\bibfnamefont{F.}~\bibnamefont{Edler}},
  \bibinfo{author}{\bibfnamefont{C.}~\bibnamefont{Tegenkamp}},
  \bibinfo{author}{\bibfnamefont{J.}~\bibnamefont{Mi}},
  \bibinfo{author}{\bibfnamefont{M.}~\bibnamefont{Bremholm}},
  \bibinfo{author}{\bibfnamefont{B.~B.} \bibnamefont{Iversen}},
  \bibinfo{author}{\bibfnamefont{C.}~\bibnamefont{Frydendahl}},
  \bibinfo{author}{\bibfnamefont{M.}~\bibnamefont{Bianchi}}, \bibnamefont{and}
  \bibinfo{author}{\bibfnamefont{P.}~\bibnamefont{Hofmann}},
  \bibinfo{journal}{Nano Letters} \textbf{\bibinfo{volume}{14}},
  \bibinfo{pages}{3755} (\bibinfo{year}{2014}).

\bibitem[{\citenamefont{Joucken et~al.}(2016)\citenamefont{Joucken, Reckinger,
  Lorcy, Avila, Chen, Lagoute, Colomer, Ghijsen, Asensio, and
  Sporken}}]{Joucken:2016aa}
\bibinfo{author}{\bibfnamefont{F.}~\bibnamefont{Joucken}},
  \bibinfo{author}{\bibfnamefont{N.}~\bibnamefont{Reckinger}},
  \bibinfo{author}{\bibfnamefont{S.}~\bibnamefont{Lorcy}},
  \bibinfo{author}{\bibfnamefont{J.}~\bibnamefont{Avila}},
  \bibinfo{author}{\bibfnamefont{C.}~\bibnamefont{Chen}},
  \bibinfo{author}{\bibfnamefont{J.}~\bibnamefont{Lagoute}},
  \bibinfo{author}{\bibfnamefont{J.-F. m.~c.} \bibnamefont{Colomer}},
  \bibinfo{author}{\bibfnamefont{J.}~\bibnamefont{Ghijsen}},
  \bibinfo{author}{\bibfnamefont{M.~C.} \bibnamefont{Asensio}},
  \bibnamefont{and} \bibinfo{author}{\bibfnamefont{R.}~\bibnamefont{Sporken}},
  \bibinfo{journal}{Phys. Rev. B} \textbf{\bibinfo{volume}{93}},
  \bibinfo{pages}{241101} (\bibinfo{year}{2016}).

\bibitem[{\citenamefont{Joucken
  et~al.}(2019{\natexlab{a}})\citenamefont{Joucken, Quezada-L\'opez, Avila,
  Chen, Davenport, Chen, Watanabe, Taniguchi, Asensio, and
  Velasco}}]{Joucken:2019ab}
\bibinfo{author}{\bibfnamefont{F.}~\bibnamefont{Joucken}},
  \bibinfo{author}{\bibfnamefont{E.~A.} \bibnamefont{Quezada-L\'opez}},
  \bibinfo{author}{\bibfnamefont{J.}~\bibnamefont{Avila}},
  \bibinfo{author}{\bibfnamefont{C.}~\bibnamefont{Chen}},
  \bibinfo{author}{\bibfnamefont{J.~L.} \bibnamefont{Davenport}},
  \bibinfo{author}{\bibfnamefont{H.}~\bibnamefont{Chen}},
  \bibinfo{author}{\bibfnamefont{K.}~\bibnamefont{Watanabe}},
  \bibinfo{author}{\bibfnamefont{T.}~\bibnamefont{Taniguchi}},
  \bibinfo{author}{\bibfnamefont{M.~C.} \bibnamefont{Asensio}},
  \bibnamefont{and} \bibinfo{author}{\bibfnamefont{J.}~\bibnamefont{Velasco}},
  \bibinfo{journal}{Phys. Rev. B} \textbf{\bibinfo{volume}{99}},
  \bibinfo{pages}{161406} (\bibinfo{year}{2019}{\natexlab{a}}).

\bibitem[{\citenamefont{Paradisi et~al.}(2015)\citenamefont{Paradisi, Biscaras,
  and Shukla}}]{Paradisi:2015aa}
\bibinfo{author}{\bibfnamefont{A.}~\bibnamefont{Paradisi}},
  \bibinfo{author}{\bibfnamefont{J.}~\bibnamefont{Biscaras}}, \bibnamefont{and}
  \bibinfo{author}{\bibfnamefont{A.}~\bibnamefont{Shukla}},
  \bibinfo{journal}{Applied Physics Letters} \textbf{\bibinfo{volume}{107}},
  \bibinfo{pages}{143103} (\bibinfo{year}{2015}).

\bibitem[{\citenamefont{Rosenzweig et~al.}(2020)\citenamefont{Rosenzweig,
  Karakachian, Marchenko, K\"uster, and Starke}}]{Rosenzweig:2020aa}
\bibinfo{author}{\bibfnamefont{P.}~\bibnamefont{Rosenzweig}},
  \bibinfo{author}{\bibfnamefont{H.}~\bibnamefont{Karakachian}},
  \bibinfo{author}{\bibfnamefont{D.}~\bibnamefont{Marchenko}},
  \bibinfo{author}{\bibfnamefont{K.}~\bibnamefont{K\"uster}}, \bibnamefont{and}
  \bibinfo{author}{\bibfnamefont{U.}~\bibnamefont{Starke}},
  \bibinfo{journal}{Phys. Rev. Lett.} \textbf{\bibinfo{volume}{125}},
  \bibinfo{pages}{176403} (\bibinfo{year}{2020}).

\bibitem[{\citenamefont{Zhang et~al.}(2008)\citenamefont{Zhang, Brar, Wang,
  Girit, Yayon, Panlasigui, Zettl, and Crommie}}]{Zhang:2008ac}
\bibinfo{author}{\bibfnamefont{Y.}~\bibnamefont{Zhang}},
  \bibinfo{author}{\bibfnamefont{V.~W.} \bibnamefont{Brar}},
  \bibinfo{author}{\bibfnamefont{F.}~\bibnamefont{Wang}},
  \bibinfo{author}{\bibfnamefont{C.}~\bibnamefont{Girit}},
  \bibinfo{author}{\bibfnamefont{Y.}~\bibnamefont{Yayon}},
  \bibinfo{author}{\bibfnamefont{M.}~\bibnamefont{Panlasigui}},
  \bibinfo{author}{\bibfnamefont{A.}~\bibnamefont{Zettl}}, \bibnamefont{and}
  \bibinfo{author}{\bibfnamefont{M.~F.} \bibnamefont{Crommie}},
  \bibinfo{journal}{Nature Physics} \textbf{\bibinfo{volume}{4}},
  \bibinfo{pages}{627} (\bibinfo{year}{2008}).

\bibitem[{\citenamefont{Joucken
  et~al.}(2019{\natexlab{b}})\citenamefont{Joucken, Avila, Ge, Quezada-Lopez,
  Yi, Le~Goff, Baudin, Davenport, Watanabe, Taniguchi et~al.}}]{Joucken:2019aa}
\bibinfo{author}{\bibfnamefont{F.}~\bibnamefont{Joucken}},
  \bibinfo{author}{\bibfnamefont{J.}~\bibnamefont{Avila}},
  \bibinfo{author}{\bibfnamefont{Z.}~\bibnamefont{Ge}},
  \bibinfo{author}{\bibfnamefont{E.~A.} \bibnamefont{Quezada-Lopez}},
  \bibinfo{author}{\bibfnamefont{H.}~\bibnamefont{Yi}},
  \bibinfo{author}{\bibfnamefont{R.}~\bibnamefont{Le~Goff}},
  \bibinfo{author}{\bibfnamefont{E.}~\bibnamefont{Baudin}},
  \bibinfo{author}{\bibfnamefont{J.~L.} \bibnamefont{Davenport}},
  \bibinfo{author}{\bibfnamefont{K.}~\bibnamefont{Watanabe}},
  \bibinfo{author}{\bibfnamefont{T.}~\bibnamefont{Taniguchi}},
  \bibnamefont{et~al.}, \bibinfo{journal}{Nano Letters}
  \textbf{\bibinfo{volume}{19}}, \bibinfo{pages}{2682}
  (\bibinfo{year}{2019}{\natexlab{b}}).

\bibitem[{\citenamefont{McCann}(2006)}]{McCann:2006ab}
\bibinfo{author}{\bibfnamefont{E.}~\bibnamefont{McCann}},
  \bibinfo{journal}{Phys. Rev. B} \textbf{\bibinfo{volume}{74}},
  \bibinfo{pages}{161403} (\bibinfo{year}{2006}).

\bibitem[{\citenamefont{Ohta et~al.}(2006)\citenamefont{Ohta, Bostwick,
  Seyller, Horn, and Rotenberg}}]{Ohta:2006aa}
\bibinfo{author}{\bibfnamefont{T.}~\bibnamefont{Ohta}},
  \bibinfo{author}{\bibfnamefont{A.}~\bibnamefont{Bostwick}},
  \bibinfo{author}{\bibfnamefont{T.}~\bibnamefont{Seyller}},
  \bibinfo{author}{\bibfnamefont{K.}~\bibnamefont{Horn}}, \bibnamefont{and}
  \bibinfo{author}{\bibfnamefont{E.}~\bibnamefont{Rotenberg}},
  \bibinfo{journal}{Science} \textbf{\bibinfo{volume}{313}},
  \bibinfo{pages}{951} (\bibinfo{year}{2006}).

\bibitem[{\citenamefont{Kipp et~al.}(1999)\citenamefont{Kipp, Ro{\ss}nagel,
  Solterbeck, Strasser, Schattke, and Skibowski}}]{Kipp:1999aa}
\bibinfo{author}{\bibfnamefont{L.}~\bibnamefont{Kipp}},
  \bibinfo{author}{\bibfnamefont{K.}~\bibnamefont{Ro{\ss}nagel}},
  \bibinfo{author}{\bibfnamefont{C.}~\bibnamefont{Solterbeck}},
  \bibinfo{author}{\bibfnamefont{T.}~\bibnamefont{Strasser}},
  \bibinfo{author}{\bibfnamefont{W.}~\bibnamefont{Schattke}}, \bibnamefont{and}
  \bibinfo{author}{\bibfnamefont{M.}~\bibnamefont{Skibowski}},
  \bibinfo{journal}{Physical Review Letters} \textbf{\bibinfo{volume}{83}},
  \bibinfo{pages}{5551} (\bibinfo{year}{1999}).

\bibitem[{\citenamefont{Gaylord et~al.}(1989)\citenamefont{Gaylord, Jeong, and
  Kevan}}]{Gaylord:1989aa}
\bibinfo{author}{\bibfnamefont{R.~H.} \bibnamefont{Gaylord}},
  \bibinfo{author}{\bibfnamefont{K.~H.} \bibnamefont{Jeong}}, \bibnamefont{and}
  \bibinfo{author}{\bibfnamefont{S.~D.} \bibnamefont{Kevan}},
  \bibinfo{journal}{Physical Review Letters} \textbf{\bibinfo{volume}{62}},
  \bibinfo{pages}{2036} (\bibinfo{year}{1989}).

\bibitem[{\citenamefont{Aebi et~al.}(1994)\citenamefont{Aebi, Osterwalder,
  Fasel, Naumovi{\'{c}}, and Schlapbach}}]{Aebi:1994ab}
\bibinfo{author}{\bibfnamefont{P.}~\bibnamefont{Aebi}},
  \bibinfo{author}{\bibfnamefont{J.}~\bibnamefont{Osterwalder}},
  \bibinfo{author}{\bibfnamefont{R.}~\bibnamefont{Fasel}},
  \bibinfo{author}{\bibfnamefont{D.}~\bibnamefont{Naumovi{\'{c}}}},
  \bibnamefont{and}
  \bibinfo{author}{\bibfnamefont{L.}~\bibnamefont{Schlapbach}},
  \bibinfo{journal}{Surface Science} \textbf{\bibinfo{volume}{307-309}},
  \bibinfo{pages}{917} (\bibinfo{year}{1994}).

\bibitem[{\citenamefont{Shirley et~al.}(1995)\citenamefont{Shirley, Terminello,
  Santoni, and Himpsel}}]{Shirley:1995aa}
\bibinfo{author}{\bibfnamefont{E.~L.} \bibnamefont{Shirley}},
  \bibinfo{author}{\bibfnamefont{L.~J.} \bibnamefont{Terminello}},
  \bibinfo{author}{\bibfnamefont{A.}~\bibnamefont{Santoni}}, \bibnamefont{and}
  \bibinfo{author}{\bibfnamefont{F.~J.} \bibnamefont{Himpsel}},
  \bibinfo{journal}{Phys. Rev. B} \textbf{\bibinfo{volume}{51}},
  \bibinfo{pages}{13614} (\bibinfo{year}{1995}).

\bibitem[{\citenamefont{Mucha-Kruczynski
  et~al.}(2008)\citenamefont{Mucha-Kruczynski, Tsyplyatyev, Grishin, McCann,
  Fal'ko, Bostwick, and Rotenberg}}]{Mucha-Kruczynski:2008aa}
\bibinfo{author}{\bibfnamefont{M.}~\bibnamefont{Mucha-Kruczynski}},
  \bibinfo{author}{\bibfnamefont{O.}~\bibnamefont{Tsyplyatyev}},
  \bibinfo{author}{\bibfnamefont{A.}~\bibnamefont{Grishin}},
  \bibinfo{author}{\bibfnamefont{E.}~\bibnamefont{McCann}},
  \bibinfo{author}{\bibfnamefont{V.~I.} \bibnamefont{Fal'ko}},
  \bibinfo{author}{\bibfnamefont{A.}~\bibnamefont{Bostwick}}, \bibnamefont{and}
  \bibinfo{author}{\bibfnamefont{E.}~\bibnamefont{Rotenberg}},
  \bibinfo{journal}{Physical Review B} \textbf{\bibinfo{volume}{77}},
  \bibinfo{eid}{195403} (\bibinfo{year}{2008}).

\bibitem[{\citenamefont{Lizzit et~al.}(2010)\citenamefont{Lizzit, Zampieri,
  Petaccia, Larciprete, Lacovig, Rienks, Bihlmayer, Baraldi, and
  Hofmann}}]{Lizzit:2010aa}
\bibinfo{author}{\bibfnamefont{S.}~\bibnamefont{Lizzit}},
  \bibinfo{author}{\bibfnamefont{G.}~\bibnamefont{Zampieri}},
  \bibinfo{author}{\bibfnamefont{L.}~\bibnamefont{Petaccia}},
  \bibinfo{author}{\bibfnamefont{R.}~\bibnamefont{Larciprete}},
  \bibinfo{author}{\bibfnamefont{P.}~\bibnamefont{Lacovig}},
  \bibinfo{author}{\bibfnamefont{E.~D.~L.} \bibnamefont{Rienks}},
  \bibinfo{author}{\bibfnamefont{G.}~\bibnamefont{Bihlmayer}},
  \bibinfo{author}{\bibfnamefont{A.}~\bibnamefont{Baraldi}}, \bibnamefont{and}
  \bibinfo{author}{\bibfnamefont{P.}~\bibnamefont{Hofmann}},
  \bibinfo{journal}{Nature Physics} \textbf{\bibinfo{volume}{6}},
  \bibinfo{pages}{345} (\bibinfo{year}{2010}).

\bibitem[{\citenamefont{Gierz et~al.}(2011)\citenamefont{Gierz, Henk, H\"ochst,
  Ast, and Kern}}]{Gierz:2011ab}
\bibinfo{author}{\bibfnamefont{I.}~\bibnamefont{Gierz}},
  \bibinfo{author}{\bibfnamefont{J.}~\bibnamefont{Henk}},
  \bibinfo{author}{\bibfnamefont{H.}~\bibnamefont{H\"ochst}},
  \bibinfo{author}{\bibfnamefont{C.~R.} \bibnamefont{Ast}}, \bibnamefont{and}
  \bibinfo{author}{\bibfnamefont{K.}~\bibnamefont{Kern}},
  \bibinfo{journal}{Phys. Rev. B} \textbf{\bibinfo{volume}{83}},
  \bibinfo{pages}{121408} (\bibinfo{year}{2011}).

\bibitem[{\citenamefont{Bostwick et~al.}(2010)\citenamefont{Bostwick, Speck,
  Seyller, Horn, Polini, Asgari, MacDonald, and Rotenberg}}]{Bostwick:2010ac}
\bibinfo{author}{\bibfnamefont{A.}~\bibnamefont{Bostwick}},
  \bibinfo{author}{\bibfnamefont{F.}~\bibnamefont{Speck}},
  \bibinfo{author}{\bibfnamefont{T.}~\bibnamefont{Seyller}},
  \bibinfo{author}{\bibfnamefont{K.}~\bibnamefont{Horn}},
  \bibinfo{author}{\bibfnamefont{M.}~\bibnamefont{Polini}},
  \bibinfo{author}{\bibfnamefont{R.}~\bibnamefont{Asgari}},
  \bibinfo{author}{\bibfnamefont{A.~H.} \bibnamefont{MacDonald}},
  \bibnamefont{and}
  \bibinfo{author}{\bibfnamefont{E.}~\bibnamefont{Rotenberg}},
  \bibinfo{journal}{Science} \textbf{\bibinfo{volume}{328}},
  \bibinfo{pages}{999} (\bibinfo{year}{2010}).

\bibitem[{\citenamefont{Walter et~al.}(2011)\citenamefont{Walter, Bostwick,
  Jeon, Speck, Ostler, Seyller, Moreschini, Chang, Polini, Asgari
  et~al.}}]{Walter:2011ac}
\bibinfo{author}{\bibfnamefont{A.~L.} \bibnamefont{Walter}},
  \bibinfo{author}{\bibfnamefont{A.}~\bibnamefont{Bostwick}},
  \bibinfo{author}{\bibfnamefont{K.-J.} \bibnamefont{Jeon}},
  \bibinfo{author}{\bibfnamefont{F.}~\bibnamefont{Speck}},
  \bibinfo{author}{\bibfnamefont{M.}~\bibnamefont{Ostler}},
  \bibinfo{author}{\bibfnamefont{T.}~\bibnamefont{Seyller}},
  \bibinfo{author}{\bibfnamefont{L.}~\bibnamefont{Moreschini}},
  \bibinfo{author}{\bibfnamefont{Y.~J.} \bibnamefont{Chang}},
  \bibinfo{author}{\bibfnamefont{M.}~\bibnamefont{Polini}},
  \bibinfo{author}{\bibfnamefont{R.}~\bibnamefont{Asgari}},
  \bibnamefont{et~al.}, \bibinfo{journal}{Phys. Rev. B}
  \textbf{\bibinfo{volume}{84}}, \bibinfo{pages}{085410}
  (\bibinfo{year}{2011}).

\bibitem[{\citenamefont{Das~Sarma and Hwang}(2013)}]{DasSarma:2013aa}
\bibinfo{author}{\bibfnamefont{S.}~\bibnamefont{Das~Sarma}} \bibnamefont{and}
  \bibinfo{author}{\bibfnamefont{E.~H.} \bibnamefont{Hwang}},
  \bibinfo{journal}{Phys. Rev. B} \textbf{\bibinfo{volume}{87}},
  \bibinfo{pages}{045425} (\bibinfo{year}{2013}).

\bibitem[{\citenamefont{Ulstrup
  et~al.}(2016{\natexlab{a}})\citenamefont{Ulstrup, Sch\"uler, Bianchi, Fromm,
  Raidel, Seyller, Wehling, and Hofmann}}]{Ulstrup:2016ab}
\bibinfo{author}{\bibfnamefont{S.}~\bibnamefont{Ulstrup}},
  \bibinfo{author}{\bibfnamefont{M.}~\bibnamefont{Sch\"uler}},
  \bibinfo{author}{\bibfnamefont{M.}~\bibnamefont{Bianchi}},
  \bibinfo{author}{\bibfnamefont{F.}~\bibnamefont{Fromm}},
  \bibinfo{author}{\bibfnamefont{C.}~\bibnamefont{Raidel}},
  \bibinfo{author}{\bibfnamefont{T.}~\bibnamefont{Seyller}},
  \bibinfo{author}{\bibfnamefont{T.}~\bibnamefont{Wehling}}, \bibnamefont{and}
  \bibinfo{author}{\bibfnamefont{P.}~\bibnamefont{Hofmann}},
  \bibinfo{journal}{Phys. Rev. B} \textbf{\bibinfo{volume}{94}},
  \bibinfo{pages}{081403} (\bibinfo{year}{2016}{\natexlab{a}}).

\bibitem[{\citenamefont{Jones et~al.}(2020)\citenamefont{Jones, Muzzio,
  Majchrzak, Pakdel, Curcio, Volckaert, Biswas, Gobbo, Singh, Robinson
  et~al.}}]{Jones:2020aa}
\bibinfo{author}{\bibfnamefont{A.~J.~H.} \bibnamefont{Jones}},
  \bibinfo{author}{\bibfnamefont{R.}~\bibnamefont{Muzzio}},
  \bibinfo{author}{\bibfnamefont{P.}~\bibnamefont{Majchrzak}},
  \bibinfo{author}{\bibfnamefont{S.}~\bibnamefont{Pakdel}},
  \bibinfo{author}{\bibfnamefont{D.}~\bibnamefont{Curcio}},
  \bibinfo{author}{\bibfnamefont{K.}~\bibnamefont{Volckaert}},
  \bibinfo{author}{\bibfnamefont{D.}~\bibnamefont{Biswas}},
  \bibinfo{author}{\bibfnamefont{J.}~\bibnamefont{Gobbo}},
  \bibinfo{author}{\bibfnamefont{S.}~\bibnamefont{Singh}},
  \bibinfo{author}{\bibfnamefont{J.~T.} \bibnamefont{Robinson}},
  \bibnamefont{et~al.}, \bibinfo{journal}{Advanced Materials} \textbf{\bibinfo{volume}{32}},
  \bibinfo{pages}{2001656} (\bibinfo{year}{2020}).

\bibitem[{\citenamefont{Luryi}(1988)}]{Luryi:1988aa}
\bibinfo{author}{\bibfnamefont{S.}~\bibnamefont{Luryi}},
  \bibinfo{journal}{Applied Physics Letters} \textbf{\bibinfo{volume}{52}},
  \bibinfo{pages}{501} (\bibinfo{year}{1988}).

\bibitem[{\citenamefont{Cheiwchanchamnangij and
  Lambrecht}(2012)}]{Cheiwchanchamnangij:2012aa}
\bibinfo{author}{\bibfnamefont{T.}~\bibnamefont{Cheiwchanchamnangij}}
  \bibnamefont{and} \bibinfo{author}{\bibfnamefont{W.~R.~L.}
  \bibnamefont{Lambrecht}}, \bibinfo{journal}{Phys. Rev. B}
  \textbf{\bibinfo{volume}{85}}, \bibinfo{pages}{205302}
  (\bibinfo{year}{2012}).

\bibitem[{\citenamefont{Cappelluti et~al.}(2013)\citenamefont{Cappelluti,
  Rold\'an, Silva-Guill\'en, Ordej\'on, and Guinea}}]{Cappelluti:2013aa}
\bibinfo{author}{\bibfnamefont{E.}~\bibnamefont{Cappelluti}},
  \bibinfo{author}{\bibfnamefont{R.}~\bibnamefont{Rold\'an}},
  \bibinfo{author}{\bibfnamefont{J.~A.} \bibnamefont{Silva-Guill\'en}},
  \bibinfo{author}{\bibfnamefont{P.}~\bibnamefont{Ordej\'on}},
  \bibnamefont{and} \bibinfo{author}{\bibfnamefont{F.}~\bibnamefont{Guinea}},
  \bibinfo{journal}{Phys. Rev. B} \textbf{\bibinfo{volume}{88}},
  \bibinfo{pages}{075409} (\bibinfo{year}{2013}).

\bibitem[{\citenamefont{Riley et~al.}(2014)\citenamefont{Riley, Mazzola,
  Dendzik, Michiardi, Takayama, Bawden, Granerod, Leandersson, Balasubramanian,
  Hoesch et~al.}}]{Riley:2014aa}
\bibinfo{author}{\bibfnamefont{J.~M.} \bibnamefont{Riley}},
  \bibinfo{author}{\bibfnamefont{F.}~\bibnamefont{Mazzola}},
  \bibinfo{author}{\bibfnamefont{M.}~\bibnamefont{Dendzik}},
  \bibinfo{author}{\bibfnamefont{M.}~\bibnamefont{Michiardi}},
  \bibinfo{author}{\bibfnamefont{T.}~\bibnamefont{Takayama}},
  \bibinfo{author}{\bibfnamefont{L.}~\bibnamefont{Bawden}},
  \bibinfo{author}{\bibfnamefont{C.}~\bibnamefont{Granerod}},
  \bibinfo{author}{\bibfnamefont{M.}~\bibnamefont{Leandersson}},
  \bibinfo{author}{\bibfnamefont{T.}~\bibnamefont{Balasubramanian}},
  \bibinfo{author}{\bibfnamefont{M.}~\bibnamefont{Hoesch}},
  \bibnamefont{et~al.}, \bibinfo{journal}{Nat Phys}
  \textbf{\bibinfo{volume}{10}}, \bibinfo{pages}{835} (\bibinfo{year}{2014}).

\bibitem[{\citenamefont{Mak et~al.}(2010)\citenamefont{Mak, Lee, Hone, Shan,
  and Heinz}}]{Mak:2010aa}
\bibinfo{author}{\bibfnamefont{K.~F.} \bibnamefont{Mak}},
  \bibinfo{author}{\bibfnamefont{C.}~\bibnamefont{Lee}},
  \bibinfo{author}{\bibfnamefont{J.}~\bibnamefont{Hone}},
  \bibinfo{author}{\bibfnamefont{J.}~\bibnamefont{Shan}}, \bibnamefont{and}
  \bibinfo{author}{\bibfnamefont{T.~F.} \bibnamefont{Heinz}},
  \bibinfo{journal}{Phys. Rev. Lett.} \textbf{\bibinfo{volume}{105}},
  \bibinfo{pages}{136805} (\bibinfo{year}{2010}).

\bibitem[{\citenamefont{Splendiani et~al.}(2010)\citenamefont{Splendiani, Sun,
  Zhang, Li, Kim, Chim, Galli, and Wang}}]{Splendiani:2010aa}
\bibinfo{author}{\bibfnamefont{A.}~\bibnamefont{Splendiani}},
  \bibinfo{author}{\bibfnamefont{L.}~\bibnamefont{Sun}},
  \bibinfo{author}{\bibfnamefont{Y.}~\bibnamefont{Zhang}},
  \bibinfo{author}{\bibfnamefont{T.}~\bibnamefont{Li}},
  \bibinfo{author}{\bibfnamefont{J.}~\bibnamefont{Kim}},
  \bibinfo{author}{\bibfnamefont{C.-Y.} \bibnamefont{Chim}},
  \bibinfo{author}{\bibfnamefont{G.}~\bibnamefont{Galli}}, \bibnamefont{and}
  \bibinfo{author}{\bibfnamefont{F.}~\bibnamefont{Wang}},
  \bibinfo{journal}{Nano Letters} \textbf{\bibinfo{volume}{10}},
  \bibinfo{pages}{1271} (\bibinfo{year}{2010}).

\bibitem[{\citenamefont{Katoch et~al.}(2018)\citenamefont{Katoch, Ulstrup,
  Koch, Moser, McCreary, Singh, Xu, Jonker, Kawakami, Bostwick
  et~al.}}]{Katoch:2018aa}
\bibinfo{author}{\bibfnamefont{J.}~\bibnamefont{Katoch}},
  \bibinfo{author}{\bibfnamefont{S.}~\bibnamefont{Ulstrup}},
  \bibinfo{author}{\bibfnamefont{R.~J.} \bibnamefont{Koch}},
  \bibinfo{author}{\bibfnamefont{S.}~\bibnamefont{Moser}},
  \bibinfo{author}{\bibfnamefont{K.~M.} \bibnamefont{McCreary}},
  \bibinfo{author}{\bibfnamefont{S.}~\bibnamefont{Singh}},
  \bibinfo{author}{\bibfnamefont{J.}~\bibnamefont{Xu}},
  \bibinfo{author}{\bibfnamefont{B.~T.} \bibnamefont{Jonker}},
  \bibinfo{author}{\bibfnamefont{R.~K.} \bibnamefont{Kawakami}},
  \bibinfo{author}{\bibfnamefont{A.}~\bibnamefont{Bostwick}},
  \bibnamefont{et~al.}, \bibinfo{journal}{Nature Physics}
  \textbf{\bibinfo{volume}{14}}, \bibinfo{pages}{355} (\bibinfo{year}{2018}).

\bibitem[{\citenamefont{Qiu et~al.}(2013)\citenamefont{Qiu, da~Jornada, and
  Louie}}]{Qiu:2013aa}
\bibinfo{author}{\bibfnamefont{D.~Y.} \bibnamefont{Qiu}},
  \bibinfo{author}{\bibfnamefont{F.~H.} \bibnamefont{da~Jornada}},
  \bibnamefont{and} \bibinfo{author}{\bibfnamefont{S.~G.} \bibnamefont{Louie}},
  \bibinfo{journal}{Phys. Rev. Lett.} \textbf{\bibinfo{volume}{111}},
  \bibinfo{pages}{216805} (\bibinfo{year}{2013}).

\bibitem[{\citenamefont{Grubi\v{s}i\'c~\v{C}abo
  et~al.}(2015)\citenamefont{Grubi\v{s}i\'c~\v{C}abo, Miwa, Gr\o{}nborg, Riley,
  Johannsen, Cacho, Alexander, Chapman, Springate, Grioni
  et~al.}}]{Antonija-Grubisic-Cabo:2015aa}
\bibinfo{author}{\bibfnamefont{A.}~\bibnamefont{Grubi\v{s}i\'c~\v{C}abo}},
  \bibinfo{author}{\bibfnamefont{J.~A.} \bibnamefont{Miwa}},
  \bibinfo{author}{\bibfnamefont{S.~S.} \bibnamefont{Gr\o{}nborg}},
  \bibinfo{author}{\bibfnamefont{J.~M.} \bibnamefont{Riley}},
  \bibinfo{author}{\bibfnamefont{J.~C.} \bibnamefont{Johannsen}},
  \bibinfo{author}{\bibfnamefont{C.}~\bibnamefont{Cacho}},
  \bibinfo{author}{\bibfnamefont{O.}~\bibnamefont{Alexander}},
  \bibinfo{author}{\bibfnamefont{R.~T.} \bibnamefont{Chapman}},
  \bibinfo{author}{\bibfnamefont{E.}~\bibnamefont{Springate}},
  \bibinfo{author}{\bibfnamefont{M.}~\bibnamefont{Grioni}},
  \bibnamefont{et~al.}, \bibinfo{journal}{Nano Letters}
  \textbf{\bibinfo{volume}{15}}, \bibinfo{pages}{5883} (\bibinfo{year}{2015}).

\bibitem[{\citenamefont{Ulstrup
  et~al.}(2016{\natexlab{b}})\citenamefont{Ulstrup, {\v C}abo, Miwa, Riley,
  Gr{\o}nborg, Johannsen, Cacho, Alexander, Chapman, Springate
  et~al.}}]{Ulstrup:2016aa}
\bibinfo{author}{\bibfnamefont{S.}~\bibnamefont{Ulstrup}},
  \bibinfo{author}{\bibfnamefont{A.~G.} \bibnamefont{{\v C}abo}},
  \bibinfo{author}{\bibfnamefont{J.~A.} \bibnamefont{Miwa}},
  \bibinfo{author}{\bibfnamefont{J.~M.} \bibnamefont{Riley}},
  \bibinfo{author}{\bibfnamefont{S.~S.} \bibnamefont{Gr{\o}nborg}},
  \bibinfo{author}{\bibfnamefont{J.~C.} \bibnamefont{Johannsen}},
  \bibinfo{author}{\bibfnamefont{C.}~\bibnamefont{Cacho}},
  \bibinfo{author}{\bibfnamefont{O.}~\bibnamefont{Alexander}},
  \bibinfo{author}{\bibfnamefont{R.~T.} \bibnamefont{Chapman}},
  \bibinfo{author}{\bibfnamefont{E.}~\bibnamefont{Springate}},
  \bibnamefont{et~al.}, \bibinfo{journal}{ACS Nano}
  \textbf{\bibinfo{volume}{10}}, \bibinfo{pages}{6315}
  (\bibinfo{year}{2016}{\natexlab{b}}).

\bibitem[{\citenamefont{Bertoni et~al.}(2016)\citenamefont{Bertoni, Nicholson,
  Waldecker, H\"ubener, Monney, De~Giovannini, Puppin, Hoesch, Springate,
  Chapman et~al.}}]{Bertoni:2016aa}
\bibinfo{author}{\bibfnamefont{R.}~\bibnamefont{Bertoni}},
  \bibinfo{author}{\bibfnamefont{C.~W.} \bibnamefont{Nicholson}},
  \bibinfo{author}{\bibfnamefont{L.}~\bibnamefont{Waldecker}},
  \bibinfo{author}{\bibfnamefont{H.}~\bibnamefont{H\"ubener}},
  \bibinfo{author}{\bibfnamefont{C.}~\bibnamefont{Monney}},
  \bibinfo{author}{\bibfnamefont{U.}~\bibnamefont{De~Giovannini}},
  \bibinfo{author}{\bibfnamefont{M.}~\bibnamefont{Puppin}},
  \bibinfo{author}{\bibfnamefont{M.}~\bibnamefont{Hoesch}},
  \bibinfo{author}{\bibfnamefont{E.}~\bibnamefont{Springate}},
  \bibinfo{author}{\bibfnamefont{R.~T.} \bibnamefont{Chapman}},
  \bibnamefont{et~al.}, \bibinfo{journal}{Phys. Rev. Lett.}
  \textbf{\bibinfo{volume}{117}}, \bibinfo{pages}{277201}
  (\bibinfo{year}{2016}).

\bibitem[{\citenamefont{Ulstrup et~al.}(2017)\citenamefont{Ulstrup,
  \ifmmode~\check{C}\else \v{C}\fi{}abo, Biswas, Riley, Dendzik, Sanders,
  Bianchi, Cacho, Matselyukh, Chapman et~al.}}]{Ulstrup:2017aa}
\bibinfo{author}{\bibfnamefont{S.}~\bibnamefont{Ulstrup}},
  \bibinfo{author}{\bibfnamefont{A.~G. c. v. a.~c.}
  \bibnamefont{\ifmmode~\check{C}\else \v{C}\fi{}abo}},
  \bibinfo{author}{\bibfnamefont{D.}~\bibnamefont{Biswas}},
  \bibinfo{author}{\bibfnamefont{J.~M.} \bibnamefont{Riley}},
  \bibinfo{author}{\bibfnamefont{M.}~\bibnamefont{Dendzik}},
  \bibinfo{author}{\bibfnamefont{C.~E.} \bibnamefont{Sanders}},
  \bibinfo{author}{\bibfnamefont{M.}~\bibnamefont{Bianchi}},
  \bibinfo{author}{\bibfnamefont{C.}~\bibnamefont{Cacho}},
  \bibinfo{author}{\bibfnamefont{D.}~\bibnamefont{Matselyukh}},
  \bibinfo{author}{\bibfnamefont{R.~T.} \bibnamefont{Chapman}},
  \bibnamefont{et~al.}, \bibinfo{journal}{Phys. Rev. B}
  \textbf{\bibinfo{volume}{95}}, \bibinfo{pages}{041405}
  (\bibinfo{year}{2017}).

\bibitem[{\citenamefont{Beyer et~al.}(2019)\citenamefont{Beyer, Rohde, Grubi{\v
  s}i{\'c}~{\v C}abo, Stange, Jacobsen, Bignardi, Lizzit, Lacovig, Sanders,
  Lizzit et~al.}}]{Beyer:2019aa}
\bibinfo{author}{\bibfnamefont{H.}~\bibnamefont{Beyer}},
  \bibinfo{author}{\bibfnamefont{G.}~\bibnamefont{Rohde}},
  \bibinfo{author}{\bibfnamefont{A.}~\bibnamefont{Grubi{\v s}i{\'c}~{\v
  C}abo}}, \bibinfo{author}{\bibfnamefont{A.}~\bibnamefont{Stange}},
  \bibinfo{author}{\bibfnamefont{T.}~\bibnamefont{Jacobsen}},
  \bibinfo{author}{\bibfnamefont{L.}~\bibnamefont{Bignardi}},
  \bibinfo{author}{\bibfnamefont{D.}~\bibnamefont{Lizzit}},
  \bibinfo{author}{\bibfnamefont{P.}~\bibnamefont{Lacovig}},
  \bibinfo{author}{\bibfnamefont{C.}~\bibnamefont{Sanders}},
  \bibinfo{author}{\bibfnamefont{S.}~\bibnamefont{Lizzit}},
  \bibnamefont{et~al.}, \bibinfo{journal}{Physical Review Letters}
  \textbf{\bibinfo{volume}{123}} (\bibinfo{year}{2019}).

\bibitem[{\citenamefont{Kutnyakhov et~al.}(2020)\citenamefont{Kutnyakhov, Xian,
  Dendzik, Heber, Pressacco, Agustsson, Wenthaus, Meyer, Gieschen, Mercurio
  et~al.}}]{Kutnyakhov:2020aa}
\bibinfo{author}{\bibfnamefont{D.}~\bibnamefont{Kutnyakhov}},
  \bibinfo{author}{\bibfnamefont{R.~P.} \bibnamefont{Xian}},
  \bibinfo{author}{\bibfnamefont{M.}~\bibnamefont{Dendzik}},
  \bibinfo{author}{\bibfnamefont{M.}~\bibnamefont{Heber}},
  \bibinfo{author}{\bibfnamefont{F.}~\bibnamefont{Pressacco}},
  \bibinfo{author}{\bibfnamefont{S.~Y.} \bibnamefont{Agustsson}},
  \bibinfo{author}{\bibfnamefont{L.}~\bibnamefont{Wenthaus}},
  \bibinfo{author}{\bibfnamefont{H.}~\bibnamefont{Meyer}},
  \bibinfo{author}{\bibfnamefont{S.}~\bibnamefont{Gieschen}},
  \bibinfo{author}{\bibfnamefont{G.}~\bibnamefont{Mercurio}},
  \bibnamefont{et~al.}, \bibinfo{journal}{Review of Scientific Instruments}
  \textbf{\bibinfo{volume}{91}}, \bibinfo{pages}{013109}
  (\bibinfo{year}{2020}).

\bibitem[{\citenamefont{Chernikov
  et~al.}(2015{\natexlab{a}})\citenamefont{Chernikov, Ruppert, Hill, Rigosi,
  and Heinz}}]{Chernikov:2015aa}
\bibinfo{author}{\bibfnamefont{A.}~\bibnamefont{Chernikov}},
  \bibinfo{author}{\bibfnamefont{C.}~\bibnamefont{Ruppert}},
  \bibinfo{author}{\bibfnamefont{H.~M.} \bibnamefont{Hill}},
  \bibinfo{author}{\bibfnamefont{A.~F.} \bibnamefont{Rigosi}},
  \bibnamefont{and} \bibinfo{author}{\bibfnamefont{T.~F.} \bibnamefont{Heinz}},
  \bibinfo{journal}{Nat Photon} \textbf{\bibinfo{volume}{9}},
  \bibinfo{pages}{466} (\bibinfo{year}{2015}{\natexlab{a}}).

\bibitem[{\citenamefont{Eickholt et~al.}(2018)\citenamefont{Eickholt, Sanders,
  Dendzik, Bignardi, Lizzit, Lizzit, Bruix, Hofmann, and
  Donath}}]{Eickholt:2018aa}
\bibinfo{author}{\bibfnamefont{P.}~\bibnamefont{Eickholt}},
  \bibinfo{author}{\bibfnamefont{C.}~\bibnamefont{Sanders}},
  \bibinfo{author}{\bibfnamefont{M.}~\bibnamefont{Dendzik}},
  \bibinfo{author}{\bibfnamefont{L.}~\bibnamefont{Bignardi}},
  \bibinfo{author}{\bibfnamefont{D.}~\bibnamefont{Lizzit}},
  \bibinfo{author}{\bibfnamefont{S.}~\bibnamefont{Lizzit}},
  \bibinfo{author}{\bibfnamefont{A.}~\bibnamefont{Bruix}},
  \bibinfo{author}{\bibfnamefont{P.}~\bibnamefont{Hofmann}}, \bibnamefont{and}
  \bibinfo{author}{\bibfnamefont{M.}~\bibnamefont{Donath}},
  \bibinfo{journal}{Phys. Rev. Lett.} \textbf{\bibinfo{volume}{121}},
  \bibinfo{pages}{136402} (\bibinfo{year}{2018}).

\bibitem[{\citenamefont{Zhang et~al.}(2015)\citenamefont{Zhang, Chen, Johnson,
  Li, Li, Mende, Feenstra, and Shih}}]{Zhang:2015aa}
\bibinfo{author}{\bibfnamefont{C.}~\bibnamefont{Zhang}},
  \bibinfo{author}{\bibfnamefont{Y.}~\bibnamefont{Chen}},
  \bibinfo{author}{\bibfnamefont{A.}~\bibnamefont{Johnson}},
  \bibinfo{author}{\bibfnamefont{M.-Y.} \bibnamefont{Li}},
  \bibinfo{author}{\bibfnamefont{L.-J.} \bibnamefont{Li}},
  \bibinfo{author}{\bibfnamefont{P.~C.} \bibnamefont{Mende}},
  \bibinfo{author}{\bibfnamefont{R.~M.} \bibnamefont{Feenstra}},
  \bibnamefont{and} \bibinfo{author}{\bibfnamefont{C.-K.} \bibnamefont{Shih}},
  \bibinfo{journal}{Nano Letters} \textbf{\bibinfo{volume}{15}},
  \bibinfo{pages}{6494} (\bibinfo{year}{2015}).

\bibitem[{\citenamefont{Bruix et~al.}(2016)\citenamefont{Bruix, Miwa,
  Hauptmann, Wegner, Ulstrup, Gr\o{}nborg, Sanders, Dendzik, Grubi\ifmmode
  \check{s}\else \v{s}\fi{}i\ifmmode \acute{c}\else \'{c}\fi{}
  \ifmmode~\check{C}\else \v{C}\fi{}abo, Bianchi et~al.}}]{Bruix:2016aa}
\bibinfo{author}{\bibfnamefont{A.}~\bibnamefont{Bruix}},
  \bibinfo{author}{\bibfnamefont{J.~A.} \bibnamefont{Miwa}},
  \bibinfo{author}{\bibfnamefont{N.}~\bibnamefont{Hauptmann}},
  \bibinfo{author}{\bibfnamefont{D.}~\bibnamefont{Wegner}},
  \bibinfo{author}{\bibfnamefont{S.}~\bibnamefont{Ulstrup}},
  \bibinfo{author}{\bibfnamefont{S.~S.} \bibnamefont{Gr\o{}nborg}},
  \bibinfo{author}{\bibfnamefont{C.~E.} \bibnamefont{Sanders}},
  \bibinfo{author}{\bibfnamefont{M.}~\bibnamefont{Dendzik}},
  \bibinfo{author}{\bibfnamefont{A.}~\bibnamefont{Grubi\ifmmode \check{s}\else
  \v{s}\fi{}i\ifmmode \acute{c}\else \'{c}\fi{} \ifmmode~\check{C}\else
  \v{C}\fi{}abo}}, \bibinfo{author}{\bibfnamefont{M.}~\bibnamefont{Bianchi}},
  \bibnamefont{et~al.}, \bibinfo{journal}{Phys. Rev. B}
  \textbf{\bibinfo{volume}{93}}, \bibinfo{pages}{165422}
  (\bibinfo{year}{2016}).

\bibitem[{\citenamefont{Riley et~al.}(2015)\citenamefont{Riley, Meevasana,
  Bawden, Asakawa, Takayama, Eknapakul, Kim, Hoesch, Mo, Takagi
  et~al.}}]{Riley:2015aa}
\bibinfo{author}{\bibfnamefont{J.~M.} \bibnamefont{Riley}},
  \bibinfo{author}{\bibfnamefont{W.}~\bibnamefont{Meevasana}},
  \bibinfo{author}{\bibfnamefont{L.}~\bibnamefont{Bawden}},
  \bibinfo{author}{\bibfnamefont{M.}~\bibnamefont{Asakawa}},
  \bibinfo{author}{\bibfnamefont{T.}~\bibnamefont{Takayama}},
  \bibinfo{author}{\bibfnamefont{T.}~\bibnamefont{Eknapakul}},
  \bibinfo{author}{\bibfnamefont{T.~K.} \bibnamefont{Kim}},
  \bibinfo{author}{\bibfnamefont{M.}~\bibnamefont{Hoesch}},
  \bibinfo{author}{\bibfnamefont{S.~K.} \bibnamefont{Mo}},
  \bibinfo{author}{\bibfnamefont{H.}~\bibnamefont{Takagi}},
  \bibnamefont{et~al.}, \bibinfo{journal}{Nat Nano}
  \textbf{\bibinfo{volume}{10}}, \bibinfo{pages}{1043} (\bibinfo{year}{2015}).

\bibitem[{\citenamefont{Chernikov
  et~al.}(2015{\natexlab{b}})\citenamefont{Chernikov, van~der Zande, Hill,
  Rigosi, Velauthapillai, Hone, and Heinz}}]{Chernikov:2015ab}
\bibinfo{author}{\bibfnamefont{A.}~\bibnamefont{Chernikov}},
  \bibinfo{author}{\bibfnamefont{A.~M.} \bibnamefont{van~der Zande}},
  \bibinfo{author}{\bibfnamefont{H.~M.} \bibnamefont{Hill}},
  \bibinfo{author}{\bibfnamefont{A.~F.} \bibnamefont{Rigosi}},
  \bibinfo{author}{\bibfnamefont{A.}~\bibnamefont{Velauthapillai}},
  \bibinfo{author}{\bibfnamefont{J.}~\bibnamefont{Hone}}, \bibnamefont{and}
  \bibinfo{author}{\bibfnamefont{T.~F.} \bibnamefont{Heinz}},
  \bibinfo{journal}{Physical Review Letters} \textbf{\bibinfo{volume}{115}}
  (\bibinfo{year}{2015}{\natexlab{b}}).

\bibitem[{\citenamefont{Young et~al.}(2012)\citenamefont{Young, Dean, Meric,
  Sorgenfrei, Ren, Watanabe, Taniguchi, Hone, Shepard, and Kim}}]{Young:2012ab}
\bibinfo{author}{\bibfnamefont{A.~F.} \bibnamefont{Young}},
  \bibinfo{author}{\bibfnamefont{C.~R.} \bibnamefont{Dean}},
  \bibinfo{author}{\bibfnamefont{I.}~\bibnamefont{Meric}},
  \bibinfo{author}{\bibfnamefont{S.}~\bibnamefont{Sorgenfrei}},
  \bibinfo{author}{\bibfnamefont{H.}~\bibnamefont{Ren}},
  \bibinfo{author}{\bibfnamefont{K.}~\bibnamefont{Watanabe}},
  \bibinfo{author}{\bibfnamefont{T.}~\bibnamefont{Taniguchi}},
  \bibinfo{author}{\bibfnamefont{J.}~\bibnamefont{Hone}},
  \bibinfo{author}{\bibfnamefont{K.~L.} \bibnamefont{Shepard}},
  \bibnamefont{and} \bibinfo{author}{\bibfnamefont{P.}~\bibnamefont{Kim}},
  \bibinfo{journal}{Phys. Rev. B} \textbf{\bibinfo{volume}{85}},
  \bibinfo{pages}{235458} (\bibinfo{year}{2012}).

\bibitem[{\citenamefont{Bostwick et~al.}(2009)\citenamefont{Bostwick,
  McChesney, Emtsev, Seyller, Horn, Kevan, and Rotenberg}}]{Bostwick:2009aa}
\bibinfo{author}{\bibfnamefont{A.}~\bibnamefont{Bostwick}},
  \bibinfo{author}{\bibfnamefont{J.~L.} \bibnamefont{McChesney}},
  \bibinfo{author}{\bibfnamefont{K.~V.} \bibnamefont{Emtsev}},
  \bibinfo{author}{\bibfnamefont{T.}~\bibnamefont{Seyller}},
  \bibinfo{author}{\bibfnamefont{K.}~\bibnamefont{Horn}},
  \bibinfo{author}{\bibfnamefont{S.~D.} \bibnamefont{Kevan}}, \bibnamefont{and}
  \bibinfo{author}{\bibfnamefont{E.}~\bibnamefont{Rotenberg}},
  \bibinfo{journal}{Physical Review Letters} \textbf{\bibinfo{volume}{103}},
  \bibinfo{eid}{056404} (pages~\bibinfo{numpages}{4}) (\bibinfo{year}{2009}).

\bibitem[{\citenamefont{Muralt and Pohl}(1986)}]{Muralt:1986aa}
\bibinfo{author}{\bibfnamefont{P.}~\bibnamefont{Muralt}} \bibnamefont{and}
  \bibinfo{author}{\bibfnamefont{D.~W.} \bibnamefont{Pohl}},
  \bibinfo{journal}{Applied Physics Letters} \textbf{\bibinfo{volume}{48}},
  \bibinfo{pages}{514} (\bibinfo{year}{1986}).

\bibitem[{\citenamefont{Zhang et~al.}(2016)\citenamefont{Zhang, Li, Chen,
  Durand, Li, and Zhang}}]{Zhang:2016ag}
\bibinfo{author}{\bibfnamefont{H.}~\bibnamefont{Zhang}},
  \bibinfo{author}{\bibfnamefont{X.}~\bibnamefont{Li}},
  \bibinfo{author}{\bibfnamefont{Y.}~\bibnamefont{Chen}},
  \bibinfo{author}{\bibfnamefont{C.}~\bibnamefont{Durand}},
  \bibinfo{author}{\bibfnamefont{A.-P.} \bibnamefont{Li}}, \bibnamefont{and}
  \bibinfo{author}{\bibfnamefont{X.-G.} \bibnamefont{Zhang}},
  \bibinfo{journal}{Review of Scientific Instruments}
  \textbf{\bibinfo{volume}{87}}, \bibinfo{pages}{083702}
  (\bibinfo{year}{2016}).

\bibitem[{\citenamefont{Tetienne et~al.}(2017)\citenamefont{Tetienne,
  Dontschuk, Broadway, Stacey, Simpson, and Hollenberg}}]{Tetienne:2017aa}
\bibinfo{author}{\bibfnamefont{J.-P.} \bibnamefont{Tetienne}},
  \bibinfo{author}{\bibfnamefont{N.}~\bibnamefont{Dontschuk}},
  \bibinfo{author}{\bibfnamefont{D.~A.} \bibnamefont{Broadway}},
  \bibinfo{author}{\bibfnamefont{A.}~\bibnamefont{Stacey}},
  \bibinfo{author}{\bibfnamefont{D.~A.} \bibnamefont{Simpson}},
  \bibnamefont{and} \bibinfo{author}{\bibfnamefont{L.~C.~L.}
  \bibnamefont{Hollenberg}}, \bibinfo{journal}{Science Advances}
  \textbf{\bibinfo{volume}{3}}, \bibinfo{pages}{e1602429}
  (\bibinfo{year}{2017}).
  
\bibitem[{\citenamefont{Voigtl{\"a}nder
  et~al.}(2018)\citenamefont{Voigtl{\"a}nder, Cherepanov, Korte, Leis, Cuma,
  Just, and L{\"u}pke}}]{Voigtlander:2018aa}
\bibinfo{author}{\bibfnamefont{B.}~\bibnamefont{Voigtl{\"a}nder}},
  \bibinfo{author}{\bibfnamefont{V.}~\bibnamefont{Cherepanov}},
  \bibinfo{author}{\bibfnamefont{S.}~\bibnamefont{Korte}},
  \bibinfo{author}{\bibfnamefont{A.}~\bibnamefont{Leis}},
  \bibinfo{author}{\bibfnamefont{D.}~\bibnamefont{Cuma}},
  \bibinfo{author}{\bibfnamefont{S.}~\bibnamefont{Just}}, \bibnamefont{and}
  \bibinfo{author}{\bibfnamefont{F.}~\bibnamefont{L{\"u}pke}},
  \bibinfo{journal}{Review of Scientific Instruments}
  \textbf{\bibinfo{volume}{89}}, \bibinfo{pages}{101101}
  (\bibinfo{year}{2018}).


%\bibitem[{\citenamefont{L{\"u}pke et~al.}(2017)\citenamefont{L{\"u}pke,
%  Eschbach, Heider, Lanius, Sch{\"u}ffelgen, Rosenbach, von~den Driesch,
%  Cherepanov, Mussler, Plucinski et~al.}}]{Lupke:2017aa}
%\bibinfo{author}{\bibfnamefont{F.}~\bibnamefont{L{\"u}pke}},
%  \bibinfo{author}{\bibfnamefont{M.}~\bibnamefont{Eschbach}},
%  \bibinfo{author}{\bibfnamefont{T.}~\bibnamefont{Heider}},
%  \bibinfo{author}{\bibfnamefont{M.}~\bibnamefont{Lanius}},
%  \bibinfo{author}{\bibfnamefont{P.}~\bibnamefont{Sch{\"u}ffelgen}},
%  \bibinfo{author}{\bibfnamefont{D.}~\bibnamefont{Rosenbach}},
%  \bibinfo{author}{\bibfnamefont{N.}~\bibnamefont{von~den Driesch}},
%  \bibinfo{author}{\bibfnamefont{V.}~\bibnamefont{Cherepanov}},
%  \bibinfo{author}{\bibfnamefont{G.}~\bibnamefont{Mussler}},
%  \bibinfo{author}{\bibfnamefont{L.}~\bibnamefont{Plucinski}},
%  \bibnamefont{et~al.}, \bibinfo{journal}{Nature Communications}
%  \textbf{\bibinfo{volume}{8}}, \bibinfo{pages}{15704} (\bibinfo{year}{2017}).

\bibitem[{\citenamefont{Ella et~al.}(2019)\citenamefont{Ella, Rozen, Birkbeck,
  Ben-Shalom, Perello, Zultak, Taniguchi, Watanabe, Geim, Ilani
  et~al.}}]{Ella:2019aa}
\bibinfo{author}{\bibfnamefont{L.}~\bibnamefont{Ella}},
  \bibinfo{author}{\bibfnamefont{A.}~\bibnamefont{Rozen}},
  \bibinfo{author}{\bibfnamefont{J.}~\bibnamefont{Birkbeck}},
  \bibinfo{author}{\bibfnamefont{M.}~\bibnamefont{Ben-Shalom}},
  \bibinfo{author}{\bibfnamefont{D.}~\bibnamefont{Perello}},
  \bibinfo{author}{\bibfnamefont{J.}~\bibnamefont{Zultak}},
  \bibinfo{author}{\bibfnamefont{T.}~\bibnamefont{Taniguchi}},
  \bibinfo{author}{\bibfnamefont{K.}~\bibnamefont{Watanabe}},
  \bibinfo{author}{\bibfnamefont{A.~K.} \bibnamefont{Geim}},
  \bibinfo{author}{\bibfnamefont{S.}~\bibnamefont{Ilani}},
  \bibnamefont{et~al.}, \bibinfo{journal}{Nature Nanotechnology}
  \textbf{\bibinfo{volume}{14}}, \bibinfo{pages}{480} (\bibinfo{year}{2019}).

\bibitem[{\citenamefont{Gudde et~al.}(2007)\citenamefont{Gudde, Rohleder,
  Meier, Koch, and Hofer}}]{Gudde:2007aa}
\bibinfo{author}{\bibfnamefont{J.}~\bibnamefont{Gudde}},
  \bibinfo{author}{\bibfnamefont{M.}~\bibnamefont{Rohleder}},
  \bibinfo{author}{\bibfnamefont{T.}~\bibnamefont{Meier}},
  \bibinfo{author}{\bibfnamefont{S.~W.} \bibnamefont{Koch}}, \bibnamefont{and}
  \bibinfo{author}{\bibfnamefont{U.}~\bibnamefont{Hofer}},
  \bibinfo{journal}{Science} \textbf{\bibinfo{volume}{318}},
  \bibinfo{pages}{1287} (\bibinfo{year}{2007}).

\bibitem[{\citenamefont{Reimann et~al.}(2018)\citenamefont{Reimann,
  Schlauderer, Schmid, Langer, Baierl, Kokh, Tereshchenko, Kimura, Lange,
  G{\"u}dde et~al.}}]{Reimann:2018aa}
\bibinfo{author}{\bibfnamefont{J.}~\bibnamefont{Reimann}},
  \bibinfo{author}{\bibfnamefont{S.}~\bibnamefont{Schlauderer}},
  \bibinfo{author}{\bibfnamefont{C.~P.} \bibnamefont{Schmid}},
  \bibinfo{author}{\bibfnamefont{F.}~\bibnamefont{Langer}},
  \bibinfo{author}{\bibfnamefont{S.}~\bibnamefont{Baierl}},
  \bibinfo{author}{\bibfnamefont{K.~A.} \bibnamefont{Kokh}},
  \bibinfo{author}{\bibfnamefont{O.~E.} \bibnamefont{Tereshchenko}},
  \bibinfo{author}{\bibfnamefont{A.}~\bibnamefont{Kimura}},
  \bibinfo{author}{\bibfnamefont{C.}~\bibnamefont{Lange}},
  \bibinfo{author}{\bibfnamefont{J.}~\bibnamefont{G{\"u}dde}},
  \bibnamefont{et~al.}, \bibinfo{journal}{Nature}
  \textbf{\bibinfo{volume}{562}}, \bibinfo{pages}{396} (\bibinfo{year}{2018}).

\bibitem[{\citenamefont{Nakayama et~al.}(2020)\citenamefont{Nakayama, Kera, and
  Ueno}}]{Nakayama:2020aa}
\bibinfo{author}{\bibfnamefont{Y.}~\bibnamefont{Nakayama}},
  \bibinfo{author}{\bibfnamefont{S.}~\bibnamefont{Kera}}, \bibnamefont{and}
  \bibinfo{author}{\bibfnamefont{N.}~\bibnamefont{Ueno}},
  \bibinfo{journal}{Journal of Materials Chemistry C}
  \textbf{\bibinfo{volume}{8}}, \bibinfo{pages}{9090} (\bibinfo{year}{2020}).

\bibitem[{\citenamefont{Bostwick et~al.}(2007)\citenamefont{Bostwick, Ohta,
  Seyller, Horn, and Rotenberg}}]{Bostwick:2007aa}
\bibinfo{author}{\bibfnamefont{A.}~\bibnamefont{Bostwick}},
  \bibinfo{author}{\bibfnamefont{T.}~\bibnamefont{Ohta}},
  \bibinfo{author}{\bibfnamefont{T.}~\bibnamefont{Seyller}},
  \bibinfo{author}{\bibfnamefont{K.}~\bibnamefont{Horn}}, \bibnamefont{and}
  \bibinfo{author}{\bibfnamefont{E.}~\bibnamefont{Rotenberg}},
  \bibinfo{journal}{Nature Physics} \textbf{\bibinfo{volume}{3}},
  \bibinfo{pages}{36} (\bibinfo{year}{2007}).

\bibitem[{\citenamefont{Zhou et~al.}(2005)\citenamefont{Zhou, Wannberg, Yang,
  Brouet, Sun, Douglas, Dessau, Hussain, and Shen}}]{Zhou:2005ab}
\bibinfo{author}{\bibfnamefont{X.}~\bibnamefont{Zhou}},
  \bibinfo{author}{\bibfnamefont{B.}~\bibnamefont{Wannberg}},
  \bibinfo{author}{\bibfnamefont{W.}~\bibnamefont{Yang}},
  \bibinfo{author}{\bibfnamefont{V.}~\bibnamefont{Brouet}},
  \bibinfo{author}{\bibfnamefont{Z.}~\bibnamefont{Sun}},
  \bibinfo{author}{\bibfnamefont{J.}~\bibnamefont{Douglas}},
  \bibinfo{author}{\bibfnamefont{D.}~\bibnamefont{Dessau}},
  \bibinfo{author}{\bibfnamefont{Z.}~\bibnamefont{Hussain}}, \bibnamefont{and}
  \bibinfo{author}{\bibfnamefont{Z.-X.} \bibnamefont{Shen}},
  \bibinfo{journal}{Journal of Electron Spectroscopy and Related Phenomena}
  \textbf{\bibinfo{volume}{142}}, \bibinfo{pages}{27} (\bibinfo{year}{2005}).

\bibitem[{\citenamefont{Passlack et~al.}(2006)\citenamefont{Passlack, Mathias,
  Andreyev, Mittnacht, Aeschlimann, and Bauer}}]{Passlack:2006aa}
\bibinfo{author}{\bibfnamefont{S.}~\bibnamefont{Passlack}},
  \bibinfo{author}{\bibfnamefont{S.}~\bibnamefont{Mathias}},
  \bibinfo{author}{\bibfnamefont{O.}~\bibnamefont{Andreyev}},
  \bibinfo{author}{\bibfnamefont{D.}~\bibnamefont{Mittnacht}},
  \bibinfo{author}{\bibfnamefont{M.}~\bibnamefont{Aeschlimann}},
  \bibnamefont{and} \bibinfo{author}{\bibfnamefont{M.}~\bibnamefont{Bauer}},
  \bibinfo{journal}{Journal of Applied Physics} \textbf{\bibinfo{volume}{100}},
  \bibinfo{pages}{024912} (\bibinfo{year}{2006}).

\bibitem[{\citenamefont{Hellmann et~al.}(2009)\citenamefont{Hellmann,
  Rossnagel, Marczynski-B\"uhlow, and Kipp}}]{Hellmann:2009aa}
\bibinfo{author}{\bibfnamefont{S.}~\bibnamefont{Hellmann}},
  \bibinfo{author}{\bibfnamefont{K.}~\bibnamefont{Rossnagel}},
  \bibinfo{author}{\bibfnamefont{M.}~\bibnamefont{Marczynski-B\"uhlow}},
  \bibnamefont{and} \bibinfo{author}{\bibfnamefont{L.}~\bibnamefont{Kipp}},
  \bibinfo{journal}{Phys. Rev. B} \textbf{\bibinfo{volume}{79}},
  \bibinfo{pages}{035402} (\bibinfo{year}{2009}).

\bibitem[{\citenamefont{Hellmann
  et~al.}(2012{\natexlab{a}})\citenamefont{Hellmann, Ott, Kipp, and
  Rossnagel}}]{Hellmann:2012ac}
\bibinfo{author}{\bibfnamefont{S.}~\bibnamefont{Hellmann}},
  \bibinfo{author}{\bibfnamefont{T.}~\bibnamefont{Ott}},
  \bibinfo{author}{\bibfnamefont{L.}~\bibnamefont{Kipp}}, \bibnamefont{and}
  \bibinfo{author}{\bibfnamefont{K.}~\bibnamefont{Rossnagel}},
  \bibinfo{journal}{Phys. Rev. B} \textbf{\bibinfo{volume}{85}},
  \bibinfo{pages}{075109} (\bibinfo{year}{2012}{\natexlab{a}}).

\bibitem[{\citenamefont{Kr{{\"o}}ger et~al.}(2001)\citenamefont{Kr{{\"o}}ger,
  Greber, Kreutz, and Osterwalder}}]{Kroger:2001aa}
\bibinfo{author}{\bibfnamefont{J.}~\bibnamefont{Kr{{\"o}}ger}},
  \bibinfo{author}{\bibfnamefont{T.}~\bibnamefont{Greber}},
  \bibinfo{author}{\bibfnamefont{T.~J.} \bibnamefont{Kreutz}},
  \bibnamefont{and}
  \bibinfo{author}{\bibfnamefont{J.}~\bibnamefont{Osterwalder}},
  \bibinfo{journal}{Journal of Electron Spectroscopy and Related Phenomena}
  \textbf{\bibinfo{volume}{113}}, \bibinfo{pages}{241} (\bibinfo{year}{2001}).

\bibitem[{\citenamefont{Ulstrup
  et~al.}(2014{\natexlab{a}})\citenamefont{Ulstrup, Johannsen, Grioni, and
  Hofmann}}]{Ulstrup:2014aa}
\bibinfo{author}{\bibfnamefont{S.}~\bibnamefont{Ulstrup}},
  \bibinfo{author}{\bibfnamefont{J.~C.} \bibnamefont{Johannsen}},
  \bibinfo{author}{\bibfnamefont{M.}~\bibnamefont{Grioni}}, \bibnamefont{and}
  \bibinfo{author}{\bibfnamefont{P.}~\bibnamefont{Hofmann}},
  \bibinfo{journal}{Review of Scientific Instruments}
  \textbf{\bibinfo{volume}{85}}, \bibinfo{eid}{013907}
  (\bibinfo{year}{2014}{\natexlab{a}}).

\bibitem[{\citenamefont{Xian et~al.}(2019)\citenamefont{Xian, Acremann,
  Agustsson, Dendzik, B{\"u}hlmann, Curcio, Kutnyakhov, Pressacco, Heber, Dong
  et~al.}}]{Xian:2019aa}
\bibinfo{author}{\bibfnamefont{R.~P.} \bibnamefont{Xian}},
  \bibinfo{author}{\bibfnamefont{Y.}~\bibnamefont{Acremann}},
  \bibinfo{author}{\bibfnamefont{S.~Y.} \bibnamefont{Agustsson}},
  \bibinfo{author}{\bibfnamefont{M.}~\bibnamefont{Dendzik}},
  \bibinfo{author}{\bibfnamefont{K.}~\bibnamefont{B{\"u}hlmann}},
  \bibinfo{author}{\bibfnamefont{D.}~\bibnamefont{Curcio}},
  \bibinfo{author}{\bibfnamefont{D.}~\bibnamefont{Kutnyakhov}},
  \bibinfo{author}{\bibfnamefont{F.}~\bibnamefont{Pressacco}},
  \bibinfo{author}{\bibfnamefont{M.}~\bibnamefont{Heber}},
  \bibinfo{author}{\bibfnamefont{S.}~\bibnamefont{Dong}}, \bibnamefont{et~al.}
  (\bibinfo{year}{2019}), \eprint{arXiv.1909.07714}.

\bibitem[{\citenamefont{Miwa et~al.}(2013)\citenamefont{Miwa, Hofmann, Simmons,
  and Wells}}]{Miwa:2013aa}
\bibinfo{author}{\bibfnamefont{J.~A.} \bibnamefont{Miwa}},
  \bibinfo{author}{\bibfnamefont{P.}~\bibnamefont{Hofmann}},
  \bibinfo{author}{\bibfnamefont{M.~Y.} \bibnamefont{Simmons}},
  \bibnamefont{and} \bibinfo{author}{\bibfnamefont{J.~W.} \bibnamefont{Wells}},
  \bibinfo{journal}{Phys. Rev. Lett.} \textbf{\bibinfo{volume}{110}},
  \bibinfo{pages}{136801} (\bibinfo{year}{2013}).

\bibitem[{\citenamefont{Suga et~al.}(2004)\citenamefont{Suga, Shigemoto,
  Sekiyama, Imada, Yamasaki, Irizawa, Kasai, Saitoh, Muro, Tomita
  et~al.}}]{Suga:2004aa}
\bibinfo{author}{\bibfnamefont{S.}~\bibnamefont{Suga}},
  \bibinfo{author}{\bibfnamefont{A.}~\bibnamefont{Shigemoto}},
  \bibinfo{author}{\bibfnamefont{A.}~\bibnamefont{Sekiyama}},
  \bibinfo{author}{\bibfnamefont{S.}~\bibnamefont{Imada}},
  \bibinfo{author}{\bibfnamefont{A.}~\bibnamefont{Yamasaki}},
  \bibinfo{author}{\bibfnamefont{A.}~\bibnamefont{Irizawa}},
  \bibinfo{author}{\bibfnamefont{S.}~\bibnamefont{Kasai}},
  \bibinfo{author}{\bibfnamefont{Y.}~\bibnamefont{Saitoh}},
  \bibinfo{author}{\bibfnamefont{T.}~\bibnamefont{Muro}},
  \bibinfo{author}{\bibfnamefont{N.}~\bibnamefont{Tomita}},
  \bibnamefont{et~al.}, \bibinfo{journal}{Phys. Rev. B}
  \textbf{\bibinfo{volume}{70}}, \bibinfo{pages}{155106}
  (\bibinfo{year}{2004}).

\bibitem[{\citenamefont{Kobayashi et~al.}(2012)\citenamefont{Kobayashi, Muneta,
  Schmitt, Patthey, Ohya, Tanaka, Oshima, and Strocov}}]{Kobayashi:2012aa}
\bibinfo{author}{\bibfnamefont{M.}~\bibnamefont{Kobayashi}},
  \bibinfo{author}{\bibfnamefont{I.}~\bibnamefont{Muneta}},
  \bibinfo{author}{\bibfnamefont{T.}~\bibnamefont{Schmitt}},
  \bibinfo{author}{\bibfnamefont{L.}~\bibnamefont{Patthey}},
  \bibinfo{author}{\bibfnamefont{S.}~\bibnamefont{Ohya}},
  \bibinfo{author}{\bibfnamefont{M.}~\bibnamefont{Tanaka}},
  \bibinfo{author}{\bibfnamefont{M.}~\bibnamefont{Oshima}}, \bibnamefont{and}
  \bibinfo{author}{\bibfnamefont{V.~N.} \bibnamefont{Strocov}},
  \bibinfo{journal}{Applied Physics Letters} \textbf{\bibinfo{volume}{101}},
  \bibinfo{pages}{242103} (\bibinfo{year}{2012}).

\bibitem[{\citenamefont{Woodruff}(2002)}]{Woodruff:2002aa}
\bibinfo{author}{\bibfnamefont{D.}~\bibnamefont{Woodruff}},
  \bibinfo{journal}{Journal of Electron Spectroscopy and Related Phenomena}
  \textbf{\bibinfo{volume}{126}}, \bibinfo{pages}{55} (\bibinfo{year}{2002}).

\bibitem[{\citenamefont{Molodtsov et~al.}(1997)\citenamefont{Molodtsov,
  Richter, Danzenb\"acher, Wieling, Steinbeck, and
  Laubschat}}]{Molodtsov:1997aa}
\bibinfo{author}{\bibfnamefont{S.~L.} \bibnamefont{Molodtsov}},
  \bibinfo{author}{\bibfnamefont{M.}~\bibnamefont{Richter}},
  \bibinfo{author}{\bibfnamefont{S.}~\bibnamefont{Danzenb\"acher}},
  \bibinfo{author}{\bibfnamefont{S.}~\bibnamefont{Wieling}},
  \bibinfo{author}{\bibfnamefont{L.}~\bibnamefont{Steinbeck}},
  \bibnamefont{and}
  \bibinfo{author}{\bibfnamefont{C.}~\bibnamefont{Laubschat}},
  \bibinfo{journal}{Phys. Rev. Lett.} \textbf{\bibinfo{volume}{78}},
  \bibinfo{pages}{142} (\bibinfo{year}{1997}).

\bibitem[{\citenamefont{Hofmann et~al.}(2002)\citenamefont{Hofmann,
  S{\o}ndergaard, Agergaard, Hoffmann, Gayone, Zampieri, Lizzit, and
  Baraldi}}]{Hofmann:2002aa}
\bibinfo{author}{\bibfnamefont{P.}~\bibnamefont{Hofmann}},
  \bibinfo{author}{\bibfnamefont{C.}~\bibnamefont{S{\o}ndergaard}},
  \bibinfo{author}{\bibfnamefont{S.}~\bibnamefont{Agergaard}},
  \bibinfo{author}{\bibfnamefont{S.~V.} \bibnamefont{Hoffmann}},
  \bibinfo{author}{\bibfnamefont{J.~E.} \bibnamefont{Gayone}},
  \bibinfo{author}{\bibfnamefont{G.}~\bibnamefont{Zampieri}},
  \bibinfo{author}{\bibfnamefont{S.}~\bibnamefont{Lizzit}}, \bibnamefont{and}
  \bibinfo{author}{\bibfnamefont{A.}~\bibnamefont{Baraldi}},
  \bibinfo{journal}{Physical Review B} \textbf{\bibinfo{volume}{66}},
  \bibinfo{pages}{245422} (\bibinfo{year}{2002}).

\bibitem[{\citenamefont{Perfetti et~al.}(2007)\citenamefont{Perfetti, Loukakos,
  Lisowski, Bovensiepen, Eisaki, and Wolf}}]{Perfetti:2007aa}
\bibinfo{author}{\bibfnamefont{L.}~\bibnamefont{Perfetti}},
  \bibinfo{author}{\bibfnamefont{P.~A.} \bibnamefont{Loukakos}},
  \bibinfo{author}{\bibfnamefont{M.}~\bibnamefont{Lisowski}},
  \bibinfo{author}{\bibfnamefont{U.}~\bibnamefont{Bovensiepen}},
  \bibinfo{author}{\bibfnamefont{H.}~\bibnamefont{Eisaki}}, \bibnamefont{and}
  \bibinfo{author}{\bibfnamefont{M.}~\bibnamefont{Wolf}},
  \bibinfo{journal}{Phys. Rev. Lett.} \textbf{\bibinfo{volume}{99}},
  \bibinfo{pages}{197001} (\bibinfo{year}{2007}).

\bibitem[{\citenamefont{Johannsen et~al.}(2013)\citenamefont{Johannsen,
  Ulstrup, Cilento, Crepaldi, Zacchigna, Cacho, Turcu, Springate, Fromm, Raidel
  et~al.}}]{Johannsen:2013ab}
\bibinfo{author}{\bibfnamefont{J.~C.} \bibnamefont{Johannsen}},
  \bibinfo{author}{\bibfnamefont{S.}~\bibnamefont{Ulstrup}},
  \bibinfo{author}{\bibfnamefont{F.}~\bibnamefont{Cilento}},
  \bibinfo{author}{\bibfnamefont{A.}~\bibnamefont{Crepaldi}},
  \bibinfo{author}{\bibfnamefont{M.}~\bibnamefont{Zacchigna}},
  \bibinfo{author}{\bibfnamefont{C.}~\bibnamefont{Cacho}},
  \bibinfo{author}{\bibfnamefont{I.~C.~E.} \bibnamefont{Turcu}},
  \bibinfo{author}{\bibfnamefont{E.}~\bibnamefont{Springate}},
  \bibinfo{author}{\bibfnamefont{F.}~\bibnamefont{Fromm}},
  \bibinfo{author}{\bibfnamefont{C.}~\bibnamefont{Raidel}},
  \bibnamefont{et~al.}, \bibinfo{journal}{Phys. Rev. Lett.}
  \textbf{\bibinfo{volume}{111}}, \bibinfo{pages}{027403}
  (\bibinfo{year}{2013}).

\bibitem[{\citenamefont{Gierz et~al.}(2013)\citenamefont{Gierz, Petersen,
  Mitrano, Cacho, Turcu, Springate, St{\"o}hr, K{\"o}hler, Starke, and
  Cavalleri}}]{Gierz:2013aa}
\bibinfo{author}{\bibfnamefont{I.}~\bibnamefont{Gierz}},
  \bibinfo{author}{\bibfnamefont{J.~C.} \bibnamefont{Petersen}},
  \bibinfo{author}{\bibfnamefont{M.}~\bibnamefont{Mitrano}},
  \bibinfo{author}{\bibfnamefont{C.}~\bibnamefont{Cacho}},
  \bibinfo{author}{\bibfnamefont{I.~C.~E.} \bibnamefont{Turcu}},
  \bibinfo{author}{\bibfnamefont{E.}~\bibnamefont{Springate}},
  \bibinfo{author}{\bibfnamefont{A.}~\bibnamefont{St{\"o}hr}},
  \bibinfo{author}{\bibfnamefont{A.}~\bibnamefont{K{\"o}hler}},
  \bibinfo{author}{\bibfnamefont{U.}~\bibnamefont{Starke}}, \bibnamefont{and}
  \bibinfo{author}{\bibfnamefont{A.}~\bibnamefont{Cavalleri}},
  \bibinfo{journal}{Nature Materials} \textbf{\bibinfo{volume}{12}},
  \bibinfo{pages}{1119} (\bibinfo{year}{2013}).

\bibitem[{\citenamefont{Ulstrup
  et~al.}(2014{\natexlab{b}})\citenamefont{Ulstrup, Johannsen, Cilento, Miwa,
  Crepaldi, Zacchigna, Cacho, Chapman, Springate, Mammadov
  et~al.}}]{Ulstrup:2014ac}
\bibinfo{author}{\bibfnamefont{S.}~\bibnamefont{Ulstrup}},
  \bibinfo{author}{\bibfnamefont{J.~C.} \bibnamefont{Johannsen}},
  \bibinfo{author}{\bibfnamefont{F.}~\bibnamefont{Cilento}},
  \bibinfo{author}{\bibfnamefont{J.~A.} \bibnamefont{Miwa}},
  \bibinfo{author}{\bibfnamefont{A.}~\bibnamefont{Crepaldi}},
  \bibinfo{author}{\bibfnamefont{M.}~\bibnamefont{Zacchigna}},
  \bibinfo{author}{\bibfnamefont{C.}~\bibnamefont{Cacho}},
  \bibinfo{author}{\bibfnamefont{R.}~\bibnamefont{Chapman}},
  \bibinfo{author}{\bibfnamefont{E.}~\bibnamefont{Springate}},
  \bibinfo{author}{\bibfnamefont{S.}~\bibnamefont{Mammadov}},
  \bibnamefont{et~al.}, \bibinfo{journal}{Phys. Rev. Lett.}
  \textbf{\bibinfo{volume}{112}}, \bibinfo{pages}{257401}
  (\bibinfo{year}{2014}{\natexlab{b}}).

\bibitem[{\citenamefont{Rohde et~al.}(2018)\citenamefont{Rohde, Stange,
  M{\"u}ller, Behrendt, Oloff, Hanff, Albert, Hein, Rossnagel, and
  Bauer}}]{Rohde:2018ab}
\bibinfo{author}{\bibfnamefont{G.}~\bibnamefont{Rohde}},
  \bibinfo{author}{\bibfnamefont{A.}~\bibnamefont{Stange}},
  \bibinfo{author}{\bibfnamefont{A.}~\bibnamefont{M{\"u}ller}},
  \bibinfo{author}{\bibfnamefont{M.}~\bibnamefont{Behrendt}},
  \bibinfo{author}{\bibfnamefont{L.-P.} \bibnamefont{Oloff}},
  \bibinfo{author}{\bibfnamefont{K.}~\bibnamefont{Hanff}},
  \bibinfo{author}{\bibfnamefont{T.}~\bibnamefont{Albert}},
  \bibinfo{author}{\bibfnamefont{P.}~\bibnamefont{Hein}},
  \bibinfo{author}{\bibfnamefont{K.}~\bibnamefont{Rossnagel}},
  \bibnamefont{and} \bibinfo{author}{\bibfnamefont{M.}~\bibnamefont{Bauer}},
  \bibinfo{journal}{Physical Review Letters} \textbf{\bibinfo{volume}{121}}
  (\bibinfo{year}{2018}).

\bibitem[{\citenamefont{Perfetti et~al.}(2006)\citenamefont{Perfetti, Loukakos,
  Lisowski, Bovensiepen, Berger, Biermann, Cornaglia, Georges, and
  Wolf}}]{Perfetti:2006aa}
\bibinfo{author}{\bibfnamefont{L.}~\bibnamefont{Perfetti}},
  \bibinfo{author}{\bibfnamefont{P.~A.} \bibnamefont{Loukakos}},
  \bibinfo{author}{\bibfnamefont{M.}~\bibnamefont{Lisowski}},
  \bibinfo{author}{\bibfnamefont{U.}~\bibnamefont{Bovensiepen}},
  \bibinfo{author}{\bibfnamefont{H.}~\bibnamefont{Berger}},
  \bibinfo{author}{\bibfnamefont{S.}~\bibnamefont{Biermann}},
  \bibinfo{author}{\bibfnamefont{P.~S.} \bibnamefont{Cornaglia}},
  \bibinfo{author}{\bibfnamefont{A.}~\bibnamefont{Georges}}, \bibnamefont{and}
  \bibinfo{author}{\bibfnamefont{M.}~\bibnamefont{Wolf}},
  \bibinfo{journal}{Phys. Rev. Lett.} \textbf{\bibinfo{volume}{97}},
  \bibinfo{pages}{067402} (\bibinfo{year}{2006}).

\bibitem[{\citenamefont{Hellmann
  et~al.}(2012{\natexlab{b}})\citenamefont{Hellmann, Rohwer, Kall{\"a}ne,
  Hanff, Sohrt, Stange, Carr, Murnane, Kapteyn, Kipp et~al.}}]{Hellmann:2012aa}
\bibinfo{author}{\bibfnamefont{S.}~\bibnamefont{Hellmann}},
  \bibinfo{author}{\bibfnamefont{T.}~\bibnamefont{Rohwer}},
  \bibinfo{author}{\bibfnamefont{M.}~\bibnamefont{Kall{\"a}ne}},
  \bibinfo{author}{\bibfnamefont{K.}~\bibnamefont{Hanff}},
  \bibinfo{author}{\bibfnamefont{C.}~\bibnamefont{Sohrt}},
  \bibinfo{author}{\bibfnamefont{A.}~\bibnamefont{Stange}},
  \bibinfo{author}{\bibfnamefont{A.}~\bibnamefont{Carr}},
  \bibinfo{author}{\bibfnamefont{M.~M.} \bibnamefont{Murnane}},
  \bibinfo{author}{\bibfnamefont{H.~C.} \bibnamefont{Kapteyn}},
  \bibinfo{author}{\bibfnamefont{L.}~\bibnamefont{Kipp}}, \bibnamefont{et~al.},
  \bibinfo{journal}{Nature Communications} \textbf{\bibinfo{volume}{3}},
  \bibinfo{pages}{1069} (\bibinfo{year}{2012}{\natexlab{b}}).

\bibitem[{\citenamefont{Leuenberger et~al.}(2013)\citenamefont{Leuenberger,
  Yanagisawa, Roth, Dil, Wells, Hofmann, Osterwalder, and
  Hengsberger}}]{Leuenberger:2013aa}
\bibinfo{author}{\bibfnamefont{D.}~\bibnamefont{Leuenberger}},
  \bibinfo{author}{\bibfnamefont{H.}~\bibnamefont{Yanagisawa}},
  \bibinfo{author}{\bibfnamefont{S.}~\bibnamefont{Roth}},
  \bibinfo{author}{\bibfnamefont{J.~H.} \bibnamefont{Dil}},
  \bibinfo{author}{\bibfnamefont{J.~W.} \bibnamefont{Wells}},
  \bibinfo{author}{\bibfnamefont{P.}~\bibnamefont{Hofmann}},
  \bibinfo{author}{\bibfnamefont{J.}~\bibnamefont{Osterwalder}},
  \bibnamefont{and}
  \bibinfo{author}{\bibfnamefont{M.}~\bibnamefont{Hengsberger}},
  \bibinfo{journal}{Phys. Rev. Lett.} \textbf{\bibinfo{volume}{110}},
  \bibinfo{pages}{136806} (\bibinfo{year}{2013}).

\bibitem[{\citenamefont{Sobota et~al.}(2014)\citenamefont{Sobota, Yang,
  Leuenberger, Kemper, Analytis, Fisher, Kirchmann, Devereaux, and
  Shen}}]{Sobota:2014aa}
\bibinfo{author}{\bibfnamefont{J.~A.} \bibnamefont{Sobota}},
  \bibinfo{author}{\bibfnamefont{S.-L.} \bibnamefont{Yang}},
  \bibinfo{author}{\bibfnamefont{D.}~\bibnamefont{Leuenberger}},
  \bibinfo{author}{\bibfnamefont{A.~F.} \bibnamefont{Kemper}},
  \bibinfo{author}{\bibfnamefont{J.~G.} \bibnamefont{Analytis}},
  \bibinfo{author}{\bibfnamefont{I.~R.} \bibnamefont{Fisher}},
  \bibinfo{author}{\bibfnamefont{P.~S.} \bibnamefont{Kirchmann}},
  \bibinfo{author}{\bibfnamefont{T.~P.} \bibnamefont{Devereaux}},
  \bibnamefont{and} \bibinfo{author}{\bibfnamefont{Z.-X.} \bibnamefont{Shen}},
  \bibinfo{journal}{Phys. Rev. Lett.} \textbf{\bibinfo{volume}{113}},
  \bibinfo{pages}{157401} (\bibinfo{year}{2014}).

\bibitem[{\citenamefont{Gerber et~al.}(2017)\citenamefont{Gerber, Yang, Zhu,
  Soifer, Sobota, Rebec, Lee, Jia, Moritz, Jia et~al.}}]{Gerber:2017aa}
\bibinfo{author}{\bibfnamefont{S.}~\bibnamefont{Gerber}},
  \bibinfo{author}{\bibfnamefont{S.-L.} \bibnamefont{Yang}},
  \bibinfo{author}{\bibfnamefont{D.}~\bibnamefont{Zhu}},
  \bibinfo{author}{\bibfnamefont{H.}~\bibnamefont{Soifer}},
  \bibinfo{author}{\bibfnamefont{J.~A.} \bibnamefont{Sobota}},
  \bibinfo{author}{\bibfnamefont{S.}~\bibnamefont{Rebec}},
  \bibinfo{author}{\bibfnamefont{J.~J.} \bibnamefont{Lee}},
  \bibinfo{author}{\bibfnamefont{T.}~\bibnamefont{Jia}},
  \bibinfo{author}{\bibfnamefont{B.}~\bibnamefont{Moritz}},
  \bibinfo{author}{\bibfnamefont{C.}~\bibnamefont{Jia}}, \bibnamefont{et~al.},
  \bibinfo{journal}{Science} \textbf{\bibinfo{volume}{357}},
  \bibinfo{pages}{71} (\bibinfo{year}{2017}).

\bibitem[{\citenamefont{Hein et~al.}(2020)\citenamefont{Hein, Jauernik, Erk,
  Yang, Qi, Sun, Felser, and Bauer}}]{Hein:2020aa}
\bibinfo{author}{\bibfnamefont{P.}~\bibnamefont{Hein}},
  \bibinfo{author}{\bibfnamefont{S.}~\bibnamefont{Jauernik}},
  \bibinfo{author}{\bibfnamefont{H.}~\bibnamefont{Erk}},
  \bibinfo{author}{\bibfnamefont{L.}~\bibnamefont{Yang}},
  \bibinfo{author}{\bibfnamefont{Y.}~\bibnamefont{Qi}},
  \bibinfo{author}{\bibfnamefont{Y.}~\bibnamefont{Sun}},
  \bibinfo{author}{\bibfnamefont{C.}~\bibnamefont{Felser}}, \bibnamefont{and}
  \bibinfo{author}{\bibfnamefont{M.}~\bibnamefont{Bauer}},
  \bibinfo{journal}{Nature Communications} \textbf{\bibinfo{volume}{11}}
  (\bibinfo{year}{2020}).

\bibitem[{\citenamefont{Reimann et~al.}(2014)\citenamefont{Reimann, G{\"u}dde,
  Kuroda, Chulkov, and H{\"o}fer}}]{Reimann:2014aa}
\bibinfo{author}{\bibfnamefont{J.}~\bibnamefont{Reimann}},
  \bibinfo{author}{\bibfnamefont{J.}~\bibnamefont{G{\"u}dde}},
  \bibinfo{author}{\bibfnamefont{K.}~\bibnamefont{Kuroda}},
  \bibinfo{author}{\bibfnamefont{E.~V.} \bibnamefont{Chulkov}},
  \bibnamefont{and}
  \bibinfo{author}{\bibfnamefont{U.}~\bibnamefont{H{\"o}fer}},
  \bibinfo{journal}{Phys. Rev. B} \textbf{\bibinfo{volume}{90}},
  \bibinfo{pages}{081106} (\bibinfo{year}{2014}).

\bibitem[{\citenamefont{R{\"o}sner et~al.}(2016)\citenamefont{R{\"o}sner,
  Steinke, Lorke, Gies, Jahnke, and Wehling}}]{Rosner:2016aa}
\bibinfo{author}{\bibfnamefont{M.}~\bibnamefont{R{\"o}sner}},
  \bibinfo{author}{\bibfnamefont{C.}~\bibnamefont{Steinke}},
  \bibinfo{author}{\bibfnamefont{M.}~\bibnamefont{Lorke}},
  \bibinfo{author}{\bibfnamefont{C.}~\bibnamefont{Gies}},
  \bibinfo{author}{\bibfnamefont{F.}~\bibnamefont{Jahnke}}, \bibnamefont{and}
  \bibinfo{author}{\bibfnamefont{T.~O.} \bibnamefont{Wehling}},
  \bibinfo{journal}{Nano Letters} \textbf{\bibinfo{volume}{16}},
  \bibinfo{pages}{2322} (\bibinfo{year}{2016}).

\bibitem[{\citenamefont{Oloff et~al.}(2014)\citenamefont{Oloff, Oura,
  Rossnagel, Chainani, Matsunami, Eguchi, Kiss, Nakatani, Yamaguchi, Miyawaki
  et~al.}}]{Oloff:2014aa}
\bibinfo{author}{\bibfnamefont{L.-P.} \bibnamefont{Oloff}},
  \bibinfo{author}{\bibfnamefont{M.}~\bibnamefont{Oura}},
  \bibinfo{author}{\bibfnamefont{K.}~\bibnamefont{Rossnagel}},
  \bibinfo{author}{\bibfnamefont{A.}~\bibnamefont{Chainani}},
  \bibinfo{author}{\bibfnamefont{M.}~\bibnamefont{Matsunami}},
  \bibinfo{author}{\bibfnamefont{R.}~\bibnamefont{Eguchi}},
  \bibinfo{author}{\bibfnamefont{T.}~\bibnamefont{Kiss}},
  \bibinfo{author}{\bibfnamefont{Y.}~\bibnamefont{Nakatani}},
  \bibinfo{author}{\bibfnamefont{T.}~\bibnamefont{Yamaguchi}},
  \bibinfo{author}{\bibfnamefont{J.}~\bibnamefont{Miyawaki}},
  \bibnamefont{et~al.}, \bibinfo{journal}{New Journal of Physics}
  \textbf{\bibinfo{volume}{16}}, \bibinfo{pages}{123045}
  (\bibinfo{year}{2014}).

\bibitem[{\citenamefont{Ulstrup et~al.}(2015)\citenamefont{Ulstrup, Johannsen,
  Cilento, Crepaldi, Miwa, Zacchigna, Cacho, Chapman, Springate, Fromm
  et~al.}}]{Ulstrup:2015ab}
\bibinfo{author}{\bibfnamefont{S.}~\bibnamefont{Ulstrup}},
  \bibinfo{author}{\bibfnamefont{J.~C.} \bibnamefont{Johannsen}},
  \bibinfo{author}{\bibfnamefont{F.}~\bibnamefont{Cilento}},
  \bibinfo{author}{\bibfnamefont{A.}~\bibnamefont{Crepaldi}},
  \bibinfo{author}{\bibfnamefont{J.~A.} \bibnamefont{Miwa}},
  \bibinfo{author}{\bibfnamefont{M.}~\bibnamefont{Zacchigna}},
  \bibinfo{author}{\bibfnamefont{C.}~\bibnamefont{Cacho}},
  \bibinfo{author}{\bibfnamefont{R.~T.} \bibnamefont{Chapman}},
  \bibinfo{author}{\bibfnamefont{E.}~\bibnamefont{Springate}},
  \bibinfo{author}{\bibfnamefont{F.}~\bibnamefont{Fromm}},
  \bibnamefont{et~al.}, \bibinfo{journal}{Journal of Electron Spectroscopy and
  Related Phenomena} \textbf{\bibinfo{volume}{200}}, \bibinfo{pages}{340 }
  (\bibinfo{year}{2015}), \bibinfo{note}{special Anniversary Issue: Volume
  200}.

\bibitem[{\citenamefont{Spiecker et~al.}(1998)\citenamefont{Spiecker, Schmidt,
  Ziethen, Menke, Kleineberg, Ahuja, Merkel, Heinzmann, and
  Sch{\"o}nhense}}]{Spiecker:1998aa}
\bibinfo{author}{\bibfnamefont{H.}~\bibnamefont{Spiecker}},
  \bibinfo{author}{\bibfnamefont{O.}~\bibnamefont{Schmidt}},
  \bibinfo{author}{\bibfnamefont{C.}~\bibnamefont{Ziethen}},
  \bibinfo{author}{\bibfnamefont{D.}~\bibnamefont{Menke}},
  \bibinfo{author}{\bibfnamefont{U.}~\bibnamefont{Kleineberg}},
  \bibinfo{author}{\bibfnamefont{R.}~\bibnamefont{Ahuja}},
  \bibinfo{author}{\bibfnamefont{M.}~\bibnamefont{Merkel}},
  \bibinfo{author}{\bibfnamefont{U.}~\bibnamefont{Heinzmann}},
  \bibnamefont{and}
  \bibinfo{author}{\bibfnamefont{G.}~\bibnamefont{Sch{\"o}nhense}},
  \bibinfo{journal}{Nuclear Instruments and Methods in Physics Research Section
  A: Accelerators, Spectrometers, Detectors and Associated Equipment}
  \textbf{\bibinfo{volume}{406}}, \bibinfo{pages}{499} (\bibinfo{year}{1998}).

\bibitem[{\citenamefont{Chernov et~al.}(2015)\citenamefont{Chernov, Medjanik,
  Tusche, Kutnyakhov, Nepijko, Oelsner, Braun, Min{\'{a}}r, Borek, Ebert
  et~al.}}]{Chernov:2015aa}
\bibinfo{author}{\bibfnamefont{S.}~\bibnamefont{Chernov}},
  \bibinfo{author}{\bibfnamefont{K.}~\bibnamefont{Medjanik}},
  \bibinfo{author}{\bibfnamefont{C.}~\bibnamefont{Tusche}},
  \bibinfo{author}{\bibfnamefont{D.}~\bibnamefont{Kutnyakhov}},
  \bibinfo{author}{\bibfnamefont{S.}~\bibnamefont{Nepijko}},
  \bibinfo{author}{\bibfnamefont{A.}~\bibnamefont{Oelsner}},
  \bibinfo{author}{\bibfnamefont{J.}~\bibnamefont{Braun}},
  \bibinfo{author}{\bibfnamefont{J.}~\bibnamefont{Min{\'{a}}r}},
  \bibinfo{author}{\bibfnamefont{S.}~\bibnamefont{Borek}},
  \bibinfo{author}{\bibfnamefont{H.}~\bibnamefont{Ebert}},
  \bibnamefont{et~al.}, \bibinfo{journal}{Ultramicroscopy}
  \textbf{\bibinfo{volume}{159}}, \bibinfo{pages}{453} (\bibinfo{year}{2015}).

\bibitem[{\citenamefont{Mad{\'e}o et~al.}(2020)\citenamefont{Mad{\'e}o, Man,
  Sahoo, Campbell, Pareek, Wong, Mahboob, Chan, Karmakar, Mariserla
  et~al.}}]{Madeo:2020aa}
\bibinfo{author}{\bibfnamefont{J.}~\bibnamefont{Mad{\'e}o}},
  \bibinfo{author}{\bibfnamefont{M.~K.~L.} \bibnamefont{Man}},
  \bibinfo{author}{\bibfnamefont{C.}~\bibnamefont{Sahoo}},
  \bibinfo{author}{\bibfnamefont{M.}~\bibnamefont{Campbell}},
  \bibinfo{author}{\bibfnamefont{V.}~\bibnamefont{Pareek}},
  \bibinfo{author}{\bibfnamefont{E.~L.} \bibnamefont{Wong}},
  \bibinfo{author}{\bibfnamefont{A.~A.} \bibnamefont{Mahboob}},
  \bibinfo{author}{\bibfnamefont{N.~S.} \bibnamefont{Chan}},
  \bibinfo{author}{\bibfnamefont{A.}~\bibnamefont{Karmakar}},
  \bibinfo{author}{\bibfnamefont{B.~M.~K.} \bibnamefont{Mariserla}},
  \bibnamefont{et~al.} (\bibinfo{year}{2020}), \eprint{arXiv.2005.00241}.

\bibitem[{\citenamefont{Kumai et~al.}(1999)\citenamefont{Kumai, Okimoto, and
  Tokura}}]{Kumai:1999aa}
\bibinfo{author}{\bibfnamefont{R.}~\bibnamefont{Kumai}},
  \bibinfo{author}{\bibfnamefont{Y.}~\bibnamefont{Okimoto}}, \bibnamefont{and}
  \bibinfo{author}{\bibfnamefont{Y.}~\bibnamefont{Tokura}},
  \bibinfo{journal}{Science} \textbf{\bibinfo{volume}{284}},
  \bibinfo{pages}{1645} (\bibinfo{year}{1999}).

\bibitem[{\citenamefont{Lang et~al.}(2002)\citenamefont{Lang, Madhavan,
  Hoffman, Hudson, Eisaki, Uchida, and Davis}}]{Lang:2002aa}
\bibinfo{author}{\bibfnamefont{K.~M.} \bibnamefont{Lang}},
  \bibinfo{author}{\bibfnamefont{V.}~\bibnamefont{Madhavan}},
  \bibinfo{author}{\bibfnamefont{J.~E.} \bibnamefont{Hoffman}},
  \bibinfo{author}{\bibfnamefont{E.~W.} \bibnamefont{Hudson}},
  \bibinfo{author}{\bibfnamefont{H.}~\bibnamefont{Eisaki}},
  \bibinfo{author}{\bibfnamefont{S.}~\bibnamefont{Uchida}}, \bibnamefont{and}
  \bibinfo{author}{\bibfnamefont{J.~C.} \bibnamefont{Davis}},
  \bibinfo{journal}{Nature} \textbf{\bibinfo{volume}{415}},
  \bibinfo{pages}{412} (\bibinfo{year}{2002}).

\bibitem[{\citenamefont{Qazilbash et~al.}(2007)\citenamefont{Qazilbash, Brehm,
  Chae, Ho, Andreev, Kim, Yun, Balatsky, Maple, Keilmann
  et~al.}}]{Qazilbash:2007aa}
\bibinfo{author}{\bibfnamefont{M.~M.} \bibnamefont{Qazilbash}},
  \bibinfo{author}{\bibfnamefont{M.}~\bibnamefont{Brehm}},
  \bibinfo{author}{\bibfnamefont{B.-G.} \bibnamefont{Chae}},
  \bibinfo{author}{\bibfnamefont{P.-C.} \bibnamefont{Ho}},
  \bibinfo{author}{\bibfnamefont{G.~O.} \bibnamefont{Andreev}},
  \bibinfo{author}{\bibfnamefont{B.-J.} \bibnamefont{Kim}},
  \bibinfo{author}{\bibfnamefont{S.~J.} \bibnamefont{Yun}},
  \bibinfo{author}{\bibfnamefont{A.~V.} \bibnamefont{Balatsky}},
  \bibinfo{author}{\bibfnamefont{M.~B.} \bibnamefont{Maple}},
  \bibinfo{author}{\bibfnamefont{F.}~\bibnamefont{Keilmann}},
  \bibnamefont{et~al.}, \bibinfo{journal}{Science}
  \textbf{\bibinfo{volume}{318}}, \bibinfo{pages}{1750} (\bibinfo{year}{2007}).

\bibitem[{\citenamefont{Lai et~al.}(2010)\citenamefont{Lai, Nakamura,
  Kundhikanjana, Kawasaki, Tokura, Kelly, and Shen}}]{Lai:2010aa}
\bibinfo{author}{\bibfnamefont{K.}~\bibnamefont{Lai}},
  \bibinfo{author}{\bibfnamefont{M.}~\bibnamefont{Nakamura}},
  \bibinfo{author}{\bibfnamefont{W.}~\bibnamefont{Kundhikanjana}},
  \bibinfo{author}{\bibfnamefont{M.}~\bibnamefont{Kawasaki}},
  \bibinfo{author}{\bibfnamefont{Y.}~\bibnamefont{Tokura}},
  \bibinfo{author}{\bibfnamefont{M.~A.} \bibnamefont{Kelly}}, \bibnamefont{and}
  \bibinfo{author}{\bibfnamefont{Z.-X.} \bibnamefont{Shen}},
  \bibinfo{journal}{Science} \textbf{\bibinfo{volume}{329}},
  \bibinfo{pages}{190} (\bibinfo{year}{2010}).

\bibitem[{\citenamefont{Torre et~al.}(2015)\citenamefont{Torre, Tomadin, Geim,
  and Polini}}]{Torre:2015aa}
\bibinfo{author}{\bibfnamefont{I.}~\bibnamefont{Torre}},
  \bibinfo{author}{\bibfnamefont{A.}~\bibnamefont{Tomadin}},
  \bibinfo{author}{\bibfnamefont{A.~K.} \bibnamefont{Geim}}, \bibnamefont{and}
  \bibinfo{author}{\bibfnamefont{M.}~\bibnamefont{Polini}},
  \bibinfo{journal}{Physical Review B} \textbf{\bibinfo{volume}{92}},
  \bibinfo{pages}{165433} (\bibinfo{year}{2015}).

\bibitem[{\citenamefont{Bandurin et~al.}(2016)\citenamefont{Bandurin, Torre,
  Kumar, Ben~Shalom, Tomadin, Principi, Auton, Khestanova, Novoselov,
  Grigorieva et~al.}}]{Bandurin:2016aa}
\bibinfo{author}{\bibfnamefont{D.~A.} \bibnamefont{Bandurin}},
  \bibinfo{author}{\bibfnamefont{I.}~\bibnamefont{Torre}},
  \bibinfo{author}{\bibfnamefont{R.~K.} \bibnamefont{Kumar}},
  \bibinfo{author}{\bibfnamefont{M.}~\bibnamefont{Ben~Shalom}},
  \bibinfo{author}{\bibfnamefont{A.}~\bibnamefont{Tomadin}},
  \bibinfo{author}{\bibfnamefont{A.}~\bibnamefont{Principi}},
  \bibinfo{author}{\bibfnamefont{G.~H.} \bibnamefont{Auton}},
  \bibinfo{author}{\bibfnamefont{E.}~\bibnamefont{Khestanova}},
  \bibinfo{author}{\bibfnamefont{K.~S.} \bibnamefont{Novoselov}},
  \bibinfo{author}{\bibfnamefont{I.~V.} \bibnamefont{Grigorieva}},
  \bibnamefont{et~al.}, \bibinfo{journal}{Science}
  \textbf{\bibinfo{volume}{351}}, \bibinfo{pages}{1055} (\bibinfo{year}{2016}).

\bibitem[{\citenamefont{Krishna~Kumar et~al.}(2017)\citenamefont{Krishna~Kumar,
  Bandurin, Pellegrino, Cao, Principi, Guo, Auton, Ben~Shalom, Ponomarenko,
  Falkovich et~al.}}]{Krishna-Kumar:2017aa}
\bibinfo{author}{\bibfnamefont{R.}~\bibnamefont{Krishna~Kumar}},
  \bibinfo{author}{\bibfnamefont{D.~A.} \bibnamefont{Bandurin}},
  \bibinfo{author}{\bibfnamefont{F.~M.~D.} \bibnamefont{Pellegrino}},
  \bibinfo{author}{\bibfnamefont{Y.}~\bibnamefont{Cao}},
  \bibinfo{author}{\bibfnamefont{A.}~\bibnamefont{Principi}},
  \bibinfo{author}{\bibfnamefont{H.}~\bibnamefont{Guo}},
  \bibinfo{author}{\bibfnamefont{G.~H.} \bibnamefont{Auton}},
  \bibinfo{author}{\bibfnamefont{M.}~\bibnamefont{Ben~Shalom}},
  \bibinfo{author}{\bibfnamefont{L.~A.} \bibnamefont{Ponomarenko}},
  \bibinfo{author}{\bibfnamefont{G.}~\bibnamefont{Falkovich}},
  \bibnamefont{et~al.}, \bibinfo{journal}{Nature Physics}
  \textbf{\bibinfo{volume}{13}}, \bibinfo{pages}{1182} (\bibinfo{year}{2017}).

\bibitem[{\citenamefont{Sulpizio et~al.}(2019)\citenamefont{Sulpizio, Ella,
  Rozen, Birkbeck, Perello, Dutta, Ben-Shalom, Taniguchi, Watanabe, Holder
  et~al.}}]{Sulpizio:2019aa}
\bibinfo{author}{\bibfnamefont{J.~A.} \bibnamefont{Sulpizio}},
  \bibinfo{author}{\bibfnamefont{L.}~\bibnamefont{Ella}},
  \bibinfo{author}{\bibfnamefont{A.}~\bibnamefont{Rozen}},
  \bibinfo{author}{\bibfnamefont{J.}~\bibnamefont{Birkbeck}},
  \bibinfo{author}{\bibfnamefont{D.~J.} \bibnamefont{Perello}},
  \bibinfo{author}{\bibfnamefont{D.}~\bibnamefont{Dutta}},
  \bibinfo{author}{\bibfnamefont{M.}~\bibnamefont{Ben-Shalom}},
  \bibinfo{author}{\bibfnamefont{T.}~\bibnamefont{Taniguchi}},
  \bibinfo{author}{\bibfnamefont{K.}~\bibnamefont{Watanabe}},
  \bibinfo{author}{\bibfnamefont{T.}~\bibnamefont{Holder}},
  \bibnamefont{et~al.}, \bibinfo{journal}{Nature}
  \textbf{\bibinfo{volume}{576}}, \bibinfo{pages}{75} (\bibinfo{year}{2019}).

\bibitem[{\citenamefont{Polini and Geim}(2020)}]{Polini:2020aa}
\bibinfo{author}{\bibfnamefont{M.}~\bibnamefont{Polini}} \bibnamefont{and}
  \bibinfo{author}{\bibfnamefont{A.~K.} \bibnamefont{Geim}},
  \bibinfo{journal}{Physics Today} \textbf{\bibinfo{volume}{73}},
  \bibinfo{pages}{28} (\bibinfo{year}{2020}).

\bibitem[{\citenamefont{de~Jong and Molenkamp}(1995)}]{Jong:1995aa}
\bibinfo{author}{\bibfnamefont{M.~J.~M.} \bibnamefont{de~Jong}}
  \bibnamefont{and} \bibinfo{author}{\bibfnamefont{L.~W.}
  \bibnamefont{Molenkamp}}, \bibinfo{journal}{Physical Review B}
  \textbf{\bibinfo{volume}{51}}, \bibinfo{pages}{13389} (\bibinfo{year}{1995}).

\bibitem[{\citenamefont{Hasan and Kane}(2010)}]{Hasan:2010aa}
\bibinfo{author}{\bibfnamefont{M.~Z.} \bibnamefont{Hasan}} \bibnamefont{and}
  \bibinfo{author}{\bibfnamefont{C.~L.} \bibnamefont{Kane}},
  \bibinfo{journal}{Rev. Mod. Phys.} \textbf{\bibinfo{volume}{82}},
  \bibinfo{pages}{3045} (\bibinfo{year}{2010}).

\bibitem[{\citenamefont{Ando}(2013)}]{Ando:2013aa}
\bibinfo{author}{\bibfnamefont{Y.}~\bibnamefont{Ando}},
  \bibinfo{journal}{Journal of the Physical Society of Japan}
  \textbf{\bibinfo{volume}{82}}, \bibinfo{pages}{102001}
  (\bibinfo{year}{2013}).

\bibitem[{\citenamefont{Benia et~al.}(2011)\citenamefont{Benia, Lin, Kern, and
  Ast}}]{Benia:2011aa}
\bibinfo{author}{\bibfnamefont{H.~M.} \bibnamefont{Benia}},
  \bibinfo{author}{\bibfnamefont{C.}~\bibnamefont{Lin}},
  \bibinfo{author}{\bibfnamefont{K.}~\bibnamefont{Kern}}, \bibnamefont{and}
  \bibinfo{author}{\bibfnamefont{C.~R.} \bibnamefont{Ast}},
  \bibinfo{journal}{Phys. Rev. Lett.} \textbf{\bibinfo{volume}{107}},
  \bibinfo{pages}{177602} (\bibinfo{year}{2011}).

\bibitem[{\citenamefont{King et~al.}(2011)\citenamefont{King, Hatch, Bianchi,
  Ovsyannikov, Lupulescu, Landolt, Slomski, Dil, Guan, Mi
  et~al.}}]{King:2011aa}
\bibinfo{author}{\bibfnamefont{P.~D.~C.} \bibnamefont{King}},
  \bibinfo{author}{\bibfnamefont{R.~C.} \bibnamefont{Hatch}},
  \bibinfo{author}{\bibfnamefont{M.}~\bibnamefont{Bianchi}},
  \bibinfo{author}{\bibfnamefont{R.}~\bibnamefont{Ovsyannikov}},
  \bibinfo{author}{\bibfnamefont{C.}~\bibnamefont{Lupulescu}},
  \bibinfo{author}{\bibfnamefont{G.}~\bibnamefont{Landolt}},
  \bibinfo{author}{\bibfnamefont{B.}~\bibnamefont{Slomski}},
  \bibinfo{author}{\bibfnamefont{J.~H.} \bibnamefont{Dil}},
  \bibinfo{author}{\bibfnamefont{D.}~\bibnamefont{Guan}},
  \bibinfo{author}{\bibfnamefont{J.~L.} \bibnamefont{Mi}},
  \bibnamefont{et~al.}, \bibinfo{journal}{Phys. Rev. Lett.}
  \textbf{\bibinfo{volume}{107}}, \bibinfo{pages}{096802}
  (\bibinfo{year}{2011}).

\end{thebibliography}
\end{document}